%% file: Doktorarbeit-p.tex
\documentclass[12pt,fleqn,a4paper]{report} 
\usepackage{epsfig,graphics,graphicx}
\usepackage{clshan-math,clshan-font,clshan-dm-dd}
\newcommand{\cheqnC}            {\cheqnX{\thechapter.}}

\newcommand{\cheqnCN}       [1] {\cheqnXN{\thechapter.}{#1}}
\newcommand{\cheqnCNx}      [2] {\cheqnXNx{\thechapter.}{#1}{#2}}
\newcommand{\cheqnCa}           {\cheqnXa{\thechapter.}}
\newcommand{\cheqnCb}           {\cheqnXb{\thechapter.}}
\newcommand{\cheqnCc}           {\cheqnXc{\thechapter.}}
\newcommand{\imageswitch}   [2] {#2}
\input{Doktorarbeit-Fig}
\begin{document}
\begin{titlepage}
\input{Doktorarbeit-Title-p}
\end{titlepage}
\setlength{\baselineskip}{3.3ex}
\newpage
\thispagestyle{empty}
\begin{center}
{\Large\bf Acknowledgments}
\vspace{1cm}
\end{center}
\input{Doktorarbeit-Ack}
\newpage
\thispagestyle{empty}
 ~~
\newpage
\thispagestyle{empty}
\begin{center}
{\Large\bf Abstract}
\vspace{1cm}
\end{center}
\input{Doktorarbeit-Abs}
\newpage
\thispagestyle{empty}
 ~~
\newpage
\renewcommand{\thepage}{\roman{page}}
\setcounter{page}{1}
\tableofcontents
\addtocontents{toc}{}
\newpage
\renewcommand{\thepage}{\arabic{page}}
\setcounter{page}{1}
\input{Doktorarbeit-Ch1-p}
\input{Doktorarbeit-Ch2}
\input{Doktorarbeit-Ch3}
\input{Doktorarbeit-Ch4}
\input{Doktorarbeit-Ch5}
\input{Doktorarbeit-Ch6}
\begin{appendix}
\setlength{\unitlength}{1.5cm}
\input{Doktorarbeit-AppA}
\input{Doktorarbeit-AppB}
\input{Doktorarbeit-AppC}
\input{Doktorarbeit-AppD}
\input{Doktorarbeit-AppE}
\end{appendix}
\input{Doktorarbeit-Ref}
\end{document}

%% file: Doktorarbeit-Fig.tex
\newsavebox{\SUSYa}
\savebox{\SUSYa}(16.4,2.6){
\begin{picture}(16.4,2.6)
\thicklines
\put( 1.2,1.8){\line   (0  ,1  ){0.3}}
\put( 1.7,1.8){\oval   (1  ,1  )[lb] }
\put( 1.7,1.3){\vector (1  ,0  ){2.6}}
\put( 1.7,0.8){\oval   (1  ,1  )[lt] }
\put( 1.2,0.5){\line   (0  ,1  ){0.3}}
\thinlines
\put( 0  ,2.1){\makebox(4.4,0.5){Special relativity}}
\put( 0  ,0  ){\makebox(4.4,0.5){Quantum mechanics}}
\put( 1.7,1.4){\makebox(2.6,0.5){Dirac}}
\put( 4.3,1.1){\makebox(6  ,0.4){Charge-conjugation symmetry}}
\put(10.3,1.1){\makebox(0.8,0.4){$\lTo$}}
\put(11.1,1.1){\makebox(5.3,0.4){Particles $\lgetsto$ Antiparticles}}
\end{picture}
}
\newsavebox{\SUSYb}
\savebox{\SUSYb}(16,2.6){
\begin{picture}(16,2.6)
\thicklines
\put( 1.2,1.8){\line   (0  ,1  ){0.3}}
\put( 1.7,1.8){\oval   (1  ,1  )[lb] }
\put( 1.7,1.3){\vector (1  ,0  ){2.6}}
\put( 1.7,0.8){\oval   (1  ,1  )[lt] }
\put( 1.2,0.5){\line   (0  ,1  ){0.3}}
\thinlines
\put( 0  ,2.1){\makebox(4.4,0.5){General relativity}}
\put( 0  ,0  ){\makebox(4.4,0.5){Quantum field theory}}
\put( 1.7,1.4){\makebox(2.6,0.5){SUSY models}}
\put( 4.3,1.1){\makebox(3.4,0.4){Supersymmetry}}
\put( 7.7,1.1){\makebox(0.8,0.4){$\lTo$}}
\put( 8.6,0.6){\makebox(0.4,1.4){$\Bigg\{$}}
\put( 9  ,1.3){\makebox(7  ,0.6){Fermions $\lgetsto$ Bosonic   superpartners}}
\put( 9  ,0.7){\makebox(7  ,0.6){Bosons   $\lgetsto$ Fermionic superpartners}}
\end{picture}
}
\newsavebox{\FigureAa}
\savebox{\FigureAa}(7.5,5.5){
\begin{picture}(7.5,5.5)
\put(0,0){\makebox(7.5,5.5){}}
\thicklines
\put(0.5,0.5){\vector (1  ,0  ){6.5}}
\put(1  ,0  ){\vector (0  ,1  ){5}}
\thinlines
\put(7  ,0.3){\makebox(0.4,0.4){$v$}}
\put(0.5,5.1){\makebox(1  ,0.4){$F_1(v)$}}
\multiput(1  ,4.2)(0.2,0){26}{\line(1,0){0.1}}
\put(0.3,4  ){\makebox(0.7,0.4){$F_{1,\infty}$}}
\multiput(1  ,1.5)(0.2,0){5}{\line(1,0){0.1}}
\put(0.4,1.3){\makebox(0.6,0.4){$F_{1,0}$}}
\multiput(2.5,0.4)(0,0.4){10}{\line(0,1){0.2}}
\put(2.2,0  ){\makebox(0.8,0.5){$\vmin$}}
\end{picture}
}
\newsavebox{\Figurekave}
\savebox{\Figurekave}(9,6){
\begin{picture}(9,6)
\put(0,0){\makebox(9,6){}}
\put(0.5,0.7){\vector (1  , 0  ){7.5}}
\put(1  ,0.5){\vector (0  , 1  ){4.7}}
\put(8  ,0.4){\makebox(0.4, 0.6){$Q$}}
\put(0.3,5.2){\makebox(1.4, 0.6){$\ln\adRdQ$}}
\put(2  ,4.4){\circle*{0.1}}
\put(4  ,3.2){\circle*{0.1}}
\put(6  ,1.2){\circle*{0.1}}
\put(2  ,4.4){\line   (5  ,-3  ){2}}
\put(4  ,3.2){\line   (1  ,-1  ){2}}
\put(3  ,4  ){\line   (5  ,-4  ){2}}
\put(5.1,2.1){\makebox(0.6, 0.4){$k_{n,\rmave}$}}
\put(2.6,3.3){\makebox(0.8, 0.4){$k_{n-1,n  }$}}
\put(4.4,1.7){\makebox(0.8, 0.4){$k_{n  ,n+1}$}}
\put(1.5,4.4){\makebox(1  , 0.4){$\ln r_{n-1}$}}
\put(3.8,3.2){\makebox(0.8, 0.4){$\ln r_{n  }$}}
\put(5.9,1.2){\makebox(1  , 0.4){$\ln r_{n+1}$}}
\put(2  ,0.6){\line   (0  , 1  ){0.2}}
\put(1.5,0.2){\makebox(1  , 0.4){$Q_{n-1}$}}
\put(4  ,0.6){\line   (0  , 1  ){0.2}}
\put(3.7,0.2){\makebox(0.6, 0.4){$Q_{n  }$}}
\put(6  ,0.6){\line   (0  , 1  ){0.2}}
\put(5.5,0.2){\makebox(1  , 0.4){$Q_{n+1}$}}
\multiput(2,0.7)(0,0.2){19}{\line(0,1){0.1}}
\multiput(4,0.7)(0,0.2){13}{\line(0,1){0.1}}
\multiput(6,0.7)(0,0.2){ 3}{\line(0,1){0.1}}
\end{picture}
}
\newsavebox{\FigurelnrnQ}
\savebox{\FigurelnrnQ}(9,6){
\begin{picture}(9,6)
\put(0,0){\makebox(9,6){}}
\put(0.5,0.7){\vector (1  , 0  ){7.5}}
\put(1  ,0.5){\vector (0  , 1  ){4.7}}
\put(8  ,0.4){\makebox(0.4, 0.6){$Q$}}
\put(0.3,5.2){\makebox(1.4, 0.6){$\ln\adRdQ$}}
\put(4  ,3.2){\circle*{0.1}}
\put(2  ,4.4){\line   (5  ,-3  ){4}}
\put(3.8,3.2){\makebox(0.8, 0.4){$\ln r_n$}}
\put(6.1,1.8){\makebox(0.6, 0.4){$k_{n,\rmave}$}}
\put(2  ,0.6){\line   (0  , 1  ){0.2}}
\put(1.5,0.1){\makebox(1  , 0.5){$\T\frac{Q_{n-1}+Q_{n  }}{2}$}}
\put(4  ,0.6){\line   (0  , 1  ){0.2}}
\put(3.7,0.1){\makebox(0.6, 0.5){$Q_n$}}
\put(6  ,0.6){\line   (0  , 1  ){0.2}}
\put(5.5,0.1){\makebox(1  , 0.5){$\T\frac{Q_{n  }+Q_{n+1}}{2}$}}
\multiput(2,0.7)(0,0.2){19}{\line(0,1){0.1}}
\multiput(4,0.7)(0,0.2){13}{\line(0,1){0.1}}
\multiput(6,0.7)(0,0.2){ 7}{\line(0,1){0.1}}
\put(3  ,3.8){\circle*{0.1}}
\put(2.6,4.2){\makebox(1.9, 0.6){$\ln r_{n,\rmave}(Q)$}}
\put(3  ,4.3){\vector (0  ,-1  ){0.4}}
\put(3  ,0.6){\line   (0  , 1  ){0.2}}
\put(2.7,0.1){\makebox(0.6, 0.5){$Q$}}
\multiput(3,0.7)(0,0.2){16}{\line(0,1){0.1}}
\end{picture}
}
\newsavebox{\Figuredeltarn}
\savebox{\Figuredeltarn}(9,6){
\begin{picture}(9,6)
\put(0,0){\makebox(9,6){}}
\put(0.5,0.7){\vector ( 1  , 0  ){7.5}}
\put(1  ,0.5){\vector ( 0  , 1  ){4.7}}
\put(8  ,0.4){\makebox( 0.4, 0.6){$Q$}}
\put(0.3,5.2){\makebox( 1.4, 0.6){$\dRdQ$}}
\put(4  ,3.2){\circle*{0.1}}
\put(4  ,3  ){\circle*{0.1}}
\put(4  ,2.8){\circle*{0.1}}
\put(4.1,3.7){\makebox( 0.6, 0.4){$r_{n,\rmave}^{\ast}$}}
\put(4.3,3.7){\vector (-1  ,-2  ){0.2}}
\put(4.5,2.8){\makebox( 0.4, 0.4){$r_n$}}
\put(4.5,3  ){\vector (-1  , 0  ){0.4}}
\put(4.1,1.7){\makebox( 1.6, 0.6){$\adRdQ_{{\rm real},Q = Q_n}$}}
\put(4.3,2.3){\vector (-1  , 2  ){0.2}}
\put(2  ,0.6){\line   ( 0  , 1  ){0.2}}
\put(1.5,0  ){\makebox( 1  , 0.6){$\frac{Q_{n-1}+Q_{n  }}{2}$}}
\put(4  ,0.6){\line   ( 0  , 1  ){0.2}}
\put(3.7,0  ){\makebox( 0.6, 0.6){$Q_n$}}
\put(6  ,0.6){\line   ( 0  , 1  ){0.2}}
\put(5.5,0  ){\makebox( 1  , 0.6){$\frac{Q_{n  }+Q_{n+1}}{2}$}}
\multiput(2,0.7)(0,0.2){18}{\line(0,1){0.1}}
\multiput(4,0.7)(0,0.2){13}{\line(0,1){0.1}}
\multiput(6,0.7)(0,0.2){10}{\line(0,1){0.1}}
\put(6.4,2.3){\makebox( 1  , 0.6)[l]{$r_{n,\rmave}(Q)$}}
\put(6.4,1.6){\makebox( 1  , 0.6)[l]{$\adRdQ_{\rm real}$}}
\put(3.7,3  ){\line   ( 1  , 0  ){0.2}}
\put(3.7,2.8){\line   ( 1  , 0  ){0.2}}
\put(3.8,3.3){\vector ( 0  ,-1  ){0.3}}
\put(3.8,2.5){\vector ( 0  , 1  ){0.3}}
\put(3  ,2.7){\makebox( 0.8, 0.4){$\Delta_{r,n}$}}
\end{picture}
}

%% file: Doktorarbeit-Title-p.tex
%
%
%
%
%
\newpage
\thispagestyle{empty}
\begin{center}
\vspace*{1.2cm}
{\LARGE\bf Theoretical Interpretation of Experimental \\ \vspace{0.3cm}
           Data from Direct Dark Matter Detection}    \\
\vspace{3cm}
{\Large
 Dissertation                                         \\ \vspace{0.2cm}
 zur                                                  \\ \vspace{0.2cm}
 Erlangung der Doktorgrades (Dr.~rer.~nat.)           \\ \vspace{0.2cm}
 der                                                  \\ \vspace{0.2cm}
 Mathematisch-Naturwissenschaftlichen Fakult\"at      \\ \vspace{0.2cm}
 der                                                  \\ \vspace{0.2cm}
 Rheinischen Friedrich-Wilhelms-Universit\"at Bonn    \\
\vspace{3cm}
 vorgelegt von                                        \\ \vspace{0.3cm}
 {\sc Chung-Lin Shan}                                 \\ \vspace{0.2cm}
 aus Taipeh                                           \\
\vspace{4.6cm}
 Bonn 2007
}
\end{center}
\newpage
\thispagestyle{empty}
 ~~
%
%
%
%
%
%
%
%
%
\newpage
\thispagestyle{empty}
 ~
\vspace*{1cm}
\hspace*{\fill} \\
{\large
 Angefertigt mit Genehmigung \vspace{0.2cm}
 der Mathematisch-Naturwissenschaftlichen Fakult\"at der Universit\"at Bonn}

\vspace{18cm}
\begin{flushleft}
\hspace{-0.35cm}
\begin{tabular}{l l}
 Referent:          & Prof. Manuel Drees    \\
 Korreferent:       & Prof. Herbert Dreiner \\
                    &                       \\
 Tag der Promotion: & 16. Oktober 2007
\end{tabular}
\end{flushleft}
\newpage
\thispagestyle{empty}
 ~~
\newpage
\thispagestyle{empty}
 ~
\vspace*{20cm}
\hspace*{\fill} \\
 Ich versichere,
 dass ich diese Arbeit selbst\"andig verfasst
 und keine anderen als die angegebenen Quellen und Hilfsmittel benutzt
 sowie die Zitate kenntlich gemacht habe.

%
\newpage
\thispagestyle{empty}
 ~~

%% file: Doktorarbeit-Ack.tex
 ~~~$\,$
 First and foremost
 I would like to express my gratitude to my advisor Prof.~Manuel Drees
 for the opportunity to work on the subject of Dark Matter.
 As a collaborator I have learned to appreciate
 his profound knowledge of the subject and intuition for the essential
 as well as his manner of researching
 and special unique point of views.

 Next I would like to thank Mitsuru Kakizaki
 for discussing about our research,
 and especially
 for his help to correct this work.

 I would also like to thank all colleagues in my office:
 Markus Bernhardt,
 Sascha Bornhauser,
 Sebastian Grab,
 Jong Soo Kim,
 Olaf Kittel,
 Ulrich Langenfeld,
 Anja Marold,
 and Federico von der Pahlen.
 I always enjoy the friendly atmosphere in our office
 and it is always being a pleasure to discuss about Physics
 together with something else with you.

 My thanks also go to
 Prof.~Herbert Dreiner,
 Prof.~Hans-Peter Nilles,
 Kin-ya Oda and his wife Naoko,
 Saul-Noe Ramos-Sanchez,
 Patrick Vaudrevange,
 and all current and former colleagues
 in the theory group in the Physikalischen Institut.
%
%
%
 Moreover,
 I particularly want to thank the very helpful secretaries:
 Dagmar Fassbender,
 Sandra Heidbrink,
 Patricia Z\"undorf,
 and our computer and all-round specialist Andreas Wisskirchen.
 I am pleased to know you and thank for all your assistances.

 My special thank goes to
 Prof. Andrzej Buras,
 Christoph Bobeth,
 Martin Gorbahn,
 Sebastian J\"ager,
 Frank Kr\"uger,
 Anton Poschenrieder,
 Andreas Weiler and his wife Laura
 as well as the other former colleagues
 in the Physics Department at the Technische Universit\"at M\"unchen.
 I was glad for the time spent with you
 at the beginning of my study in Germany.

 Furthermore
 it is a great pleasure to thank
 the Deutschem Akademischem Austausch Dienst (DAAD),
 especially Ruth Eberlein,
 Petra Hagemann,
 Heike Sch\"adlich,
 Elena Schmid,
 and all other members of the Referat 423,
 not only for the financial support,
 but also their thoughtful caring about the study and life.
 
 In addition,
 I would also like to thank the following friends:
 Simone M\"uller,
 Dieter Rupp,
 Ritschi Spindler
 and Li-Chi Wu in Munich
 as well as
 the family of Kunkel,
 Herv\'e Barreteau,
 Chun-yi Chen,
 Piyada Dok-angkab,
 Marie Kubota,
 Hsiao-Yun Lee,
 Yin-Chen Lin,
 Yuka Nakajima,
 Naoko Teranishi,
 and Hsin-Tzu Wang in Bonn.
 You make my life in Germany more interesting and colorful.

 My parents always respect my decisions,
 especially the long-time study in Germany,
 and support me for the whole time.
 I am very grateful to them.

%% file: Doktorarbeit-Abs.tex
 ~~~$\,$
 Weakly Interacting Massive Particles (WIMPs)
 are one of the leading candidates for Dark Matter.
 Currently,
 the most promising method to detect many different WIMP candidates
 is the direct detection of the recoil energy deposited
 in a low-background laboratory detector due to elastic WIMP-nucleus scattering.
 So far the usual procedure has been
 to predict the event rate of direct detection of WIMPs
 based on some model(s) of the Galactic halo from cosmology
 and of WIMPs from elementary particle physics.

 The aim of this work is to invert this process.
 In this thesis
 I will present methods
 which allow to extract information
 on the WIMP velocity distribution function as well as on the WIMP mass
 from the recoil energy spectrum
 as well as from experimental data directly.

 At first
 I will derive
 the expression that allow to reconstruct
 the normalized one-dimensional velocity distribution function of WIMPs
 from the recoil spectrum.
 I will also derive the formulae
 for determining the moments of the velocity distribution function.
 All these expressions are independent of
 the as yet unknown WIMP density near the Earth
 as well as of the WIMP-nucleus cross section.
 The only information about the nature of WIMPs which one needs is the WIMP mass.

 Then I will present methods
 that allow to apply the expressions directly to experimental data,
 without the need to fit the recoil spectrum to a functional form.
 These methods are independent of the Galactic halo model.
 The reconstruction of the velocity distribution function
 will be further extended to take into account the annual modulation of the event rate.

 Moreover,
 I will present
 a method for reconstructing
 the amplitude of the annual modulation of the velocity distribution.
 The only information which one needs
 is the measured recoil energies and their measuring times.
 An alternative, better way for confirming the annual modulation of the event rate
 will also be given.

 Finally,
 I will present
 a method for determining the WIMP mass
 by combining two (or more) experiments with different detector materials.
 This method is not only independent of the model of Galactic halo
 but also of that of WIMPs.
 In addition,
 some meaningful information on the WIMP mass can already be extracted
 from less than one hundred events.

%% file: Doktorarbeit-Ch1-p.tex
\chapter{Dark Matter}
 ~~~$\,$
 One of the most fundamental open questions
 in cosmology and elementary particle physics today is
 {\em what is the nature of Dark Matter}.
 Earlier the question was whether Dark Matter actually exists.
 But nowadays we have some strong evidence to believe that
 {\em something which we do not know exists}.

 As introduction I review briefly
 the history of the discovery of (the existence of) Dark Matter in the Universe.
 It will be seen that,
 according to some astronomical observations and measurements,
 more than 80\% of the total mass content of the Universe consists of Dark Matter.
 I will also present some models of Dark Matter halo in this chapter.
\section{Evidence for Dark Matter}
\label{DMevidence}
 ~~~$\,$
 We call such {\em something} ``dark''
 because it {\em (almost) neither emits nor absorbs electromagnetic radiation}.
 Historically the observational evidence for the existence of Dark Matter
 came only from galactic dynamics and are gravitational \cite{SUSYDM96}.
 The following discussions in this section show that
 {\em the observed luminous objects}
 (stars, gas clouds, globular clusters, or even entire galaxies)
 {\em can not have enough mass to support the observed gravitational effects}
 \cite{SUSYDM96}.
\subsection{Clusters of galaxies}
 ~~~$\,$
 Clusters of galaxies are the largest gravitationally-bound objects in the Universe.
 For example,
 our Milky Way and the M31 galaxy belong to the Local Group of Galaxies and
 are part of the Virgo Supercluster of Galaxies.

 At the beginning of the 1930s,
 F.~Zwicky and other astronomers measured the total mass of a few clusters of galaxies
 and the masses of the luminous objects in these clusters of galaxies
 \cite{Zwicky33}, \cite{Smith36}.
 Their measurements showed that
 {\em the masses of these clusters of galaxies required to gravitationally bind their galaxies
 are much larger than the sum of the luminous masses of their individual galaxies}.
%
%
\subsection{Rotation curves of spiral galaxies}
\label{rotationcurves}
 ~~~$\,$
 The most convincing evidence for the existence of Dark Matter
 came from the {\em measurement of the rotation curves of spiral galaxies} in the 1970s
 by V.~C.~Rubin and other astronomers \cite{Rubin70}-\cite{Rubin80}.

 According to Newton's Second Law,
 the rotational velocity $v$ of an object
 on a stable orbit with radius $r$ from the center of galaxy is
\footnote{
 Here the galaxy is assumed to be spherical symmetric.}
\beq
  \frac{v^2(r)}{r}
= \frac{G_N M(r)}{r^2}
\~,
\label{eqn1010201}
\eeq
 namely,
\beq
         v(r)
 \propto \sfrac{M(r)}{r}
\~,
\eeq
 where $M(r)$ is the mass inside the orbit.
 For an object outside the visible part of the galaxy,
 one would expect that
\beq
         v(r)
 \propto \frac{1}{\sqrt{r}}
\~.
\eeq
\begin{figure}[t]
\begin{center}
\imageswitch{
\begin{picture}(14,10.2)
\put(0,0){\framebox(14,10.2){}}
\end{picture}}
{\includegraphics[width=14cm]{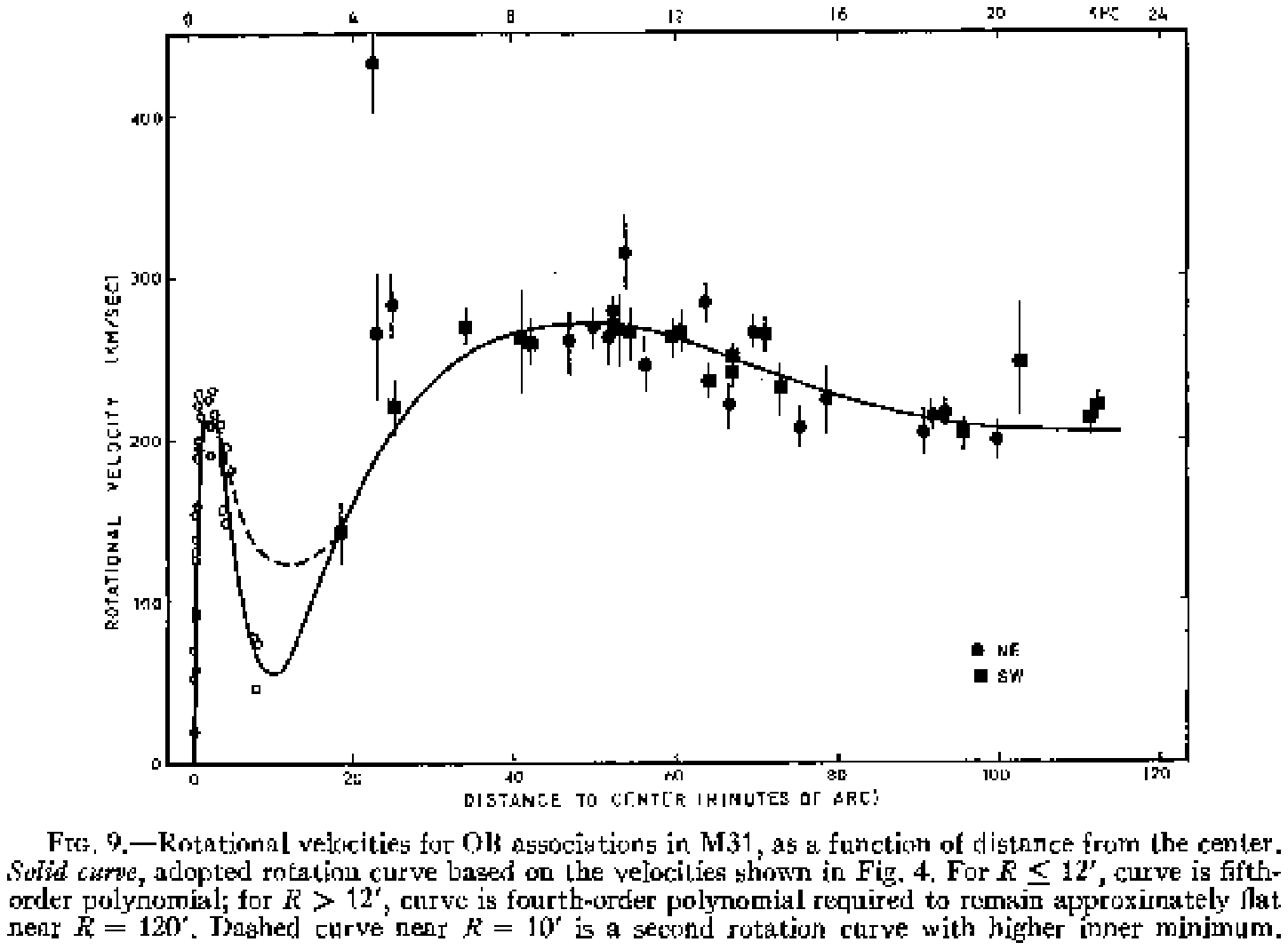}}
\end{center}
\caption{
 Rotation curve for OB associations in M31,
 as a function of distance from the galaxy center
 (figure from \cite{Rubin70}).}
\label{fig1010201}
\end{figure}
\begin{figure}[p]
\begin{center}
\imageswitch{
\begin{picture}(15.9,20.5)
\put(0,0){\framebox(15.9,20.5){}}
\end{picture}}
{\includegraphics[width=15.9cm]{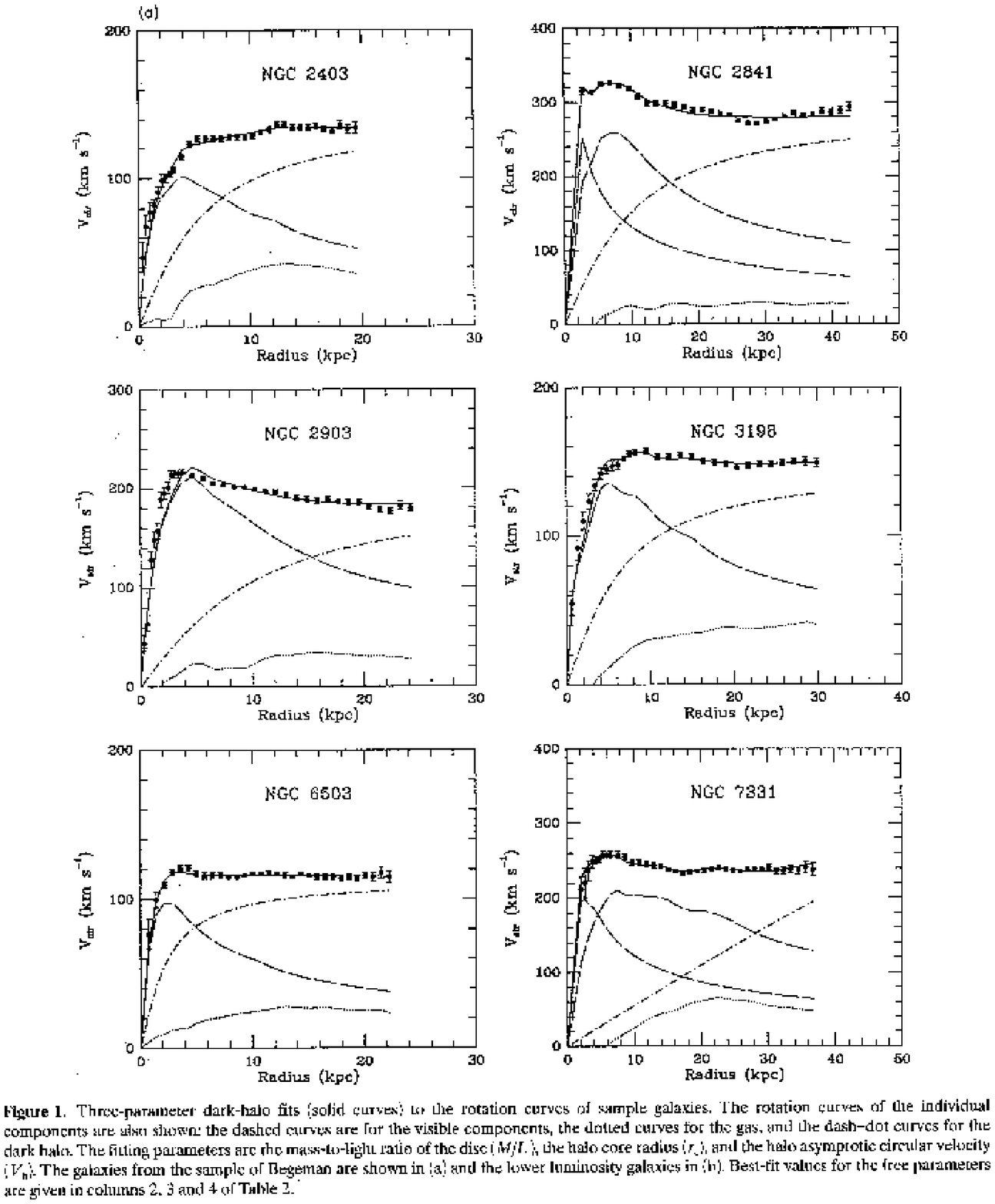}}
\end{center}
\renewcommand{\thefigure}{\thechapter.\arabic{figure}a}
\caption{
 Rotation curves for some simple spiral galaxies.
 The rotation curves of the individual components: visible component, gas, and dark halo,
 are also shown (figure from \cite{Begeman91}).}
\label{fig1010202a}
\end{figure}
\begin{figure}[t!]
\begin{center}
\imageswitch{
\begin{picture}(15.9,14.4)
\put(0,0){\framebox(15.9,14.4){}}
\end{picture}}
{\includegraphics[width=15.9cm]{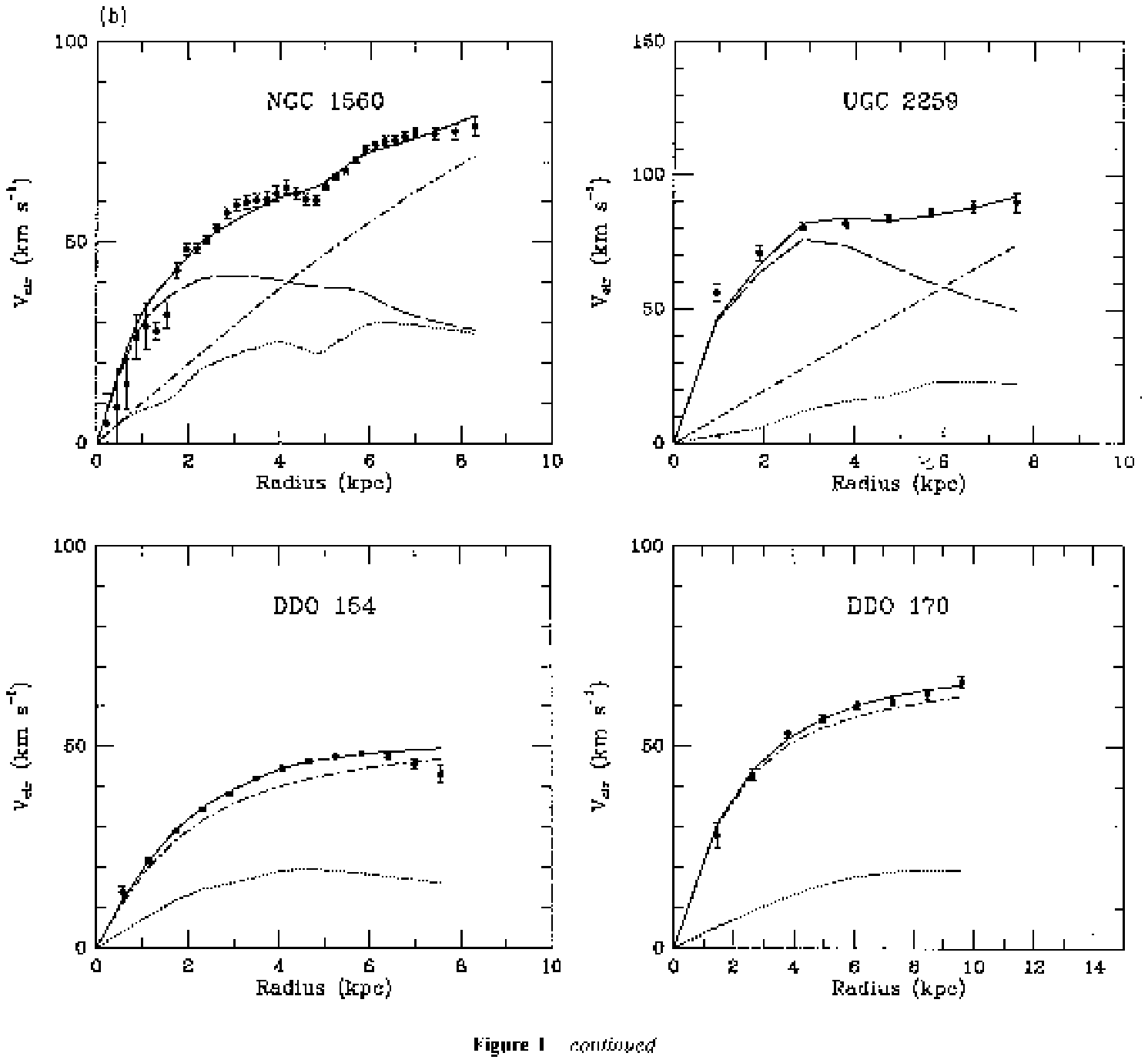}}
\end{center}
\addtocounter{figure}{-1}
\renewcommand{\thefigure}{\thechapter.\arabic{figure}b}
\caption{
 Rotation curves for some lower luminosity galaxies.
 The rotation curves of the individual components: visible component, gas, and dark halo,
 are also shown (figure from \cite{Begeman91}).}
\label{fig1010202b}
\end{figure}
\renewcommand{\thefigure}{\thechapter.\arabic{figure}}
 However,
 measurements of the circular velocities of clouds of neutral hydrogen in galaxies
 by using their 21-cm emission \cite{SUSYDM96} showed that
 the {\em rotation curves of spiral galaxies are flat}
 (see Figs.~\ref{fig1010201} and \ref{fig1010202a})
 {\em or even rising} (see Fig.~\ref{fig1010202b})
 {\em at distances far away from their stellar and gaseous components}
 \cite{Rubin70}-\cite{Olling00}.
 This implies the {\em existence of a ``dark halo'' around the galaxy}
 with a total mass profile:
\beq
         M(r)
 \propto r
\~,
\eeq
 i.e., the profile of the mass density should be
\beq
         \rho(r)
 \propto \frac{1}{r^2}
\~,
\label{eqn1010202}
\eeq
 since
\beq
   M(r)
 = 4 \pi \int_0^r r'^2 \rho(r') \~ dr'
\~.
\label{eqn1010203}
\eeq
\subsection{Escape velocity from the Milky Way}
 ~~~$\,$
 The escape velocity from the Milky Way at the position of our Solar system
 has been estimated as \cite{Fich91}, \cite{SUSYDM96}
\beq
                 v_{\rm esc}^{\rm Galaxy}
 \amssyasymc{38}~450~{\rm km/s}
\~.
\label{eqn1010301}
\eeq
 It is {\em much larger than can be accounted for by the luminous matter in our Galaxy}.
 It is not difficult to understand why this result so surprising
 if one thinks about the huge difference
 between the escape velocity from the Sun's surface \cite{SUSYDM96}:
\beq
       v_{\rm esc}^{\odot}
 \cong 617.5~{\rm km/s}
\~,
\eeq
 and that from the Solar system at the position of the Earth \cite{TaschenbuchPhysik}:
\beq
       v_{\rm esc}^{\rm solar}
 \cong 42.1~{\rm km/s}
\~.
\eeq
\begin{figure}[t]
\begin{center}
\imageswitch{
\begin{picture}(9,8.2)
\put(0,0){\framebox(9,8.2){}}
\end{picture}}
{\includegraphics[width=9cm]{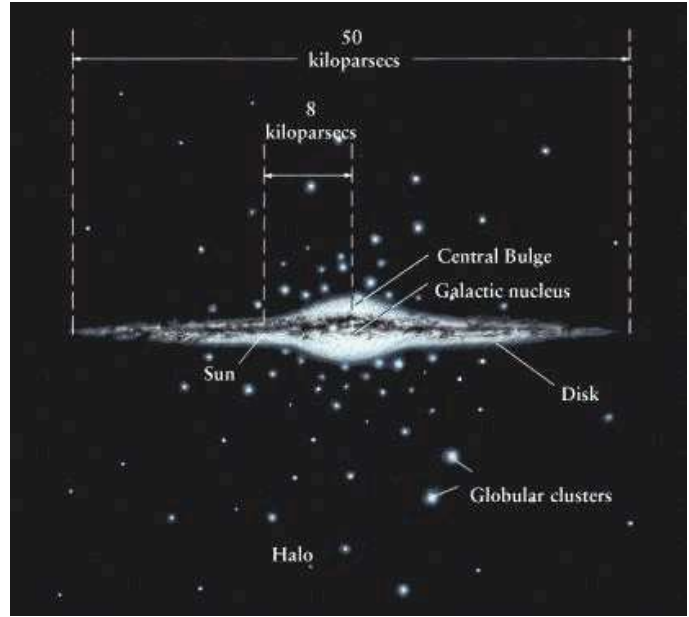}}
\end{center}
\caption{
 The position of the Sun in the Milky Way.
 The visible (luminous) component has been shown.
 It can be seen that
 our Solar system is already out of the Central Bulge of the Galaxy.}
\label{fig1010301}
\end{figure}
 Recall that
 the gravitational well in our Solar system is essentially
 only caused by the Sun's mass
 which dominates the total mass of the Solar system.
 If the mass of the luminous matter in our Galaxy
 would also dominate the total mass of the Galaxy
 (see Fig.~\ref{fig1010301}),
 the escape velocity from our Galaxy at the position of our Solar system
 would also be reduced (at least) one order of magnitude.
\section{Cosmological density parameters}
\label{Omegai}
 ~~~$\,$
 The cosmological density parameter of a given component of the total energy of the Universe $i$
 has been defined as
 the density of this component averaged over the Universe, $\rho_i$,
 in units of the {\em critical energy density} of the Universe, $\rho_{\rm crit}$,
\beq
        \Omega_i
 \equiv \frac{\rho_i}{\rho_{\rm crit}}
\~.
\label{eqn1020001}
\eeq
 The critical energy density of the Universe is the value
 that makes the geometry of the Universe flat
 (a more detailed explanation about the ``flat Universe''
  will be given in Subsec.~\ref{CMBRanisotropy})
 \cite{RPP06AP}:
\footnote{
 Note that $\rho_{\rm crit}$ here is the critical ``energy density''.
 However,
 the factor $c^2$ in the expression has been usually neglected.}
\beqn
           \rho_{\rm crit}
 =         \frac{3 H_0^2}{8 \pi G_N}
 \eqnsimeq 2.775 \times 10^{11}  h^2~M_{\odot}/{\rm Mpc}^3
           \non\\
 \eqnsimeq 1.878 \times 10^{-29} h^2~{\rm g/cm}^3
\~.
\label{eqn1020002}
\eeqn
 Here $H$ is the expansion rate of the Universe (the time dependent Hubble parameter) defined as
\beq
        H
 \equiv \frac{\dot{a}}{a}
\label{eqn1020003}
\eeq
 with the scale factor of the Universe, $a(t)$,
 and $H_0$ denotes the expansion rate of the Universe
 ``at the present epoch'' (redshift $z = 0$),
\beq
        H_0
 \equiv 100 h~{\rm km/s/Mpc}
\~,
\eeq
 with the dimensionless Hubble constant $h$.
 Moreover,
 the Newtonian gravitational constant, the mass of the Sun, and the parsec (pc) are given as
 \cite{RPP06AP}
\beq
   G_N
 = 6.674 \times 10^{-11}~{\rm m^3/kg/s^2}
\~,
\eeq
\beq
   M_{\odot}
 = 1.988 \times 10^{30}~{\rm kg}
\~,
\label{eqn1020004}
\eeq
\beq
        1~{\rm pc}
 \equiv \frac{1~{\rm AU}}{1~{\rm arc~sec}}
 \simeq 3.0857 \times 10^{16}~{\rm m}
 \simeq 3.262~c\cdot{\rm yr}
\~,
\eeq
 where the astronomical unit (AU), i.e., the mean distance between the Earth and the Sun,
 and the speed of light, $c$, are given as \cite{RPP06AP}
\beq
   1~{\rm AU}
 = 1.4960 \times 10^{11}~{\rm m}
\~,
\eeq
\beq
        c
 \equiv 2.9979 2458 \times 10^{8}~{\rm m/s}
\~.
\eeq

 In the rest of this section
 I present briefly some important astronomical measurements and their current results,
 by which the cosmological density parameters of different components of our Universe
 can be determined pretty exactly
 (to one or even two significant figure accuracy \cite{RPP06CP}).
 Particularly prominent are the measurement of
 the anisotropy of the cosmic microwave background (CMB) radiation,
 led by the three-year results from the Wilkinson Microwave Anisotropy Probe (WMAP) \cite{WMAP3}.
 In the last subsection we will see that
 the cosmological density parameters also show
 the evidence for (or the necessity of) the existence of Dark Matter
 (and, more exactly, also of {\em Dark Energy},
  both of them are ``something which we do not know'').
 More details about theoretical explanations and experimental results of these measurements
 can be found in e.g., Refs.~\cite{EarlyUniverse}, \cite{Turner99},
 \cite{RPP06CP}, and \cite{RPP06BBC}-\cite{RPP06CMB}.
\subsection{Cosmic microwave background (CMB)}
 ~~~$\,$
 The cosmic microwave background radiation (CMBR or CBR) discovered in 1965
 provides the fundamental evidence for the hot Big-Bang model of the early Universe
 \cite{EarlyUniverse}.
 The spectrum of the CBR can be described very well
 by a blackbody function with the temperature $T$ \cite{RPP06CMB}.
 The energy density of ``CMB photons'' can then be obtained directly as \cite{RPP06AP}
\beq
   \rho_{\gamma}
 = \frac{\pi^2}{15} \frac{(k_B T)^4}{(\hbar c)^3}
\~,
\label{eqn1020101}
\eeq
 where the Boltzmann's constant, $k_B$, and the Planck's constant, $\hbar$, have been given as
 \cite{EarlyUniverse}
\beq
   k_B
 = 1.3807 \times 10^{-23}~{\rm J/K}
\~,
\eeq
\beq
   \hbar
 = 1.0546 \times 10^{-34}~{\rm J \~ s} 
\~.
\eeq
 The present (mean) CBR temperature has been measured as \cite{RPP06AP}
\beq
   T_0
 = (2.725 \pm 0.001)~{\rm K}
\~.
\label{eqn1020102}
\eeq
\subsection{Anisotropy of the CMB radiation}
\label{CMBRanisotropy}
\begin{figure}[t]
\begin{center}
\imageswitch{
\begin{picture}(14,7)
\put(0,0){\framebox(14,7){}}
\end{picture}}
{\includegraphics[width=14cm]{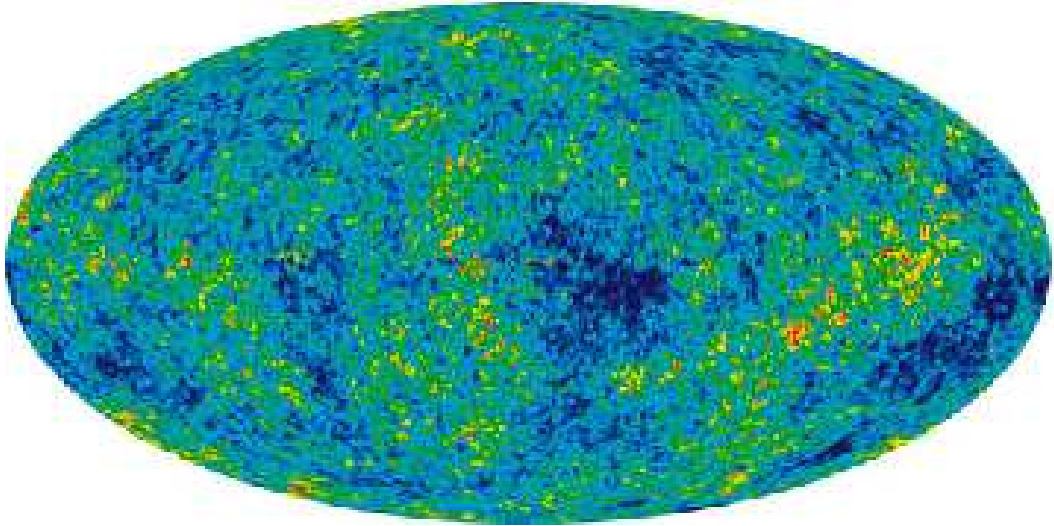}}
\end{center}
\caption{
 Anisotropy of the CMB radiation.
 The detailed, all-sky picture of the infant Universe from three years of WMAP data.
 The image reveals 13.7 billion year old temperature fluctuations
 (shown as color differences)
 which correspond to the seeds that grew to become the galaxies.
 The signal from our Galaxy was subtracted using the multi-frequency data.
 This image shows a temperature range of $\pm 200~\mu$K
 (figure from NASA/WMAP Science Team).}
\label{fig1020201}
\end{figure}

 ~~~$\,$
 Another important observable quantity from the CMB is its anisotropy:
 tiny temperature difference
 (of order of $10^{-5}$ of the magnitude of the mean temperature $T$ \cite{RPP06CMB})
 between two points on the sky (see Fig.~\ref{fig1020201}).
 The measurement of the anisotropy of the CBR can be expanded in spherical harmonics as:
\beq
   \delta T(\theta,\phi)
 = \sum_{l,m} a_{lm} Y_{lm}(\theta,\phi)
\~.
\eeq
 Here the multipole number $l$ is given as
\beq
        l
 \simeq \frac{200^{\circ}}{\theta}
\~,
\eeq
 and a useful quantity $C_l$ has been defined as
\beq
        C_l
 \equiv \expv{|a_{lm}|^2}
 =      \frac{1}{2l+1} \sum_{m = -l}^{l} |a_{lm}|^2
\~.
\eeq
 The anisotropy of the CBR offers the best means for
 determining the curvature of the Universe, $R_{\rm curv}$, \cite{Turner99}
 and thereby the ``total matter/energy density'' of the Universe, $\Omega_0$,
 according to the Friedmann equation \cite{Turner99}, \cite{RPP06CP}:
\beq
   \Omega_0-1
 = \frac{k}{R_{\rm curv}^2 H_0^2}
\~,
\eeq
 where $k$ is a curvature constant which can be chosen to take only three discrete values:
 $\pm 1$ and 0.
 According to the Friedmann equation,
 when the total matter/energy density of our Universe is equal to 1,
 the Universe is ``spatially flat'' ($R_{\rm curv} = \infty$, or, equivalently, $k = 0$).
 While,
 for $\Omega_0 > 1$ ($\Omega_0 < 1$),
 the constant $k$ should be $+1$ ($-1$) and we call the Universe ``closed'' (``open'')
 \cite{RPP06BBC}.
\begin{figure}[t]
\begin{center}
\imageswitch{
\begin{picture}(14,9.9)
\put(0,0){\framebox(14,9.9){}}
\end{picture}}
{\includegraphics[width=14cm]{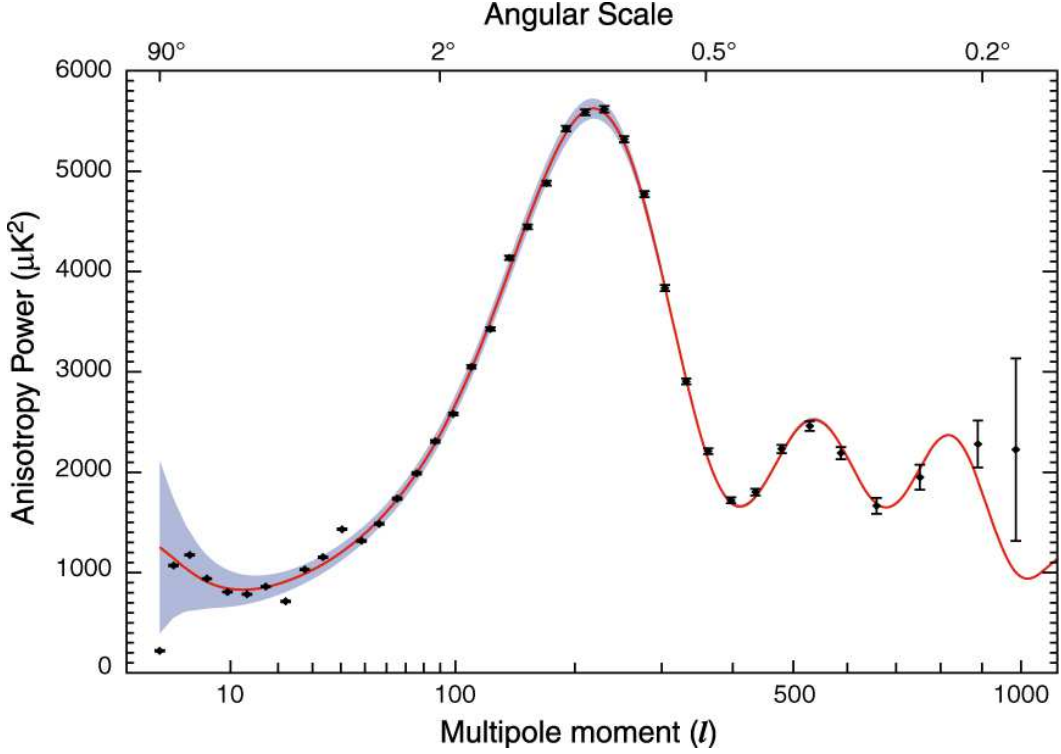}}
\end{center}
\caption{
 The angular power spectrum of the CMB temperature from three-year data of the WMAP satellite.
 The solid curve is the prediction from the best-fitting $\Lambda$CDM model.
 The error bars on the data points (which are tiny for most of them)
 indicate the observational errors,
 while the shaded region indicates the statistical uncertainty
 from being able to observe only one microwave sky,
 known as cosmic variance,
 which is the dominant uncertainty on large angular scales \cite{RPP06CP}.
 The first peak around $l \sim 200$ corresponds to $\theta \sim 1^{\circ}$
 (figure from NASA/WMAP Science Team).}
\label{fig1020202}
\end{figure}
 In Fig.~\ref{fig1020202} one can find that
 the anisotropy power,
 sometimes shown as $l(l+1) \~ C_l/2\pi$,
 oscillates (the so-called ``gravity-driven acoustic oscillations'')
 with some ``acoustic peaks''.
 Roughly speaking,
 the angular position of these peaks
 is a sensitive probe of the spatial curvature of the Universe:
 if our Universe is open (close),
 these peaks should lie at higher (lower) $l$ \cite{RPP06CMB}.

 Moreover,
 according to standard Big-Bang Cosmology,
 the higher the primordial matter density,
 the shorter the duration of the epoch of structure formation
 and thereby the larger fluctuations in the CBR \cite{SUSYDM96},
 or, equivalently,
 the stronger these acoustic oscillations \cite{Turner99}.
 Hence,
 the relative height of the first acoustic peak
 can be used to determine the ``primordial matter density''.

 More details about the physics and the analyses of anisotropy of CBR
 can be found in Ref.~\cite{RPP06CMB}. 
\subsection{Age of the Universe}
 ~~~$\,$
 As mentioned in the previous subsection,
 the higher the primordial matter density,
 the faster the Universe expanded
 and thus the shorter the age of the Universe reaching its present size.
 Hence,
 the measurements of the age of the Universe, $T_{\rm U}$,
 and the expansion rate of the Universe, $h$,
 can give the upper and lower limits on the ``matter density'' in the Universe.

 According to WMAP results combined with other astronomical measurements,
 the age of the Universe has been estimated as \cite{WMAP3}, \cite{RPP06AP}
\beq
    T_{\rm U}
 = 13.7^{+0.1}_{-0.2}~{\rm Gyr}
\~.
\eeq
\subsection{Present expansion rate of the Universe}
 ~~~$\,$
 According to the Hubble law \cite{Turner99}:
\beq
   H_0
 = \frac{v}{d}
\~.
\eeq
 Here the velocity $v$ can be determined by the redshift,
 thus the most accurate direct methods for measuring distances to distant objects $d$
 can be used to estimate the Hubble parameter $H_0$ \cite{Primack02}.
 Currently,
 there are two methods for measuring extra galactic distances \cite{Primack02}:
 time delays between luminosity variations
 in different gravitationally lensed images of distant quasars and
 the Sunyaev-Zel'dovich effect:
 Compton scattering of the CMB by the hot electrons in clusters of galaxies.
 Note that
 the error on the estimates of the Hubble parameter
 is dominated by one systematic uncertainty:
 the distance from out Galaxy to the Large Magellanic Cloud (LMC),
 which has been used to calibrate
 the Cepheid period-luminosity relationship \cite{Primack02}.

 The dimensionless Hubble constant has been estimated as \cite{WMAP3}, \cite{RPP06AP}
\beq
   h
 = 0.73^{+0.03}_{-0.04}
\~,
\eeq
 and the present expansion rate of the Universe can then be given as
\beq
   H_0
 = 73^{+3}_{-4}~{\rm km/s/Mpc}
\~.
\eeq
\subsection{Abundances of the light elements}
 ~~~$\,$
\begin{figure}[p]
\begin{center}
\imageswitch{
\begin{picture}(15.5,19.5)
\put(0,0){\framebox(15.5,19.5){}}
\end{picture}}
{\includegraphics[width=15.5cm]{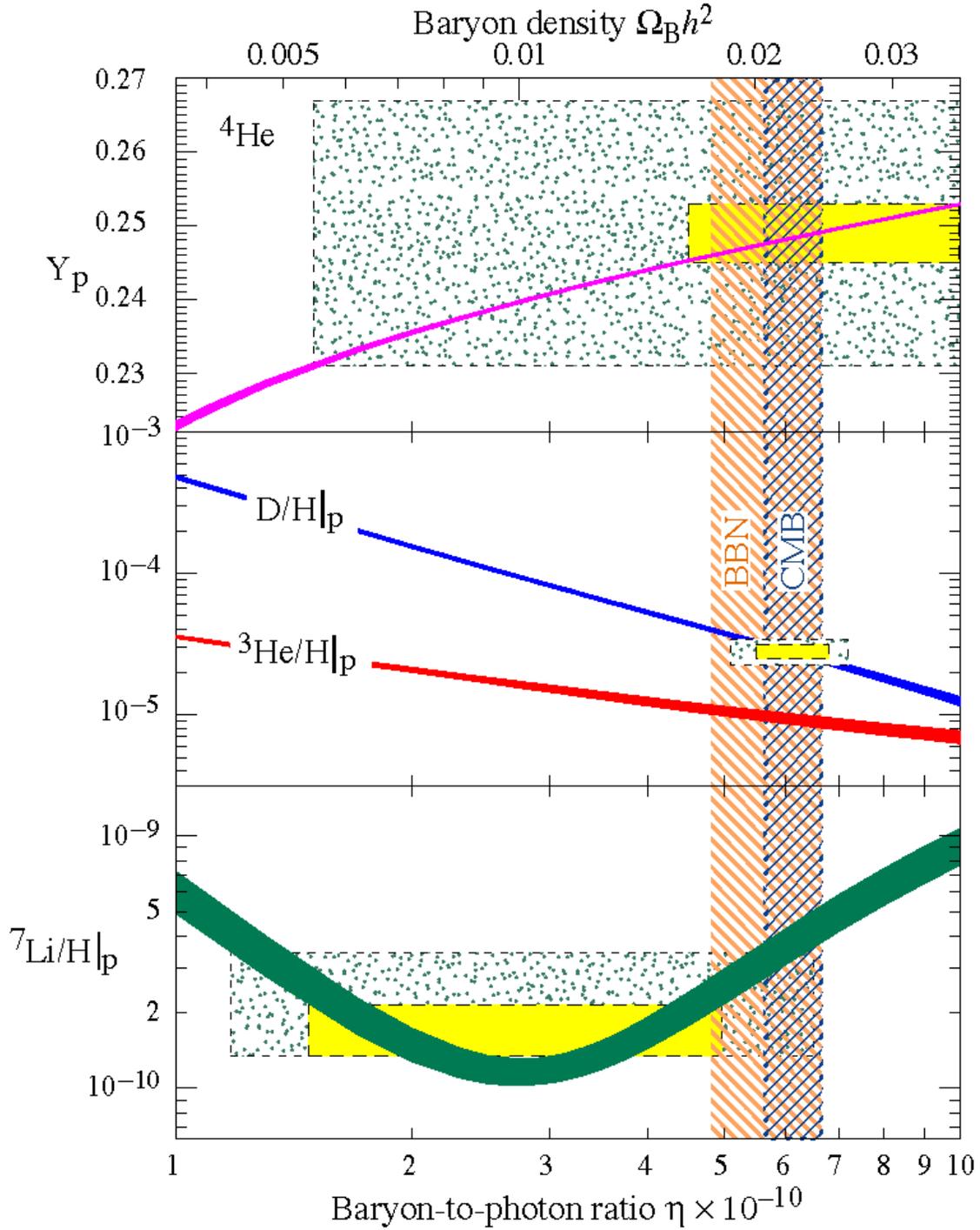}}
\end{center}
\caption{
 The predicted abundances of $\rmXA{He}{4}$ (mass fraction),
 D, $\rmXA{He}{3}$, and $\rmXA{Li}{7}$ (number relative to hydrogen)
 by the standard model of the BBN as a function of the baryon density.
 ${\rm Y_p} \equiv 2 n_{\rm n} n_{\rm p}/(n_{\rm n}+n_{\rm p}) \simeq 0.25$,
 where $n_{\rm n}$ and $n_{\rm p}$ are the neutron and proton number densities.
 Widths of the curves indicate $2 \sigma$ theoretical uncertainty.
 Boxes indicate the observed light element abundances
 (smaller boxes: $2 \sigma$ statistical errors;
  larger boxes: $\pm 2 \sigma$ statistical and systematic errors).
 The narrow vertical band indicates the CMB measurement of the cosmic baryon density
 (figure from \cite{RPP06BBN}).}
\label{fig1020501}
\end{figure}
 BBN predicts the primordial abundances of the light elements.
 Thus measurements of the primordial abundances of the light elements
 produced in the Big Bang,
 such as deuterium (D), helium ($\rmXA{He}{3}$ and $\rmXA{He}{4}$),
 and lithium ($\rmXA{Li}{7}$),
 can also give the upper and lower limits of the ``baryon density'' in the Universe.

 Moreover,
 among these four light elements,
 because the primordial abundance of deuterium depends strongly on the baryon density
 ($\propto \rho_{\rm b}^{-1.7}$),
 and it can only be destroyed by the astrophysical processes,
 deuterium becomes the most powerful "baryometer" \cite{Turner99}.

 Fig.~\ref{fig1020501} shows the theoretically predicted abundances of the four lightest elements
 and the observational results. 
\subsection{Gas-to-total mass ratio}
\label{gastotalmass}
 ~~~$\,$
 The clusters of galaxies formed due to density perturbations
 with a co-moving size of the order of 10 Mpc
 and gathered material from such a large region of the Universe \cite{Turner99}.
 Meanwhile,
 most of the baryons in the clusters of galaxies reside not in the galaxies themselves
 but in form of hot intercluster, x-ray emitting gas \cite{Turner99}.
 Hence,
 by measuring the gas-to-total mass ratio of the cluster, $f_{\rm gas/total}$,
 and combining with the (measured) baryon density in the Universe, $\Omega_{\rm b}$,
 we can determine the ``matter density'' in the Universe \cite{Turner99}:
\beq
   \Omega_{\rm m}
 = \frac{\Omega_{\rm b}}{f_{\rm gas/total}}
\~.
\eeq

 There are two methods for determining the mass of the intercluster gas:
 the x-ray flux emitted from the intercluster gas
 or the Sunyaev-Zel'dovich CBR distortion caused by CMB photons
 scattering off hot electrons in the intracluster gas \cite{Turner99}.
 While,
 there are also three independent methods for estimating the total mass of a cluster:
 the motions of cluster galaxies with the virial theorem,
 assuming that the gas is in hydrostatic equilibrium and
 using it to infer the underlying mass distribution,
 or mapping the cluster mass directly by gravitational lensing \cite{Turner99}, \cite{Tyson00}.
 Within their uncertainties and where comparisons can be made,
 the two methods for determining the mass of the intercluster gas
 and the three methods for estimating the total mass of a cluster
 are consistent with each other, respectively \cite{Turner99}.
\subsection{Mass-to-light ratio}
 ~~~$\,$
 One other method for estimating the ``matter density'' of the Universe
 is using the mass-to-light ratios \cite{Turner99}:
\beq
   \rho_{\rm m}
 = \afrac{M}{L} {\cal L}
\~,
\label{eqn1020701}
\eeq
 where $\cal L$ is the averaged luminosity density of the Universe
 \cite{SUSYDM96}, \cite{Turner99}.
 In V-band \cite{SUSYDM96} and in B-band \cite{Turner99},
 we have,
 respectively,
\cheqnCa
\beq
   {\cal L}_{\rm V}
 = (1.7 \pm 0.6) \times 10^8 h~L_{\odot}/{\rm Mpc}^3
\~,
\label{eqn1020702a}
\eeq
 and
\cheqnCb
\beq
   {\cal L}_{\rm B}
 = 2.4 \times 10^8 h~L_{\odot}/{\rm Mpc}^3
\~,
\label{eqn1020702b}
\eeq
\cheqnC
 where $L_{\odot}$ is the luminosity of the Sun \cite{RPP06AP},
\beq
   L_{\odot}
 = (3.846 \pm 0.008) \times 10^{26}~{\rm W}
\~.
\label{eqn1020703}
\eeq
 Once we have estimated the mass-to-light ratios of some systems, i.e.,
\beq
        \Upsilon_{\rm x}
 \equiv \frac{M}{L_{\rm x}}
\~,
        ~~~~~~~~~~~~~~~~~~~~~~~~~~~~~~ 
\rm
        x
 =      V,~B
.
\eeq
 Then,
 combining Eqs.(\ref{eqn1020001}), (\ref{eqn1020002}), (\ref{eqn1020701})
 and (\ref{eqn1020702a}) or (\ref{eqn1020702b}),
 the total matter density can be obtained as
\beq
   \Omega_{\rm m,x}
 = \frac{C_{\rm x}}{10^{4} h} \afrac{\Upsilon_{\rm x}}{\Upsilon_{\odot}}
\~,
   ~~~~~~~~~~~~~~~ 
\rm
   x
 = V,~B
.
\eeq
 Here
\beq
   C_{\rm V}
 = 6.1
\~,
   ~~~~~~~~~~~~~~~~~~~~~~~~~~~~~~ 
   C_{\rm B}
 = 8.6
\~,
\eeq
 and $\Upsilon_{\odot}$ is the mass-to-light ratio of the Sun,
\beq
   \Upsilon_{\odot}
 = 5.169 \times 10^{3}~{\rm kg/W}
\~,
\eeq
 where I have used Eqs.(\ref{eqn1020004}) and (\ref{eqn1020703}).
\subsection{Supernovae type Ia (SNe Ia) at high-redshift}
 ~~~$\,$
 If we could measure the present extra galactic distances $d_0$ and velocities $v_0$,
 they should obey the Hubble law \cite{Turner99}:
\beq
   \frac{v_0}{d_0}
 = H_0
\~,
\eeq
 since the expansion of the Universe is simply a rescaling.
 But what we can actually measure
 are the distances $d_z$ and velocities $v_z$ at an earlier time (redshift $z$).
 If we suppose that
 the expansion of our Universe should slow down due to the attractive force of gravity,
 i.e., $H_z > H_0$,
 the measured galactic velocities should be larger than that expected by the Hubble law:
\beq
   v_z
 = d_z H_z
 > d_z H_0
\~,
\eeq
 or,
 equivalently,
 for the galaxies with known velocities,
 their distances should be shorter than that expected by the Hubble law:
\beq
   d_z
 = \frac{v_z}{H_z}
 < \frac{v_z}{H_0}
\~.
\eeq

 In 1998
 two groups: the Supernova Cosmology Project and the High-z Supernova Search Team
 have published their
 ``magnitude-redshift (Hubble) diagram for fifty-some type Ia supernovae (SNe Ia)
 out to redshifts of nearly 1'' \cite{Turner99}.
 By using SNe Ia as standard candles for estimating the distances to faraway galaxies,
 the two groups concluded that
 {\em distant galaxies are moving slower than predicted by the Hubble law}
 and that {\em this implies an accelerated expansion of our Universe} \cite{Turner99}.

 In order to explain this observational indication,
 i.e., in order to find the discrepancy 
 between the measured total matter(/energy) density, $\Omega_0$,
 and the matter density, $\Omega_{\rm m}$, (data given in the next subsection),
 a new term ``Dark Energy''
\footnote{
 Dark Energy is beyond the area of my research and thus will not be discussed in this work.
 Short reviews and summaries can be found
 in e.g., \cite{Turner99}, \cite{RPP06CP}, \cite{RPP06BBC}, and \cite{Peebles03}.}
 (or sometimes also called ``quintessence'') has been introduced
 \cite{Turner99}.
\subsection{Cosmological densities of different components}
\label{Omegailist}
 ~~~$\,$
 According to the various astronomical measurements described above
 (and other measurements,
  e.g., the peculiar velocities of galaxies,
  the shape of the present power spectrum of density perturbations,
  and the opacity of the Lyman-$\alpha$ forest toward high-redshift quasars),
 we can conclude today the cosmological densities of different components as follows.

 The total matter/energy density is \cite{WMAP3}, \cite{RPP06AP}
\beq
   \Omega_0
 = 1.003^{+0.013}_{-0.017}
\~.
\label{eqn1020901}
\eeq
 It can be separated into Dark Energy \cite{RPP06AP}:
\beq
        \Omega_{\Lambda}
 \equiv \frac{\Lambda}{3 H_0^2}
 =      0.76^{+0.04}_{-0.06}
\~,
\label{eqn1020902}
\eeq
 where we have used
\beq
   \rho_{\Lambda}
 = \frac{\Lambda}{8 \pi G_N}
\~,
\eeq
 and total matter \cite{WMAP3}, \cite{RPP06AP}:
\beq
   \Omega_{\rm m}
 = 0.127^{+0.007}_{-0.009}~h^{-2}
 = 0.24 ^{+0.03 }_{-0.04 }
\~.
\label{eqn1020903}
\eeq
 The total matter in the Universe can also be separated into baryons
 \cite{WMAP3}, \cite{RPP06AP}:
\beq
   \Omega_{\rm b}
 = 0.0223^{+0.0007}_{-0.0009}~h^{-2}
 = 0.042 ^{+0.003 }_{-0.005 }
\~,
\label{eqn1020904}
\eeq
 and {\em non-baryonic Dark Matter} \cite{RPP06AP}:
\beq
   \Omega_{\rm DM}
 = \Omega_{\rm m}-\Omega_{\rm b}
 = 0.105^{+0.007}_{-0.010}~h^{-2}
 = 0.20 ^{+0.02 }_{-0.04 }
\~.
\label{eqn1020905}
\eeq
 Among the baryons in the Universe there is luminous matter with a density of
 \cite{SUSYDM96} \cite{Jesus04}:
\beq
        \Omega_{\rm lum}
 \simeq 0.01
\~,
\label{eqn1020906}
\eeq
 including the density of the stars \cite{Frenk02}:
\beq
   \Omega_{\rm stars}
 = (0.0023 \sim 0.0041) \pm 0.0004
\~.
\label{eqn1020907}
\eeq
 While,
 the non-baryonic Dark Matter can be separated into {\em Cold Dark Matter} (CDM)
 and {\em Hot Dark Matter} (HDM)
 (the definitions and some discussions about the CDM and HDM
  will be given in Secs.~\ref{CDM} and \ref{HDM},
  respectively).
 Finally,
 among the relativistic particles (see Sec.~\ref{HDM}),
 the density of the CMB photons can be estimated directly
 by inserting $T_0$ in Eq.(\ref{eqn1020102}) into Eq.(\ref{eqn1020101}) as
 \cite{RPP06AP}
\beq
   \Omega_{\gamma}
 = (2.471 \pm 0.004) \times 10^{-5}~h^{-2}
 = (4.6   \pm 0.5  ) \times 10^{-5}
\~,
\label{eqn1020910}
\eeq
 and the density of the neutrinos has been estimated as \cite{RPP06AP}
\beq
   \Omega_{\nu}
 < 0.014~(95\%~{\rm C.~L.})
\~.
\label{eqn1020911}
\eeq
\section{Galactic halo models}
\label{halomodels}
\begin{figure}[p]
\begin{center}
\imageswitch{
\begin{picture}(13,20)
\put(0,0){\framebox(13,20){}}
\end{picture}}
{\includegraphics[width=13cm]{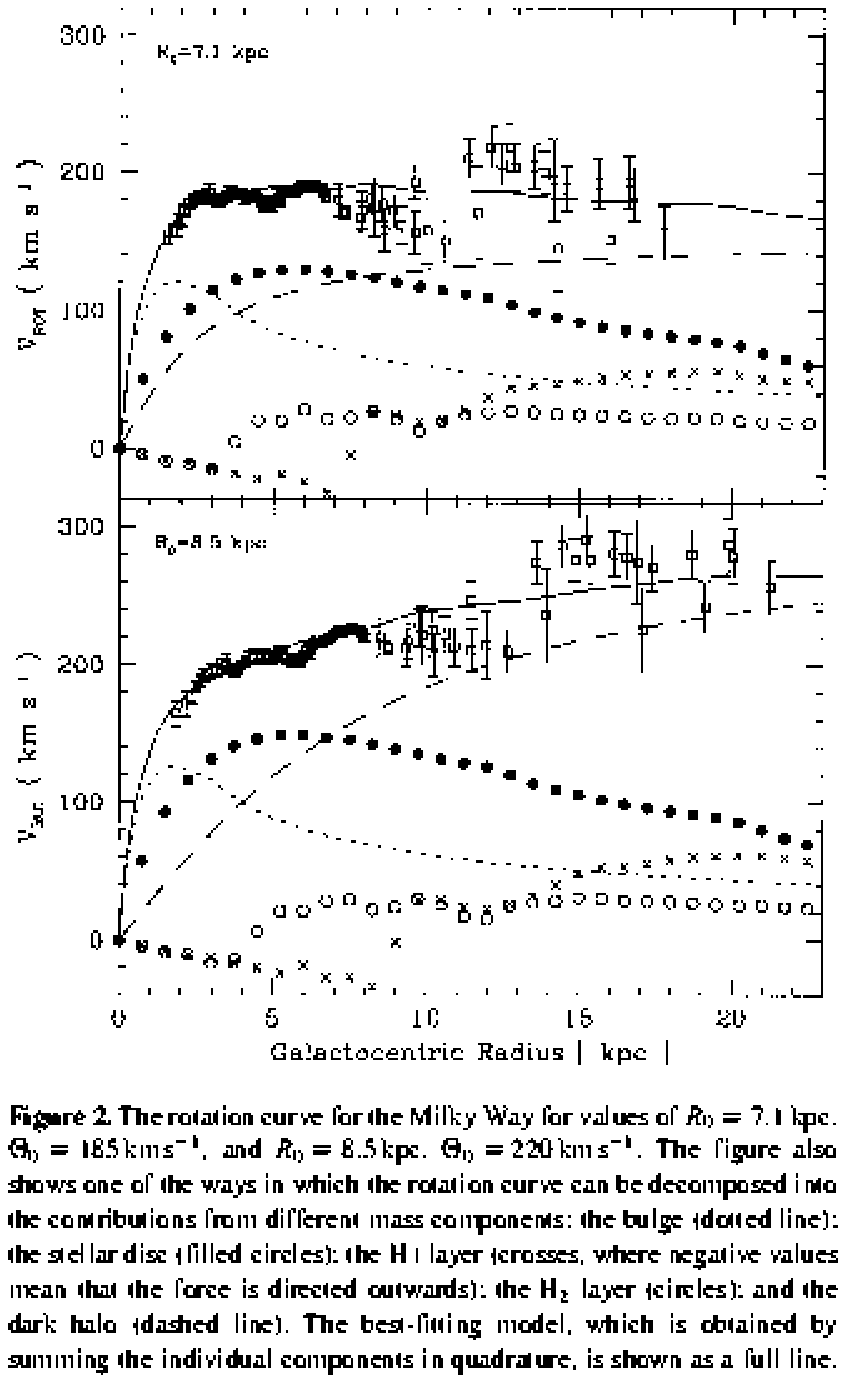}}
\end{center}
\caption{
 Rotation curve for the Milky Way,
 as a function of distance from the Galactic center,
 with two different assumptions for the Sun's distance
 from the Galactic center, $r_0$,
 and the rotation velocity at $r_0$, $v_{\rm rot}(r_0)$
 (figure from \cite{Olling00}).}
\label{fig1030001}
\end{figure}

 ~~~$\,$
 In this section
 I present some simple halo models.
 For estimating some characteristics of halo Dark Matter,
 such as the velocity dispersion of Dark Matter particles, $\bar{v}$,
 and the local Dark Matter density, $\rho_0$,
 the rotation curve of our Galaxy is the most important observational quantity,
 since it measures the change in density
 and sets the scale for the depth of the Galactic potential well \cite{SUSYDM96}.
 Essentially,
 all important direct information which has been obtained about the halo
 is provided by the rotation curve \cite{Kamionkowski98}.
 However,
 due to our location inside the Milky Way,
 it is more difficult to measure the accurate rotation curve of our own Galaxy
 than those of other galaxies (see Fig.~\ref{fig1030001}).
 In addition,
 the ``disk contribution'' to the rotation curve
 must be known to infer the halo contribution,
 but precise determination of the disk contribution is also difficult.
\subsection{Standard assumptions of Dark Matter halo}
\label{haloDM}
 ~~~$\,$
 The velocity dispersion of Dark Matter particles
\footnote{
\label{expvv2}
 Strictly speaking,
 the velocity dispersion should be $\expv{{\bf v}^2}-\expv{{\bf v}}^2$.
 However,
 since the major component of Dark Matter should be cold
 (a detailed discussion will be given in Sec.~\ref{CDM})
 and assumed to have negligibly small velocity average $\expv{\bf{v}}$
 in the Galactic rest frame,
 its rms velocity $\expv{v^2}$ has been called sometimes simply
 as the velocity dispersion.}
 in the Solar neighborhood has been assumed as
\beq
        \bar{v}
 =      \expv{v^2}^{1/2}
 \simeq 270~{\rm km/s}
\~.
\label{eqn1030101}
\eeq
 And the IAU standard value for the rotational velocity
 at the Sun's distance from the Galactic center is \cite{RPP06AP}
\beq
        v_0
 \equiv v_{\rm rot}(r_0)
 \simeq (220 \pm 20)~{\rm km/s}
\~,
\label{eqn1030102}
\eeq
 where the distance from the Sun to the Galactic center is \cite{RPP06AP}
\beq
         r_0
 \simeq (8.0 \pm 0.5)~{\rm kpc}
\~.
\label{eqn1030103}
\eeq
 On the other hand,
 the local Dark Matter density (the Dark Matter density near the Solar system) is given by
 \cite{SUSYDM96}
\beqn
            \rho_0
 \equiv     \rho(r_0)
 \eqnapprox 0.3~{\rm GeV}/c^2/{\rm cm^3}
            \non\\
 \eqnapprox 5 \times 10^{-25}~{\rm g/cm^3}
\~,
\label{eqn1030104}
\eeqn
 with an uncertainty of slightly less than a factor of 2 \cite{SUSYDM96}, \cite{Turner99}.
 Here I have used \cite{RPP06P}
\beq
   1~{\rm GeV}/c^2
 = 1.7827 \times 10^{-24}~{\rm g}
\~.
\eeq
\subsection{Canonical isothermal spherical halo model}
\label{ishalo}
 ~~~$\,$
 The simplest halo model is an isothermal spherical halo.
 An empirically plausible radial profile for a spherical galactic halo
 is constrained only by its contribution to the galactic rotation curve.
 This means that
 the radial profile should approach to a constant near its core
 so that it gives rise to a linearly rising rotation curve at small radii,
 and it should fall as $1/r^2$ or eventually faster at large radii
 to provide a flat rotation curve \cite{Kamionkowski98}.
 The density profile of the cored isothermal spherical halo is given by \cite{SUSYDM96}
\beq
   \rho_{\rm IS}(r)
 = \rho_0 \afrac{r_c^2+r_0^2}{r_c^2+r^2}
\~,
\label{eqn1030201}
\eeq
 where $\rho_0$ is the local halo density
 and $r_c$ is the core radius of the isothermal spherical halo,
 within which the density $\rho_{\rm IS}(r)$ behaves no longer as $1/r^2$,
 but goes to a constant as $r$ approaches 0.

 Substituting this expression into Eq.(\ref{eqn1010203}) and using Eq.(\ref{eqn1010201}),
 the rotational velocity at a radius $r$ from the halo center can be found as
\beqn
     v_{\rm IS}^2(r)
 \= 4 \pi G_N \cdot \frac{1}{r} \int_0^r r'^2 \rho(r') \~ dr'
    \non\\
 \= 4 \pi G_N \rho_0 \abrac{r_c^2+r_0^2} \bbrac{1-\afrac{r_c}{r} \tan^{-1}\afrac{r}{r_c}}
\~,
\label{eqn1030202}
\eeqn
 where I have used
\beqN
   \int \frac{x^2 \~ dx}{a^2+x^2}
 = x-a \~ \tan^{-1}\afrac{x}{a}
\~.
\eeqN
 Define $v_{\infty}$ as the (measured) rotational velocity as $r \to \infty$.
 One can find that
\beq
   v_{\infty}^2
 = v_{\rm IS}^2(r \to \infty)
 = 4 \pi G_N \rho_0 \abrac{r_c^2+r_0^2}
\~,
\label{eqn1030203}
\eeq
 thus the local halo density $\rho_0$ in Eq.(\ref{eqn1030201}) can be expressed as
\beq
   \rho_0
 = \frac{v_{\infty}^2}{4 \pi G_N \abrac{r_c^2+r_0^2}}
\~.
\label{eqn1030204}
\eeq
\begin{figure}[p]
\begin{center}
\imageswitch{
\begin{picture}(15,21)
\put(0  , 0  ){\framebox(15,21  ){}}
\put(1  ,12  ){\framebox(13, 9  ){}}
\put(7  ,11  ){\makebox ( 1, 1  ){(a)}}
\put(1  , 1  ){\framebox(13, 9  ){}}
\put(7  , 0  ){\makebox ( 1, 1  ){(b)}}
\end{picture}}
{\includegraphics[width=13cm]{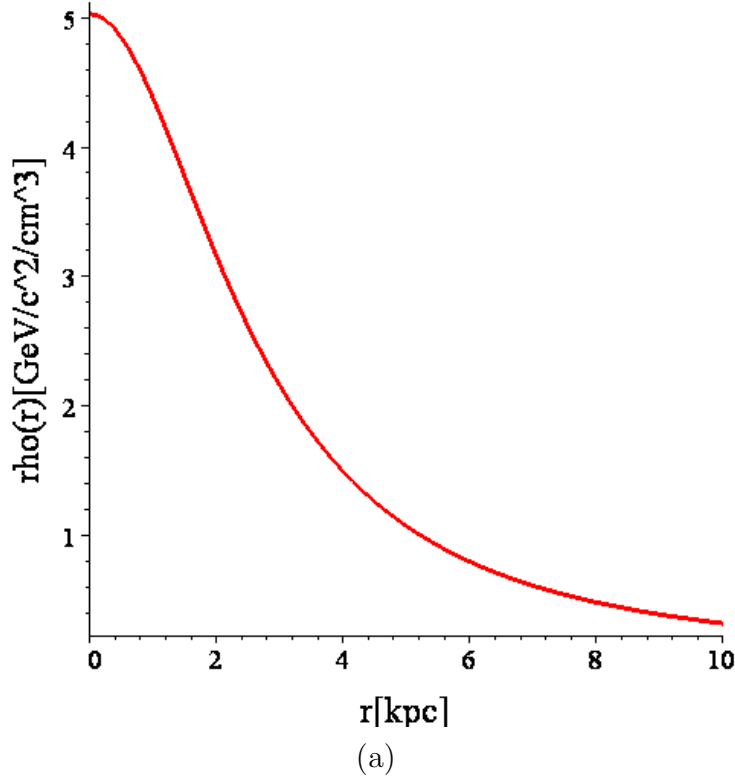} \\ (a) \\ \vspace{1cm}
 \includegraphics[width=13cm]{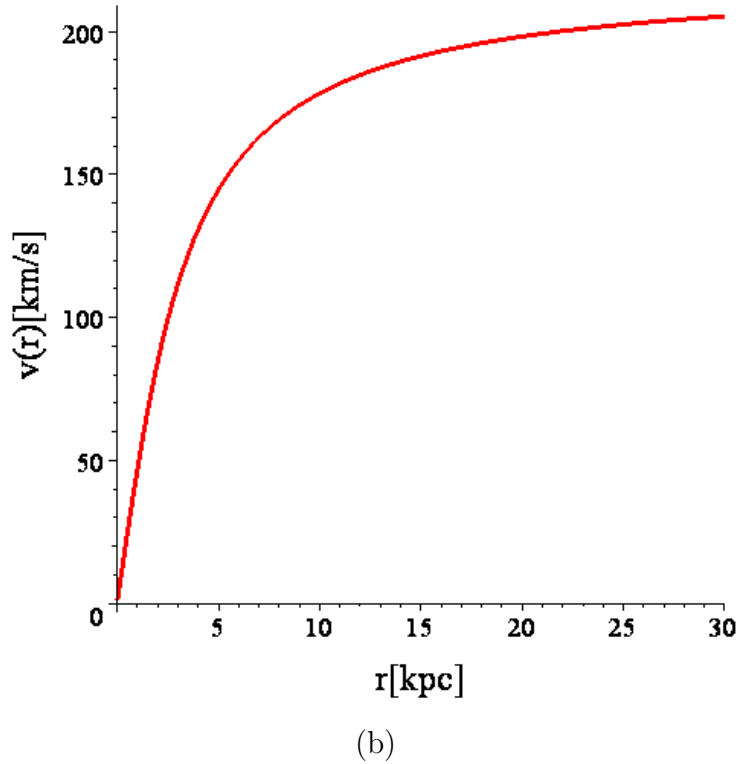}   \\ (b)}
\end{center}
\caption{
 (a) The radial density profile $\rho_{\rm IS}(r)$ given in Eq.(\ref{eqn1030201}') and
 (b) the rotation curve $v_{\rm IS}(r)$ given in Eq.(\ref{eqn1030202}')
 of the canonical isothermal spherical halo model.
 Here I have used $v_{\infty} = 220~{\rm km/s}$ \cite{Kamionkowski98}
 and $r_c = 2.6$ kpc (see Subsec.~\ref{Evanshalo}).}
\label{fig1030201}
\end{figure}
 Meanwhile,
 combining Eqs.(\ref{eqn1030202}) and (\ref{eqn1030203}),
 the core radius of the isothermal spherical halo in unit of $r_0$, i.e., $r_c/r_0$,
 can be solved (numerically) by \cite{SUSYDM96}:
\beq
   \afrac{r_c}{r_0} \tan^{-1}\afrac{r_0}{r_c}
 = 1-\frac{v_{\rm IS}^2(r_0)}{v_{\infty}^2}
\~.
\label{eqn1030205}
\eeq
 Eqs.(\ref{eqn1030204}) and (\ref{eqn1030205}) show
 how we can estimate the local halo density and the halo core radius
 in the isothermal spherical halo model
 once the rotational velocities $v_0$ and $v_{\infty}$ have been measured.

 Finally,
 substituting Eq.(\ref{eqn1030204}) into Eqs.(\ref{eqn1030201}) and (\ref{eqn1030202})
 the density profile and the rotation curve of the isothermal spherical halo model
 can be rewritten as
\cheqnrefp{eqn1030201}
\beq
   \rho_{\rm IS}(r)
 = \frac{v_{\infty}^2}{4 \pi G_N} \afrac{1}{r_c^2+r^2}
\~,
\label{eqn1030201p}
\eeq
 and
\cheqnrefp{eqn1030202}
\beq
    v_{\rm IS}(r)
 = v_{\infty} \bbrac{1-\afrac{r_c}{r} \tan^{-1}\afrac{r}{r_c}}^{1/2}
\~,
\label{eqn1030202p}
\eeq

 Figs.~\ref{fig1030201} show
 the radial density profile (upper frame) and the rotation curve (lower frame)
 of the canonical isothermal spherical halo model
 given in Eqs.(\ref{eqn1030201}') and (\ref{eqn1030202}'),
 respectively.
\subsection{Alternative isothermal spherical halo model}
\label{aishalo}
 ~~~$\,$
 An alternative density profile for the isothermal spherical halo
 has been given by \cite{Kamionkowski98}
\cheqnCN{-2}
\beq
   \rho_{\rm AIS}(r)
 = \rho_0 \afrac{r_c+r_0}{r_c+r}^2
\~,
\label{eqn1030301}
\eeq
 where $\rho_0$ is again the local halo density
 and $r_c$ is the core radius of this alternative isothermal spherical halo mode.
\footnote{
 Here,
 for simplicity,
 I use the same notation as in Eq.(\ref{eqn1030201}),
 but $r_c$ for these two halo models are not the same.
 The determination of these core radii will be given in Subsec.~\ref{Evanshalo}}
 Using Eqs.(\ref{eqn1010203}) and (\ref{eqn1010201}),
 the rotation curve of this alternative isothermal spherical halo can be found as
\beq
   v_{\rm AIS}^2(r)
 = 4 \pi G_N \rho_0 \abrac{r_c+r_0}^2
   \bbrac{1+\frac{r_c}{r_c+r}-2 \afrac{r_c}{r} \~ \ln\afrac{r_c+r}{r_c}}
\~,
\label{eqn1030302}
\eeq
 where I have used
\beqN
   \int \frac{x^2 \~ dx}{(ax+b)^2}
 = \frac{1}{a^3} \bbrac{(ax+b)-\frac{b^2}{ax+b}-2 b \~ \ln(ax+b)}
\~.
\eeqN
 Meanwhile,
 the core radius of this alternative halo model in unit of $r_0$
 can be solved (numerically) by :
\beq
   2 \alpha_c \~ \ln\afrac{\alpha_c+1}{\alpha_c}-\frac{\alpha_c}{\alpha_c+1}
 = 1-\frac{v_{\rm AIS}^2(r_0)}{v_{\infty}^2}
\~.
\label{eqn1030303}
\eeq
 where I have defined
\beq
        \alpha_c
 \equiv \frac{r_c}{r_0}
\~.
\label{eqn1030304}
\eeq
 Finally,
 from Eq.(\ref{eqn1030302}),
 one has
\beq
   v_{\infty}^2
 = v_{\rm AIS}^2(r \to \infty)
 = 4 \pi G_N \rho_0 \abrac{r_c+r_0}^2
\~,
\label{eqn1030305}
\eeq
 the density profile and the rotation curve of the alternative isothermal spherical halo model
 in Eqs.(\ref{eqn1030301}) and (\ref{eqn1030302}) can be rewritten as
\cheqnrefp{eqn1030301}
\beq
   \rho_{\rm AIS}(r)
 = \frac{v_{\infty}^2}{4 \pi G_N} \afrac{1}{r_c+r}^2
\~,
\label{eqn1030301p}
\eeq
 and
\cheqnrefp{eqn1030302}
\beq
   v_{\rm AIS}(r)
 = v_{\infty}
   \bbrac{1+\frac{r_c}{r_c+r}-2 \afrac{r_c}{r} \~ \ln\afrac{r_c+r}{r_c}}^{1/2}
\~.
\label{eqn1030302p}
\eeq
\cheqnCN{-2}

\begin{figure}[t]
\begin{center}
\imageswitch{
\begin{picture}(14,10.1)
\put(0,0){\framebox(14,10.1){}}
\end{picture}}
{\includegraphics[width=13.5cm]{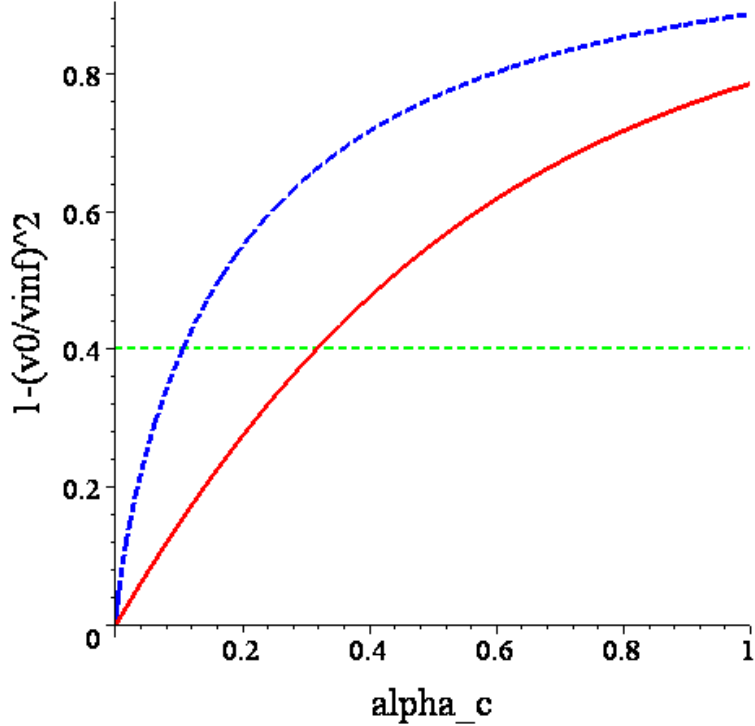}}
\end{center}
\caption{
 The equations for solving
 the ratios of the core radii of two cored isothermal spherical halo models
 to the distance from the Sun to the Galactic center,
 $\alpha_c \equiv r_c/r_0$.
 The solid (red) line shows Eq.(\ref{eqn1030205})
 and the dash-dotted (blue) line shows Eq.(\ref{eqn1030303}).
 Here the dash (green) line denotes
 $v_{\rm IS}(r_0) = v_{\rm AIS}(r_0) = v_{\rm halo}(r_0) = 170$ km/s
 \cite{SUSYDM96}, \cite{Kamionkowski98}
 and $v_{\infty} = 220$ km/s.}
\label{fig1030301}
\end{figure}

 Fig.~\ref{fig1030301} shows
 the equations for solving
 the ratios of the core radii of two cored isothermal spherical halo models
 to the distance from the Sun to the Galactic center,
 $\alpha_c \equiv r_c/r_0$,
 given in Eqs.(\ref{eqn1030205}) and (\ref{eqn1030303}).
\subsection{Evans' power-law halo model}
\label{Evanshalo}
 ~~~$\,$
 Even if we consider only spherical halo distributions,
 there are still some latitudes in our choice
 for the precise form of the radial density profile of a halo model.
 Meanwhile,
 N-body simulations of gravitational collapse produce axisymmetric or triaxial halos
 \cite{Green01},
 and other spiral galaxies appear to have flattened halos \cite{Sackett94}, \cite{Green01}.

 The equipotentials of elliptical galaxies and the halos of spiral galaxies
 could be roughly stratified on similar concentric spheroids \cite{Evans94}.
 This suggests a useful approximation to their gravity field as \cite{Evans94}:
\beq
     \psi(r,z)
 =   \frac{\psi_a r_c^{\beta}}{(r_c^2+r^2+z^2/q^2)^{\beta/2}}
\~,
     ~~~~~~~~~~~~~~~~~~~~ 
     \beta
 \ne 0
\~,
\label{eqn1030401}
\eeq
 where $\psi_a$ is the central potential,
 $r_c$ is the core radius of the halo,
 and $q$ is the axis ratio of the equipotentials or the so-called flattening parameter.
 Note that I use here the cylindrical coordinates
 and $r$ denotes the distance from the point which one considers to the rotation axis.
 The potential is just a power of the spheroidal radius $r$,
 thus this model has been called the power-law halo model \cite{Evans94}.

 Using the Poisson Equation:
\footnote{
 Note that
 the Poisson Equation holds actually for
 the ``total'' potential and the ``total'' density distribution,
 i.e., for luminous baryonic matter and Dark Matter together,
 not for each component separately.}
\beq
   \Lap \psi
 =-4 \pi G_N \rho
\~,
\eeq
 the density distribution of the gravitational potential described in Eq.(\ref{eqn1030401})
 can be obtained as \cite{Evans94}
\beq
   \rho(r,z)
 = \frac{v_a^2}{4 \pi G_N} \afrac{r_c^{\beta}}{q^2}
   \bfrac{(2 q^2+1) r_c^2+(1-\beta q^2) r^2+(2q^2-\beta-1) z^2/q^2}
         {(r_c^2+r^2+z^2/q^2)^{\beta/2+2}}
\~,
\label{eqn1030402}
\eeq
 where
\beq
        v_a^2
 \equiv \beta \psi_a
\~.
\label{eqn1030403}
\eeq
 Meanwhile,
 the velocity of the circular orbit in the equatorial plane with radius $r$
 can be obtained as \cite{Evans94}
\beq
   v_{\rm circ}(r)
 = v_a \bfrac{r_c^{\beta} r^2}{(r_c^2+r^2)^{\beta/2+1}}^{1/2}
\~,
\label{eqn1030404}
\eeq
 since the central force
\beq
   F_{\rm cen}(r,z=0)
 =-\Grad \psi(r,z) \cdot \h{r} \Big|_{z=0}
 = \frac{v_{\rm circ}^2(r)}{r}
\~.
\eeq
 The rotational velocity in Eq.(\ref{eqn1030404}) is asymptotically falling if $\beta > 0$
 and rising if $\beta < 0$ \cite{Evans94}.

 On the other hand,
 the model with spheroidal equipotentials and a completely flat rotation curve at large radii
 is well known as an axisymmetric logarithmic potential
 \cite{Binney81}, \cite{Evans93}, \cite{Evans94}:
\beq
   \psi(r,z)
 =-\frac{v_a^2}{2} \ln\abrac{r_c^2+r^2+\frac{z^2}{q^2}}
\~,
\label{eqn1030405}
\eeq
 where $v_a$ is the rotational velocity at large radii
 (i.e., $v_{\infty}$ used in the previous two subsections).
 Using the Poisson Equation,
 the density distribution can be found as \cite{Evans93}
\beq
   \rho(r,z)
 = \frac{v_a^2}{4 \pi G_N} \bfrac{(2 q^2+1) r_c^2+r^2+(2q^2-1) z^2/q^2}{q^2 (r_c^2+r^2+z^2/q^2)^2}
\~.
\label{eqn1030406}
\eeq
 Comparing this expression with the expression in Eq.(\ref{eqn1030402}),
 it can be seen that
 the logarithmic potential given in Eq.(\ref{eqn1030405}) has the properties
 corresponding to the missing $\beta = 0$ case in Eq.(\ref{eqn1030401}) \cite{Evans94}.
 The velocity of the circular orbit in the equatorial plane with radius $r$
 due to the potential given in Eq.(\ref{eqn1030405}) can be obtained as
\beq
   v_{\rm circ}(r)
 = v_a \afrac{r^2}{r_c^2+r^2}^{1/2}
\~.
\label{eqn1030407}
\eeq

 Furthermore,
 let $q = 1$ and use the replacement:
\beq
     r^2+z^2
 \to r^2
\~,
\eeq
 we can rewrite the potential in Eq.(\ref{eqn1030405})
 to the potential for a spherical halo as
 \cite{Evans93}
\beq
   \psi(r)
 =-\frac{v_a^2}{2} \ln\abrac{r_c^2+r^2}
\~.
\label{eqn1030408}
\eeq
 The density distribution and the velocity of the circular orbit
 of this spherical Evans model with radius $r$ can be obtained as,
 respectively \cite{Evans93}, \cite{Kamionkowski98},
\beq
   \rho_{\rm PL}(r)
 = \frac{v_a^2}{4 \pi G_N} \bfrac{3 r_c^2+r^2}{(r_c^2+r^2)^2}
\~,
\label{eqn1030409}
\eeq
 and
\beq
   v_{\rm circ}(r)
 = v_a \afrac{r^2}{r_c^2+r^2}^{1/2}
\~.
\label{eqn1030410}
\eeq
 As done in Subsecs.~\ref{ishalo} and \ref{aishalo},
 the core radius of the halo in unit of $r_0$ can be solved by means of the following equation
 \cite{Kamionkowski98}:
\beq
   \frac{r_c}{r_0}
 = \bbrac{\frac{v_a^2}{v_{\rm circ}^2(r_0)}-1}^{1/2}
\~.
\label{eqn1030411}
\eeq

\begin{figure}[p]
\begin{center}
\imageswitch{
\begin{picture}(15,20.5)
\put(0  , 0  ){\framebox(15,20.5){}}
\put(1  ,11.5){\framebox(13, 9  ){}}
\put(7  ,10.5){\makebox ( 1, 1  ){(a)}}
\put(1  , 1  ){\framebox(13, 9  ){}}
\put(7  , 0  ){\makebox ( 1, 1  ){(b)}}
\end{picture}}
{\includegraphics[width=12.5cm]{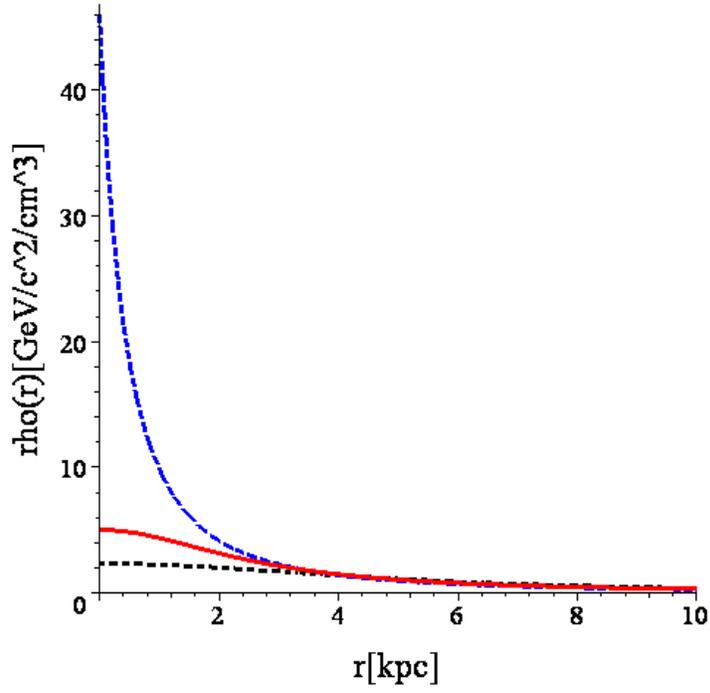} \\ (a) \\ \vspace{0.5cm}
 \includegraphics[width=12.5cm]{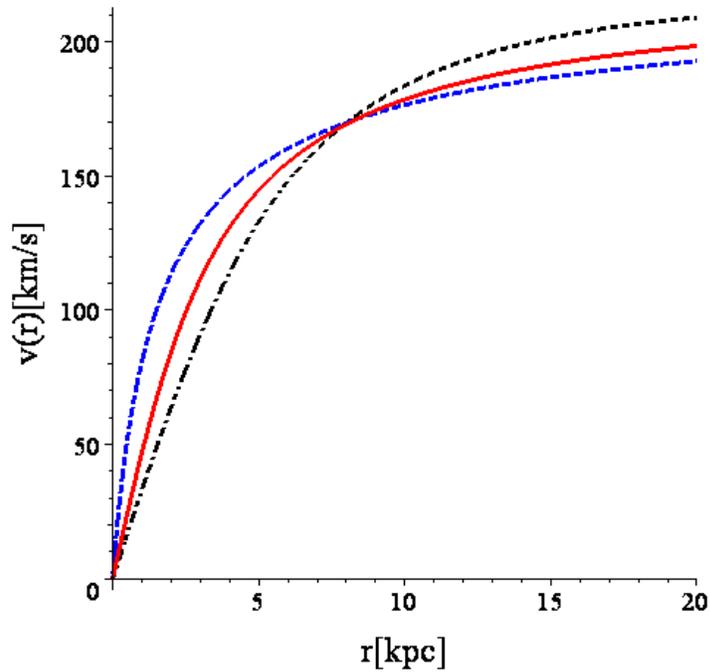}   \\ (b)}
\end{center}
\caption{
 (a) The radial density profiles and (b) the rotation curves
 of different isothermal spherical halo models.
 The solid (red) lines indicate
 the canonical halo model in Eq.(\ref{eqn1030201}') with $r_c \simeq 2.6~{\rm kpc}$,
 the dash (blue) lines indicate
 the alternative halo model in Eq.(\ref{eqn1030301}') with $r_c \simeq 0.86~{\rm kpc}$,
 and the dash-dotted (black) lines indicate
 the spherical Evans' halo model in Eq.(\ref{eqn1030409})
 with $r_c \simeq 6.6~{\rm kpc}$.
 Here I have used $v_{\infty} = 220~{\rm km/s}$ \cite{Kamionkowski98}.}
\label{fig1030401}
\end{figure}
 Finally,
 using the local rotational velocity given in Eq.(\ref{eqn1030102}):
\cheqnrefp{eqn1030102}
\beq
        v_{\rm rot}(r_0)
 \simeq 220~{\rm km/s}
\~,
\eeq
\cheqnCN{-1}
 and the assumption for the disk contribution to the rotational velocity at $r = r_0$
 \cite{SUSYDM96}, \cite{Kamionkowski98}:
\beq
   v_{\rm disk}(r_0)
 = 140~{\rm km/s}
\~,
\eeq
 the local halo contribution can be found as
\beq
        v_{\rm halo}(r_0)
 \simeq 170~{\rm km/s}
\~,
\eeq
 since,
\beq
   v_{\rm rot}(r_0)
 = \sqrt{v_{\rm disk}^2(r_0)+v_{\rm halo}^2(r_0)}
\~.
\eeq
 Then the core radii for the canonical, alternative, and Evans spherical halo models
 given in Eqs.(\ref{eqn1030201}'), (\ref{eqn1030301}'), and (\ref{eqn1030409})
 can be found by means of Eqs.(\ref{eqn1030205}), (\ref{eqn1030303}), and (\ref{eqn1030411}) as
\beq
        r_{c,{\rm IS}}
 \simeq 2.6~{\rm kpc}
\~,
\eeq
\beq
        r_{c,{\rm AIS}}
 \simeq 0.86~{\rm kpc}
\~,
\eeq
 and
\beq
        r_{c,{\rm PL}}
 \simeq 6.6~{\rm kpc}
\~,
\eeq
 where I used
\cheqnrefp{eqn1030103}
\beq
        r_0
 \simeq 8.0~{\rm kpc}
\~,
\label{eqn1030103p}
\eeq
\cheqnCN{-1}
 and the rotational velocity at large radii as \cite{Kamionkowski98}
\beq
   v_{\infty}
 = v_a
 = 220 ~{\rm km/s}
\~.
\eeq

 Figs.~\ref{fig1030401} show
 the radial density profiles (upper frame) and the rotation curves (lower frame)
 of different isothermal spherical halo models
 with the core radii obtained above.
\subsection{NFW halo model}
\label{NFWhalo}
 ~~~$\,$
 Besides the three spherical halo models presented in the previous three subsections,
 there is also the well-known NFW density profile given as
 \cite{Navarro96}-\cite{Gentile07}:
\beq
   \rho_{\rm NFW}(r)
 = \frac{\rho_s}{(r/r_s) (1+r/r_s)^2}
\~.
\label{eqn1030501}
\eeq
 Here $\rho_s$ is the characteristic density \cite{Gentile07}:
\beq
   \rho_s
 = \rho_{\rm crit} \afrac{\lambda_s}{3} \bfrac{c_s^3}{\ln(1+c_s)-c_s/(1+c_s)}
\~,
\eeq
 with the virial overdensity $\lambda_s$ and the halo concentration parameter $c_s$,
 and $r_s$ is the characteristic radius \cite{Gentile07}:
\beq
   r_s
 = \frac{r_{\rm vir}}{c_s}
 = \frac{1.2 \times 10^{2}}{c_s} \afrac{M_{\rm vir}}{10^{11} M_{\odot}}^{1/3} {\rm kpc}
\~.
\label{eqn1030502}
\eeq
 The virial radius, $r_{\rm vir}$, defined as the radius,
 inside which the average overdensity is
 $\lambda_s$ times the critical density of the Universe,
 and the virial mass, $M_{\rm vir}$ is then the total mass
 within this virial radius $r_{\rm vir}$ \cite{Burkert99}.
 Usually the average overdensity $\lambda_s$ has been chosen as $\approx$ 200,
 and the virial radius and the virial mass for $\lambda_s = 200$
 have been then specially labeled as $r_{200}$ and $M_{200}$,
 respectively \cite{Burkert99}.
 However,
 for a flat Universe, i.e., $\Omega_0 = 1$,
 the average overdensity $\lambda_s$ has been found to be \cite{Lokas99}
\beq
        \lambda_s
 =      18 \pi^2
 \simeq 178
\~.
\eeq
 Note that
 there is a good correlation between $c_s$ and $M_{\rm vir}$ in Eq.(\ref{eqn1030502}),
 which results from the fact that
 dark halo densities reflect the density of the Universe
 at the epoch of their formation
 and that halos of a given mass are preferentially
 assembled over a narrow range of redshifts \cite{Burkert99}.
 Hence,
 as lower mass halos form earlier,
 at times when the Universe was significantly denser,
 they are more centrally concentrated \cite{Burkert99}.

 Although the NFW density profile given in Eq.(\ref{eqn1030501})
 doesn't approach the $1/r^2$ form for large $r$,
 high-resolution N-body simulations of structure formation and some observational results
 have shown that
 the NFW profile indeed provides
 a good description of the density distribution in clusters
 \cite{Lokas99}, \cite{Burkert99}.

 Moreover,
 the NFW density profile given in Eq.(\ref{eqn1030501})
 has been expanded to the following form \cite{Kuwabara02}:
\beq
   \rho_{\rm NFWE}(r)
 = \frac{\rho_s}{(r/r_s)^{\alpha} (1+r/r_s)^{\nu}}
\~.
\eeq
 For the original NFW profile, one has $\alpha = 1$, $\nu = 2$;
 for a modified NFW profile, one can use $\alpha = \nu = 3/2$;
 and for a so-called Hernquist profile one has $\alpha = 1$, $\nu = 3$
 \cite{Hernquist90}, \cite{Kuwabara02}.

 On the other hand,
 A. Burkert has suggested a Burkert density profile \cite{Burkert99}:
\beq
   \rho_{\rm B}(r)
 = \frac{\rho_b}{(1+r/r_b) \bbrac{1+(r/r_b)^2}}
\~.
\eeq
 Some analyses show that
 density profiles of dwarfs and low surface brightness galaxies
 can be fitted by the Burkert profile much better than the NFW profile \cite{Lokas99}.
\subsection{Bulk rotation}
\label{buklrotation}
 ~~~$\,$
 So far the halo models presented above are only classified
 by their density profiles or the gravity field,
 namely, their mass distribution.
 Because,
 as mentioned in the beginning of this section,
 what has been measured is just the rotation curve \cite{Kamionkowski98}.
 However,
 the rotation curve is determined by the halo mass distribution and
 is insensitive to its velocity distribution \cite{Kamionkowski98}.
 Thus,
 even though there are some theoretical arguments
 against a rotation-dominated velocity distribution,
 there is no empirical evidence to rule out a halo with bulk rotation \cite{Kamionkowski98}.
 Note that
 such bulk rotation can also affect the velocity distribution of WIMPs
 seen near the Earth \cite{Kamionkowski98}.
\footnote{
 The velocity distribution of WIMPs will be discussed in Chap.~3.}

 Halo models with bulk rotation can be constructed by
 taking linear combination of the velocity distribution function \cite{Green01}:
\beqn
    f_{\rm rot}({\bf v})
 \= a_{\rm rot} f_{+}({\bf v})+(1-a_{\rm rot}) f_{-}({\bf v})
    \non\\
 \= \cleft{\renewcommand{\arraystretch}{1.5}
           \begin{array}{l l}
               a_{\rm rot}  f({\bf v}) \~ , &  {\rm for}~v_{\phi} > 0 \~ , \\
            (1-a_{\rm rot}) f({\bf v}) \~ , &  {\rm for}~v_{\phi} < 0 \~ .
           \end{array}}
\eeqn
 A non-rotating halo has $a_{\rm rot} = 0.5$,
 whereas a counter-rotating or a co-rotating one has
 $0 \leq a_{\rm rot} < 0.5$ or $0.5< a_{\rm rot} \leq 1$,
 respectively \cite{Kamionkowski98}, \cite{Green01}.

 Moreover,
 $a_{\rm rot}$ is related to a dimensionless spin parameter $\lambda_{\rm rot}$,
 which usually has been used to quantify the galactic angular momentum \cite{Kamionkowski98}:
\beq
   \lambda_{\rm rot}
 = 0.36 \vbrac{a_{\rm rot}-0.5}
\~.
\eeq
 Numerical studies of galaxy formation find that $|\lambda_{\rm rot}| < 0.05$,
 corresponding to $0.36 < a_{\rm rot} < 0.64$ \cite{Green01}.
%
%
%
%

%% file: Doktorarbeit-Ch2.tex
\chapter{Candidates for Dark Matter}
 ~~~$\,$
 As defined in the Introduction,
 Dark Matter (almost) neither emits nor absorbs electromagnetic radiation.
 It is thus {\em non-luminous}.

 Meanwhile,
 as described in Secs.~\ref{DMevidence} and \ref{Omegai},
 so far we can ``observe'' (or, actually, ``feel'') the existence of Dark Matter
 only through its gravitational effects.
 Moreover,
 according to the observational results for the rotation curves of spiral galaxies
 (described in Subsec.~\ref{rotationcurves}),
 Dark Matter forms halos with an approximately spherical distribution around galaxies.
 Hence,
 Dark Matter {\em (almost) does not interact with ordinary matter}
 and is {\em collisionless}.
 Otherwise,
 if Dark Matter could interact with ordinary matter,
 it would dissipate its kinetic energy after the interactions,
 fall onto galaxies,
 settle deep into the galactic gravitational wells,
 and thus form the ``galactic disks'' with ordinary matter.

 On the other hand,
 (the major part of) Dark Matter particles
 should {\em moved non-relativistically} in the early Universe
 or, equivalently,
 {\em have sufficiently low primordial velocity dispersion},
 in order to allow it to merge to galactic scale structures
 (e.g., galaxies and clusters of galaxies).
 In contrast,
 although neutrinos are also collisionless,
 they moved relativistically
 and have thus too large velocity dispersion
 to build galactic scale structures.

 Therefore,
 Dark Matter should be some
 ``non-luminous, non-baryonic, non-relativistic, and collisionless''
 elementary particles have not yet been discovered.
 In addition,
 the candidates for Dark Matter must satisfy the following cosmological conditions:
 they must be stable on cosmological time scales 
 and have the right relic cosmological density \cite{Drees06}.

 In this chapter
 I present some most motivated and studied candidates for Dark Matter.
 Most of them are non-baryonic and non-relativistic particles.
 However,
 some relativistic particles and baryonic objects could also be (part of) Dark Matter.
\section{Cold Dark Matter (CDM)}
\label{CDM}
 ~~~$\,$
 ``Cold'' has been used here to indicate that
 such Dark Matter particles moved {\em non-relativistically}
 at the matter-radiation decoupling time in the early Universe \cite{Jesus04},
 i.e., at the time in which galaxies could just start to form.
 Due to their {\em relatively slower velocities}
 Cold Dark Matter would first form some relatively small galactic scale structures;
 large galaxies and clusters of galaxies are formed
 through ``hierarchical merging'' of these smaller structures.
\subsection{Minimal Supersymmetric Standard Model (MSSM)}
\label{MSSM}
 ~~~$\,$
 {\em Supersymmetry} has been considered to solve the hierarchy problem
 in the Standard Model (SM) of particle physics:
 Why is the electroweak scale ($E_{\rm EW} \simeq {\cal O}(100~{\rm GeV})$)
 so small compared to the other known scales
 such as the grand unification scale ($E_{\rm GUT} \simeq 10^{16}~{\rm GeV}$ \cite{SUSYDM96})
 or the Planck scale ($E_{\rm Pl} \simeq 10^{19}~{\rm GeV}$ \cite{SUSYDM96})?
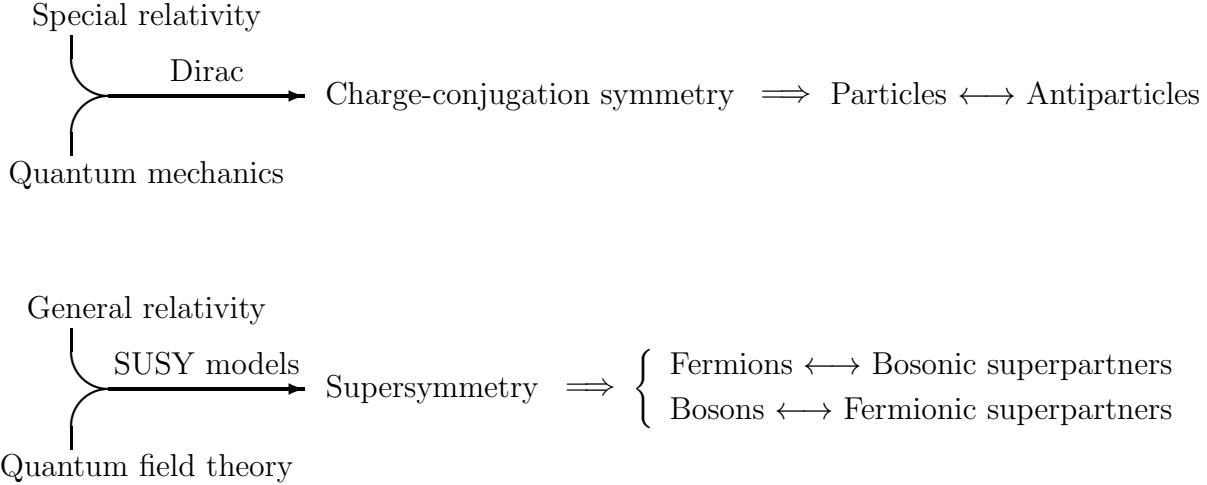
\begin{figure}[t]
\begin{center}
\begin{picture}(15.9,6.5)
\put(-0.5,3.9){\usebox{\SUSYa}}
\put(-0.5,0  ){\usebox{\SUSYb}}
\end{picture}
\end{center}
\caption{
 Extension of the Standard Model of particle physics to supersymmetric models}
\label{fig2010101}
\end{figure}

 As shown in Fig.~\ref{fig2010101},
 supersymmetry provides a natural framework
 for discussing theories with large hierarchies of scales
 and unification with gravity \cite{SUSYDM96}. 
 In supersymmetric models,
 {\em for every fermionic degree of freedom
  there is a bosonic degree of freedom and vice versa}.
 This means also that,
 for each ``normal'' particle,
 there will be a supersymmetric partner.
 Hence,
 the particle spectrum has been greatly extended in the MSSM.

\newcommand{\bl}  [1]{\multicolumn{2}{| l |}{#1}}
\newcommand{\mbxc}[2]{\makebox[#1 cm][c]{#2}}
\begin{table}[t]
\begin{center}
\renewcommand{\arraystretch}{1.4}
\small
\hspace*{-1cm}
\begin{tabular}{|| l | l || l | l | l ||}
\hline
\hline
 \multicolumn{2}{|| c ||}{Normal particles} & \multicolumn{3}{| c ||}{SUSY partners}     \\
\hline
\hline
 \mbxc{3.2}{Name} & \mbxc{1.8}{Symbol} & \multicolumn{2}{| c |}{Name} & \mbxc{4}{Symbol} \\
\hline
\hline
 up-quarks   & $q = u,~c,~t$                        & \bl{up-squarks}   &
 $\Td{u}_{L},~\Td{u}_{R},~\Td{c}_{L},~\Td{c}_{R},~\Td{t}_{L},~\Td{t}_{R}$                \\
 down-quarks & $q = d,~s,~b$                        & \bl{down-squarks} &
 $\Td{d}_{L},~\Td{d}_{R},~\Td{s}_{L},~\Td{s}_{R},~\Td{b}_{L},~\Td{b}_{R}$                \\
\hline
 leptons     & $e,~\mu,~\tau$                       & \bl{sleptons}     &
 $\Td{e}_{L},~\Td{e}_{R},~\Td{\mu}_{L},~\Td{\mu}_{R},~\Td{\tau}_{L},~\Td{\tau}_{R}$      \\
 neutrinos   & $\nu_e,~\nu_{\mu},~\nu_{\tau}$       & \bl{sneutrinos}   &
 $\Td{\nu}_e,~\Td{\nu}_{\mu},~\Td{\nu}_{\tau}$                                           \\
\hline
 gluons      & $g$                                  & \bl{gluinos}      & $\Td{g}$       \\
\hline
\hline
 photon      & $\gamma$ & \makebox[4.1cm][l]{photino $\Td{\gamma}$}     & \mbxc{1.8}{} & \\
 Z boson     & $Z^0$    & Z-ino $\Td{Z}$            &                   &                \\
\cline{1-3}
 light scalar Higgs     & $h^0$     &    & & \\
 heavy scalar Higgs     & $H^0$     &
 \raisebox{2.4ex}[0pt]{neutral higgsinos $\Td{h}^0,~\Td{H}^0$} &
 \raisebox{7.2ex}[0pt]{neutralinos}                            &
 \raisebox{7.2ex}[0pt]{$\Td{\chi}_1^0,~\Td{\chi}_2^0,~\Td{\chi}_3^0,~\Td{\chi}_4^0$}     \\
\hline
 pseudoscalar Higgs     & $A^0$     & \bl{}                    &           \\
\hline
 charged Higgs          & $H^{\pm}$ & charged higgsinos $\Td{H}^{\pm}$ & & \\
 W bosons               & $W^{\pm}$ & gauginos, W-inos  $\Td{W}^{\pm}$ &
 \raisebox{2.4ex}[0pt]{charginos}   &
 \raisebox{2.4ex}[0pt]{$\Td{\chi}_1^{\pm},~\Td{\chi}_2^{\pm}$}             \\
\hline
\hline
 graviton & $G$ & \bl{gravitino} & $\Td{G}$ \\
\hline
 axion    & $a$ & \bl{axino}     & $\Td{a}$ \\
\hline
\hline
\end{tabular}
\end{center}
\caption{Particles of typical supersymmetric models}
\label{tab2010101}
\end{table}
 The particles of typical supersymmetric models are given
 in Table \ref{tab2010101}.
 The spectrum of the normal particles is specified
 in the same manner as in non-supersymmetric models.
 Quark mass matrices determine the masses and the mixing angles,
 which are encoded in the Cabibbo-Kobayashi-Maskawa (CKM) matrix.
 The pattern of gauge-symmetry breaking is unchanged from the Standard Model,
 and gives the same tree-level relation
 between the masses of the $W^{\pm}$ and $Z^0$ bosons.
 Quarks in the Standard Model have spin $\frac{1}{2}$,
 while their superpartners, squarks, are scalars \cite{SUSYDM96}.
 There are two squarks (left-hand and right-hand) for each quark.
 In some models there is no mixing between different flavors,
 and each squark is associated with a given quark \cite{SUSYDM96},
 for example, $\Td{u}_{L}$ and $\Td{u}_{R}$, $\Td{d}_{L}$ and $\Td{d}_{R}$.
 However,
 generally
 the three up-quarks can mix among themselves and similarly for the three down-quarks,
 so there are totally six up-squarks and six down-squarks in the particle spectrum
 \cite{SUSYDM96}.
 Similarly for the leptons.
 In these models,
 left-right sfermion mixing is proportional to
 the corresponding fermion mass \cite{SUSYDM96}.
 Thus there is little left-right mixing
 for $u$, $d$, and $s$ squarks or selectrons or smuons,
 but mixing of staus and $c$, $b$, and especially $t$ squarks
 can be substantial \cite{SUSYDM96}.

 A most important technical difference between the Standard Model and the MSSM
 occurs in the Higgs sector.
 Two weak isospin Higgs doublet fields are required in the MSSM,
 whereas only one is required in the SM \cite{SUSYDM96}.
 This enrichment of the Higgs sector gives rise to five physical states and
 provides an important phenomenological window.
 The superpartners of the $W^{\pm}$ and charged Higgs bosons,
 the gauginos and the charged higgsinos,
 carry the same $SU(3) \times U(1)$ quantum numbers.
 Thus they will generally mix after electroweak-symmetry breaking,
 and the two resulting mass eigenstates
 are linear combinations known as charginos \cite{SUSYDM96}.
 Similarly for the superpartners of
 the photon, $Z^0$ boson, and neutral Higgs bosons.
 These fields generally mix to create four mass eigenstates
 called neutralinos \cite{SUSYDM96}.
 In many supersymmetric models,
 constraints on the Higgs-higgsino sector
 are therefore an important area of supersymmetric phenomenology \cite{SUSYDM96}.

 In Table \ref{tab2010101},
 the tilde $\sim$ has been used to denote a supersymmetric particle.
 However,
 the tildes for neutralinos and charginos are sometimes omitted
 since there is no ambiguity for such particles.
 Moreover,
 the lightest neutralino is in most models
 the lightest supersymmetric particle (LSP)
 and usually just to be called as ``the neutralino''.
 Note here that,
 although the lightest neutralino is the most studied LSP
 and some authors even use the LSP to indicate it,
 there are also some other candidates for LSP,
 e.g., the gravitino.

 Furthermore,
 a $R$-parity should be also presented here \cite{SUSYDM96}:
\beq
   R
 = (-1)^{3 (B-L)+2 S}
\~,
\eeq
 where $B$ and $L$ are the baryon and lepton number,
 $S$ is the spin.
 For ordinary particles $R = +1$ (even) and
 for supersymmetric particles $R = -1$ (odd).
 If the $R$-parity is conserved,
 one SUSY particle (with $R = -1$) could only decay to
 a lighter SUSY particle (with $R = -1$)
 and any number of ordinary particles (with $R = +1$).
 Certainly,
 such decay can not happen with the ``lightest'' SUSY particle
 since no SUSY particle can be lighter than the lightest one.
 Hence,
 in models with strict $R$-parity conservation,
 the LSP must be absolutely stable and
 is then the best candidate for Dark Matter \cite{SUSYDM96}.

 In contrast,
 if $R$-parity is broken,
 there is no special selection rule to
 prevent the decays of the supersymmetric particles
 in the spectrum with masses of order of a few GeV or larger \cite{SUSYDM96}.
 In particular,
 there were no natural candidate for Cold Dark Matter \cite{SUSYDM96}.
 Theories with broken $R$-parity also
 possess baryon- and lepton-number violating interactions
 with strengths controlled by the scale of $R$-parity violation \cite{SUSYDM96}.
\subsection{Weakly Interacting Massive Particles (WIMPs)}
\label{WIMPs}
 ~~~$\,$
 Weakly Interacting Massive Particles (WIMPs) $\chi$
 arise e.g., in supersymmetric extensions
 of the Standard Model of electroweak interactions
 and are the leading non-baryonic candidates for Cold Dark Matter \cite{SUSYDM96}.
 They are {\em stable} particles
 and {\em interact with ordinary matter only via weak interactions}.
 Typically their masses have been presumed to be between 10 GeV and a few TeV
 \cite{SUSYDM96}, \cite{Drees06}.
 They could include neutralinos,
 sneutrinos,
 heavy fourth-generation Dirac and Majorana neutrinos,
 and non-minimal neutralinos
 (neutralinos in non-minimal supersymmetric models) \cite{SUSYDM96}.

 Relic elementary particles are left over from the Big Bang.
 Stable or long-lived particles with very weak interactions
 can remain in sufficient numbers
 to account for a significant fraction of critical density \cite{Turner99}.
 Very weak interactions are necessary
 for their annihilations to cease before their numbers are too small \cite{Turner99}.

 WIMPs exist in thermal equilibrium and in abundance in the early Universe,
 when the temperature of the Universe $T$ exceeds their masses $m_{\chi}$
 ($T~\amssyasyma{38}~m_{\chi}$) \cite{SUSYDM96}.
 The equilibrium abundance is maintained
 by pair annihilation of WIMPs with their antiparticles $\bar{\chi}$
 into lighter particles $l$
 (quarks and leptons, or even gauge- and Higgs-bosons if their masses are heavy enough)
 and vice versa ($\chi \bar{\chi} \getsto l \bar{l}$) \cite{SUSYDM96}.
 The rate of this reaction is proportional to
 the product of the WIMP number density $n_{\chi}$
 and the WIMP pair annihilation cross section into SM particles $\sigma_A$ times
 the relative velocity between the two WIMPs in their center-of-mass system $v$
 \cite{SUSYDM96}, \cite{Drees06}:
\beq
   \Gamma_{\chi}
 = n_{\chi} \expv{\sigma_A v}
\~,
\label{eqn2010201}
\eeq
 where $\expv{\cdots}$ indicates the thermal averaging.

 As the Universe cools to a temperature less than the masses of WIMPs ($T < m_{\chi}$),
 the equilibrium abundance (number density) of WIMPs drops exponentially
 until the rate for the annihilation reaction ($\chi \bar{\chi} \to l \bar{l}$)
 becomes smaller than the Hubble expansion rate of the Universe
 ($\Gamma_{\chi}~\amssyasyma{46}~H$).
 At this point the interactions which maintain thermal equilibrium ``freeze out''
 and the WIMPs cease to annihilate
 and drop out of thermal equilibrium \cite{Drees06}.
 Hence,
 a relic cosmological abundance ``freezes in'' \cite{SUSYDM96},
 i.e., the density of the co-moving WIMPs remains essentially constant.

 The time evolution of the number density of WIMPs $n_{\chi}(t)$
 can be described by the Boltzmann equation \cite{SUSYDM96}:
\beq
   \Dd{n_{\chi}}{t}+3 H n_{\chi}
 =-\expv{\sigma_A v} \bbrac{n_{\chi}^2-\abrac{n_{\chi}^{\rm eq}}^2}
\~.
\label{eqn2010202}
\eeq
 Here
 $n_{\chi}^{\rm eq}$ is the number density of WIMPs in thermal equilibrium.
 The second term on the left-hand side accounts for the expansion of the Universe,
 the first term in the brackets on the right-hand side
 accounts for the depletion of WIMPs due to their pair-annihilation,
 and the second term arises from creation of WIMPs from the inverse reaction \cite{SUSYDM96}.
 In the absence of number-changing interactions,
 the right-hand side would be zero and we would find $n \propto 1/a^3(t)$ \cite{SUSYDM96},
 where $a(t)$ is the scale factor of the Universe in Eq.(\ref{eqn1020003}).

 Note that
 Eq.(\ref{eqn2010202}) describes both Dirac particles
\footnote{
 $n_{\chi}^2$ on the right-hand side
 should be modified to $n_{\chi} n_{\bar{\chi}}$
 for Dirac particles,
 but often $n_{\chi} = n_{\bar{\chi}}$ has been assumed.}
 as well as Majorana particles
 \cite{SUSYDM96},
 which are particles that themselves are also their antiparticles,
 such as neutralinos ($\chi = \bar{\chi}$) \cite{SUSYDM96}.
 For the case of Majorana particles
 (they are so-called ``self-annihilating''),
 the annihilation rate in Eq.(\ref{eqn2010201}) should be modified to \cite{SUSYDM96}
\beq
   \Gamma_{\chi}
 = \afrac{n_{\chi}}{2} \expv{\sigma_A v} 
\~.
\eeq
 However,
 in each annihilation,
 two particles are removed and the factor of 2 can be canceled.
 For Dirac particles with no particle-antiparticle asymmetry,
 i.e., $n_{\chi} = n_{\bar{\chi}}$,
 Eq.(\ref{eqn2010201}) is true.
 But the total number of particles plus antiparticles must be $2 n_{\chi}$ \cite{SUSYDM96}.
 In the case of Dirac particles with a particle-antiparticle asymmetry,
 the relic abundance is generally that given by the asymmetry \cite{SUSYDM96}.
 For example,
 the relic proton density is fixed by the proton-antiproton asymmetry,
 i.e., the baryon number of the Universe \cite{SUSYDM96}.

 The early Universe is radiation dominated and
 the Hubble expansion rate falls with temperature as \cite{SUSYDM96}
\beq
   H(T)
 = 1.66 \afrac{g_{\ast}^{1/2} \~ T^2}{M_{\rm Pl}}
\~.
\label{eqn2010203}
\eeq
 Here $M_{\rm Pl}$ is the Planck mass
 \cite{RPP06AP}
\beq
        M_{\rm Pl}
 \equiv \sfrac{\hbar c}{G_{\rm N}}
 =      1.2209 \times 10^{19}~{\rm GeV}/c^2
 =      2.1764 \times 10^{-5}~{\rm g}
\~,
\eeq
 and the quantity $g_{\ast}$ is the effective number of relativistic degrees of freedom.
 It is approximately equal to the number of bosonic relativistic degrees of freedom
 plus $\frac{7}{8}$ times the number of fermionic relativistic degrees of freedom
 \cite{SUSYDM96}.
 While,
 for very high temperature ($T~\amssyasyma{38}~m_{\chi}$) \cite{SUSYDM96},
\beq
         n_{\chi}^{\rm eq}
 \propto T^3
\~.
\eeq
 Hence,
 the expansion rate $H(T)$ in Eq.(\ref{eqn2010203})
 decreases less rapidly than the number density of WIMPs.
 This means that,
 at early times,
 the expansion term in Eq.(\ref{eqn2010202}), $3 H n_{\chi}$,
 is negligible compared with the right-hand side,
 and the number density tracks its equilibrium abundance \cite{SUSYDM96}.

 However,
 at later times or at low temperatures ($T~\amssyasyma{46}~m_{\chi}$),
 the right-hand side in Eq.(\ref{eqn2010202}) becomes negligible compared with the expansion term,
 and the co-moving abundance of WIMPs remains unchanged \cite{SUSYDM96}.
 It can be found that \cite{SUSYDM96}
\beq
   n_{\chi}^{\rm eq}
 = g \afrac{m_{\chi} T}{2 \pi}^{3/2} e^{-m_{\chi}/T}
\~,
\eeq
 where $g$ is the number of internal degrees of freedom of the WIMPs
 and thus their density is Boltzmann suppressed \cite{SUSYDM96}.
 If the expansion of the Universe were so slow that thermal equilibrium was always maintained,
 the number of WIMPs today would be exponentially suppressed
 (essentially, there would be no WIMPs) \cite{SUSYDM96}.
 The temperature $T_F$ at which the WIMPs freeze out
 is given by $\Gamma_{\chi}(T_F) = H(T_F)$
 \cite{SUSYDM96}.
 Using typical weak-scale numbers,
 the freeze-out temperature turns out to be \cite{SUSYDM96}
\beq
        T_F
 \simeq \frac{\mchi}{20}
\~.
\eeq
 There is a small logarithmic dependence on the mass and annihilation cross section here
 \cite{SUSYDM96}.
 As stated above,
 after freeze out,
 the abundance of WIMPs per co-moving volume remains constant.

 Finally,
 the present relic density of WIMPs is then approximately given by
 (ignoring logarithmic corrections) \cite{SUSYDM96}, \cite{Drees06}
\beq
        \Omega_{\chi} h^2
 \simeq const. \times \frac{T_0^3}{M_{\rm Pl}^3 \expv{\sigma_A v}}
 \simeq \frac{0.1 c~{\rm pb}}{\expv{\sigma_A v}}
 =      \frac{3 \times 10^{-27}~{\rm cm^3/s}}{\expv{\sigma_A v}}
\~,
\label{eqn2010204}
\eeq
 where $T_0$ is the current CMB temperature given in Eq.(\ref{eqn1020102}),
 and $c$ is the speed of light.
 It is inversely proportional to the annihilation cross section of WIMPs.
 Hence,
 as the annihilation cross section is increased,
 the WIMPs stay in equilibrium longer,
 and we are left with a smaller relic abundance \cite{SUSYDM96}.
 The annihilation cross section is generally
 expected to decrease as the WIMP mass is increased,
 so the relic abundance should be also increased \cite{SUSYDM96}.
 Therefore,
 heavier WIMPs should be more likely to
 contribute too much to the mass of the Universe,
 and then be cosmologically inconsistent \cite{SUSYDM96}.
\subsection{Neutralinos}
 ~~~$\,$
 As introduced in Subsec.~\ref{MSSM},
 neutralinos are linear combinations of photino, Z-ino and neutral higgsinos
 (the supersymmetric partners of the photon, $Z^0$ and neutral Higgs bosons,
  see Table \ref{tab2010101}):
\beq
   \Td{\chi}_i^0
 = a_i \Td{\gamma}+b_i \Td{Z}+c_i \Td{h}^0+d_i \Td{H}^0
\~,
   ~~~~~~~~~~~~~~~~~~~~ 
   i
 = 1,~2,~3,~4
.
\eeq
 In most SUSY models,
 the lightest neutralino
 is the lightest supersymmetric particle \cite{Drees06}
 and therefore the best motivated and the most widely studied candidate
 for WIMP Dark Matter,
 but not the unique candidate for LSP (e.g., sneutrinos) \cite{Drees06}.

 There are some theoretical reasons to believe that
 the lightest neutralino should be the LSP.
 First,
 suppose a charged uncolored SUSY particle,
 such as a chargino or a slepton,
 were the LSP.
 The relic number density of such particles can be given as
 roughly $10^{-6} n_{\rm b} M/{\rm GeV}$ \cite{SUSYDM96},
 where $n_{\rm b}$ is the baryon number density and $M$ is the mass of such particles.
 Then they would show up in searches for anomalously heavy protons \cite{SUSYDM96}.
 Null results from such searches rule out such charged particles over a broad mass range
 \cite{SUSYDM96}.
 Moreover,
 grand unified models predict relations between the masses of the SUSY particles .
 In most models the gluino is more massive than the neutralino,
 and the squarks are also heavier than the sleptons \cite{SUSYDM96}.
 In addition,
 some detailed calculations show that
 the lightest neutralino has the desired thermal relic density,
 $\Omega_{\rm DM}$ in Eq.(\ref{eqn1020905}),
 in at least four distinct regions of parameter space \cite{Drees06}.
\subsection{Sneutrinos}
 ~~~$\,$
 Sneutrinos are the spin-0 supersymmetric partner of the neutrinos
 (see Table \ref{tab2010101}).
 There are some reasons to rule out sneutrinos to be good candidate for Dark Matter.
 First,
 in most models,
 there is a slepton with mass similar to, but slightly smaller than, the sneutrino mass
 \cite{SUSYDM96}.
 Meanwhile,
 their masses would have to exceed several hundred GeV for them
 to make good Dark Matter candidates.
 This is uncomfortably heavy for the lightest sparticle \cite{Drees06}.
 On the other hand,
 the annihilation cross sections of sneutrinos are expected to be quite large \cite{Drees06}.
 Hence,
 the negative outcome of various WIMP searches
 rules out ordinary sneutrinos as primary component of the Dark Matter halo of our Galaxy
 \cite{Drees06}.
 However,
 in models with gauge-mediated SUSY breaking
 the lightest messenger sneutrino could be a good candidate \cite{Drees06}.
\subsection{Heavy fourth-generation Dirac and Majorana neutrinos}
\label{forthgenerationnu}
 ~~~$\,$
 They are the first proposed WIMP candidates for CDM \cite{SUSYDM96}.
 They are heavy, but stable particles and
 assumed to have (weak) interactions with ordinary matter
 though Standard Model coupling to the $Z^0$ boson \cite{SUSYDM96}.
 Such neutrinos could annihilate into light fermions
 via s-channel exchange of a $Z^0$ boson.
 For $m \ll M_Z$
 the cross section is proportional to the square of their mass \cite{SUSYDM96}.
 Because
 their interactions are fixed by gauge symmetry,
 the only adjustable scale is then their masses \cite{SUSYDM96}.
 The cosmological abundances of the heavy Dirac and Majorana neutrinos
 have been given as \cite{SUSYDM96}
\beq
        \Omega_{\nu,{\rm D}} h^2
 \simeq \afrac{m_{\nu,{\rm D}}}{2~{\rm GeV}}^{-2}
\~,
\eeq
 and
\beq
        \Omega_{\nu,{\rm M}} h^2
 \simeq \afrac{m_{\nu,{\rm M}}}{5~{\rm GeV}}^{-2}
\~,
\eeq
 for neutrino masses in the range
 ${\cal O}(1~{\rm GeV})~\amssyasyma{46}~m_{\nu} \ll m_{Z} = 91.19~{\rm GeV}$.

 However,
 there is no obvious reason
 why such massive neutrinos should not be allowed to decay \cite{Drees06}.
 Moreover,
 an SU(2) doublet neutrino will have a too small relic density
 if its mass exceeds $M_{\rm z}/2$,
 as required by LEP data \cite{Drees06}.

 On the other hand,
 for such neutrinos with masses greater than the electroweak gauge-boson masses,
 annihilations into gauge- and/or Higgs-boson pairs could occur.
 However,
 the cross section would not decrease as the neutrino mass increases \cite{SUSYDM96},
 so the relic abundance of neutrinos with masses of the order of 100 GeV
 remains too small to account for the Dark Matter
 in the Galactic halo \cite{SUSYDM96}.

 Dirac neutrinos interact with nuclei through
 a coherent vector interaction \cite{SUSYDM96}
 (some details about the vector interaction with nuclei
  will be given in Subsec.~\ref{SIcrosssection}).
 Thus the Dirac-neutrino-nucleus cross section
 is expected to be quite substantial \cite{SUSYDM96},
 and this would lead to a significant event rate in a direct detection experiment.
 Null results from such experiments have ruled out
 Dirac neutrinos with masses in the range
 12 GeV \amssyasyma{46} $m_{\nu,{\rm D}}$ \amssyasyma{46} 1.4 TeV
 as the primary component of the Dark Matter halo \cite{SUSYDM96}.

 Meanwhile,
 Majorana neutrinos interact with nuclei
 only via an axial-vector interaction \cite{SUSYDM96}
 (some details about the axial-vector interaction with nuclei
  will be given in Subsec.~\ref{SDcrosssection}),
 and are therefore difficult to detect directly.
 However,
 such neutrinos would be captured in the Sun by scattering from hydrogen therein
 and their pair annihilations in the Sun would produce energetic neutrinos from the Sun
 \cite{SUSYDM96}.
 Null results from searches for energetic neutrinos at e.g., Kamiokande
 have also ruled out Majorana neutrinos
 with mass less than a few hundred GeV \cite{SUSYDM96}.
\subsection{Axions}
 ~~~$\,$
 Axion $a$ is also one of the leading candidates for CDM.
 Axions have been introduced  by Peccei and Quinn to solve
 the strong CP (charge-conjugation and parity) violation problem of QCD.
 They are pseudo Nambu-Goldstone bosons
 associated with the spontaneous breaking of a new global Peccei-Quinn (PQ) U(1) symmetry
 at scale $f_a$ \cite{Drees06}.

 The present relic density of the axions can be given as \cite{Drees06}
\beq
   \Omega_a
 = \kappa_a \afrac{f_a}{10^{12}~{\rm GeV}}^{1.175} \theta_a^2
\~,
\eeq
 where $\kappa_a$ is a numerical factor lying roughly between 0.5 and a few,
 $\theta_a$ is a ``misalignment angle'' which parameterizes the axion field.
 Suppose $\theta_a \sim {\cal O}(1)$,
 axions will have the required cosmological energy density in Eq.(\ref{eqn1020905})
 to be Dark Matter,
 if $f_a \sim {\cal O}(10^{11}~{\rm GeV})$.
 It is pretty comfortably above the laboratory and astrophysical constraints
 and this would correspond to an axion mass $m_a \sim 10^{-4}$ eV \cite{Drees06}.

 Axions could be detected by looking for
 their conversion to microwave photons, $a \to \gamma$,
 in a strong magnetic field \cite{Drees06}.
 Such a conversion could proceed through the loop-induced $a \gamma \gamma$ coupling,
 whose strength $g_{a \gamma \gamma}$ is thus an important parameter of axion models
 \cite{Drees06}.
 Moreover,
 the conversion rate can be enhanced in a high quality cavity on resonance and,
 due to the equation $m_a c^2 = \hbar \omega_{\rm res}$,
 varying this resonance frequency can give a range of $m_a$,
 or, equivalently, $f_a$ \cite{Drees06}.
\subsection{Other possible SUSY candidates}
\label{otherSUSY}
 ~~~$\,$
 Besides the neutralinos and the sneutrinos,
 some other supersymmetric particles are also (theoretically) possible
 to be candidates for Dark Matter.
\footnote{
 Besides the different supersymmetric extensions of the Standard Model,
 there are also some theories based on ``flat universal extra dimensions (UED)''.
 The most studied candidate for CDM in these extra-dimension models
 is the first Kaluza-Klein (KK) mode of the hypercharge gauge boson
 (the lightest KK particle, LKP)
 $\gamma_{(1)}$.}

 Axino is the spin-$\frac{1}{2}$ superpartner of the axion.
 It may be the LSP or the next-lightest supersymmetric particle (NLSP)
 and may decay to the LSP
 \cite{SUSYDM96}.
 When the axino is the lightest supersymmetric particle and has a mass of a few keV,
 it can be a good candidate for {\em Warm Dark Matter} (WDM) \cite{SUSYDM96}.

 While,
 gravitino, the spin-$\frac{3}{2}$ superpartner of graviton,
 is also a possible candidate for Dark Matter.
 The gravitinos will decouple at temperatures of order of the Planck scale
 ($E_{\rm Pl} \simeq 10^{19}~{\rm GeV}$ \cite{SUSYDM96}).
 Thus the physics of the gravitinos must be considered
 at energies and temperatures right up to this scale \cite{SUSYDM96}.
 In addition,
 if gravitinos behave as standard stable thermal relics
 with an abundance determined by consideration of their decoupling,
 the mass of gravitinos should be less than a few keV \cite{SUSYDM96}.
 However,
 in some models with gravitinos as LSP,
 the NLSP should decay to a gravitino plus ordinary particles \cite{SUSYDM96}.
 Since the coupling to gravitinos is so weak,
 this NLSP will be very long-lived
 and the products of its decay will contain $\gamma$-ray with high energies \cite{SUSYDM96}.
\section{Hot Dark Matter (HDM)}
\label{HDM}
 ~~~$\,$
 ``Hot'' has been used here to indicate that
 such Dark Matter particles moved {\em relativistically}
 in the early Universe.
 Due to their {\em fast velocities},
 they would cover great distances and then form some very large scale structures.
 This means that
 the Hot Dark Matter forms the structure of our Universe from the top down,
 with superclusters fragmenting into clusters and galaxies \cite{Turner99}.
 It is in contrast to the observational evidence which indicates that
 the structure of our Universe has been formed {\em from the bottom up}
 by merging dust to galactic scale structures \cite{Turner99}.

 However,
 there are still some suggestions in which part of Dark Matter is hot and the rest is cold.
 In these models the bulk of the Dark Matter (especially in galactic halos) is still cold.
%
%
\subsection{Massive neutrinos}
 ~~~$\,$
 The leading candidates for Hot Dark Matter are the massive neutrinos.
 As shown in Subsec.~\ref{Omegailist},
 WMAP results combined with other astronomical measurements
 lead to a contribution for light (but massive) neutrino species \cite{RPP06AP}:
\cheqnref{eqn1020911}
\beq
   \Omega_{\nu}
 < 0.014
\~.
\eeq
\cheqnCN{-1}
 They could include the electron-, muon-, and tauon-neutrinos in the Standard Model
 with non-zero masses
\footnote{
 At present
 we know from $\nu$ oscillations that
 at least two of these three SM neutrinos
 have small, but non-vanishing masses.}
 as well as the forth-generation Dirac and Majorana neutrinos
 (described in Subsec.~\ref{forthgenerationnu})
 with extremely light masses.
\section{Dark baryons}
 ~~~$\,$
 As mentioned above,
 some CDM particles, e.g., neutralinos and axions,
 could form galactic scale structures such as galaxies and clusters of galaxies,
 while,
 some HDM particles, e.g., massive neutrinos,
 could form larger structures of the Universe.
 This means that
 {\em on different scales Dark Matter might consist of different materials}
 \cite{SUSYDM96}.
\footnote{
 However,
 some recent researches indicate that
 there should be only one species of Dark Matter in the Universe.}

 Moreover,
 in this chapter I presented some theoretically predicted (SUSY) particles
 as candidates for Dark Matter.
 However,
 {\em until now there is no direct accelerator evidence for the existence of supersymmetry}
 \cite{SUSYDM96}.
 Actually,
 {\em it is not absolutely certain that Dark Matter is neither baryons nor neutrinos}
 \cite{SUSYDM96}.
 There are also some conservative cosmological models
 which describe the Universe only in terms of baryons and perhaps neutrinos \cite{SUSYDM96}.

 On the other hand,
 as shown in Subsec.~\ref{Omegailist},
 the baryonic matter density in the Universe is
\cheqnrefp{eqn1020904}
\beq
       \Omega_{\rm b}
 \simeq 0.042
\~,
\label{eqn1020904p}
\eeq
 but only around 25\% of the baryonic matter are luminous: 
\cheqnref{eqn1020906}
\beq
        \Omega_{\rm lum}
 \simeq 0.01
\~.
\eeq
\cheqnCN{-2}
 Although,
 as mentioned in Subsec.~\ref{gastotalmass},
 most of the baryons in the clusters of galaxies
 reside not in the galaxies themselves,
 but in form of hot intercluster, x-ray emitting gas \cite{Turner99},
 such hot gas in the clusters of galaxies
 only accounts for around 10\% of the baryons in the Universe \cite{Turner99}.
 Hence,
 there should (must) be some {\em baryonic Dark Matter}.
\footnote{
 However,
 some recent results of the measurement of
 the opacity of the Lyman-$\alpha$ forest toward high-redshift quasars
 indicate that
 there are probably enough baryons at $z \ge 3$,
 but it is not clear where they are now.}

 Two most promising possibilities for such dark baryons
 are diffuse hot gas and dark stars,
 which include white dwarfs, neutron stars, black holes,
 or objects with masses around or below the hydrogen-burning limit \cite{Turner99}.
\subsection{Massive astrophysical compact halo objects (MACHOs)}
 ~~~$\,$
 Massive astrophysical compact halo objects include,
 for example,
 brown dwarfs
 which are balls of hydrogen and helium with masses $< 0.08 M_{\odot}$
 and therefore never begin nuclear fusion of hydrogen \cite{SUSYDM96}
 (but they do burn deuterium),
 jupiters
 which are similar to brown dwarfs but have masses $\sim 0.001 M_{\odot}$ \cite{SUSYDM96}
 and do not burn anything,
 and white dwarfs \cite{SUSYDM96}.
 Actually,
 objects with masses around or below the hydrogen-burning limit
 could be baryonic Dark Matter \cite{Turner99}.

 Meanwhile,
 neutron stars and stellar black-hole remnants are also candidates for baryonic Dark Matter
 \cite{SUSYDM96}.
 Black holes with masses $\sim 100 M_{\odot}$ could be remnants of an early generation of stars
 which were massive enough
 so that not many heavy elements were dispersed when they underwent their supernova explosions
 \cite{SUSYDM96}.
 Primordial black holes which formed before the era of Big Bang
 could be counted for non-baryonic Dark Matter rather than baryonic one \cite{Drees06}.
 However,
 such an early creation of a large number of black holes is possible
 only in certain somewhat contrived cosmological models \cite{Drees06}.

 MACHOs might represent a large part of the galactic Dark Matter and
 could be detected through the microlensing effect \cite{Drees06}.
 The MACHO, EROS, OGLE collaborations have performed programs of observation of such objects
 by monitoring the luminosity of millions of stars in the Large and Small Magellanic Clouds
 \cite{Drees06}.
 They concluded that
 MACHOs contribute \amssyasyma{46} 40\% (MACHO) or even \amssyasyma{46} 20\% (EROS)
 to the mass of the galactic halo \cite{Drees06}.

%% file: Doktorarbeit-Ch3.tex
\chapter{Direct Detection of WIMPs}
 ~~~$\,$
 By definition,
 Dark Matter could neither emit nor absorb electromagnetic radiation.
 However,
 as described in Subsec.~\ref{WIMPs},
 WIMPs would have annihilated to some ordinary matter,
 e.g., quarks and leptons,
 in the early Universe.
 Otherwise,
 they would have unacceptable large abundance today.
 According to the crossing symmetry,
 the amplitude for WIMP annihilation to, for example, quarks is related to
 the amplitude for elastic scattering of WIMPs from quarks \cite{SUSYDM96}.
 Therefore,
 {\em WIMPs should have some small, but non-zero couplings to ordinary matter}.

 Due to this coupling to nucleus (through the coupling to quarks),
 WIMPs could scatter elastically from target nuclei of the detector material
 and produce nuclear recoils which deposit energy in the detector.
 Hence,
 one of the most promising methods of detecting Galactic Dark Matter is
 the direct detection of WIMPs \cite{Goodman85}-\cite{Smith96}.
 Note that,
 although the lightest neutralino is the leading candidate for Dark Matter,
 such WIMP direct searches are not specialized to detect the neutralino
 but any particle with similar generic properties,
 e.g., a mass between a few GeV and a few TeV and weakly interacting with ordinary matter
 \cite{Ramachers02}.
\footnote{
 The lightest Kaluza-Klein particle arising in the extra-dimension models
 (mentioned as footnote in Subsec.~\ref{otherSUSY})
 can also scatter elastically from the detector nuclei
 through KK-quark $q_{(1)}$ and Higgs exchange \cite{Akerib06}.
 Thus such particles could also be detected from direct detection experiments.
 A brief description about the interaction between $q_{(1)}$ and Higgs
 and the analysis using recent experimental results
 can be found in Ref.~\cite{Akerib06}.}
\section{Elastic WIMP-nucleus scattering}
\label{elasticscattering}
 ~~~$\,$
 Using the standard assumption of the WIMP density near the Earth
\cheqnref{eqn1030104}
\beq
         \rho_0
 \approx 0.3~{\rm GeV}/c^2/{\rm cm^3}
\~,
\eeq
\cheqnCN{-1}
 and assuming a WIMP mass
\beq 
         \mchi
 \approx 100~{\rm GeV}/c^2
\~,
\label{eqn3010001}
\eeq
 the number density of WIMP can be found as
\beq
   n
 = \frac{\rho_0}{\mchi}
 \approx 3 \times 10^{-3}~{\rm cm}^{-3}
\~.
\label{eqn3010002}
\eeq
 Meanwhile,
 by the assumption that
 the halo WIMPs are gravitationally bound to the Galaxy and its halo,
 the average velocity of WIMP wind
 is then approximately equal to the stellar velocity in the Solar neighborhood:
\footnote{
 Note that $\expv{v} \equiv \expv{|\bf v|}$ is ${\cal O}(v_0)$
 even though $\expv{\bf v} = 0$
 (see footnote on p.\pageref{expvv2}).}
\beq
         \expv{v}
 \approx 250~{\rm km/s}
\~.
\label{eqn3010003}
\eeq
 Therefore,
 the WIMP flux is $\sim 10^5$ WIMPs
 per square centimeter of the Earth's surface per second.

 However,
 the very low cross section of WIMPs on ordinary material
 makes the elastic WIMP-nucleus scattering very rare.
 In typical SUSY models with neutralino WIMPs,
 WIMP-nucleus cross section is about $10^{-6} \sim 10^{-4}$ pb
 ($10^{-42} \sim 10^{-40}~{\rm cm^2}$)
\footnote{
 1 barn = $10^{-24}~{\rm cm^2}$,
 1 pb   = $10^{-36}~{\rm cm^2}$}
 and the expected event rate is then {\em at most 1 event/kg/day} \cite{Drees06},
 in some models it is even {\em less than 1 event/ton/yr} \cite{Sanglard05}.

 With expected WIMP mass in the range 10 GeV$/c^2$ to 10 TeV$/c^2$ \cite{Drees06},
 typical nuclear recoil energies are {\em of order of 1 to 100 keV}.
 However,
 as we can see in Fig.~\ref{fig4020101} in Subsec.~\ref{dRdQkn},
 the event rate drops approximately exponentially
 and most events should be with energies less than 40 keV
 (a simple theoretical estimate will be given in Subsec.~\ref{Qestimate}).

 On the other hand,
 in the energy range from a few to a couple hundred keV,
 typical background noise due to cosmic rays and ambient radioactivity is much larger.
 Thus a underground laboratory and extensive shielding around the detector
 to protect against cosmic-ray induced backgrounds,
 and selection of extremely radiopure materials are necessary and important \cite{Drees06}
 (more details about background and its discrimination
  will be given in Sec.~\ref{background}).

 The event rate of elastic WIMP-nucleus scattering depends on various parameters
 coming from astrophysics, particle physics and nuclear physics:
 the WIMP density near the Solar system $\rho_0$,
 the WIMP-nucleus cross section,
 the WIMP mass $\mchi$,
 and the velocity distribution of the incident WIMPs $f(v)$ in the Galactic halo near the Earth.
 However,
 by some standard assumptions about the halo model,
 e.g., the WIMP density profiles (have been presented in Sec.~\ref{halomodels})
 and velocity distributions (will be discussed in Subsecs.~\ref{fGau} and \ref{fsh}),
 the expected event rate mainly depends on two unknowns:
 the mass of the incident WIMPs and the WIMP-nucleus cross section.
 Hence the experimental observable is usually expressed
 as a contour in the WIMP mass-cross section plane
 (e.g., Figs.~\ref{fig3080601} and \ref{fig3100001}),
 although it is basically the scattering rate and is a function of energy
 \cite{Drees06}.
\subsection{Rate for elastic WIMP-nucleus scattering}
\label{differentialrate}
 ~~~$\,$
 The direct detection experiment measures
 the number of events per unit time per unit mass of detector material
 as a function of the energy deposited in the detector $Q$.
 Qualitatively,
 the event rate of direct detection, $R$, can be simply expressed as \cite{SUSYDM96}
\beq
         R
 \approx \frac{n \~ \expv{v} \~ \sigma}{\mN}
\~,
\label{eqn3010101}
\eeq
 where $\expv{v}$ is the average velocity of the incident WIMPs relative to the Earth frame
 (i.e., to the target),
 $\sigma$ is the WIMP-nucleus cross section,
 and $\mN$ is the mass of the target nucleus.
 Here we multiply the factor $1/\mN$
 to get the number of target nuclei per unit mass of the detector material.

 More accurately,
 one should take into account the fact that
 the WIMPs move in the halo with velocities
 determined by their velocity distribution function $f(v)$,
 and that the differential cross section depends on $f(v)$
 through an elastic nuclear form factor $F(q)$
 \cite{SUSYDM96}:
\beq
   d \sigma
 = \frac{1}{v^2} \afrac{\sigma_0}{4 \mr^2} F^2(q) \~ dq^2
\~.
\label{eqn3010102}
\eeq
 Here $\sigma_0$ is the total cross section ignoring the form factor suppression,
\beq
   \mr
 = \frac{\mchi \mN}{\mchi+\mN}
\label{eqn3010103}
\eeq
 is the reduced mass,
 $q$ is the transferred 3-momentum:
\beq
   q
 = \sqrt{2 \mN Q}
\~.
\label{eqn3010104}
\eeq
 Therefore,
 in general,
 the differential scattering event rate (per unit detector mass) should be written as
 \cite{SUSYDM96}
\beqn
    dR
 \= \frac{\rho_0}{\mchi \mN} \int v f_1(v) \~ d\sigma \~ dv
    \non\\
 \= \afrac{\rho_0 \sigma_0}{2 \mchi \mr^2} F^2(Q) \int \bfrac{f_1(v)}{v} \~ dv \~ dQ
\~,
\label{eqn3010105}
\eeqn
 where $f_1(v)$ is the one-dimensional velocity distribution function of WIMPs
 impinging on the detector,
 $v$ is the absolute value of the WIMP velocity in the Earth rest frame,
 and we have to integrate over all possible incoming velocities.
 By means of classical mechanics,
 the transferred momentum $q$ can be expressed as
\beq
   q
 = 2 \bbrac{\mN \afrac{\mchi v}{\mchi+\mN}} \sin{\T\afrac{\theta_{\rm CM}}{2}}
 = 2 \mr v \sfrac{1-\cos\theta_{\rm CM}}{2}
\~,
\eeq
 where $\theta_{\rm CM}$ is the scattering angle in the center-of-momentum frame.
 Since
\beqN
     0
 \le 1-\cos\theta_{\rm CM}
 \le 2
\~,
\eeqN
 for a given deposited energy $Q$, we have
\beqN
   Q
 = \frac{\abrac{2 \mr \vmin}^2}{2 \mN}
 = \frac{2 \mr^2 \vmin^2}{\mN}
\~,
\eeqN
 i.e.,
 the minimal incoming velocity of incident WIMPs
 that can deposit the energy $Q$ in the detector can be expressed as
\beq
   \vmin(Q)
 = \sfrac{\mN}{2 \mr^2} \sqrt{Q}
 = \alpha \sqrt{Q}
\~,
\label{eqn3010106}
\eeq
 where I have defined
\beq
        \alpha
 \equiv \sfrac{\mN}{2 \mr^2}
\~.
\label{eqn3010107}
\eeq
 Then the differential event rate for elastic WIMP-nucleus scattering, Eq.(\ref{eqn3010105}),
 can be rewritten as
\beq
   \dRdQ
 = \calA \FQ \intvmin \bfrac{f_1(v)}{v} dv
\~,
\label{eqn3010108}
\eeq
 where the constant coefficient $\calA$ is defined as
\beq
        \calA
 \equiv \frac{\rho_0 \sigma_0}{2 \mchi \mr^2}
\~.
\label{eqn3010109}
\eeq
 Note that,
 first,
 $\alpha$ defined in Eq.(\ref{eqn3010107}) depends only on the WIMP mass $\mchi$
 (and the mass of the target material, $\mN$, which we can choose).
 The two as yet unknown parameters,
 i.e., the WIMP density $\rho_0$ and the total WIMP-nucleus cross section $\sigma_0$,
 have been collected in the coefficient $\calA$ defined in Eq.(\ref{eqn3010109}).
 Second,
 I assumed here that the detector essentially only consists of nuclei of a single isotope.
 If the detector contains several different nuclei
 (e.g., NaI as in the DAMA detector \cite{DAMA}),
 the right-hand side of Eq.(\ref{eqn3010108}) has to be replaced by a sum of terms,
 each term describing the contribution of one isotope.
 For simplicity,
 in the remainder of this work I will focus on mono-isotopic detectors.

 Finally,
 the total event rate per unit time per unit mass of detector material can be expressed as
\beq
   R
 = \int_{\Qthre}^{\infty} \adRdQ dQ
\~,
\label{eqn3010110}
\eeq
 where $\Qthre$ is the threshold energy of the detector.
\subsection{Nuclear form factor (for spin-independent coupling)}
\label{formfactor}
 ~~~$\,$
 Here I present two most commonly used parameterizations
 of the squared nuclear form factor $\FQ$ in Eq.(\ref{eqn3010108})
 for spin-independent coupling,
 which usually dominates the event rate
 (more details about WIMP-nucleus couplings
  will be given in Sec.~\ref{targetmaterial}).
 Moreover,
 the form factors for spin-dependent coupling are still only poorly understood.

 The simplest form factor is the exponential one,
 first introduced by Ahlen {\it et al.} \cite{Ahlen87} and Freese {\it et al.} \cite{Freese88}:
\beq
   F_{\rm ex}^2(Q)
 = e^{-Q/Q_0}
\~,
\label{eqn3010201}
\eeq
 where $Q$ is the recoil energy transferred from the incident WIMP to the target nucleus,
\cheqnCa
\beq
   Q_0
 = \frac{1.5}{\mN R_0^2}
\eeq
 is the nuclear coherence energy and
\cheqnCb
\beq
   R_0
 = \bbrac{0.3+0.91 \afrac{\mN}{\rm GeV}^{1/3}}~{\rm fm}
\eeq
\cheqnC
 is the radius of the nucleus.
 The exponential form factor implies that
 the radial density profile of the nucleus has a Gaussian form.
 This Gaussian density profile is simple,
 but not very realistic.
 Engel has therefore suggested a more accurate form factor \cite{Engel91},
 inspired by the Woods-Saxon nuclear density profile,
\beq
   F_{\rm WS}^2(Q)
 = \bfrac{3 j_1(q R_1)}{q R_1}^2 e^{-(q s)^2}
\~.
\label{eqn3010202}
\eeq
 Here $j_1(x)$ is a spherical Bessel function,
 $q$ is the transferred 3-momentum given in Eq.(\ref{eqn3010104}), and
\cheqnCa
\beq
   R_1
 = \sqrt{R_A^2-5 s^2}
\eeq
 with
\cheqnCb
\beq
        R_A
 \simeq 1.2 \~ A^{1/3}~{\rm fm}
\~,
        ~~~~~~~~~~~~~~~~ 
        s
 \simeq 1~{\rm fm}
\~,
\eeq
\cheqnC
 where $A$ is the atomic mass number of the nucleus.
\subsection{Simple isothermal Maxwellian halo}
\label{fGau}
 ~~~$\,$
 For the simplest halo model presented in Subsec.~\ref{ishalo},
 the canonical isothermal spherical halo,
 with the assumption that
 the WIMPs trapped in the galactic field have attained thermal equilibrium
 with a Maxwellian velocity distribution \cite{Gascon05},
 the velocity distribution function is given by \cite{SUSYDM96}
\beq
   f_\Gau(v)
 = \afrac{1}{\pi^{3/2} v_0^3} e^{-v^2/v_0^2}
\~,
\eeq
 where $v_0$ is the orbital velocity of the Sun in the Galactic frame:
\cheqnrefp{eqn1030102}
\beq
        v_0
 \simeq 220~{\rm km/s}
\~,
\label{eqn1030102p}
\eeq
\cheqnCN{-1}
 which characterizes the velocity of all virialized objects in the Solar vicinity.
 Then,
 since
\beqN
   d^3 v
 = v^2 dv \~ d\Omega
 = 4\pi v^2 dv
\~,
\eeqN
 the normalized one-dimensional velocity distribution function has been obtained as
 \cite{SUSYDM96}
\beq
   f_{1,\Gau}(v)
 = \frac{4}{\sqrt{\pi}} \afrac{v^2}{v_0^3} e^{-v^2/v_0^2}
\~.
\label{eqn3010301}
\eeq
 According to Eq.(\ref{eqn3010108}),
 the scattering spectrum of the simplest theoretical velocity distribution
 given in Eq.(\ref{eqn3010301}) can be obtained as
 (a detailed calculation will be given in App.~\ref{dRdQGau})
\beq
   \adRdQ_\Gau
 = \calA \afrac{2}{\sqrt{\pi} v_0} \FQ \~ e^{-\alpha^2 Q/v_0^2}
\~.
\label{eqn3010302}
\eeq
 Meanwhile,
 the mean velocity and velocity dispersion of the halo WIMPs can be obtained as
 (detailed calculations will be given in App.~\ref{vGau}),
 respectively,
\beq
   \expv{v}_\Gau
 = \intz v f_{1,\Gau}(v) \~ dv
 = \afrac{2}{\sqrt{\pi}} v_0
\~,
\label{eqn3010303}
\eeq
 and
\beq
   \expv{v^2}_\Gau
 = \intz v^2 f_{1,\Gau}(v) \~ dv
 = \afrac{3}{2} v_0^2
\~.
\label{eqn3010304}
\eeq
 For light WIMPs,
 the effect due to the form factor introduced in Eq.(\ref{eqn3010102}) can be neglected
 and we can use $\FQ \approx 1$ \cite{SUSYDM96}.
 Then the total event rate in Eq.(\ref{eqn3010110}) can be found directly as
\beq
   R_\Gau(\Qthre)
 = \frac{\rho_0 \sigma_0}{\mchi \mN} \afrac{2 v_0}{\sqrt{\pi}} e^{-\alpha^2 \Qthre/v_0^2}
\~.
\label{eqn3010305}
\eeq
 For the case of $\Qthre = 0$,
 this result can be reduced to
\beq
   R_\Gau(\Qthre = 0)
 = \frac{\rho_0 \sigma_0 \expv{v}_\Gau}{\mchi \mN}
\~,
\label{eqn3010306}
\eeq
 which is exactly the naive estimate in Eq.(\ref{eqn3010101}).
 On the other hand,
 with the exponential form factor $F_{\rm ex}^2(Q)$ given in Eq.(\ref{eqn3010201}),
 one can find that
 (a detailed calculation will be given in App.~\ref{RexGau})
\beq
   R_{\Gau,\rm ex}(\Qthre)
 = \frac{\rho_0 \sigma_0 \expv{v}_\Gau}{\mchi \mN}
   \abigg{\beta^2 \~ e^{-\alpha^2 \Qthre/v_0^2 \beta^2}}
\~,
\label{eqn3010307}
\eeq
 and then
\beq
   R_{\Gau,\rm ex}(\Qthre = 0)
 = \frac{\rho_0 \sigma_0 \expv{v}_\Gau}{\mchi \mN} \cdot \beta^2
\~,
\label{eqn3010308}
\eeq
 where I have defined
\beq
        \beta
 \equiv \abrac{1+\frac{v_0^2}{\alpha^2 Q_0}}^{-1/2}
\~.
\label{eqn3010309}
\eeq
 It can be seen that,
 for the case that the exponential form factor $F_{\rm ex}^2(Q)$ can be neglected,
 or, equivalently, $Q_0 \to \infty$,
 i.e., $\beta \to 1$,
 $R_{\Gau,\rm ex}(\Qthre)$ in Eqs.(\ref{eqn3010307}) and (\ref{eqn3010308})
 will reduce to Eqs.(\ref{eqn3010305}) and (\ref{eqn3010306}).
\section{Annual modulation of the event rate}
\label{annualmodulation}
 ~~~$\,$
\begin{figure}[t]
\begin{center}
\imageswitch{
\begin{picture}(10.7,8.3)
\put(0,0){\framebox(10.7,8.3){}}
\end{picture}}
{\includegraphics[width=10.5cm]{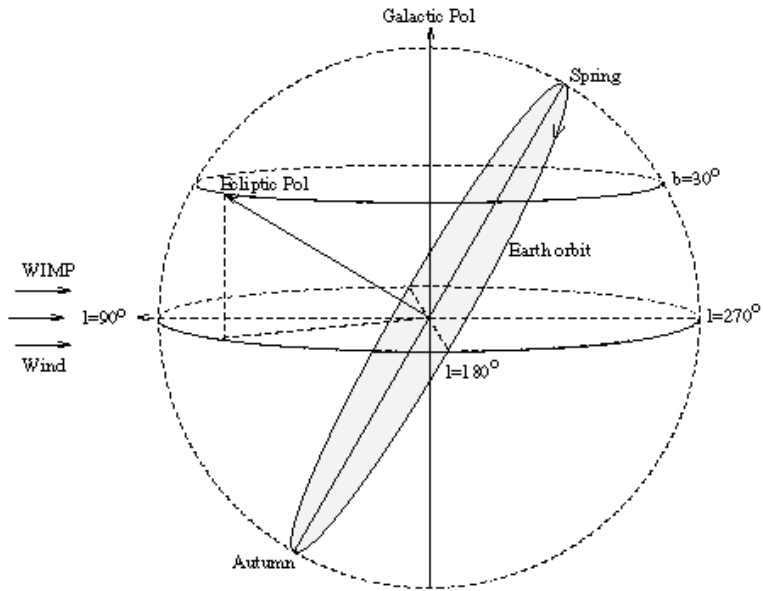}}
\end{center}
\caption{
 The Earth orbit in Galactic coordinates.
 The Sun moves to the left with about $220~{\rm km/s}$,
 inducing a WIMP wind (figure from \cite{Ramachers02}).}
\label{fig3020001}
\end{figure}
 The one-dimensional velocity distribution function $f_{1,\Gau}(v)$
 given in Eq.(\ref{eqn3010301})
 has been considered in the Galactic rest frame.
 More realistically,
 the orbital motion of the Solar system in the Galaxy
 as well as the motion of the Earth around the Sun
 must be considered \cite{Drukier86}-\cite{Spergel88}.
 As shown in Fig.~\ref{fig3020001},
 since the speed of the Earth adds to or subtracts from the speed of the Sun,
 the event rate for a given recoil energy (or energy range) in Eq.(\ref{eqn3010108})
 should be a cosinusoidal function with a one-year period
 (see Figs.~\ref{fig3070301} and \ref{fig5010001})
 and a peak around June 2nd \cite{Freese88}.
 The expected amplitude of this annual modulation is around $5\%$.
\footnote{
 The ratio of a theoretically expected amplitude of this annual modulation
 to the time-averaged scattering spectrum
 as function of the recoil energy
 will be given in Fig.~\ref{fig5020101} in Subsec.~\ref{DeltaQt}.
 It can be seen that,
 for recoil energy between 0 and 50 keV,
 the modulated amplitude is around $-4\% \sim 5\%$.
 Some detailed discussions about the annual modulation of event rate
 will be given in Secs.~\ref{dRdQt} and \ref{DeltaQt}.}

 Originally,
 such an annual modulation was only expected for the WIMP signal,
 not for the background.
 Thus this effect might serve a method to distinguish
 the WIMP signal from the background.
 And actually,
 the DAMA collaboration \cite{DAMA} has claimed that
 they have observed this annual modulation of the event rate
 \cite{Bernabei00}-\cite{Bernabei03b}
 (more details about the DAMA result will be given in Subsec.~\ref{DAMA}).
 However,
 {\em the much larger background might also be subject to modulation}
 \cite{Drees06}.
 For example,
 the dependence of the cosmic muon flux
 on the atmospherical temperature,
 or the dependence of the background neutron flux
 on water in rock and concrete.
 Hence,
 the signal identification should also be performed.
\subsection{Shifted Maxwellian halo}
\label{fsh}
 ~~~$\,$
 When we take into account
 the orbital motion of the Solar system around the Galaxy,
 as well as that of the Earth around the Sun,
 the velocity distribution function in Eq.(\ref{eqn3010301})
 should be modified to \cite{SUSYDM96}
\beq
   f_{1,\sh}(v,v_e)
 = \frac{1}{\sqrt{\pi}} \afrac{v}{v_e v_0} \bbigg{e^{-(v-v_e)^2/v_0^2}-e^{-(v+v_e)^2/v_0^2}}
\label{eqn3020101}
\eeq
 with
\beq
   v_e(t)
 = v_0 \bbrac{1.05+0.07 \cos\afrac{2 \pi (t-t_p)}{1~{\rm yr}}}
\~,
\label{eqn3020102}
\eeq
 where $t_p \simeq$ June 2nd is the date
 on which the velocity of the Earth relative to the WIMP halo is maximal \cite{Freese88}.
 Eq.(\ref{eqn3020102}) includes the effect of the rotation of the Earth around the Sun
 (second term),
 but does not allow for the possibility that the halo itself might rotate
 (some discussions about such bulk rotation
  has been given in Subsec.~\ref{buklrotation}).

 Substituting the shifted Maxwellian velocity distribution function in Eq.(\ref{eqn3020101})
 into Eq.(\ref{eqn3010108}),
 the theoretically expected scattering spectrum can be obtained as
 (a detailed calculation will be given in App.~\ref{dRdQsh})
\beq
   \adRdQ_\sh
 = \calA \afrac{1}{2 v_e} \FQ
   \bbigg{\erf{\T\afrac{\alpha \sqrt{Q}+v_e}{v_0}}-\erf{\T\afrac{\alpha \sqrt{Q}-v_e}{v_0}}}
\~.
\label{eqn3020103}
\eeq
 Here $\erf(x)$ is the error function, defined as
\beqN
   \erf(x)
 = \frac{2}{\sqrt{\pi}} \int_0^{x} e^{-t^2} dt
\~.
\eeqN
 Meanwhile,
 the mean velocity and velocity dispersion in Eqs.(\ref{eqn3010303}) and (\ref{eqn3010304})
 should be modified to (detailed calculations will be given in App.~\ref{vsh}),
 respectively,
\beq
   \expv{v}_\sh
 = \afrac{v_0}{\sqrt{\pi}} e^{-v_e^2/v_0^2}+\abrac{\frac{v_0^2}{2 v_e}+v_e} \erf\afrac{v_e}{v_0}
\~,
\label{eqn3020104}
\eeq
 and
\beq
   \expv{v^2}_\sh
 = \afrac{3}{2} v_0^2+v_e^2
\~.
\label{eqn3020105}
\eeq
 As what I did in Subsec.~\ref{fGau},
 considering the light-WIMP case and using $\FQ \approx 1$,
 the total event rate for the shifted Maxwellian distribution in Eq.(\ref{eqn3020101})
 can be found as
 (a detailed calculation will be given in App.~\ref{dRdQsh})
\beqn
 \conti R_\sh(\Qthre)
        \non\\
 \=     \frac{\rho_0 \sigma_0}{\mchi \mN} \afrac{v_0^2}{2 v_e}
        \cbrac{ \abrac{\frac{1}{2}-S_{+} S_{-}} \bbigg{\erf(S_{+})-\erf(S_{-})}
               +\frac{1}{\sqrt{\pi}} \abigg{S_{+} e^{-S_{-}^2}-S_{-} e^{-S_{+}^2}}}
\~,
        \non\\
\label{eqn3020106}
\eeqn
 where I have defined
\beq
        S_{\pm}
 \equiv \frac{\alpha \sqrt{\Qthre} \pm v_e}{v_0}
\~.
\label{eqn3020107}
\eeq
 For the case of $\Qthre = 0$,
 $R_\sh(\Qthre)$ in Eq.(\ref{eqn3020106}) can be reduced directly to
\beqn
     R_\sh(\Qthre = 0)
 \= \frac{\rho_0 \sigma_0}{\mchi \mN}
    \bbrac{ \abrac{\frac{v_0^2}{2 v_e}+v_e} \erf\afrac{v_e}{v_0}
           +\afrac{v_0}{\sqrt{\pi}} e^{-v_e^2/v_0^2}}
    \non\\
 \= \frac{\rho_0 \sigma_0 \expv{v}_\sh}{\mchi \mN}
\~,
\label{eqn3020108}
\eeqn
 where I have used
\beqN
   \erf(-x)
 =-\erf(x)
\~.
\eeqN
 Moreover,
 for the case with the exponential form factor $F_{\rm ex}^2(Q)$
 given in Eq.(\ref{eqn3010201}),
 I have
 (a detailed calculation will be given in App.~\ref{Rexsh})
\beqn
        R_{\sh,\rm ex}(\Qthre)
 \=     \frac{\rho_0 \sigma_0}{\mchi \mN} \afrac{v_0^2}{2 v_e} \afrac{\beta^2}{1-\beta^2}
        \non\\
 \conti ~~~~~~ \times 
        \cBiggl{ e^{-\abrac{1-\beta^2} \alpha^2 \Qthre/v_0^2 \beta^2} 
                 \bbigg{\erf(S_{+})-\erf(S_{-})}}
        \non\\
 \conti ~~~~~~~~~~~~~~~~~~~~~~~~~~~ 
        \cBiggr{-\beta e^{-\abrac{1-\beta^2} v_e^2/v_0^2}
                 \bbigg{\erf(T_{+})-\erf(T_{-})}}
\~,
\label{eqn3020109}
\eeqn
 where I have defined
\beq
        T_{\pm}
 \equiv \frac{\alpha \sqrt{\Qthre} \pm \beta^2 v_e}{v_0 \beta}
\~.
\label{eqn3020110}
\eeq
 For the case of $\Qthre = 0$,
 $R_{\sh,\rm ex}(\Qthre)$ in Eq.(\ref{eqn3020109}) can be reduce to
\beqn
 \conti R_{\sh,\rm ex}(\Qthre = 0)
        \non\\
 \=     \frac{\rho_0 \sigma_0}{\mchi \mN} \afrac{v_0^2}{v_e} \afrac{\beta^2}{1-\beta^2}
        \bbrac{ \erf\afrac{v_e}{v_0}
               -\beta e^{-\abrac{1-\beta^2} v_e^2/v_0^2} \~ \erf\afrac{\beta v_e}{v_0}}
\~.
\label{eqn3020111}
\eeqn
 It is not difficult to check that
 $R_{\sh,\rm ex}(\Qthre)$ in Eqs.(\ref{eqn3020109}) and (\ref{eqn3020111})
 can be reduced to Eqs.(\ref{eqn3020106}) and (\ref{eqn3020108})
 when one neglects the form factor $F_{\rm ex}^2(Q)$,
 i.e., let $\beta \to 1$.
 On the other hand,
 as $v_e \ll v_0$,
 one can also prove that
 the results in Eqs.(\ref{eqn3020104}) to (\ref{eqn3020106}), and (\ref{eqn3020108})
 can be reduced to Eqs.(\ref{eqn3010303}) to (\ref{eqn3010306}).
\section{Diurnal modulation of the event rate}
\label{diurnalmodulation}
 ~~~$\,$
 Similar to the annual modulation
 caused by the orbital motion of the Earth around the Sun,
 due to the rotation of the Earth,
 the event rate for a given energy (or energy range)
 should have a diurnal modulation \cite{Ramachers02}, \cite{Jesus04}.

\begin{figure}[p]
\begin{center}
\imageswitch{
\begin{picture}(15,21.5)
\put(0  , 0  ){\framebox(15,22  ){}}
\put(2  ,12  ){\framebox(11, 9.5){Shielding of the detector}}
\put(7  ,11  ){\makebox ( 1, 1  ){(a)}}
\put(3  , 1  ){\framebox( 9, 9  ){Directionality}}
\put(7  , 0  ){\makebox ( 1, 1  ){(b)}}
\end{picture}}
{\includegraphics[width=11cm]{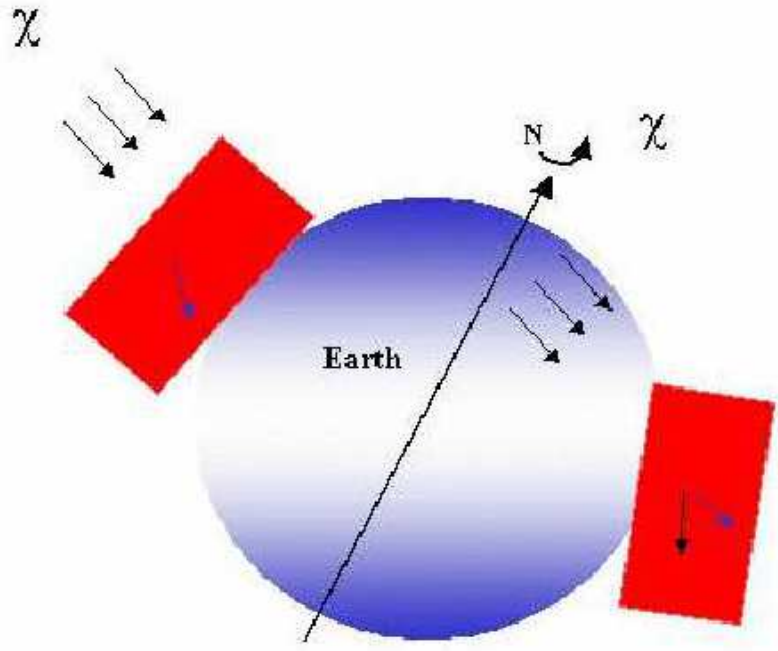} \\ (a) \\ \vspace{1cm}
 \includegraphics[width= 9cm]{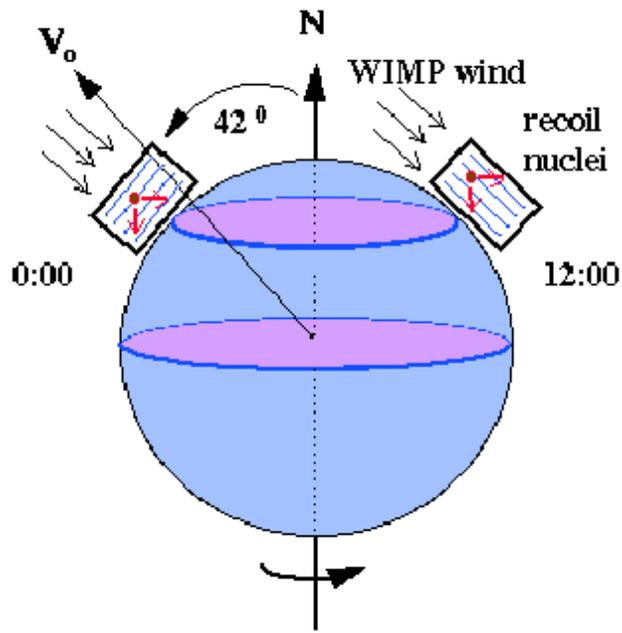} \\ (b)}
\end{center}
\caption{
 Two effects caused by the diurnal modulation:
 (a) shielding of the detector of the incident WIMP flux by the Earth
     (figure from \cite{Jesus04}),
 and
 (b) directionality of the WIMP wind
     (figure from \cite{Ramachers02}).}
\label{fig3030001}
\end{figure}
 There are two different effects caused by this diurnal modulation.
 The first one is the shielding of the detector of the incident WIMP flux by the Earth
 \cite{Jesus04} (illustrated in Figs.~\ref{fig3030001}(a)).
 Some authors claimed that,
 for WIMP masses close to 50 GeV$/c^2$ and under certain assumptions,
 this diurnal modulation due to the shielding of the WIMP flux
 could be larger than the annual modulation \cite{Spergel88}, \cite{Jesus04}.
 However,
 this requires a large WIMP-nucleus cross section
 and recent experimental results have (almost) excluded this possibility.
 Moreover,
 practically,
 it is still impossible for the detectors nowaday
 and should also be very difficult for the next-generation ones
 to get more than a few singles per day to prove this effect
 (more details about the status of the operated experiments and their results
  will be given in Secs.~\ref{cyrogenic} to \ref{SDD}).  

 On the other hand,
 the second effect due to the rotation of the Earth
 is the directionality of the WIMP wind:
 a daily forward/backward asymmetry of the nuclear recoil direction
 (illustrated in Figs.~\ref{fig3030001}(b)).
 A gaseous detector (e.g., DRIFT \cite{UKDMC})
 or anisotropic response scintillators
 should have the ability to measure this recoil direction
 \cite{Ramachers02}, \cite{Drees06}.
\section{Target material dependence}
\label{targetmaterial}
 ~~~$\,$
 The WIMP-nucleus cross section $\sigma_0$ in Eqs.(\ref{eqn3010102}) and (\ref{eqn3010109})
 depends on the nature of the WIMP couplings to nucleons.
 For non-relativistic WIMPs,
 one in general has to distinguish
 spin-independent (SI) and spin-dependent (SD) couplings \cite{Drees06}.
\subsection{Spin-independent (SI) cross section}
\label{SIcrosssection}
 ~~~$\,$
 The total cross section for ``scalar'' coupling can be expressed as \cite{SUSYDM96}
\beq
   \sigma_{\rm 0,scalar}
 = \frac{4 \mr^2}{\pi} \bBig{Z f_{\rm p}+(A-Z) f_{\rm n}}^2
\~.
\label{eqn3040101}
\eeq
 Here $\mr$ is the reduced mass of the WIMP and the target nucleus in Eq.(\ref{eqn3010103}),
 $Z$ is the atomic number,
 i.e., the number of protons,
 $A-Z$ is then the number of neutrons,
 $f_{\rm p}$ and $f_{\rm n}$ are the effective couplings of WIMPs to protons and neutrons,
 respectively.

 Here we have to sum over the couplings to each nucleon before squaring
 because the wavelength associated with the momentum transfer
 is comparable to or larger than the size of the nucleus \cite{Gascon05},
 the so-called ``coherence effect".
 In most cases,
 the couplings to protons and neutrons are approximately equal \cite{SUSYDM96},
\beq
        f_{\rm n}
 \simeq f_{\rm p}
\~.
\eeq
 Then the cross section for scalar interaction in Eq.(\ref{eqn3040101}) can be reduced to
\beq
         \sigma_{\rm 0,scalar}
 \propto A^2
\~.
\label{eqn3040102}
\eeq
 This means that,
 due to the coherence effect with the entire nucleus,
 the cross section for scalar interaction scales approximately as
 the square of the atomic mass of the target nucleus.
 Hence,
 {\em higher mass nuclei, e.g., Ge or Xe,
 are preferred for the search for the scalar interaction} \cite{Drees06}.

 On the other hand,
 WIMPs could also have a ``vector'' coupling to protons and neutrons \cite{SUSYDM96}:
\beq
   \sigma_{\rm 0,vector}
 = \frac{\mr^2}{64 \pi} \bBig{2 Z b_{\rm p}+(A-Z) b_{\rm n}}^2
\~,
\label{eqn3040103}
\eeq
 where $b_{\rm p}$ and $b_{\rm n}$ are the effective couplings to protons and neutrons.
 However,
 for Majorana WIMPs ($\chi = \bar{\chi}$), e.g., the neutralino,
 there is no such vector interaction \cite{Drees06}.
\subsection{Spin-dependent (SD) cross section}
\label{SDcrosssection}
 ~~~$\,$
 WIMPs could also couple to the spin of the target nucleus,
 an ``axial-vector'' (spin-spin) interaction.
 For this spin-spin coupling,
 only unpaired nucleons contribute significantly to the interaction,
 as the spins of the $A$ nucleons in a nucleus are systematically anti-aligned \cite{Gascon05}.
 And it is obvious that
 this spin-dependent interaction exists
 only if the incident WIMPs carry spin \cite{Drees06}.

 The total cross section for spin coupling can be expressed as  \cite{SUSYDM96}
\beq
   \sigma_{\rm 0,axial}
 = \frac{32 \mr^2}{\pi} \bBig{\Lambda^2 J (J+1)}
\~,
\label{eqn3040201}
\eeq
 where $J$ is the total angular momentum of the nucleus
 and $\Lambda$ ($\propto 1/J$) depends on the axial couplings of WIMPs to the quarks.

 Because of the dependence on the nuclear spin factor,
 {\em the useful target nuclei for search for spin interaction
 are $\XA{F}{\it 19}$ and $\XA{I}{\it 127}$} \cite{Drees06}.
\subsection{Comparison of the SI and SD cross sections}
\label{SISDcrosssection}
 ~~~$\,$
 Generally speaking,
 a WIMP could have both scalar and spin-dependent interactions with the nucleus.
 Thus the WIMP-nucleus cross section $\sigma_0$
 in Eqs.(\ref{eqn3010102}) and (\ref{eqn3010109})
 should be the sum of the scalar cross section, $\sigma_{\rm 0,scalar}$,
 in Eq.(\ref{eqn3040101})
 and the spin cross section, $\sigma_{\rm 0,axial}$,
 in Eq.(\ref{eqn3040201}).

 For the scalar interaction,
 an analytic nuclear form factor,
 e.g., the exponential and the Woods-Saxon form factors presented in Subsec.~\ref{formfactor},
 can be used.
 For the spin interaction,
 the form factor will differ from nucleus to nucleus
 and no simple analytic form factor can provide a very good approximation \cite{SUSYDM96}.
 Fortunately,
 {\em for nuclei with $A~\amssyasyma{38}~\it 30$,
 the scalar interaction almost always dominates the spin interaction} \cite{SUSYDM96}.
\subsection{Target mass}
 ~~~$\,$
 The scattering event rate depends
 also on the atomic mass of the target material directly.

 First,
 according to Eq.(\ref{eqn3010106}),
 the smaller $\alpha$ the lower the incoming velocity
 with which the incident WIMPs can deposit energy larger than the threshold energy.
 Meanwhile,
 according to the definitions of $\alpha$ and $\mr$
 in Eqs.(\ref{eqn3010107}) and (\ref{eqn3010103}),
 it can be found that,
 for WIMPs with a given mass and detector with a given threshold energy,
 $\alpha$ will be smallest
 if the mass of the target nucleus $\mN$ is equal to the WIMP mass $\mchi$.

 Second,
 for a given total mass of detector material,
 a larger target mass means also a smaller number density of the nucleus
 which can interact with the incident WIMPs.
 It will certainly reduce the total event number.
\section{Measurement of recoil energy}
\label{Qmeasurement}
\subsection{A simple estimate}
\label{Qestimate}
 ~~~$\,$
 As an example,
 we assume a WIMP mass
\cheqnref{eqn3010001}
\beq
         \mchi
 \approx 100~{\rm GeV}/c^2
\~,
\eeq
 and use the standard theoretical WIMP rms velocity
\cheqnref{eqn1030101}
\beq
        \expv{v^2}^{1/2}
 \simeq 270~{\rm km/s}
\~.
\eeq
\cheqnCN{-2}
 then the average kinetic energy of the incident WIMPs can be estimated as
\beq
         \expv{E_{\chi}}
 \approx 40~{\rm keV}
\~.
\eeq
 On the other hand,
 by means of classical mechanics,
 the recoil energy of the target nucleus due to the elastic scattering can be expressed as
\beq
   Q
 = \bbrac{\frac{4 \mchi \mN}{(\mchi+\mN)^2} \~ \cos^2\theta_{\rm Lab}} E_{\chi}
\~,
\label{eqn3050101}
\eeq
 where $\theta_{\rm Lab}$ is the recoil angle in the laboratory frame.
 This expression shows that the maximum recoil energy is obtained when $\mN = \mchi$.
 This is also why
 this search should be more efficient
 for target material with a mass comparable to the WIMP mass.
\subsection{Induced signals}
\label{inducedsignals}
 ~~~$\,$
 When a WIMP scatters off a nucleus,
 the nucleus will at first obtain a few tens of keV kinetic energy
 and then dissipate this energy in the detector via three main processes:
 the electrons can be stripped by the scattered nucleus
 and an ionized nucleus-electrons system will be produced,
 this electronic activity can emit light,
 and the movement of the recoiling nucleus in the lattice
 can also induce vibrational phonons.
 Moreover,
 the ionization and scintillation energy will convert into phonons
 that will eventually thermalize
 and produce a tiny elevation of the temperature in the detector.

 Hence,
 generally speaking,
 due to the elastic WIMP-nucleus scattering,
 the nuclear recoil can induce three different signals:
 ionization (charges),
 scintillation (light),
 and heat (phonons).
%
%
\subsection{Quenching factor}
\label{quenchingfactor}
 ~~~$\,$
 When a photon with energy between keV and MeV enters a detector,
 it will induce an electron recoil with a range of the order of the $\mu$m
 and transfer most of its energy to the electron.
 However,
 the range of a nuclear recoil is only of the order of the nm
 and the nucleus will lose a substantial part of its energy directly into phonons
 associated with atom vibrations as the nucleus is stopped in the lattice \cite{Gascon05}.

 Hence,
 the quenching factor (the nuclear recoil relative efficiency)
 for the ionization detectors
 has been defined as the ratio of the number of charge carriers
 produced by a nuclear recoil due to the WIMP interaction
 to that produced by an electron recoil with the same kinetic energy
 (energy calibrated with a $\gamma$-source, called ``electron equivalent energy'' or ``eee'').
 Meanwhile,
 for scintillating detectors,
 the quenching factor is defined as
 the ratio between the light produced by a nuclear recoil and by an electron recoil.
 For conventional detectors,
 this factor is usually lower than 0.3 \cite{Jesus04}, \cite{Gascon05}:
 $\sim 0.3$ for Ge or Si,
 $\sim 0.25$ for Na,
 $\sim 0.09$ for I,
 and $\sim 0.2$ for Xe.
 While,
 for cryogenic detectors measuring heat,
 the quenching factor has been measured to be around one
 for recoiling nuclei independently of the energy \cite{Jesus04}.

 Note that,
 due to this quenching factor,
 the measured recoil energies are often quoted practically in $\keVee$
 instead of true recoil energies in unit of keV.
%
%
\subsection{Heat}
\label{heat}
 ~~~$\,$
 Basically a cryogenic detector has been made of a crystal
 with a thermometer glued on it,
 and operated at very low temperature (around 20 mK).

 When the detectors have been cooled to the operating temperature
 in a dilution refrigerator,
 the heat capacity ($\propto T^3$) is so low
 that even a few keV of deposited energy
 raises the temperature of one of the detectors
 by a measurable amount,
 allowing the amount of energy deposited to be determined \cite{SUSYDM96}.

 Moreover,
 a superconducting-normal phase transition due to the elevation of the temperature
 has been used by the CRESST collaboration \cite{CRESST}.
 A thin film of tungsten (W) can be grown on a silicon detector
 and held just below the critical temperature.
 Phonons created by a WIMP-nucleus scattering would heat the superconducting film,
 causing it to go normal,
 and the change in resistance could be measured \cite{SUSYDM96}.
 A very low threshold energy ($\simeq$ 500 eV) of such detector were reached
 by the CRESST-I experiment with a 262 g sapphire detector
 \cite{Jesus04}, \cite{Altmann01}
 (more details about the CRESST experiments
  and their results will be given in Subsec.~\ref{CRESST}).

 Similarly,
 it is also possible
 to use some small superconducting granules in a magnetic field as detector,
 when one of such detectors is heated by a nuclear recoil,
 it would go normal and thereby cause a measurable change in the magnetic flux
 \cite{SUSYDM96}.
\subsection{Ionization}
\label{ionization}
 ~~~$\,$
 A small voltage is placed across the crystal of the detector,
 and when several atoms have been ionized,
 the freed electrons will drift to one side,
 the collected charges can be used as a measure of the energy deposited in ionization
 \cite{SUSYDM96}.

 Germanium $\rmXA{Ge}{76}$ used initially
 in the neutrino-less double-$\beta$ ($0 \nu 2 \beta$) decay experiments
 has been used as the first detector material for direct WIMP detection experiments
 by the Heidelberg-Moscow (HDMS) collaboration \cite{HDMS}
 (more details about the HDMS experiments
  and their results will be given in Subsec.~\ref{HDMS}).
 Thanks to the high intrinsic purity achieved by the semiconductor industries
 and the technique developed for the $0 \nu 2 \beta$ decay experiments,
 Ge ionization detectors
 have nowaday very low thresholds and very good resolutions
 ($\Qthre \simeq 4 \sim 10~\keVee$,
  equivalent to $\simeq 15 \sim 30$ keV recoil energy,
  for HDMS \cite{Jesus04}).
 Moreover,
 silicon (Si) has also been used by e.g., CDMS collaboration \cite{CDMS} as detector material
 (more details about the CDMS experiments
  and their results will be given in Subsec.~\ref{CDMS}).
 However, the size of such ionization detectors are limited.
\subsection{Scintillation}
\label{scintillation}
 ~~~$\,$
 Scintillation detectors,
 e.g., sodium iodine (NaI) or liquid xenon (LXe),
 are the solution to accumulate large mass of detector material
 ($\approx 100$ kg).
 However,
 it is more difficult to achieve radiopurity comparable to Ge detectors
 \cite{Gascon05}.

 Moreover,
 as mentioned in Subsec.~\ref{inducedsignals},
 scintillation detectors do not measure the elevation of the temperature in the crystal,
 but the light emitted by the electrons produced due to the ionization,
 thus the energy threshold for these detectors
 may be substantially higher than the thermal calorimeters \cite{SUSYDM96}.

 Meanwhile,
 the NaI-based experiments,
 such as DAMA \cite{DAMA}, NaIAD \cite{UKDMC}, and ELEGANT,
 originally attempted to use a pulse shape discrimination
 to statistically identify the WIMP signals from their observed events
 (detailed discussions about the background discrimination
  will be given in Sec.~\ref{background}).
 It was found that
 the low number of detected scintillation photons per keV of incident energy
 (called ``photo-electron per keV'' or ``p.e./$\keVee$'')
 restricts the usefulness of this method at low energy \cite{Gascon05}.
 This means that the background may be problematic.
 The technique is now being investigated for CsI or CaF scintillator,
 where the difference in time constants
 between scintillations induced by electron- and nuclear-recoils
 are larger than in the NaI detectors \cite{Gascon05}.
\subsection{Combinations of two different signals}
\label{signalcombinations}
 ~~~$\,$
 Actually,
 most of the direct WIMP detection experiments
 use detectors with mixed techniques and measure simultaneously two signals.
 For example,
 for cryogenic detectors,
 the CDMS and EDELWEISS collaborations \cite{CDMS}, \cite{EDELWEISS}
 investigate the heat-ionization signals
 (more details will be given in Subsecs.~\ref{CDMS} and \ref{EDW}),
 and the CRESST collaboration \cite{CRESST} explores the heat-scintillation channel
 (more details will be given in Subsec.~\ref{CRESST}).

 Combining information measuring from two different channels
 can offer a powerful event-by-event rejection method
 for the background discrimination
 down to 5 to 10 keV recoil energy \cite{Drees06}.
 As mentioned in Subsec.~\ref{quenchingfactor},
 due to the quenching effect of the detector material,
 the ratio of the ionization or the scintillation signal to the heat signal
 is significantly different for the nuclear recoils and for the electron recoils.
 Similarly,
 nuclear recoils due to WIMP or neutron interactions
 have a much higher characteristic light over charge ratio than
 electron recoils due to electron and $\gamma$-ray interactions \cite{Kaufmann06a}.
 Thus simultaneous measurements of two of
 the heat, the ionization, or the scintillation signals
 can be used to distinguish nuclear recoils induced by WIMPs
 from electron recoils induced by electron or $\gamma$-ray interactions
 (more details about different methods for background discrimination
  will be given in Sec.~\ref{background}).
\section{Background and background discrimination}
\label{background}
 ~~~$\,$
 As mentioned in the beginning of this chapter,
 due to the very low cross section of WIMPs on ordinary material,
 the event number of the elastic WIMP-nucleus scattering is very rare
 and the backgrounds coming from different sources are much larger.

 For example,
 cosmic rays and cosmic-ray induced $\gamma$-rays with energies in the keV to MeV range,
 radioactive isotopes in and around the detector (in the equipment)
 should be considered.
 Moreover,
 neutrons induced by cosmic muons
 can produce nuclear recoil events similar to the real events induced by WIMPs.
 And electron recoils
 from photons (x-ray and $\gamma$-ray radiations) and electrons
 are also a major background.
%
%
\subsection{Cosmic muons and underground laboratories}
 ~~~$\,$
 At ground level,
 approximately $10^3$ cosmic muons
 pass through per square centimeter of the Earth's surface per day \cite{Gascon05}.
 They can induce nuclear transmutations to unstable isotopes
 throughout the detector volume \cite{Gascon05}.

 In order to protect from the penetrating cosmic muon flux,
 it is necessary to place the detector in deep underground.
 In underground laboratories such as
 the Soudan Underground Laboratory (the CDMS collaboration) in Minnesota in the USA,
 the Gran Sasso National Laboratory (the CRESST and DAMA collaborations) in Italy,
 or the Laboratoire Souterrain de Modane (LSM, the EDELWEISS collaboration)
 in the Fr\'ejus Tunnel in the French-Italian Alps,
 the muon flux can be reduced by a factor of $10^5 \sim 10^7$ \cite{Gascon05}.
\subsection{External natural radioactivity and (passive) shielding}
 ~~~$\,$
 External sources of radioactivity mean the radioactive isotopes
 in the rock around the underground laboratory and in the walls of the laboratory.

 A shielding from external natural radioactivity can be achieved
 by surrounding the detector with thick absorbing material \cite{Gascon05}:
 high-$Z$ materials like lead
 are very effective for stopping $\gamma$-rays with MeV energy,
 while low-$Z$ materials
 are sufficient for stopping low energy $\gamma$-rays
 as well as $\alpha$- and $\beta$-radiations.
\subsection{Internal natural radioactivity and radiopure materials}
 ~~~$\,$
 Beyond a thickness of 15 to 25 cm of lead shielding \cite{Gascon05},
 one has to consider the internal radioactivity
 of the equipment,
 of the contamination near the detector or in the target material,
 and even of the lead shielding itself.

 Internal radioactivity can be reduced very well
 by using detectors (and the other experimental equipment) made of radiopure materials.
 Archeological lead has also often been used
 since it has already been shielded from cosmic rays for 2000 years.
\subsection{Active background rejection}
 ~~~$\,$
 Except passive shielding around the detector,
 most experiments use also some different techniques for active background rejection.

 Generally there are two different types of background discrimination.
 Statistical rejections are used to ascertain
 which fraction of the total event sample comes from a well-defined type of background,
 but cannot tell for one individual event \cite{Gascon05}.
 Moreover,
 this kind of rejections depends strongly on the theoretical predictions
 about the true signals induced by WIMPs and the background events.

 On the other hand,
 the event-by-event rejections
 check each recorded event in the detector independently of the others (``blind'')
 and can be used to reject background events with an almost 100\% certainty.
 Note that,
 however,
 in practice there is always a small probability
 that some background event may fake the signals induced by WIMPs.
\subsection{Neutron induced nuclear recoils}
 ~~~$\,$
 Cosmic muons can induce neutrons in the inner lead shielding
 and such fast neutrons can induce nuclear recoils similar to those induced by WIMPs.

 Fast neutron shielding consists of moderators
 made of material with a high density of hydrogen,
 such as polyethylene or water \cite{Gascon05}.
\subsection{Multiple-scatter events and array of detectors}
\label{multiplescatter}
 ~~~$\,$
 The interaction between WIMPs and ordinary material is too weak,
 or, equivalently,
 the mean free path of a WIMP in ordinary matter is too long
 (of the order of a light-year \cite{Gascon05}),
 so that WIMPs could never interact more than once in a single detector
 or two adjacent detectors.
 In contrast,
 the mean free path of a neutron or a high energy photon is of the order of cm,
 thus multiple-scatter events produced by neutrons are more common \cite{Gascon05}.

 Hence,
 an array of closely packed detectors
 (e.g., the tower with six detectors used by the CDMS experiment,
  see Subsec.~\ref{CDMS} and their web page \cite{CDMS})
 can efficiently identify these multiple-scatter events \cite{Gascon05}.
\subsection{Electron recoils}
\label{electronrecoils}
 ~~~$\,$
 Theoretically
 WIMPs interact only with the nuclei (through the coupling to quarks)
 and produce nuclear recoils,
 while,
 due to the electromagnetic interaction,
 the dominant radioactive backgrounds interact usually with the electrons
 and produce electron recoils.
 Therefore,
 the experiments
 which can discriminate between
 the events due to nuclear recoils and events due to electron recoils 
 can reject most radioactive background \cite{Schnee06}.

 There are three ways to distinguish nuclear recoils from electron recoils.
 First,
 as mentioned in Subsec.~\ref{quenchingfactor},
 due to the quenching effect of the detector material,
 the ratio of the ionization or the scintillation signal to the heat signal
 is significantly different for nuclear recoils and for electron recoils.
 Thus one can measure simultaneously
 the heat signal and the ionization or the scintillation signal
 to distinguish the nuclear recoil events from the electron recoil events.
 
 Second,
 the decay times of pulses for nuclear recoils
 may be different than that for electron recoils \cite{Schnee06}.
 Thus some experiments use only a scintillation detector,
 but measure also the timing of the signals \cite{Schnee06}.
 However,
 due to the limited resolution and discrimination power of this pulse shape analysis
 at low energies,
 this effect allows only a statistical background rejection \cite{Drees06}.
 This technique has been used by e.g.,
 the DAMA/NaI and NaIAD experiments (NaI(Tl) detector) \cite{DAMA}, \cite{UKDMC}
 and ZEPLIN-I experiment (liquid xenon detector) \cite{UKDMC}.

 Third,
 as mentioned in Subsec.~\ref{quenchingfactor},
 the range of a electron recoil is of the order of $\mu$m
 and that of a nuclear recoil is only of the order of nm.
 Thus the nuclear recoils have a much larger energy loss per unit length, $dQ/dx$,
 or, equivalently, produce a much higher energy density,
 than the electron recoils.
 Therefore,
 some experiments are actually immune to electron recoils
 because the energy they deposit is not dense enough to trigger \cite{Schnee06}.
\subsection{Surface events and self shielding}
\label{surfaceevents}
 ~~~$\,$
 Due to their very long mean free path,
 WIMPs will interact uniformly throughout the detector volume.
 In contrast,
 due to the short mean free path of the high-energy photons and neutrons,
 for a detector with a large volume,
 the interactions induced by radiations
 originating from the surrounding material and surface contamination
 will occur mostly at the detector surface \cite{Gascon05}.

 This ``self-shielding'' effect leads to
 the incentive of building large position-sensitive detectors
 in order to reject the surface events.
 Moreover,
 low-energy photons, $\alpha$- and $\beta$-rays have very short mean free path ($<$ mm),
 and can be rejected even if the position sensitivity is limited \cite{Gascon05}.
\subsection{Incomplete charge collection}
 ~~~$\,$
 Some electromagnetic events occurring very near the detector surface
 can also mimic nuclear recoils
 because they produce less ionizations than expected from electron recoils
 \cite{Ramachers02}.
 But such surface events can also be rejected by the self-shielding effect.
\subsection{Shape of the recoil energy spectrum}
 ~~~$\,$
 The shape of the recoil spectrum $dR/dQ$
 for a given WIMP mass in some simple halo models
 can be predicted numerically or even analytically
 (e.g., $(dR/dQ)_{\Gau}$ in Eq.(\ref{eqn3010302})
  and $(dR/dQ)_{\sh}$ in Eq.(\ref{eqn3020103})).
 The measured recoil spectrum should be consistent with the expectation.

 However,
 first,
 the overall shape of the expected spectrum is (approximately) exponential,
 as is the case for many background sources;
 second,
 different velocity distribution functions in different halo models
 could predict totally different recoil spectrum
 (cf.~$(dR/dQ)_{\Gau}$ in Eq.(\ref{eqn3010302})
  and $(dR/dQ)_{\sh}$ in Eq.(\ref{eqn3020103})).
 Moreover,
 the expected signal events
 measured by the currently operated detectors and even next-generation ones
 are at most only a few per day.
 Hence,
 as we will see in Sec.~\ref{Qni},
 at present a meaningful reconstruction of the recoil spectrum
 with a small statistical error is actually impossible.
\section{Cyrogenic detectors}
\label{cyrogenic}
 ~~~$\,$
 As discussed above,
 a WIMP detector is constrained by three important requirements:
 low threshold, (ultra) low background, and high detector material mass \cite{Jesus04}.

 In the following I will present some important collaborations worldwide
 and summarize their recent results and plans in the near future.
 More details about these collaborations and their experiments can be found in the references.
\subsection{CDMS}
\label{CDMS}
 ~~~$\,$
 The Cryogenic Dark Matter Search (CDMS) collaboration \cite{CDMS} uses
 the Berkeley Center for Particle Astrophysics (CfPA) germanium cryogenic detector
 \cite{SUSYDM96}.
 Their first test run was at the Stanford University Underground Facility \cite{SUSYDM96},
 and now moved to the deep Soudan Underground Laboratory (Soudan mine) in Minnesota in the USA
 \cite{Akerib05a}. 
 The Soudan mine
 has 780 m rock overburden (2090 meters water equivalent, m.w.e.) \cite{Akerib05a},
 the surface muon flux is then reduced by a factor of $5 \times 10^4$ \cite{Akerib05a},
 and the neutron background is also reduced by a factor of 400 \cite{Jesus04}
 ($\sim 4 \times 10^{-4}$/kg/day \cite{Brink05}).

 It was the first experiment to operate a detector
 measuring simultaneously ionization and heat signals with a germanium/silicon crystal
 as the target material \cite{Jesus04}.
 They developed Z(depth)-sensitive
 Ionization and Phonon (ZIP) detectors.
 The principle of their ZIP detector
 is basically the same as that discussed in Subsec.~\ref{ionization},
 except that the heat sensor is replaced by a thin film sensor
 and thus able to detect phonons before their complete thermalization \cite{Gascon05}.

 Their ``tower(s)'' with mixed Ge and Si detectors
 are powerful for subtracting the neutron background \cite{Akerib05a}.
 Except the neutron multiple-scatter events discussed in Subsec.~\ref{multiplescatter},
 while Ge and Si have similar scattering rates per nucleon for neutrons,
 the WIMP-nucleon scattering rate is expected to be 5-7 times greater in Ge than in Si
 for all but the lowest-mass WIMPs.
 Moreover,
 the kinematics of neutron elastic scattering gives a recoil energy spectrum
 scaled in energy by a factor of $\sim$ 2 in Si compared to Ge,
 whereas the factor would be $\sim$ 1 or less for WIMP elastic scattering.
 All of these three methods can be used (together),
 in conjunction with Monte Carlo simulations,
 to statistically subtract any neutron background.

 In addition,
 because the athermal phonons from electron recoils
 are faster than those from nuclear recoils,
 particularly if the electron recoils occur near a detector surface,
 by collecting such fast, athermal phonons with thousands of thin-film sensors,
 their ZIP detector can discriminate very well
 against the surface electron recoils \cite{Schnee06}.

 According to Ref.~\cite{Akerib05a},
 for recoil energies above 10 keV,
 events due to the background photons can be almost perfectly rejected,
 and more than 96\% of the incomplete charge collection events can also be rejected
 by using additional information from the shape, timing,
 and energy partition of the phonon pulses
 (namely,
  only events with both slow phonon pulses and low ionization
  have been accepted),
 while over half of the nuclear-recoil events should be kept \cite{Schnee06}.

 In the first Soudan run of the CDMS-II experiment
 (from October 11, 2003 to January 11, 2004) \cite{Akerib05a},
 one tower with 4 Ge (each 250 g) and 2 Si (each 100 g) ZIP detectors has been operated
 for 52.6 live days,
 the recoil energy thresholds of these six detectors were between 10 and 20 keV,
 {\em only one candidate event} with a recoil energy of 64 keV
 in one Ge detector has been measured.
 For a WIMP mass of 60 GeV/$c^2$,
 a $4 \times 10^{-7}$ pb upper limit on the spin-independent WIMP-nucleon cross section
\footnote{
 Here the cross-section $\sigma_0$ shown in Eqs.(\ref{eqn3010102}) and (\ref{eqn3010109})
 is normalized to a single nucleon $\sigma_{\chi {\rm n}}$
 in order to allow comparisons between different target nuclei.
}
 from Ge as been achieved.
 Meanwhile,
 thanks to the $\rmXA{Ge}{73}$ (spin-$\frac{9}{2}$)
 and $\rmXA{Si}{29}$ (spin-$\frac{1}{2}$)
 content of natural germanium and silicon,
 a 0.2 pb upper limit on the spin-dependent WIMP-neutron cross section
 for a WIMP mass of 50 GeV/$c^2$ has also been achieved.
 These were the world's lowest limits on the WIMP-nucleon cross-section
 in the case of spin-independent interactions
 and spin-dependent interactions with neutrons.

 In the second Soudan run (from March 25 to August 8, 2004)
 \cite{Akerib05b}, \cite{Akerib05c}, and \cite{Akerib06},
 two towers
 (one tower with 4 Ge and 2 Si detectors and the other one with 2 Ge and 4 Si detectors)
 have been operated for 74.5 live days
 and the recoil energy thresholds have been improved to be only 7 keV,
 {\em one more candidate event} with a recoil energy of 10.5 keV
 in one Ge detector has been measured.
 For a WIMP mass of 60 GeV/$c^2$,
 the upper limit on the spin-independent WIMP-nucleon cross section
 has been given as 1.6 $\times 10^{-7}$ pb from Ge
 and 3.4 $\times 10^{-6}$ pb from Si
 (see Fig.~\ref{fig3080601}).
 These limits are a factor of 6 lower than
 those given by the ZEPLIN-I experiment \cite{UKDMC}
 (see Subsec.~\ref{ZEPLIN})
 and an order of magnitude lower than those of the CRESST and EDELWEISS collaborations
 \cite{Schnee06}.
 Moreover,
 their results excluded the overlap between the CDMS and DAMA/NaI allowed regions
 at WIMP masses $\amssyasymc{38}~25~{\rm GeV}/c^2$,
 though compatible regions at lower masses remain \cite{Akerib05c}
 (see Fig.~\ref{fig3080601}).

 Now the CDMS collaboration is preparing for
 five towers with totally 19 Ge (4.75 kg) and 11 Si (1.1 kg) ZIP detectors,
 and will improve their sensitivity a factor of $\approx$ 10 \cite{Akerib06}.
 Furthermore,
 they also planned a SuperCDMS project
 which will start with a total mass of 25 kg (Phase A, each detector will be 640 g)
 and be improved to 150 kg (Phase B) and eventually 1000 kg (Phase C) \cite{Akerib06},
 in order to achieve $\sim 10^{-9}$ pb sensitivity
 (for a WIMP mass of 60 GeV/$c^2$, Phase A, see Fig.~\ref{fig3100001}) \cite{Akerib06},
 corresponding to ${\cal O}(10^{-4})$ events/kg/day event rate
 in the energy range between 15 and 45 keV
 \cite{Brink05}.
 They will also move to the SNO Underground Laboratory at the Sudbury mine in Canada.
 The $\sim$ 6000 m.w.e. overburden at this site
 results in over two orders of magnitude suppression
 in the neutron background compared to Soudan \cite{Brink05}.
\subsection{CRESST}
\label{CRESST}
 ~~~$\,$
 The Cryogenic Rare Event Search using Superconducting Phase Transition Thermometers (CRESST)
 collaboration \cite{CRESST}
 uses heat-scintillation detectors
 with $\rm Ca W O_4$ crystal
 in the Gran Sasso National Laboratory in Italy.
 Their detector provides a good rejection of surface events as of photons
 due to the much larger light yield from all electron recoils relative to nuclear recoils
 \cite{Schnee06}.

 As mentioned in Subsec.~\ref{heat},
 their detector uses
 the superconducting-normal phase transition due to the difference of the temperature.
 A thin superconducting film of tungsten (W) has be grown on a silicon detector
 and held just below the critical temperature.
 Heat produced by WIMP-nucleus scatterings will change the film to its normal state
 and the change in resistance could be measured \cite{SUSYDM96}.
 However,
 as mentioned in Subsec.~\ref{scintillation},
 the threshold energy of such a scintillation detector
 is relatively higher than for an ionization detector.
 Thus a disadvantage of the CRESST heat-scintillation detector is that
 an event measured by the phonon channel
 but producing no light may mimic a WIMP signal \cite{Schnee06}.

 In 2003 CRESST ran two prototype detectors
 for a couple of months without neutron shielding.
 A significant neutron background on the oxygen
 in their $\rm Ca W O_4$ detectors was observed \cite{Schnee06}
 and the light yield for W recoils is significantly less than for Ca or O recoils.
 This result indicates that
 WIMPs are expected to interact primarily with W nuclei,
 while neutrons will interact relatively more often with O and Ca nuclei \cite{Gascon05}.

 In early 2004
 they operated two 300 g CRESST-II prototype detector modules,
 {\em 16 events} have been recorded
 and a rate for nuclear recoil energy between 12 and 40 keV of $(0.87 \pm 0.22)$ events/kg/day
 has been obtained \cite{Angloher04}.
 However,
 this is compatible with the rate expected from neutron background,
 and most of these events lie in the region of the phonon-light plane
 anticipated for neutron-induced recoils \cite{Angloher04}.
 Moreover,
 a particularly strong limit for WIMPs with coherent scattering results from
 selecting a region of the phonon-light plane corresponding to tungsten recoils,
 where the best module shows zero events \cite{Angloher04}.
 The sensitivity achieved by the CRESST-II experiment
 is given in Fig.~\ref{fig3080601}.

 Now they are preparing the scientific run of the CRESST-II experiment
 with 33 detectors (each 300 g) and totally $\sim$ 10 kg target material
 (see Fig.~\ref{fig3100001}).
%
%
\subsection{DAMA}
\label{DAMA}
 ~~~$\,$
 The DArk MAtter (DAMA) collaboration \cite{DAMA}
 uses a scintillation detector with $\sim$ 100 kg NaI(Tl)
 in the Gran Sasso National Laboratory (Laboratori Nazionali del Gran Sasso, LNGS) in Italy
 \cite{Bernabei00}.
 With 1400 m rock overburden (3500 m.w.e.),
 the total muon flux is reduced to $\sim 1/{\rm m^2/hr}$
 (one order of magnitude lower than that of CDMS),
 the external $\gamma$-ray flux is reduced by a factor of $10^5$.

\begin{figure}[p]
\begin{center}
\imageswitch{
\begin{picture}(15,21.1)
\put(0  , 0  ){\framebox(15,21.1){}}
\put(1.5,14.6){\framebox(12, 6.5){DAMA-4yrs}}
\put(7  ,13.6){\makebox ( 1, 1  ){(a)}}
\put(3  , 1  ){\framebox( 9,12  ){DAMA-7yrs}}
\put(7  , 0  ){\makebox ( 1, 1  ){(b)}}
\end{picture}}
{\includegraphics[width=12cm]{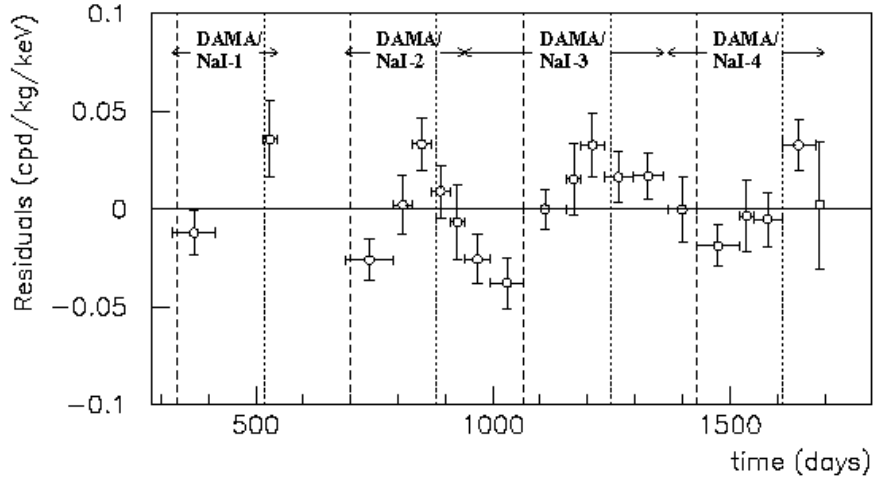} \\ (a) \\ \vspace{0.8cm}
 \includegraphics[width= 9cm]{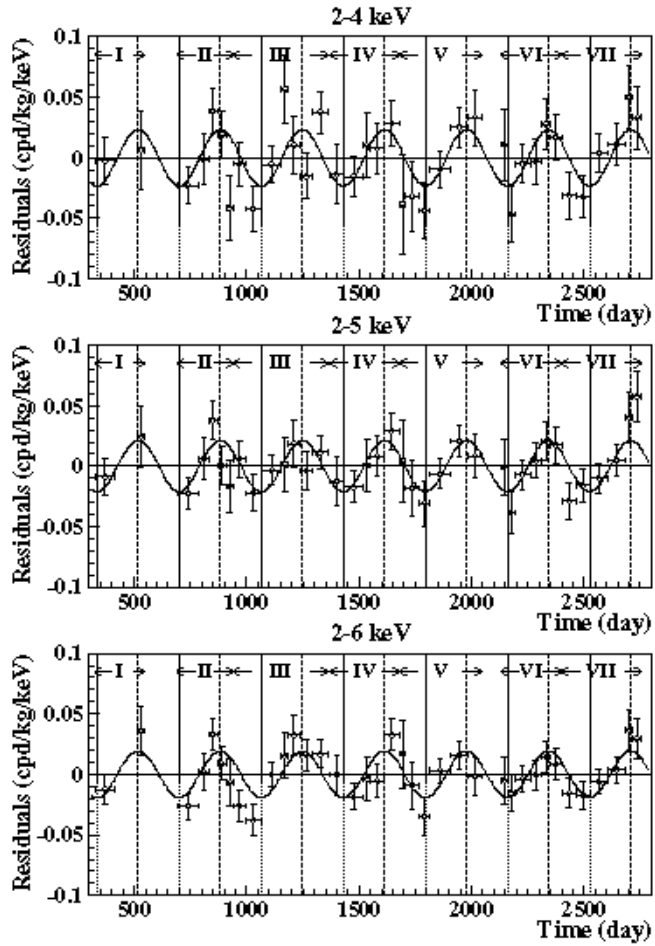} \\ (b)}
\end{center}
\caption{
 The published results of the DAMA/NaI experiment.
 (a) In the 2-6 $\keVee$ cumulative energy interval over 4 annual cycles,
     since January 1st of the first year of data taking.
     Theoretically expected minimum (dashed line), maximum (dotted line)
     (figure from \cite{Bernabei03a}).
 (b) In the 2-4 $\keVee$, 2-5 $\keVee$, and 2-6 $\keVee$ cumulative energy intervals
     over 7 annual cycles and end of data taking in July 2002
     (figure from \cite{Bernabei03b}).}
\label{fig3070301}
\end{figure}

 They are the only collaboration which claimed to detect the signal of halo Dark Matter
 due to the annual modulation effect discussed in Sec.~\ref{annualmodulation}.
 Figs.~\ref{fig3070301} show the 4-year and the 7-year results
 of the DAMA/NaI experiment \cite{Bernabei03a}, \cite{Bernabei03b},
 their threshold energy is about 2 $\keVee$,
 corresponding to approximately 22 keV recoil energy \cite{Gascon05}.
 Meanwhile,
 they published a WIMP mass $\mchi \simeq 52~{\rm GeV}/c^2$ and
 a WIMP-proton cross section $\sigma_{\chi {\rm p}} \simeq 7.2 \times 10^{-6}$ pb
 \cite{Bernabei00}
 under the standard assumptions of WIMP halo described in Subsec.~\ref{haloDM}
 (see Fig.~\ref{fig3080601}).

 However,
 the DAMA collaboration uses the pulse shape discrimination (PSD) technique
 (see Subsec.~\ref{electronrecoils})
 to statistically (not event-by-event) discriminate the measured events \cite{Gascon05}.
 On the other hand,
 the CDMS results \cite{Abusaidi00}, \cite{Akerib05a}, \cite{Akerib05b} and \cite{Akerib05c}
 are clearly incompatible with the signal claimed by DAMA
 under the standard assumptions of the WIMP halo and spin-independent WIMP-nucleus coupling
 \cite{Schnee06}.

 But these two experiments might still be compatible in some exotic scenarios.
 One possibility is to postulate rather light ($\mchi < 10~{\rm GeV}/c^2$)
 and fast WIMPs with large scattering cross section \cite{Gondolo05}.
 Since for this case $m_{\rm I} \gg \mchi$,
 the maximal recoil energy induced by the scattering on iodine
 will then be smaller than the threshold energy of the DAMA's detector
 and the recorded events have thus been induced by the scattering on sodium
 (see Eq.(\ref{eqn3050101})).
 Similarly,
 the Ge nuclei used by the CDMS experiment
 could also be too heavy to deposit recoil energy large enough to be measured.
 However,
 the null results from the Si detector used by CDMS (see Subsec.~\ref{CDMS})
 should (almost) rule out this possibility.
 Another possible way out is to postulate that the detected events are actually inelastic,
 leading to the production of a second particle
 that is almost, but not exactly, degenerate with the WIMP \cite{TuckerSmith05}.

 They are running now the DAMA/LIBRA
 (Large sodium Iodide Bulk for RAre processes) experiment
 with totally $\sim$ 250 kg NaI(Tl) \cite{Bernabei03a}.
%
%
\subsection{EDELWEISS}
\label{EDW}
 ~~~$\,$
 The Exp\'erience pour DEtecter Les WIMPs En Site Souterrain
 (EDELWEISS, EDW) collaboration \cite{EDELWEISS}
 is in the Laboratoire Souterrain de Modane (LSM)
 in the Fr\'ejus Tunnel in the French-Italian Alps.
 With $\sim$ 1800 m rock overburden ($\sim$ 4800 m.w.e.),
 the muon flux can be reduced to $\sim 4/{\rm m}^2$/day,
 the fast neutron flux can be reduced to $\sim 1.6 \times 10^{-6}/{\rm cm}^2$/s
 \cite{Sanglard06}.

 The EDELWEISS-I experiment has used 3 (each 320 g)
 cryogenic heat-and-ionization Ge detector
 \cite{Gascon05}.
 Their calibrations
 indicate that the larger ionization/recoil energy ratio of electron recoils
 results in very good discrimination ability against photon backgrounds
 down to the 20 keV threshold energy \cite{Schnee06}.

 In 2000 to 2002 \cite{Sanglard05},
 the EDELWEISS-I experiment has been operated for an exposure of 13.6 kg-day.
 The energy threshold was 13 keV and {\em no event} has been recorded.
 In 2003 \cite{Sanglard05},
 the second run of the EDELWEISS-I has been operated for an exposure of 48.4 kg-day
 (totally 62 kg-day)
 and {\em 40 nuclear recoil candidate events} have been recorded
 in the energy range 15 to 200 keV:
 18 events between  15 and  20 keV,
 16 events between  20 and  30 keV,
  3 events between  30 and 100 keV,
  3 events between 100 and 200 keV,
 and {\em more 19 events} have been observed below 15 keV;
 most likely due to remaining background neutrons and surface electrons.
 According to these results,
 they gave exclusion limits for WIMP masses above 25 GeV/$c^2$
 \cite{Sanglard05} (see Fig.~\ref{fig3080601}).

 Now they are preparing the EDELWEISS-II experiment
 with 120 detectors \cite{Sanglard06} (see Fig.~\ref{fig3100001}).
\subsection{Heidelberg-Moscow (HDMS)}
\label{HDMS}
 ~~~$\,$
 The Heidelberg-Moscow collaboration \cite{HDMS}
 uses a $\sim$ 2 kg $\rmXA{Ge}{76}$ semiconductor ionization detector
 in the Gran Sasso National Laboratory in Italy
 and achieved a very low background count rate ($<$ 0.2 event/kg/day)
 in the interval 10 $\sim$ 40 keV,
 and a threshold energy $\Qthre \simeq 4 \sim 10~\keVee$
 (equivalent to $\simeq 15 \sim 30$ keV recoil energy)
 \cite{Jesus04}.

 They produced the first limits on WIMP searches
 and until recently had the best performance.
 However,
 without positive identification of nuclear recoil events,
 these experiments could only set limits,
 e.g., excluding sneutrinos as major component of the galactic halo \cite{Drees06}.
\subsection{KIMS}
\label{KIMS}
 ~~~$\,$
 The Korea Invisible Mass Search (KIMS) collaboration \cite{DMRC}
 in the Yangyang Underground Laboratory (Y2L, $\sim$ 700 m rock overburden) in South Korea
 developed a CsI(Tl) crystal scintillation detector
 and uses an improved pulse shape discrimination
 to discriminate nuclear recoil events induced by WIMPs
 from electron recoil events induced by $\gamma$-ray background 
 \cite{HSLee07}.

 They have operated 4 detectors with an exposure of 3407 kg-day
 in the temperature $0^\circ$ C
 and {\em null signals} have been observed \cite{HSLee07}.
 Due to their $\rmXA{Cs}{133}$ and $\rmXA{I}{127}$ target nuclei
 this result has been used to give the lowest upper limits
 on the spin-dependent WIMP-proton cross section
 for WIMP masses $\amssyasymc{38}~30~{\rm GeV}/c^2$ \cite{HSLee07}
 (the lowest upper limit on the spin-dependent WIMP-neutron cross section
  has been obtained by the CDMS collaboration,
  see Subsec.~\ref{CDMS}).
 Moreover,
 although several experiments have already given exclusion limits
 rejecting the DAMA signal region,
 it is the first time that
 such a limit obtained by the crystal scintillation detector containing $\rmXA{I}{127}$,
 which has been usually assumed to be the dominant nucleus
 for the spin-independent interaction in the NaI(Tl) crystal.
%
%
\subsection{PICO-LON}
\label{PICOLON}
 ~~~$\,$
 The Planar Inorganic Crystals Observatory for LOw-background Neutr(al)ino (PICO-LON)
 collaboration in Japan
 uses detector with multilayer (3 layers for PICO-LON-0 and 16 layers for PICO-LON-I)
 thin NaI(Tl) crystals \cite{Fushimi05a}, \cite{Fushimi06}.

 A special advantage of their 0.05 cm thin and 5 cm $\times$ 5 cm wide area NaI(Tl) crystals
 is the position sensitivity of the recoil events \cite{Fushimi06}.
 The position resolution for the thinner directions is as good as 0.05 cm
 due to the segmentation of the detector \cite{Fushimi06};
 while,
 the position information in the wider area was obtained by analyzing
 the ratio of the number of photons collected on the opposite sides of the detector
 \cite{Fushimi06}
 and a circle with $\simeq$ 0.5 cm radius has been reached \cite{Fushimi05b}.
 Moreover,
 they have also claimed that
 a very low threshold energy $\sim 2~\keVee$ has been measured
 \cite{Fushimi05a}, \cite{Fushimi05b}.

 The PICO-LON-0 experiment has been run at the surface laboratory at Tokushima,
 and the PICO-LON detector will be installed into Oto Cosmo Observatory
 (100 km south from Osaka)
 covered by thick rock with $\sim$ 1200 m.w.e. \cite{Fushimi06}.
\section{Liquid noble gas detectors}
\label{nobleliquid}
 ~~~$\,$
 Liquid noble gas detectors have the advantage of easier scaling to large masses
 since it is based on liquids \cite{Schnee06}.
 They can also be allowed to operate in higher temperatures:
 165 K for xenon, 87 K for argon, and 27 K for neon \cite{Benetti07}.

 Due to its high-$A$ value,
 liquid xenon (LXe) has been used by the ZEPLIN collaboration \cite{UKDMC}
 as the first liquid noble gas detector material
 (more details about the ZEPLIN experiments will be given Subsec.~\ref{ZEPLIN}).
 Recoils in the liquid noble gas such as Xe 
 can induce both ionization and excitation of Xe atoms.
 An excitation produced by a nuclear recoil
 usually induces emission with a single photon,
 whereas that reduced by an electron recoil
 emits photons in form of a slow triplet,
 thus nuclear recoils have a faster pulse shape than electron recoils
 \cite{Schnee06}, \cite{Alner07}.

 Besides xenon,
 argon and neon are also suitable detector materials.
 The effect of faster pulse shape is particularly clear in Ar and Ne,
 leading to their extremely good discrimination based on timing \cite{Schnee06}.
 In addition,
 due to the form factor effect introduced in Eq.(\ref{eqn3010102})
 (two nuclear form factors for spin-independent cross section
  have been given in Subsec.~\ref{formfactor}),
 the event rate in e.g., argon is less sensitive to the energy threshold than in xenon
 \cite{Kaufmann06a}.

 Moreover,
 as discussed in Subsec.~\ref{signalcombinations},
 the scintillation (light) over ionization (charge) ratio can be used additionally
 to discriminate the nuclear recoils from the electron recoils
 due to electron and $\gamma$-ray interactions.

 However,
 discrimination of the radioactive contamination in the detector material,
 such as $\rmXA{Kr}{85}$ in liquid Xe (25 ppm Kr in natural Xe)
 or especially $\rmXA{Ar}{39}$ in liquid Ar \cite{Schnee06}
 as well as of the surface radioactivity from the liquid container \cite{Gascon05},
 and relatively larger threshold energies
 could be primary challenges for detectors using liquid noble gases.
\subsection{ArDM}
\label{ArDM}
 ~~~$\,$
 The Argon Dark Matter (ArDM) experiment at CERN \cite{ArDM}
 uses a ton-scale detector with liquid argon (LAr),
 which measures simultaneously the scintillation and the ionization signals
 \cite{Rubbia05}, \cite{Kaufmann06a}, and \cite{Kaufmann06b}.

 With a recoil energy threshold of 30 keV
 and a sensitivity of $10^{-6}$ pb WIMP-nucleon cross section,
 the ArDM experiment has been expected to yield
 approximately 100 events per day per ton \cite{Kaufmann06a}.
 By improving the background rejection power and further limiting the background sources,
 a sensitivity of $10^{-8}$ pb (1 event per day per ton)
 would become reachable \cite{Kaufmann06a}.
\subsection{WARP}
\label{WARP}
 ~~~$\,$
 Similar to the ArDM experiment,
 the WIMP ARgon Programme (WARP) experiment \cite{WARP}
 uses also a dual-phase (gas-liquid) argon detector \cite{Schnee06}
 in the Gran Sasso Underground Laboratory in Italy \cite{Benetti07}.
 By using a strong electric field,
 ionization electrons are drifted out of the liquid argon into gaseous phase,
 where they are detected via the secondary luminescence
 \cite{Gascon05}, \cite{Schnee06}.
 The discrimination technique is based on both of
 the pulse shape of the photon emissions
 and on the ratio of the scintillation to ionization energies \cite{Schnee06}
 described in Subsec.~\ref{signalcombinations}.

 Their first run of a 2.6 kg (1.87 $\ell$) prototype with a 96.5 kg-day exposure
 resulted in {\em no candidate events} above the threshold energy 55 keV
 \cite{Benetti07} (see Fig.~\ref{fig3080601}).
 Later they will upgrade the detector to totally 140 kg ($>$ 100 $\ell$)
 \cite{Benetti07}.
\subsection{XENON}
\label{XENON}
 ~~~$\,$
 The XENON collaboration \cite{XENON} uses also
 a dual-phase xenon time projection chamber (XeTPC) with 3D position sensitivity
 in the Gran Sasso Underground Laboratory in Italy
 \cite{Ni06}.

 The XENON10 experiment uses 15 kg liquid xenon,
 the background rate from $\rmXA{Kr}{85}$ contamination
 is reduced by a factor of 5000
 by using a commercially available low-Kr (5 ppb) xenon,
 the self-shielding effect (see Subsec.~\ref{surfaceevents}) of LXe
 can reduce the background events in the central region of the LXe target
 by more than one order of magnitude for recoil energy below 50 $\keVee$,
 (5 to 15 $\keVee$ corresponds to roughly 15 to 40 keV recoil energy) \cite{Ni06}.

 By measuring simultaneously
 the scintillation and the ionization
 produced by radiation in pure liquid xenon,
 the detector can discriminate signals from background down to 4.5 $\keVee$
 \cite{Angle07}.
 Between October 6, 2006 and February 14, 2007
 the XENON experiment has been run for 58.6 live days
 and {\em 10 events} have been observed in the energy range 4.5 to 26.9 $\keVee$.
 However,
 none of these events are likely WIMP interactions. \cite{Angle07}

 Their newest result gives
 a 90\% C.~L.~upper limit on the spin-independent WIMP-nucleon cross section of
 $8.8 \times 10^{-8}$ pb for a WIMP mass of 100 GeV/$c^2$ \cite{Angle07},
 a factor of 2.3 lower than the limit achieved by CDMS-II experiment
 (see Subsec.~\ref{CDMS}).
 For a WIMP mass of 30 GeV/$c^2$, the limit is $4.5 \times 10^{-8}$ pb \cite{Angle07}
 (see Fig.~\ref{fig3080601}).

 The XENON10 experiment will be upgraded to $\sim$ 100 kg \cite{Gascon05}
 and a WIMP-nucleon sensitivity of $2 \times 10^{-8}$ pb in 2008 \cite{Baudis07}
 (see Fig.~\ref{fig3100001}).
\subsection{XMASS}
\label{XMASS}
 ~~~$\,$
 The Xenon Neutrino MASS Detector (XMASS) experiment
 uses a 100 kg (intend to ultimately 800 kg) \cite{Drees06}
 single-phase Xe detector \cite{Schnee06}
 at the SuperKamiokande site in Japan.

 Because of their 100 kg total mass of target material,
 XMASS has a good position sensitivity and
 has demonstrated the self-shielding effect (see Subsec.~\ref{surfaceevents})
 to reduce the background events
 induced by multiple scattering (see Subsec.~\ref{multiplescatter})
 and surface contamination \cite{Drees06}.

 Besides the XMASS experiment with liquid Xe,
 CLEAN (Cryogenic Low-Energy Astrophysics with Neon)
 and DEAP are also single-phase experiments.
 They use Ar or Ne as detector material
 in oder to take advantage of the much larger timing difference
 between nuclear recoils and electron recoils described above
 \cite{Schnee06}, \cite{Spooner07}.
\subsection{ZEPLIN}
\label{ZEPLIN}
 ~~~$\,$
 The Zoned Proportional Scintillation in Liquid Noble Gases (ZEPLIN) collaboration \cite{UKDMC}
 first used a liquid xenon scintillation detector
 in the Boulby Laboratory (1070 m underground) in the United Kingdom.
%
\begin{figure}[t]
\begin{center}
\imageswitch{
\begin{picture}(15,12.2)
\put(0  , 0  ){\framebox(15,12.2){}}
\put(2.5, 2.2){\framebox(10,10  ){}}
\put(2.5, 0  ){\framebox(10, 2.1){}}
\end{picture}}
{\includegraphics[width=10cm]{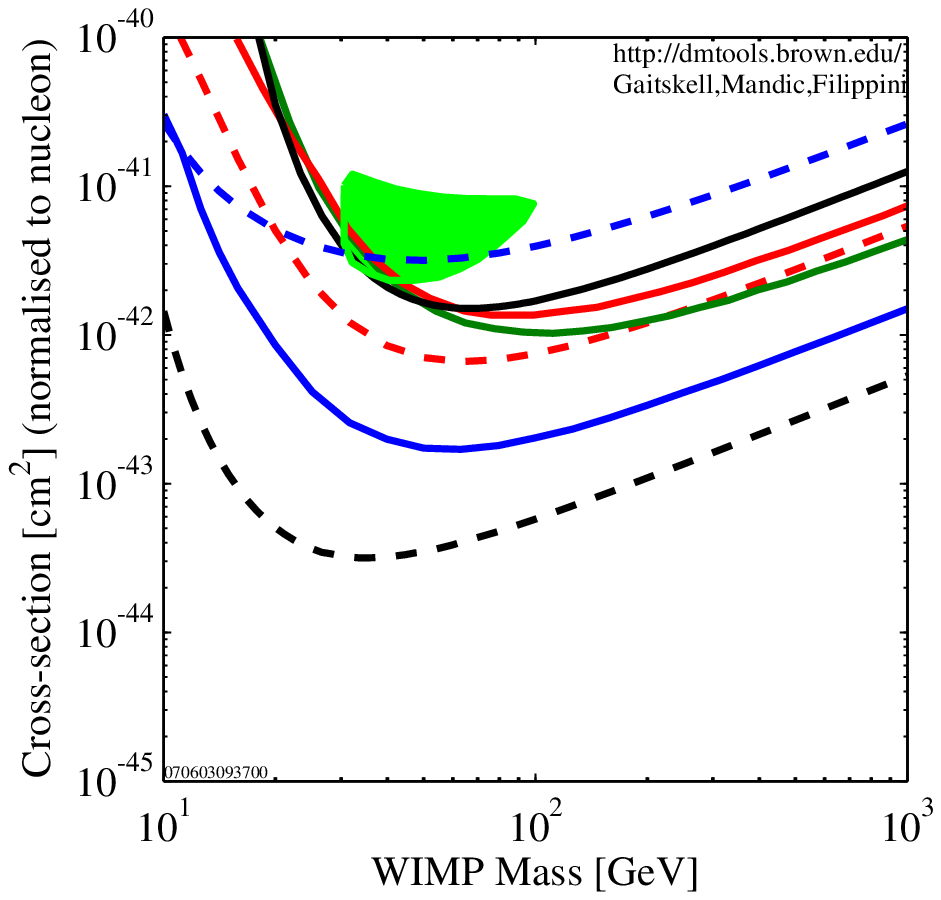}  \\
 \includegraphics[width=10cm]{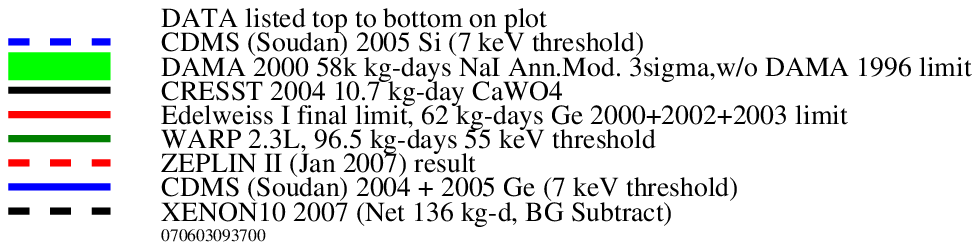}}
\end{center}
\caption{
 The curves show the sensitivities of (the exclusion limits on)
 the spin-independent WIMP-nucleon cross section versus WIMP mass
 achieved by
 the CDMS       \cite{Akerib05b}
 (the blue solid and the blue dashed lines, for Ge and Si, respectively),
 the CRESST     \cite{Angloher04} (the black solid line),
 the DAMA       \cite{Bernabei00} (the green area),
 the EDELWEISS  \cite{Sanglard05} (the red  solid line),
 the WARP       \cite{Benetti07}  (the dark green solid line),
 the XENON      \cite{Angle07}    (the black dashed line),
 and the ZEPLIN \cite{Alner07}    (the red dashed line) collaborations
 (plot generated by {\tt http://dmtools.berkeley.edu/limitplots/}).}
\label{fig3080601}
\end{figure}

 ZEPLIN-I was a single-phase experiment with 6 kg (3.1 kg fiducial) Xe,
 so it could not collect the ionization signals,
 but depended solely on the pulse shape discrimination \cite{Schnee06}.
 In an exposure of 293 kg-day,
 {\em no excess consistent with nuclear recoils} was seen \cite{Schnee06}.
 However,
 the published limits are somewhat controversial,
 because their calibration results of the neutron recoil discrimination
 do not appear to be convincing enough to consider the limits set on the WIMP signal
 to be as reliable as the ones set by the cryogenic experiments \cite{Drees06}.
 Actually,
 with only 1.5 photo-electron per keV and a three-fold coincidence,
 searching for the WIMP signal in the 2-10 $\keVee$ region is for ZEPLIN-I quite challenging
 \cite{Drees06}.

 ZEPLIN-II has been upgraded to a dual-phase experiment with 31 kg Xe \cite{Alner07}.
 In the first run with 225 kg-day exposure,
 {\em 29 events} have been observed in an acceptance window
 defined between 5 $\keVee$ and 20 $\keVee$.
 With a summed expectation of 28.6 $\pm$ 4.3
 $\gamma$-ray and radon progeny induced background events,
 these figures provide
 a 90\% C.~L.~upper limit to the number of nuclear recoils of {\em 10.4 events}
 in this acceptance window,
 which converts to a spin-independent WIMP-nucleon cross-section
 of $6.6 \times 10^{-7}$ pb
 for a WIMP mass of 65 GeV$/c^2$ (see Fig.~\ref{fig3080601}).
 For the second run a sensitivity of $2 \times 10^{-7}$ pb has been expected
 (see Fig.~\ref{fig3100001}).
%
%
\section{Superheated droplet detectors (SDD)}
\label{SDD}
 ~~~$\,$
 Metastable liquid droplets immersed in a gel expand (explode)
 due to a phase transition to the gaseous phase,
 when a particle or nucleus with sufficiently high energy density
 (energy deposited over unit length, $dQ/dx$)
 interacts in the liquid \cite{Ramachers02}.
 A broad range of detector materials could be used and inexpensively scaled to large masses
 \cite{Schnee06}.

 The main (best) advantage of such integrating detectors is that,
 by tuning pressure and temperature,
 the threshold energy of the detector can be adjusted so that
 the detector could be insensitive to the low energy density \cite{Schnee06}.
 Thus electron recoils and $\alpha$-radiation events can be rejected automatically.
\subsection{DRIFT}
\label{DRIFT}
 ~~~$\,$
 The Directional Recoil Identification From Tracks (DRIFT) experiment \cite{UKDMC}
 uses a low pressure Xe-$\rm CS_2$ gas mixture TPC \cite{Jesus04}.
 Using the negative $\rm CS_2$ ions instead of usual $e^{-}$ as charge carriers
 can reduce the diffusion and thus achieve millimetric track resolution \cite{Jesus04}.
 The ionization tracks will be measured with a multi-wire proportional chamber
 in a low-pressure gas \cite{Ramachers02},
 this offers the most convincing technique
 to measure the direction of nuclear recoils \cite{Drees06}, \cite{Spooner07}.

 However,
 one disadvantage of the DRIFT's detector is
 the very low target mass and/or the need of a huge detector volume.
\subsection{MIMAC-He3}
\label{MIMAC}
 ~~~$\,$
 The MIcro-tpc (temporal projection chambers) MAtrix
 of Chambers of Helium 3 (MIMAC-He3) experiment \cite{MIMAC}
 uses an ultra cold pure $\rmXA{He}{3}$ detector \cite{Drees06}.
 The use of $\rmXA{He}{3}$ as target material is motivated
 by its privileged features for Dark Matter search
 compared with other target nuclei.
 First,
 $\rmXA{He}{3}$ being a spin-$\frac{1}{2}$ nucleus with a single neutron,
 a detector made of such material will be
 sensitive to the ``neutron spin-dependent'' interaction,
 leading to a natural complementarity to most existing or planned Dark Matter detectors
 as well as proton based spin-dependant detectors \cite{Santos07}.
 Moreover,
 the mass of the $\rmXA{He}{3}$ atom is 2.81 GeV/$c^2$,
 the recoil energy range is expected less than 10 keV \cite{Moulin05}.
 Hence,
 MIMAC-He3 could be used to measure events with low recoil energies
 as well as detect low-mass WIMPs
 ($6~{\rm GeV}/c^2 \le \mchi \le 40~{\rm GeV}/c^2$ \cite{Santos07}).

 There are several more advantages for a detector using $\rmXA{He}{3}$
 \cite{Moulin05}, \cite{Santos07}:
 first,
 there are no intrinsic x-rays;
 second,
 a very low Compton cross section to $\gamma$-ray
 (two orders of magnitude weaker than in Ge)
 can reduce the natural radioactive background by several orders of magnitude;
 third,
 the neutron signature due to the capture process:
 n + $\rmXA{He}{3}$ $\to$ p + $\rmXA{H}{3}$ + 764 keV,
 will be very useful for discrimination of neutron background.
 On the other hand,
 the double detection of the ionization energy
 and the track projection by means of the TPC chambers
 can assure the electron-recoil discrimination.

 However,
 similar to the DRIFT experiment,
 one disadvantage of their detector is
 the very low target mass and/or the need of a huge detector volume.
\subsection{PICASSO}
\label{PICASSO}
 ~~~$\,$
 The Project In CAnada to Search for Supersymmetric Objects (PICASSO) experiment
 in the Sudbury Neutron Observatory (SNO) in Canada
 uses $\rmXA{F}{19}$ (spin-$\frac{1}{2}$ isotope) as detector material
 and search for spin-dependent Dark Matter particles 
 (see Subsec.~\ref{SDcrosssection})\cite{Jesus04}.

 The principle of the PICASSO detector is as follows:
 Small superheated freon droplets imbedded in a gel matrix at room temperature.
 The nuclear recoil of $\rmXA{F}{19}$ induces the explosion of a droplet,
 creating an acoustic shock wave which will be measured with piezoelectric transducers
 \cite{Jesus04}.
\subsection{SIMPLE}
\label{SIMPLE}
 ~~~$\,$
 The Superheated Instrument for Massive ParticLe Experiments (SIMPLE) collaboration
 in the Laboratoire Souterrain \`a Bas Bruit (LSBB, $\sim$ 1500 m.w.e.) in Frence
 uses $\rm C_2 Cl F_5$ and $\rm C F_3 I$
 and searches also for spin-dependent Dark Matter interaction
 \cite{Girard05a}, \cite{Morlet07}.
 Their results exclude the spin-dependent WIMP-proton cross section above 1.14 pb
 for a WIMP mass of 50 GeV$/c^2$ \cite{Girard05b}
 (an upper limit of 0.2 pb on the spin-dependent WIMP-neutron cross section
  has been given by the CDMS collaboration,
  see Subsec.~\ref{CDMS}).
\section{Prospects}
\label{prospects}
 ~~~$\,$
 So far we did not obtain any convincing signal
 from experiments searching for Dark Matter particles.
 In the future,
 detector technique,
 better sensitivities
 as well as better background discrimination,
 will be improved.
 We need also some new ideas for detector building
 as well as application of experimental data.

\begin{figure}[t]
\begin{center}
\imageswitch{
\begin{picture}(15,12.4)
\put(0  , 0  ){\framebox(15,12.4){}}
\put(2.5, 2.4){\framebox(10,10  ){}}
\put(2.5, 0  ){\framebox(10, 2.3){}}
\end{picture}}
{\includegraphics[width=10cm]{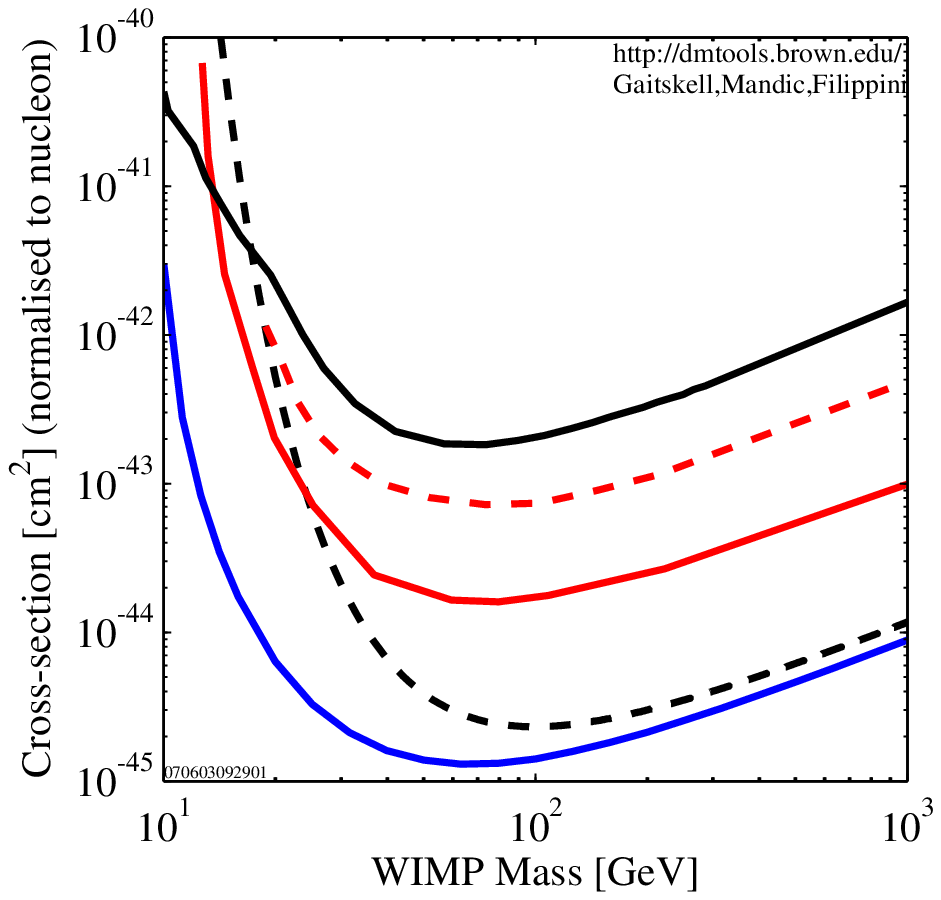}  \\
 \includegraphics[width=10cm]{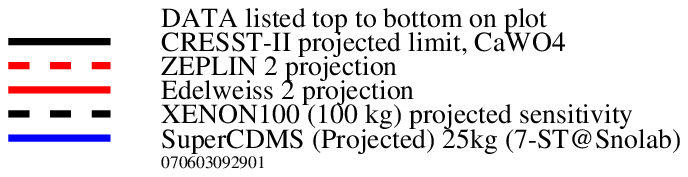}}
\end{center}
\caption{
 The curves show the sensitivities of
 the spin-independent WIMP-nucleon cross section versus WIMP mass
 projected by
 the SuperCDMS (25 kg),
 the CRESST-II (with $\rm Ca W O_4$),
 the EDELWEISS-II,
 the XENON100 (100 kg),
 and the ZEPLIN-II experiments.
 The indications of the lines as in Fig.~\ref{fig3080601}
 (plot generated by {\tt http://dmtools.berkeley.edu/limitplots/}).}
\label{fig3100001}
\end{figure}

 As described in Subsec.~\ref{CDMS},
 the CDMS collaboration has achieved
 a (so far the best) sensitivity of $\sim 10^{-7}$ pb
 for spin-independent WIMP-nucleon cross section
 and of $\sim 10^{-1}$ pb for spin-dependent WIMP-neutron cross section.
 Together with the other collaborations described above,
 direct WIMP detection experiments have started to probe
 some possible regions in the parameter space predicted by some supersymmetric models.
 For next-generation detectors,
 sensitivities will be upgraded down to $\sim 10^{-8}$ pb
 (see Fig.~\ref{fig3100001}) and,
 in long term, even $\sim 10^{-10}$ pb,
 and the corresponding WIMP-nucleus scattering event rate is then
 {\em $\sim$ 5 events/ton/yr} for Ge \cite{Drees06},
 as needed to probe large regions of MSSM parameter space.
 The total mass of detector material will also be improved to the {\em ton scale}.
 For example,
 the $\sim$ 100 kg Xe detector of the XENON and XMASS collaborations.
 The CDMS collaboration is also preparing for
 their SuperCDMS projects with maximum 1100 kg target mass,
 while the CRESST and the EDELWEISS collaborations
 will also build to a larger collaboration ``EURECA''
 (European Underground Rare Event search with Calorimeter Array).
%

 A nearly perfect background discrimination capability
 for next-generation detectors
 is also necessary.
 The ultimate neutron background will only be identified
 by the multiple interactions
 in a finely segmented or multiple interaction sensitive detector,
 and/or by operating detectors containing
 different target materials within the same setup
 \cite{Drees06}.

 Furthermore,
 the measurement of the recoil directions of the target nuclei
 would provide additional information
 on the distribution of WIMPs in our Galaxy \cite{Schnee06}.
 As described in Subsec.~\ref{DRIFT},
 the most convincing way for determining the recoil direction
 is by drifting negative ions in a temporal projection chamber
 to accurately record the tiny recoil distance.
 The DRIFT experiment has provided a proof of the principle,
 but it remains to be seen
 if such gas detectors with enough target material can detect some signals
 \cite{Schnee06}.

 By the way,
 in order to present the WIMP-nucleon cross sections
 and the detector sensitivities in the future
 more conveniently and also suitably,
 we may consider to use ``zepto'' ($10^{-21}$) or even ``yocto'' ($10^{-24}$)
 \cite{RPP06SI} barn
 instead of $10^{-9}$ pb ($10^{-45}~{\rm cm^2}$)
 or $10^{-12}$ pb ($10^{-48}~{\rm cm^2}$).

%% file: Doktorarbeit-Ch4.tex
\chapter{Reconstruction of the Velocity Distribution of WIMPs}
 ~~~$\,$
 So far most theoretical analyses of direct detection of halo WIMPs,
 as discussed in Subsecs.~\ref{fGau} and \ref{fsh},
 have predicted the detection rate for a given (class of) WIMPs,
 based on a specific model of the galactic halo
 (e.g., \cite{Kamionkowski98}, \cite{Green01}, and \cite{Ling04}).
 The goal of my work is to invert this process.
 That is,
 I wish to study,
 as {\em model-independently} as possible,
 {\em what future direct detection experiments can teach us about the WIMP halo}.

 In this chapter I will use a {\em time-averaged recoil spectrum},
 and assume that {\em no directional information exists}.
 One can thus only reconstruct the (time-averaged)
 {\em one-dimensional velocity distribution}, $f_1(v)$.
 In the first section
 I will show how to derive the velocity distribution of WIMPs
 from the functional form of the recoil spectrum;
 the assumption here is that
 this functional form has been determined by fitting the data of some (future) experiment(s).
 I will also derive formulae for moments of the velocity distribution function,
 such as the mean velocity and the velocity dispersion of WIMPs,
 which can be compared with model predictions.

 Then I will present the method for
 reconstructing the velocity distribution of WIMPs
 {\em directly from recorded signal events}.
 This allows statistically meaningful tests
 of predicted distribution functions.
 I will also discuss
 how to estimate the moments of the velocity distribution directly from these data.

 Finally,
 I will show how to determine the mass of halo WIMPs,
 which one needs for the reconstruction of (the moments of) the velocity distribution,
 from experimental data directly.
 This allows also a useful comparison of the detected WIMPs
 with the new particle(s) produced at colliders,
 e.g., the Large Hadron Collider (LHC).
\section{From the scattering spectrum}
\label{dRdQ}
 ~~~$\,$
 In this section
 I will start with the differential rate for elastic WIMP-nuclues scattering
 given in Eq.(\ref{eqn3010108})
\cheqnref{eqn3010108}
\beq
   \dRdQ
 = \calA \FQ \intvmin \bfrac{f_1(v)}{v} dv
\~,
\eeq
\cheqnCN{-1}
 and show how to
 find an expression for the one-dimensional velocity distribution function $f_1(v)$
 for given (as yet only hypothetical) measured recoil spectrum $dR/dQ$.
 To that end,
 I first define
\beq
   \Dd{F_1(v)}{v}
 = \frac{f_1(v)}{v}
\~,
\label{eqn4010001}
\eeq
 i.e., $F_1(v)$ is the primitive of $f_1(v)/v$.
 Then Eq.(\ref{eqn3010108}) can be rewritten as
\beq
   \frac{1}{\calA \FQ} \adRdQ
 = \intvmin \bfrac{f_1(v)}{v} dv
 = F_1(v \to \infty)-F_1(\vmin)
\~.
\label{eqn4010002}
\eeq
 Since WIMPs (as candidate for CDM) in today's Universe move quite slowly,
 $f_1(v)$ must vanish as $v$ approaches infinity, 
\beq
     f_1(v \to \infty)
 \to 0
\~.
\label{eqn4010003}
\eeq
 Hence
\beq
     \Dd{F_1(v)}{v} \Bigg|_{v \to \infty}
 =   \frac{f_1(v)}{v} \Bigg|_{v \to \infty}
 \to 0
\~.
\label{eqn4010004}
\eeq
 This means that $F_1(v \to \infty)$ approaches a finite value.
\footnote{
 The other properties of $F_1(v)$ will be discussed in App.~\ref{F1(v)}.}
 Differentiating both sides of Eq.(\ref{eqn4010002}) with respect to $\vmin$
 and using Eq.(\ref{eqn3010106}),
 one can find that
 (detailed derivations will be given in App.~\ref{f1(v)})
\beqn
    \Dd{F_1(\vmin)}{\vmin}
 \=-\frac{1}{\calA} \cbrac{\dd{\vmin} \bdRdQoFQ\Qvmina} 
    \non\\
    \non\\
 \= \frac{1}{\vmin} \cdot \frac{1}{\calA} \cbrac{-2 Q \cdot \ddRdQoFQdQ}\Qvmina
\~.
\label{eqn4010005}
\eeqn
 Since this expression holds for arbitrary $\vmin$,
 one can write down the following result directly: 
\beq
   \frac{f_1(v)}{v}
 = \Dd{F_1(v)}{v}
 = \frac{1}{v} \cdot \frac{1}{\calA} \cbrac{-2 Q \cdot \ddRdQoFQdQ}\Qva
\~.
\label{eqn4010006}
\eeq
 The right-hand side of Eq.(\ref{eqn4010006})
 depends on the as yet unknown constant $\calA$.
 Recall,
 however,
 that $f_1(v)$ is the {\em normalized} velocity distribution,
 i.e.,
 it is defined to satisfy
\beq
   \intz f_1(v) \~ dv
 = 1
\~.
\label{eqn4010007}
\eeq
 Therefore,
 the normalized one-dimensional velocity distribution function can be expressed as 
\beq
   f_1(v)
 = \calN \cbrac{-2 Q \cdot \ddRdQoFQdQ}\Qva
\~,
\label{eqn4010008}
\eeq
 with the normalization constant $\calN$
 (a detailed derivation will be given in App.~\ref{calN})
\beq
   \calN
 = \frac{2}{\alpha} \cbrac{\intz \frac{1}{\sqrt{Q}} \bdRdQoFQ dQ}^{-1}
\~.
\label{eqn4010009}
\eeq
 Note that the integral here starts at $Q = 0$.

 In the next step I want to compute the moments of the velocity distribution function:
\beq
   \expv{v^n}
 = \int_{\vmin(\Qthre)}^\infty v^n f_1(v) \~ dv
\~.
\label{eqn4010010}
\eeq
 Here I have introduced a threshold energy $\Qthre$.
 This is needed experimentally,
 since at very low recoil energies,
 the signal is swamped by electronic noise.
 Moreover,
 later we will meet expressions that (formally) diverge as $Q \to 0$.
 $\vmin(\Qthre)$ is calculated as in Eq.(\ref{eqn3010106}).
 Inserting Eq.(\ref{eqn4010008}) into Eq.(\ref{eqn4010010}) and integrating by parts,
 one can find that
 (a detailed derivation will be given in App.~\ref{calN})
\beq
   \expv{v^n}
 = \calN \afrac{\alpha^{n+1}}{2}
   \bbrac{\frac{2 \~ \Qthre^{(n+1)/2}}{\FQthre} \afrac{dR}{dQ}_{Q = \Qthre}+(n+1) I_n(\Qthre)}
\~,
\label{eqn4010011}
\eeq
 with
\beq
   I_n(\Qthre)
 = \int_{\Qthre}^\infty Q^{(n-1)/2} \bdRdQoFQ \~ dQ
\~.
\label{eqn4010012}
\eeq
 Physically,
 $\expv{v}$ is the average WIMP velocity,
 while $\expv{v^2}$ gives the velocity dispersion.
\footnote{
 The dispersion of the function $f_1(v)$ in the statistical sense
 is given by $\expv{v^2}-\expv{v}^2$.}
 One emphasis here is that
 {\em Eqs.(\ref{eqn4010011}) and (\ref{eqn4010012}) can be evaluated directly
 once the recoil spectrum is known;
 knowledge of the functional form of $f_1(v)$ in Eq.(\ref{eqn4010008}) is not required}.
 Note that the first term in Eq.(\ref{eqn4010011}) vanishes for $n \geq 0$ if $\Qthre \to 0$.
 In the same limit,
 $\expv{v^0} \to \calN \alpha I_0(0)/2 \to 1$ by virtue of Eq.(\ref{eqn4010009}).
 On the other hand,
 as written in Eqs.(\ref{eqn4010008}) and (\ref{eqn4010009}),
 the velocity distribution integrated over the experimentally accessible range of WIMP velocities
 gives a value smaller than unity.
 Using only quantities
 that can be measured in the presence of a non-vanishing energy threshold $\Qthre$,
 Eq.(\ref{eqn4010009}) can be replaced by
\beq
   \calN(\Qthre)
 = \frac {2}{\alpha}
   \bbrac{\frac{2 \~ \Qthre^{1/2}}{\FQthre} \afrac{dR}{dQ}_{Q = \Qthre}+I_0(\Qthre)}^{-1}
\~.
\label{eqn4010013}
\eeq
 Using $\calN(\Qthre)$ in Eq.(\ref{eqn4010008}) ensures that
 the velocity distribution integrated over $v \geq \vmin(\Qthre)$ gives unity.

 From Eqs.(\ref{eqn4010008}), (\ref{eqn4010009}), and (\ref{eqn4010011}) to (\ref{eqn4010013}),
 it can be found that 
 the (unrealistic) ``reduced'' spectrum
 (i.e., the recoil spectrum divided by the squared nuclear form factor)
 is more useful than the recoil spectrum itself.
 Meanwhile,
 note that the reduced spectrum from $(dR/dQ)_\Gau$ in Eq.(\ref{eqn3010302}):
\cheqnrefp{eqn3010302}
\beq
         \frac{1}{\FQ} \adRdQ_\Gau
 \propto e^{-\alpha^2 Q/v_0^2}
\label{eqn3010302p}
\eeq
 is exactly exponential.
 This remains approximately true for the potentially quasi-realistic spectrum
 in Eq.(\ref{eqn3020103}) as well:
\cheqnrefp{eqn3020103}
\beq
         \frac{1}{\FQ} \adRdQ_\sh
 \propto \erf{\T \afrac{\alpha \sqrt{Q}+v_e}{v_0}}-\erf{\T \afrac{\alpha \sqrt{Q}-v_e}{v_0}}
\~.
\label{eqn3020103p}
\eeq
\cheqnCN{-2}
 In order to test these formulae,
 I have substituted the spectra in Eqs.(\ref{eqn3010302}') and (\ref{eqn3020103}')
 into Eqs.(\ref{eqn4010008}) and (\ref{eqn4010011}),
 taking $\Qthre = 0$.
 They reproduced the normalized velocity distribution functions
 in Eqs.(\ref{eqn3010301}) and (\ref{eqn3020101}),
 as well as the results in Eqs.(\ref{eqn3010303}), (\ref{eqn3010304}),
 (\ref{eqn3020104}), and (\ref{eqn3020105}).
 The detailed calculations will be given in Apps.~\ref{vGaudRdQ} and \ref{vshdRdQ},
 respectively.

 One emphasis here is that
 {\em the final results in Eqs.(\ref{eqn4010008}) and (\ref{eqn4010011})
 are independent of the as yet unknown quantity $\calA$ defined in Eq.(\ref{eqn3010109})}.
 They do,
 however,
 depend on the WIMP mass $\mchi$ through the coefficient $\alpha$
 defined in Eq.(\ref{eqn3010107}).
 This mass can be extracted from a {\em single} recoil spectrum
 {\em only if} one makes some assumptions about the velocity distribution $f_1(v)$.
 In the kind of model-independent analysis pursued here,
 $\mchi$ can be determined by requiring that
 the recoil spectra {\em using two different target nuclei}
 lead to the same moments of the velocity distribution, $\expv{v^n}$,
 through Eq.(\ref{eqn4010011}).
 This method will be discussed in Sec.~\ref{mchi}.
 Note that this can be done {\em independent of the detailed particle physics model},
 which determine the value of $\sigma_0$ for two target nuclei.
 But,
 one will need to know both form factors of the target nuclei,
 which strongly depend on whether spin-dependent or spin-independent interactions dominate
 (see Subsecs.~\ref{SIcrosssection} to \ref{SISDcrosssection}).
 On the other hand,
 within a given particle physics model,
 the best determination of $\mchi$ should eventually come from experiments
 at high-energy colliders.
 However,
 we need also an alternative method as cross-check to prove
 whether the particle produced at colliders is the same particle
 detected by the direct WIMP detection.
\section{From experimental data directly}
\label{Qni}
 ~~~$\,$
 In the previous section
 I have derived formulae for
 the normalized one-dimensional velocity distribution function of WIMPs, $f_1(v)$,
 and for its moments $\expv{v^n}$,
 from the recoil-energy spectrum, $dR/dQ$.
 In order to use these expressions,
 one would need a functional form for $dR/dQ$.
 In practice this might result from a fit to experimental data.
 Note that the expression for $f_1(v)$ in Eq.(\ref{eqn4010008})
 requires knowledge not only of $dR/dQ$,
 but also of its derivative with respect to $Q$,
 i.e., we need to know both the spectrum and its slope.
 This will complicate the error analysis considerably,
 if $dR/dQ$ is the result of a fit.

 In this section I therefore go one step further,
 and derive expressions that allow to reconstruct $f_1(v)$ and its moments
 {\em directly from the data}.
 The assumption we have to make is that
 the sample to be analyzed only contains signal events,
 i.e., is free of background.
 This should be possible for the modern (next-generation) detectors
 (detailed discussions about the background and its discrimination
  have been given in Sec.~\ref{background}).
 Having a sample of pure signal events,
 we can proceed with a complete statistical analysis of the precision
 with which we can reconstruct $f_1(v)$ and its moments.

 However,
 in the absence of a true experimental sample of this kind,
 I had to resort to Monte Carlo experiments with an unweighted event generator
 written by M.~Drees.
 Since detectors without directional sensitivity have been assumed,
 a single event is uniquely characterized by the measured recoil energy $Q$.
 Existing experiments such as CDMS \cite{CDMS} and CRESST \cite{CRESST}
 can determine the recoil energy quite accurately
 (some details about their detectors and experiments
  have been given in Subsecs.~\ref{CDMS} and \ref{CRESST}).
 We will see later that the statistical errors
 on the reconstructed velocity distribution $f_1(v)$
 will be quite sizable even for next-generation experiments,
 given existing bounds on the scattering rate.
 It should therefore be a good approximation
 to ignore the error of $Q$ in the analyses.

 In the following
 I do not distinguish between the recoil spectrum $dR/dQ$
 and the actual differential counting rate $dN/dQ$.
 Since $dR/dQ$ is usually given per unit detector mass and unit time,
 the two quantities differ only by a multiplicative constant.
 This constant can be canceled in Eq.(\ref{eqn4010008}),
 since it will also appear in the normalization constant $\calN$
 in Eq.(\ref{eqn4010009}).
\subsection{Exponential ansatz for $dR/dQ$}
\label{dRdQkn}
 ~~~$\,$
 I divide the total energy range into $B$ bins with central points $Q_n$ and widths $b_n$,
 $n = 1,~2,~\cdots,~B$.
 In each bin,
 $N_n$ signal events will be recorded.
 The simulated data set can therefore be described by
\beq
     {\T Q_n-\frac{b_n}{2}}
 \le \Qni
 \le {\T Q_n+\frac{b_n}{2}}
\~,
     ~~~~~~~~~~~~ 
     i
 =   1,~2,~\cdots,~N_n
,
     ~
     n
 =   1,~2,~\cdots,~B
.
\label{eqn4020101}
\eeq
 The standard estimate for $dR/dQ$ at $Q=Q_n$ is then
\beq
   r_n
 = \frac{N_n}{b_n}
\~,
   ~~~~~~~~~~~~~~~~~~~ 
   n
 = 1,~2,~\cdots,~B
.
\label{eqn4020102}
\eeq
 The squared statistical error on $dR/dQ$ is accordingly
\beq
   \sigma^2(r_n)
 = \frac{N_n}{b_n^2}
\~,
\label{eqn4020103}
\eeq
 since
\beq
   \sigma^2(N_n)
 = N_n
\~.
\label{eqn4020104}
\eeq
\begin{figure}[t]
\begin{center}
\imageswitch{
\begin{picture}(14,10.1)
\put(0,0){\framebox(14,10.1){}}
\end{picture}}
{\includegraphics[width=13.5cm]{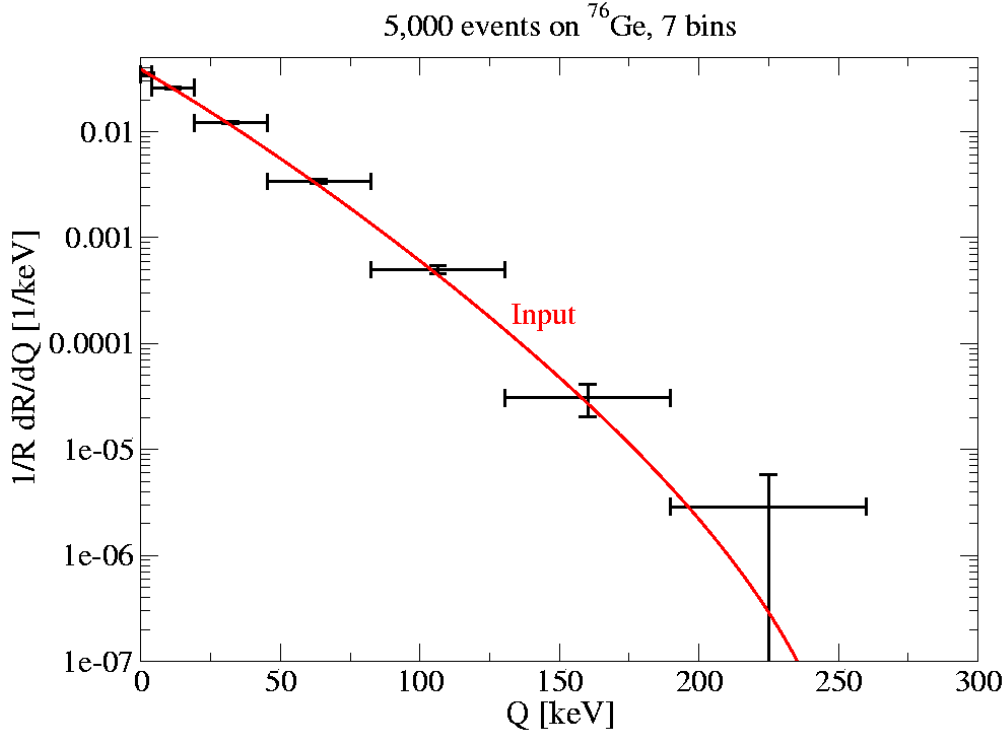}}
\end{center}
\caption{
 The curve shows the theoretical predicted recoil energy spectrum
 for the shifted Maxwellian velocity distribution
 $f_{1,\sh}(v,v_e)$ in Eq.(\ref{eqn3020101})
 with the Woods-Saxon form factor $F_{\rm WS}^2(Q)$ in Eq.(\ref{eqn3010202}).
 The data points with error bars show
 simulated experimental data produced from this spectrum
 (5,000 total events,
  $\mchi = 100~{\rm GeV}/c^2$, $\mN = 70.6~{\rm GeV}/c^2$ for $\rmXA{Ge}{76}$,
  $v_0 = 220$ km/s, $v_e = 231$ km/s,
 and the Galactic escape velocity $v_{\rm esc} = 700$ km/s
 as the cut-off velocity of the velocity distribution in Eq.(\ref{eqn3010108})).
 The vertical error bars show the statistical uncertainties of the measurements,
 while the horizontal error bars indicate the bin widths.} 
\label{fig4020101}
\end{figure}

 As mentioned at the end of the previous section,
 the predicted recoil spectrum resembles a falling exponential
 (see Eqs.(\ref{eqn3010302}') and (\ref{eqn3020103}') in Sec.~\ref{dRdQ}).
 This is confirmed in Fig.~\ref{fig4020101},
 which shows the predicted recoil spectrum of a 100 GeV$/c^2$ WIMP on $\rmXA{Ge}{76}$
 by means of the shifted Maxwellian velocity distribution $f_{1,\sh}(v,v_e)$
 in Eq.(\ref{eqn3020101})
 and the Woods-Saxon form factor $F_{\rm WS}^2(Q)$ in Eq.(\ref{eqn3010202}).
 This figure also shows the result of a simulated experiment,
 where the exposure time and cross section are chosen such that
 the expected number of events is 5,000;
 these have been collected in seven bins in recoil energy.
 Note that,
 in practice the velocity distribution in Eq.(\ref{eqn3010108})
 should be cut off at a velocity $v_{\rm esc}$,
 since WIMPs with $v > v_{\rm esc}$ can escape the gravitational well of our Galaxy.
 The cut-off velocity has been taken to be $v_{\rm esc} = 700$ km/s.

 While an approximately exponential function can be approximated
 by a linear ansatz only over a narrow range,
 i.e., for small bin widths,
 the {\em logarithm} of this function can be approximated
 by a linear ansatz for much wider bins
 (some detailed discussions about linear approximations
  can be found in App.~\ref{lineardRdQn}).
 This corresponds to the ansatz
\beq
        \adRdQ_n
 \equiv \adRdQ_{Q \simeq Q_n}
 \simeq \trn \~ e^{k_n (Q-Q_n)}
 \equiv r_n  \~ e^{k_n (Q-Q_{s,n})}
\~.
\label{eqn4020105}
\eeq
 Here $r_n$ is the standard estimator for $dR/dQ$ at $Q = Q_n$ given in Eq.(\ref{eqn4020102}),
 $\trn$ is the real value of the recoil spectrum at the point $Q = Q_n$,
\beq
        \trn
 \equiv \adRdQ_{Q = Q_n}
\~,
\label{eqn4020106}
\eeq
 and $k_n$ is the {\em logarithmic slope} of the recoil spectrum in the $n$-th $Q$-bin.

 Now our task is to find estimators for $\trn$ and $k_n$ in Eq.(\ref{eqn4020105})
 using (simulated) data.
 Note that,
 for $k_n \neq 0$,
 $\trn \neq r_n = N_n/b_n$ and cannot be estimated from the number of events $N_n$
 in the $n$-th bin alone.
 Instead,
 from the first part of Eq.(\ref{eqn4020105}),
 one has
\beq
   N_n
 = \intQnbn \adRdQ_n dQ
 = \intQnbn \trn \~ e^{k_n (Q-Q_n)} dQ
 = b_n \trn \afrac{\sinh\kappa_n}{\kappa_n}
\~, 
\label{eqn4020107}
\eeq
 where,
 for simplicity,
 I have introduced the dimensionless quantities
\beq
        \kappa_n
 \equiv \frac{b_n k_n}{2}
\~.
\label{eqn4020108}
\eeq
 Hence,
 it can be found that
\beq
   \trn
 = \frac{N_n}{b_n} \afrac{\kappa_n}{\sinh\kappa_n}
\label{eqn4020109}
\eeq
 depends on $k_n$.
 Moreover,
 using the first and second moments of the recoil spectrum in the $n$-th bin,
 one can find that
\beq
   \bQn
 = \frac{1}{N_n} \intQnbn (Q-Q_n) \adRdQ_n dQ
 = \frac{b_n}{2} \abrac{\coth\kappa_n-\frac{1}{\kappa_n}}
\~,
\label{eqn4020110}
\eeq
 and
\beqn
    \bQQn
 \= \frac{1}{N_n} \intQnbn (Q-Q_n)^2 \adRdQ_n dQ
    \non\\
 \= \afrac{b_n}{2}^2 \bbrac{1-2 \afrac{\coth\kappa_n}{\kappa_n}+\frac{2}{\kappa_n^2}}
\~,
\label{eqn4020111}
\eeqn
 where $\Bar{\cdots}|_n$ denotes the average value in the $n$-th bin.
 $\kappa_n$, or, equivalently, $k_n$ can not be solved analytically
 by only using Eq.(\ref{eqn4020110}).
 They can,
 however,
 be found numerically once
\beq
   \bQn
 = \frac{1}{N_n} \sumiNn \abrac{\Qni-Q_n}
\label{eqn4020112}
\eeq
 is known.
 Alternatively,
 multiplying both sides of Eq.(\ref{eqn4020110}) with $b_n/\kappa_n$
 and adding to Eq.(\ref{eqn4020110}),
 one can calculate the logarithmic slopes as
\beq
   k_n
 = \frac{8 \~ \bQn}{b_n^2-4 \~ \bQQn}
\~,
\label{eqn4020113}
\eeq
 where
\beq
   \bQQn
 = \frac{1}{N_n} \sumiNn \abrac{\Qni-Q_n}^2
\label{eqn4020114}
\eeq
 is now estimated from the data directly.
 Note that $k_n$ determined either
 from Eq.(\ref{eqn4020110}) or from Eq.(\ref{eqn4020113})
 is independent of the normalization $r_n$ or $\trn$.

 On the other hand,
 the second, equivalent expression in Eq.(\ref{eqn4020105}) means that
 we can still use the quantities $r_n = N_n/b_n$ as normalization.
 However,
 the logarithmic slopes, $k_n$, solved by
 either Eq.(\ref{eqn4020110}) or Eq.(\ref{eqn4020113})
 describes the spectrum $dR/dQ$ at the {\em shifted points} $Q_{s,n}$.
 Equivalence of the two expressions in Eq.(\ref{eqn4020105}) implies
\beq
   Q_{s,n}
 = Q_n+\frac{1}{k_n} \ln\afrac{\sinh\kappa_n}{\kappa_n}
\~.
\label{eqn4020115}
\eeq
 Note that,
 while $Q_n$ is simply the midpoint of the $n$-th $Q$-bin
 and can thus be chosen at will,
 $Q_{s,n}$ here is a derived quantity and
 depends on the logarithmic slope $k_n$.
 However,
 the second expression in Eq.(\ref{eqn4020105})
 combined with $Q_{s,n}$ in Eq.(\ref{eqn4020115}) has two advantages.
 First,
 the prefactor $r_n$ can be computed more easily
 than $\trn$ in Eq.(\ref{eqn4020109}).
 Second,
 according to a detailed analysis \cite{DMDD},
 it has been found that,
 for a given bin width,
 one can minimize the leading systematic error
 by interpreting the estimator of $k_n$ as logarithmic slope of the recoil spectrum,
 not at the center of the bin $Q_n$,
 but at the shifted point $Q_{s,n}$.
 Note that $Q_{s,n}$ itself depends on $k_n$.
 However,
 this {\em does not} introduce any additional error,
 if we simply interpret Eq.(\ref{eqn4020115}) as
 an - admittedly somewhat complicated - prescription
 for the determination of the $Q$-values
 where we wish to estimate the logarithmic slope of the recoil spectrum.

 In the rest of this section
 I use only $\bQn$ from Eq.(\ref{eqn4020110}) to estimate the logarithmic slope $k_n$,
 since it simplifies the error analysis somewhat.
 Note that,
 for using both $\bQn$ and $\bQQn$ from Eq.(\ref{eqn4020113}) to estimate $k_n$,
 the statistical errors of them are correlated.
\footnote{
 In contrast,
 I will use Eq.(\ref{eqn4020113}) in Subsec.~\ref{f1un1(v)}
 due to some other reasons.}
 Using standard error propagation,
 we have
\beq
   \sigma^2(k_n)
 = \bDd{k_n}{\bQn}^2 \sigma^2\abrac{\bQn}
\~.
\label{eqn4020116}
\eeq
 The first factor above can be calculated straightforwardly
 from Eq.(\ref{eqn4020110}) as
\beq
   \Dd{\bQn}{k_n}
 = \frac{1}{k_n^2} \bbrac{1-\afrac{\kappa_n}{\sinh\kappa_n}^2}
 = \frac{g(\kappa_n)}{k_n^2}
\~,
\label{eqn4020117}
\eeq
 where I have defined the auxiliary function
\beq
        g(x)
 \equiv 1-\afrac{x}{\sinh x}^2
\~.
\label{eqn4020118}
\eeq
 The error on the average energy transfer $\sigma^2\abrac{\bQn}$
 in Eq.(\ref{eqn4020116}) can be estimated directly from the data,
 using
\beq
   \sigma^2\abrac{\bQn}
 = \frac{1}{N_n-1} \bbigg{\bQQn-\bQn^2}
\~.
\label{eqn4020119}
\eeq
\subsection{Windowing the data set}
\label{windowing}
 ~~~$\,$
 From two naive linear approximations discussed in App.~\ref{lineardRdQn},
 one can find an important observation that
 the statistical error of both estimators of the slope of the recoil spectrum
 given in Eqs.(\ref{eqnD040001}) and (\ref{eqnD040005})
 scale like the bin width to the power $-1.5$
 (see Eqs.(\ref{eqnD040006}) and (\ref{eqnD040007})).
 This can intuitively be understood from the argument that
 the variation of $dR/dQ$ will be larger for larger bins
 (see Fig.~\ref{fig4020101}).
 Moreover,
 a detailed analysis for the relation
 between the statistical error of $k_n$ given in Eq.(\ref{eqn4020116})
 and the bin width $b_n$ \cite{DMDD}
 shows that,
 for small bins,
 the expected error again scales like $b_n^{-1.5}$
 and if the bin width is large,
 the statistical error decreases even faster with increasing bin width. 
 One would therefore naively conclude that
 the errors of the estimated slopes would be minimized by choosing very large bins.

 However,
 as mentioned above and discussed in Ref.~\cite{DMDD},
 neither a linear approximation of the recoil spectrum
 nor the linear ansatz of the logarithm of the spectrum
 can describe the real (but as yet unknown) recoil spectrum exactly.
 The neglected terms of higher powers of $Q-Q_n$
 will certainly induce some uncontrolled systematic errors
 which increase quickly with increasing bin width $b_n$.

 Using large bins has a second, obvious disadvantage:
 the number of bins scales inversely with their size,
 i.e., by using large bins we would be able to estimate $f_1(v)$
 only at a small number of velocities.
 Fortunately,
 this can be alleviated by using overlapping bins,
 or, equivalently,
 {\em by combining several relatively small bins into overlapping ``windows''}.
 This means that
 a given data point $\Qni$ may well contribute to several different windows,
 and hence to the estimate of $f_1(v)$ at several values of $v$.
 This can increase the total amount of information about $f_1(v)$
 since the only information we use about the data points in a given window
 is encoded in the average recoil energy in this window
 (through the estimating of $k_n$ by Eq.(\ref{eqn4020110})).
 This averaging effectively destroys information.
 By letting a given data point contribute to several overlapping windows,
 this loss of information can be reduced.

 One other obvious disadvantage of using large bins or windows is that
 it would lead to a quite large minimal value of $v$ where $f_1(v)$ can be reconstructed,
 simply because the central value $Q_1$,
 and also the shifted point $Q_{s,1}$,
 of a large first bin would be quite large.
 This can be again be alleviated by first collecting our data in relatively small bins,
 and then combining varying numbers of bins into overlapping windows.
 In particular,
 the first window would be identical with the first bin.

 A final consideration concerns the size of the bins.
 Choosing fixed bin sizes,
 and therefore also mostly fixed window sizes,
 would lead to errors on the estimated logarithmic slopes,
 and hence also on the estimates of $f_1(v)$,
 which increase quickly with increasing $Q$ or $v$.
 This is due to the essentially exponential form of the recoil spectrum,
 which would lead to a quickly falling number of events in equal-sized bins.
 A try-error analysis shows that
 the errors are roughly equal in all bins if the bin widths increase linearly
 (some different variations of binning of the data set
  will be given in App.~\ref{binning}).

 These considerations motivate the following set-up for the mock experimental analysis.
 One starts by binning the data,
 as in Eq.(\ref{eqn4020101}),
 where the bin widths satisfy
\beq
   b_n
 = b_1+(n-1) \delta
\~,
\label{eqn4020200a}
\eeq
 i.e.,
\beq
   Q_n
 = Q_{\rm min}+\abrac{n-\frac{1}{2}} b_1+\bfrac{(n-1)^2}{2} \delta
\~.
\label{eqn4020200b}
\eeq
 Here the increment $\delta$ satisfies
\beq
   \delta
 = \frac{2}{B (B-1)} \aBig{Q_{\rm max}-Q_{\rm min}-B b_1}
\~,
\label{eqn4020200c}
\eeq
 $B$ being the total number of bins,
 and $Q_{\rm max,min}$ being the (kinematical or instrumental) extrema of the recoil energy.
 Then I collect up to $n_W$ bins into a window,
 with smaller windows at the borders of the range of $Q$.
 In the rest of this section and the next chapter
 I use Latin indices $n,~m,~\cdots$ to label bins,
 and Greek indices $\mu,~\nu,~\cdots$ to label windows.
 For $1 \leq \mu \leq n_W$,
 the $\mu$-th window simply consists of the first $\mu$ bins;
 for $n_W \leq \mu \leq B$,
 the $\mu$-th window consists of bins $\mu-n_W+1,~\mu-n_W+2,~\cdots,~\mu$;
 and for $B \leq \mu \leq B+n_W-1$,
 the $\mu$-th window consists of the last $n_W-\mu+B$ bins.
 This can also be described by introducing the indices $n_{\mu-}$ and $n_{\mu+}$
 which label the first and last bin contributing to the $\mu$-th window,
 with
\cheqnCa
\beq
\renewcommand{\arraystretch}{1.3}
   n_{\mu-}
 = \cleft{\begin{array}{l l}
           1,         & \mu \leq n_W \\
           \mu-n_W+1, & \mu \geq n_W
          \end{array}},
\label{eqn4020201a}
\eeq
 and
\cheqnCb
\beq
\renewcommand{\arraystretch}{1.3}
   n_{\mu+}
 = \cleft{\begin{array}{l l}
           \mu, & \mu \leq B \\
           B,   & \mu \geq B
          \end{array}}
.
\label{eqn4020201b}
\eeq
\cheqnC
 The total number of windows defined through Eqs.(\ref{eqn4020201a}) and (\ref{eqn4020201b})
 is evidently $W = B+n_W-1$,
 i.e., $1 \leq \mu \leq B+n_W-1$.

 As shown in the previous subsection,
 the basic observables needed for the reconstruction of $f_1(v)$ in Eq.(\ref{eqn4010008})
 are the number of events in the $n$-th bin, $N_n$,
 as well as the averages $\bQn$ in Eq.(\ref{eqn4020112}).
 Once $N_n$ and $\bQn$ can be obtained,
 we can then use Eqs.(\ref{eqn4020102}), (\ref{eqn4020110}) and (\ref{eqn4020115})
 to get $r_n$, $k_n$ and $Q_{s,n}$
 as well as Eq.(\ref{eqn4020116}) to get the statistical error of $k_n$.
 One can easily calculate the number of events per window as
\beq
   N_{\mu}
 = \sum_{n=n_{\mu-}}^{n_{\mu+}} N_n
\~,
\label{eqn4020202}
\eeq
 as well as the averages
\beq
   \Bar{Q-Q_{\mu}}|_{\mu}
 = \frac{1}{N_{\mu}} \abrac{\sum_{n=n_{\mu-}}^{n_{\mu+}} N_n \Bar{Q}|_{n}}-Q_{\mu}
\~,
\label{eqn4020203}
\eeq
 where $Q_{\mu}$ is the central point of the $\mu$-th window.

 One drawback of the use of overlapping windows in the analysis is that
 the observables defined in Eqs.(\ref{eqn4020202}) and (\ref{eqn4020203})
 are all correlated (for $n_W \neq 1$).
 The slope in the $\mu$-th window, $k_{\mu}$,
 will again be calculated as in Eq.(\ref{eqn4020110})
 with ``bin'' quantities replaced by ``window'' quantities.
 We thus need the covariance matrix for the $\Bar{Q-Q_{\mu}}|_{\mu}$,
 which follows directly from the definition in Eq.(\ref{eqn4020203})
 (detailed deviations for the covariances in this subsection
  will be given in App.~\ref{covbQnrn}):
\beqn
 \conti {\rm cov}\abrac{\Bar{Q-Q_{\mu}}|_{\mu},\Bar{Q-Q_{\nu}}|_{\nu}}
        \non\\
 \=     \frac{1}{N_{\mu} N_{\nu}}
        \sum_{n=n_{\nu-}}^{n_{\mu+}}
        \bbigg{ N_n   \abrac{\Bar{Q}|_n-\Bar{Q}|_{\mu}} \abrac{\Bar{Q}|_n-\Bar{Q}|_{\nu}}
               +N_n^2 \sigma^2\abrac{\bQn}}
\~,
\label{eqn4020204}
\eeqn
 where $\sigma^2\abrac{\bQn}$ is defined as in Eq.(\ref{eqn4020119}).
 In Eq.(\ref{eqn4020204}) I have assumed $\mu \leq \nu$;
 the covariance matrix is, of course, symmetric.
 Moreover,
 the sum is understood to vanish if the two windows $\mu$, $\nu$ do not overlap,
 i.e., if $n_{\mu+} < n_{\nu-}$.

 The ansatz in Eq.(\ref{eqn4020105}) is now assumed to hold over an entire window.
 We again can estimate the prefactor as
\beq
   r_{\mu}
 = \frac{N_{\mu}}{w_{\mu}}
\~,
\label{eqn4020205}
\eeq
 $w_{\mu}$ being the width of the $\mu$-th window.
 This implies
\beq
   {\rm cov}(r_{\mu},r_{\nu})
 = \frac{1}{w_{\mu} w_{\nu}} \sum_{n=n_{\nu-}}^{n_{\mu+}} N_n
\~,
\label{eqn4020206}
\eeq
 where I have again taken $\mu \leq \nu$.
 Finally,
 the mixed covariance matrix is given by
\beq
   {\rm cov}\abrac{r_{\mu},\Bar{Q-Q_{\nu}}|_{\nu}}
 = \frac{1}{w_{\mu} N_{\nu}} \sum_{n=n_{-}}^{n_{+}} N_n \abrac{\Bar{Q}|_n-\Bar{Q}|_{\nu}}
\~.
\label{eqn4020207}
\eeq
 Note that this sub-matrix is not symmetric under the exchange of $\mu$ and $\nu$.
 In the definition of $n_{-}$ and $n_{+}$ we therefore have to distinguish two cases:
\beqn
 {\rm If}~\mu \leq \nu \&:\&~n_{-} = n_{\nu-},~n_{+} = n_{\mu+}
\~;
 \non\\
 {\rm If}~\mu \geq \nu \&:\&~n_{-} = n_{\mu-},~n_{+} = n_{\nu+}
\~.
\label{eqn4020208}
\eeqn
 As before,
 the sum in Eq.(\ref{eqn4020207}) is understood to vanish if $n_- > n_+$.

 The covariance matrices involving the estimators of the logarithmic slopes $k_{\mu}$,
 derived from Eq.(\ref{eqn4020110}) with $n \to \mu$ everywhere,
 can be calculated in terms of the covariance matrices
 in Eqs.(\ref{eqn4020204}) and (\ref{eqn4020207}):
\beq
   {\rm cov}\abrac{k_{\mu},k_{\nu}}
 = \bfrac{k_{\mu}^2 k_{\nu}^2}{g(\kappa_{\mu}) g(\kappa_{\nu})}
   {\rm cov}\abrac{\Bar{Q-Q_{\mu}}|_{\mu},\Bar{Q-Q_{\nu}}|_{\nu}}
\~,
\label{eqn4020209}
\eeq
 and
\beq
   {\rm cov}\abrac{r_{\mu},k_{\nu}}
 = \bfrac{k_{\nu}^2}{g(\kappa_{\nu})} {\rm cov}\abrac{r_{\mu},\Bar{Q-Q_{\nu}}|_{\nu}}
\~,
\label{eqn4020210}
\eeq
 where $\kappa_{\mu}$ is as in Eq.(\ref{eqn4020108}) with $n \to \mu$,
 and the function $g(x)$ has been defined in Eq.(\ref{eqn4020118}).
\subsection{Reconstructing the velocity distribution}
\label{f1r(v)}
 ~~~$\,$
 Now we are ready to put all pieces together to reconstruct the velocity distribution
 and compute its statistical error.
 Substituting the ansatz in Eq.(\ref{eqn4020105}) (with the replacement $n \to \mu$)
 into Eq.(\ref{eqn4010008}),
 the reconstructed normalized one-dimensional velocity distribution function can be expressed as
\beq
   f_{1,r}(v_{s,\mu})
 = \calN \bBigg{\frac{2 Q_{s,\mu} r_{\mu}}{F^2(Q_{s,\mu})}}
         \bbrac{\dd{Q} \ln \FQ \bigg|_{Q = Q_{s,\mu}}-k_{\mu}}
\~,
\label{eqn4020301}
\eeq
 for $\mu = 1,~2,~\cdots,~B+n_W-1$.
 Here $Q_{s,\mu}$ is given by Eq.(\ref{eqn4020115}) with $n \to \mu$,
 and,
\beq
   v_{s,\mu}
 = \alpha \sqrt{Q_{s,\mu}}
\~,
\label{eqn4020302}
\eeq
 as in Eq.(\ref{eqn3010106}).
 Finally,
 the normalization constant $\calN$ defined in Eq.(\ref{eqn4010009})
 can be estimated directly from the data:
\beq
   \calN
 = \frac{2}{\alpha} \bbrac{\sum_{a} \frac{1}{\sqrt{Q_a} \~ F^2(Q_a)}}^{-1}
\~,
\label{eqn4020303}
\eeq
 where the sum runs over all events in the sample.

 Since neighboring windows overlap,
 the estimates of $f_1(v)$ at adjacent values of $v_{\mu}$ are correlated.
 This is described by the covariance matrix
\beqn
 \conti {\rm cov}\aBig{f_{1,r}(v_{s,\mu}),f_{1,r}(v_{s,\nu})}
        \non\\
 \=     \bfrac{f_{1,r}(v_{s,\mu}) f_{1,r}(v_{s,\nu})}{r_{\mu} r_{\nu}}
        {\rm cov}\abrac{r_{\mu},r_{\nu}}
       +\abrac{2 \calN}^2
        \bfrac{Q_{s,\mu} Q_{s,\nu} r_{\mu} r_{\nu}}{F^2(Q_{s,\mu}) F^2(Q_{s,\nu})}
        {\rm cov}\abrac{k_{\mu},k_{\nu}}
        \non\\
\conti ~~~~~~~~ 
       -\calN
        \cbrac{ \bfrac{f_{1,r}(v_{s,\mu})}{r_{\mu}}
                \bfrac{2 Q_{s,\nu} r_{\nu} }{F^2(Q_{s,\nu})} {\rm cov}\abrac{r_{\mu},k_{\nu}}
               +\aBig{\mu \lgetsto \nu}}
\~,
\label{eqn4020304}
\eeqn
 where the covariance matrices involving
 the normalized counting rates $r_{\mu}$ and logarithmic slopes $k_{\mu}$
 have been given in Eqs.(\ref{eqn4020206}), (\ref{eqn4020209}), and (\ref{eqn4020210}).
 In principle
 Eq.(\ref{eqn4020304}) should also include contributions involving
 the statistical error of the estimator for $\calN$ in Eq.(\ref{eqn4020303}).
 However,
 this error and its correlations with the errors of the $r_{\mu}$ and $k_{\mu}$
 has been found to be negligible compared to the errors included in Eq.(\ref{eqn4020304}).

\begin{figure}[p]
\begin{center}
\imageswitch{
\begin{picture}(15,20)
\put(0  , 0  ){\framebox(15,20  ){}}
\put(0.5,10.5){\framebox(14, 9.5){}} 
\put(0.5, 0  ){\framebox(14, 9.5){}}
\end{picture}}
{\includegraphics[width=13.5cm]{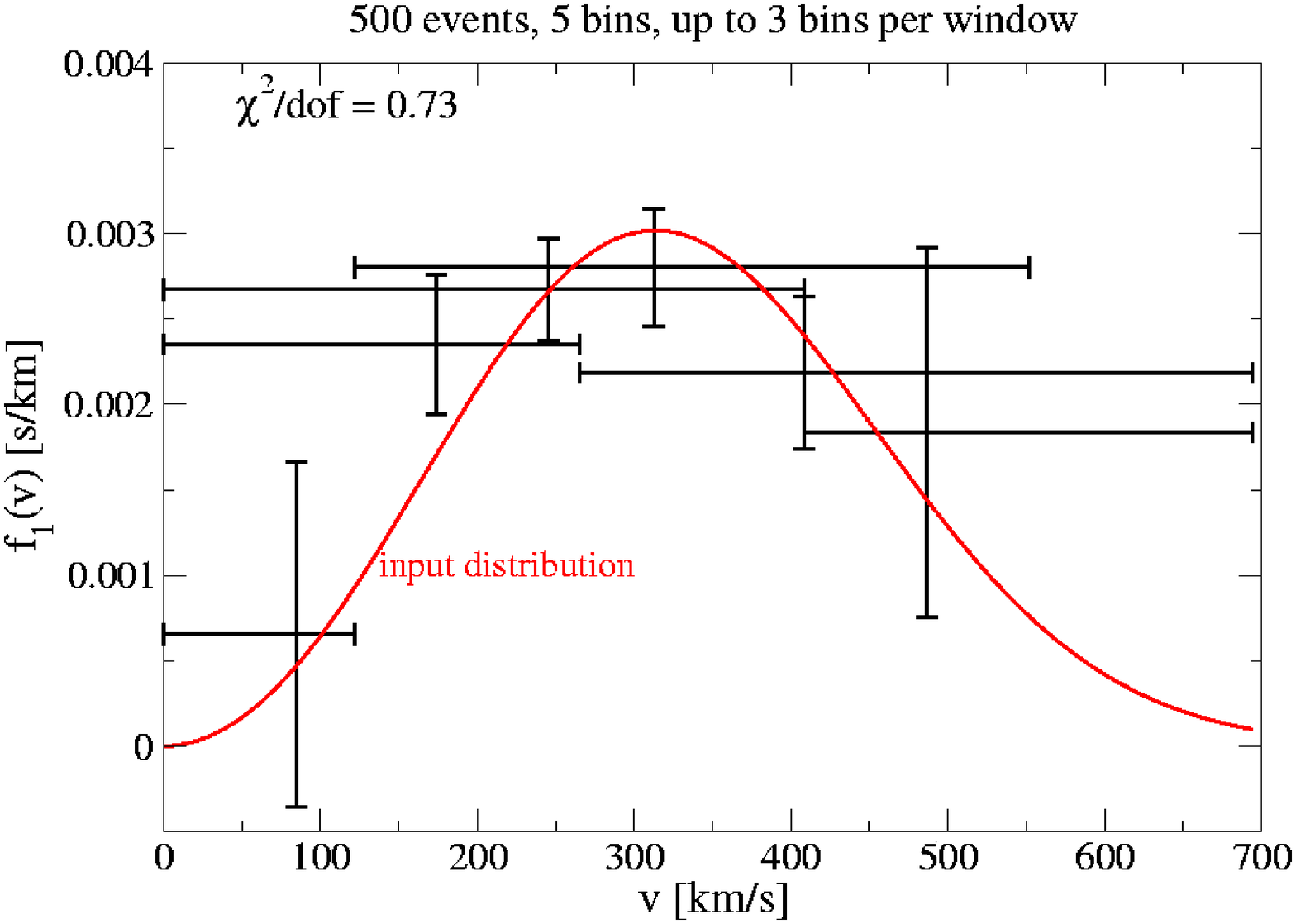} \\ \vspace{1cm}
 \includegraphics[width=13.5cm]{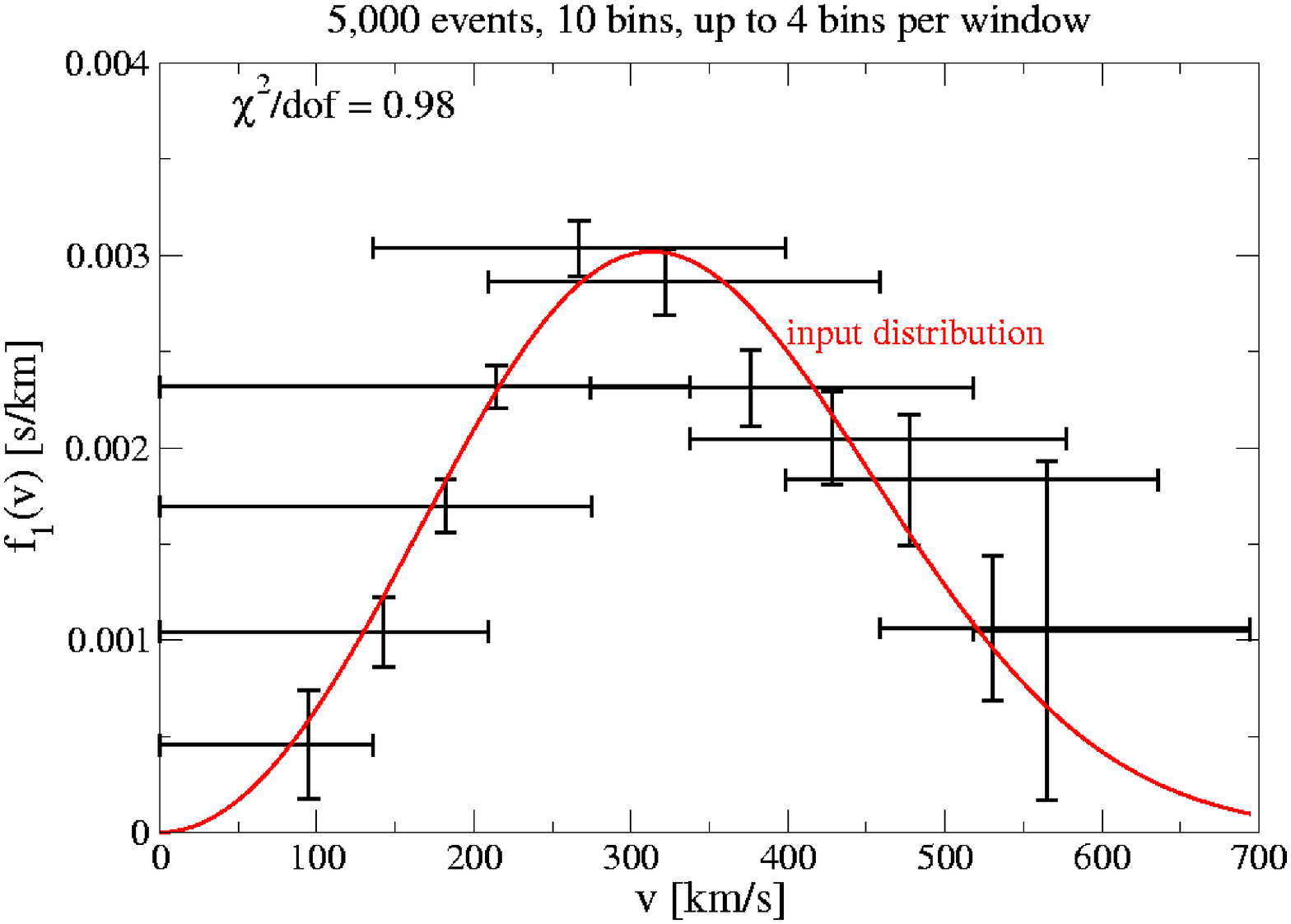}}
\end{center}
\caption{
 The WIMP velocity distribution reconstructed
 from a ``typical'' experiment with 500 (top) and 5,000 (bottom) events.
 The smooth curves show the input distributions,
 which are based on Eq.(\ref{eqn3020101}).
 The vertical error bars show the square roots of the diagonal entries of the covariance matrix
 given in Eq.(\ref{eqn4020304});
 the horizontal bars show the size of the window
 used in deriving the given value of $f_{1,r}$.
 The overlap of these horizontal bars
 thus shows the range over which the values of $f_{1,r}$ are correlated.
 Parameters as in Fig.~\ref{fig4020101}.}
\label{fig4020301}
\end{figure}
 Figs.~\ref{fig4020301} are results for the reconstructed velocity distribution,
 for ``typical'' simulated experiments with 500 (top) and 5,000 (bottom) events.
 In the top frame $B = 5$ bins has been chosen,
 the first bin having a width $b_1 = 8$ keV,
 and up to three bins have been combined into a window.
 Since the last bin is in fact empty,
 this leaves us with $W = 6$ windows,
 i.e., we can determine $f_{1,r}$ for six discrete values of the WIMP velocity $v$;
 recall that these ``measurements'' of $f_{1,r}$ are correlated,
 as indicated by the horizontal bars in the figure.
 In the lower frame $B = 10$ bins with $b_1 = 10$ keV have been chosen,
 and up to four bins have been combined into one window.
 The bins are thus significantly smaller than in the upper frame.
 As a result,
 the last two bins are now (almost) empty,
 leaving us with $W = 11$ windows.
 Figs.~\ref{fig4020301} indicate that
 one will need {\em at least a few hundred events}
 for a meaningful direct reconstruction of $f_1(v)$.

 Furthermore,
 a $\chi_f^2$ distributions has been defined by
\beq 
        \chi_f^2
 \equiv \frac{1}{W}
        \sum_{\mu,\nu}
        {\cal C}_{\mu \nu}
        \bBig{f_{1,r}(v_{s,\mu})-f_{1,\rm th}(v_{s,\mu})}
        \bBig{f_{1,r}(v_{s,\nu})-f_{1,\rm th}(v_{s,\nu})}
\~.
\label{eqn4020305}
\eeq
 Here $f_{1,r}$ is the estimate in Eq.(\ref{eqn4020301}) of the velocity distribution,
 $f_{1,\rm th}$ is a theoretical predicted velocity distribution
 (e.g., the input distributions in Figs.~\ref{fig4020301}),
 and $\cal C$ is the inverse of the covariance matrix of Eq.(\ref{eqn4020304}).
 This $\chi_f^2$ distribution allows
 statistically meaningful tests of the predicted velocity distribution function.

 More details about this $\chi_f^2$ distribution
 and some applications can be found in Ref.~\cite{DMDD}.
\subsection{Determining moments of the velocity distribution}
\label{expvvn}
 ~~~$\,$
 As mentioned in the previous subsection,
 a direct reconstruction of the WIMP velocity distribution $f_1(v)$
 will only be possible once several hundred nuclear recoil events have been collected.
 This is a tall order,
 given that not a single such event has so far been detected
 (barring the possible DAMA observation and a few candidate signals,
  see Secs.~\ref{cyrogenic} to \ref{SDD}).
 The basic reason for the large required event sample is that,
 $f_1(v)$ being a normalized distribution,
 only information on the {\em shape} of $f_1(v)$ is meaningful.
 In order to obtain such shape information via direct reconstruction,
 we have to separate the events into several bins or windows.
 Moreover,
 each window should contain sufficiently many events to allow
 an estimate of the {\em slope} of the recoil spectrum in this window.

 On the other hand,
 at the end of Sec.~\ref{dRdQ}
 I have also given expressions for the {\em moments} of $f_1(v)$
 in Eqs.(\ref{eqn4010011}) to (\ref{eqn4010013}).
 With the exception of the moment with $n = -1$,
 these are entirely inclusive quantities,
 i.e., each moment is sensitive to the entire data set;
 no binning is required,
 nor do we need to determine any slope
 (with one possible minor exception; see below).
 It thus seems reasonable to expect that
 one can obtain meaningful information about these moments with fewer events.

 The experimental implementation of Eq.(\ref{eqn4010011}) is quite straightforward.
 For $\Qthre = 0$,
 the normalization $\calN$ has already been given in Eq.(\ref{eqn4020303}).
 The case of non-vanishing threshold energy $\Qthre$ can be treated straightforwardly,
 using Eq.(\ref{eqn4010013}).
 To that end one needs to estimate the recoil spectrum at the threshold energy.
 One possibility would be to choose an artificially high value of $\Qthre$,
 and simply count the events in a bin centered on $\Qthre$.
 However,
 in this case the events with $Q < \Qthre$
 would be left out of the determination of the moments.
 We therefore should keep $\Qthre$ as small as experimentally possible,
 and to estimate the counting rate at threshold using the ansatz in Eq.(\ref{eqn4020105}).
 Since we need the recoil spectrum only at this single point,
 we only have to determine the quantities $r_1$ and $k_1$ parameterizing $dR/dQ$ in the first bin;
 this can be done as described in Subsec.~\ref{dRdQkn},
 without the need to distinguish between bins and ``windows''.
 Introduce the shorthand notation
\beq
        \rthre
 \equiv \afrac{dR}{dQ}_{Q = \Qthre}
 =      r_1 e^{k_1 (\Qthre-Q_{s,1})}
\~.
\label{eqn4020401}
\eeq
 Then,
 combining Eqs.(\ref{eqn4010011}) and (\ref{eqn4010013}),
 the $n$-th moment of the velocity distribution function can be rewritten as
\beq
   \expv{v^n}
 = \alpha^n
   \bbrac{\frac{2 \Qthre^{    1/2} \rthre}{\FQthre}+      I_0}^{-1}
   \bbrac{\frac{2 \Qthre^{(n+1)/2} \rthre}{\FQthre}+(n+1) I_n}
\~,
\label{eqn4020402}
\eeq
 where the integral $I_n$ defined in Eq.(\ref{eqn4010012}) can be estimated through the sum:
\beq
   I_n
 = \sum_a \frac{Q_a^{(n-1)/2}}{F^2(Q_a)}
\~,
\label{eqn4020403}
\eeq
 as Eq.(\ref{eqn4020303}).
 Since all $I_n$ are determined from the same data,
 they are correlated with
\beq
   {\rm cov}(I_n,I_m)
 = \sum_a \frac{Q_a^{(n+m-2)/2}}{F^4(Q_a)}
\~.
\label{eqn4020404}
\eeq
 This can e.g., be seen by writing Eq.(\ref{eqn4020403}) as a sum over narrow bins,
 such that the recoil spectrum within each bin can be approximated by a constant.
 Each term in the sum would then have to be multiplied with the number of events in this bin;
 Eq.(\ref{eqn4020404}) then follows from standard error propagation.
 Note that,
 when re-converted into an integral,
 the expression for ${\rm cov}(I_0,I_0)$ will diverge logarithmically for $\Qthre \to 0$.
 Equivalently,
 the numerical estimate of this entry can become very large
 if the sample contains events with very small $Q$-values.
 But,
 according to some numerical simulations,
 there should be no problem for samples with $\Qthre > 1$ keV.
 Many existing experiments in fact require significantly larger energy transfers
 in their definition of a WIMP signal. 

 In order to calculate the statistical error of $\expv{v^n}$
 in Eq.(\ref{eqn4020402}),
 one needs at first the error of $\rthre$
 which can be obtained from Eq.(\ref{eqn4020401}) as
\beq
    \sigma^2(\rthre)
  = \rthre^2
    \cbrac{ \frac{\sigma^2(r_1)}{r_1^2}
           +\bbrac{\Qthre-Q_{s,1}-k_1 \aPp{Q_{s,1}}{k_1}}^2 \sigma^2(k_1)}
\~.
\label{eqn4020405}
\eeq
 Here the squared errors for $r_1$ and $k_1$ are simply
 the corresponding diagonal entries of the respective covariance matrices
 given in Eqs.(\ref{eqn4020206}) and (\ref{eqn4020209}),
 and the definition of $Q_{s,1}$ in Eq.(\ref{eqn4020115}) implies
\beq
   Q_{s,1}+k_1 \aPp{Q_{s,1}}{k_1}
 = Q_1-\frac{1}{k_1}+\afrac{b_1}{2} \coth\afrac{b_1 k_1}{2}
\~,
\label{eqn4020406}
\eeq
 where $Q_1$ is the central $Q$-value in the first bin.
 It should be noted that
 the first term in Eq.(\ref{eqn4010011}) is negligible
 for all $n \geq 1$ if $\Qthre \simeq 1$ keV.
 However,
 even for this low threshold energy
 it contributes significantly to the normalization constant $\calN$,
 as described by Eq.(\ref{eqn4010013}).
 Of course,
 the first term in Eq.(\ref{eqn4010011}) always dominates for $n = -1$.
 This is not surprising,
 since the very starting point of this analysis, Eq.(\ref{eqn3010108}),
 already shows that
 the counting rate at $\Qthre$ is proportional to
 the ``minus first'' moment of the velocity distribution.

 One needs also the correlation
 between the errors on the estimate of the recoil spectrum at $Q = \Qthre$
 and the integrals $I_n$.
 It is clear that these quantities are correlated,
 since the former is estimated from all events in the first bin,
 which of course also contribute to the latter.
 These correlations can be estimated by using the ansatz in Eq.(\ref{eqn4020105}),
 which makes the following prediction
 for the contribution of the first bin to the integrals:
\beq
   I_{n,1}
 = r_1 \int_{\Qthre}^{\Qthre+b_1} \bfrac{Q^{(n-1)/2}}{\FQ} e^{k_1 (Q-Q_{s,1})} \~ dQ
\~.
\label{eqn4020407}
\eeq
 This immediately implies
\cheqnCa
\beq
   \Pp{I_{n,1}}{r_1}
 = \frac{I_{n,1}}{r_1}
\~,
\label{eqn4020408a}
\eeq
 and
\cheqnCb
\beq
   \Pp{I_{n,1}}{k_1}
 = I_{n+2,1}-\bbrac{Q_{s,1}+k_1 \aPp{Q_{s,1}}{k_1}} I_{n,1}
\~.
\label{eqn4020408b}
\eeq
\cheqnC
 Note that $I_{n,1}$ and $I_{n+2,1}$ in Eqs.(\ref{eqn4020408a}) and (\ref{eqn4020408b})
 can be evaluated as in Eq.(\ref{eqn4020403}),
 with the sum extending only over events in the first bin:
\beq
   I_{n,1}
 = \sum_{i=1}^{N_1} \frac{Q_{1,i}^{(n-1)/2}}{F^2(Q_{1,i})}
\~.
\label{eqn4020409}
\eeq
 The correlation between $\rthre$ and $I_n$ is then given by
\beqn
 \conti {\rm cov}(\rthre,I_n)
        \non\\
 \=     \rthre \~ I_{n,1}
        \cleft{ \frac{\sigma^2(r_1)}{r_1^2}
               +\bbrac{\Qthre-Q_{s,1}-k_1 \aPp{Q_{s,1}}{k_1}}}
        \non\\
 \conti ~~~~~~~~~~~~~~~~~~~~~~~~~~~~~~~~~ \times 
        \cright{\bbrac{\frac{I_{n+2,1}}{I_{n,1}}-Q_{s,1}-k_1 \aPp{Q_{s,1}}{k_1}}\! \sigma^2(k_1)}
\~.
\label{eqn4020410}
\eeqn

 Finally,
 these ingredients allow us to compute the covariance matrix
 for the estimates of the moments of the velocity distribution: 
\beqn
\conti {\rm cov}\aBig{\expv{v^n},\expv{v^m}}
       \non\\
\=     \calN_{\rm m}^2
       \bbiggl{ \expv{v^n} \expv{v^m} {\rm cov}(I_0,I_0)
               +\alpha^{n+m} (n+1) (m+1) {\rm cov}(I_n, I_m) 
       \non\\
\conti ~~~~~~~~ 
               -\alpha^m (m+1) \expv{v^n} {\rm cov}(I_0,I_m)
               -\alpha^n (n+1) \expv{v^m} {\rm cov}(I_0,I_n)\bigg.
       \non\\
\conti ~~~~~~~~~~~~ 
               + D_n D_m \sigma^2(\rthre)
               -\aBig{D_m \expv{v^n} + D_n \expv{v^m}} {\rm cov}(\rthre,I_0)}\Bigg.
        \non\\
\conti ~~~~~~~~~~~~~~~~ 
       \bbiggr{+\alpha^m (m+1) D_n {\rm cov}(\rthre,I_m)
               +\alpha^n (n+1) D_m {\rm cov}(\rthre,I_n)}
\~.
\label{eqn4020411}
\eeqn
 Here I have introduced the modified normalization constant:
\beq
        \calN_{\rm m}
 \equiv \afrac{\alpha}{2} \calN
\~,
\label{eqn4020412}
\eeq
 which exploits the partial cancellation of the $\alpha$ factors
 between Eqs.(\ref{eqn4010011}) and (\ref{eqn4010013}),
 and the quantities
\beq
        D_n
 \equiv \frac{1}{\cal N_{\rm m}} \aPp{\expv{v^n}}{\rthre} 
 =      \frac{2}{\FQthre} \abrac{\alpha^n \Qthre^{(n+1)/2}-\sqrt{\Qthre} \~ \expv{v^n}}
\~.
\label{eqn4020413}
\eeq
 Note that,
 in practice,
 one can determine $\expv{v^n}$ by a single experiment with a large number of events,
 or by averaging over many experiments with a relatively small number of events.
 However,
 numerical simulations \cite{DMDD} show that
 in the second case the average values of the reconstructed moments
 do not exactly converge to the input (exact) values.
 In order to understand this,
 consider the simple case $\Qthre = 0$.
 The moments are then proportional to the ratio $I_n / I_0$ (see Eq.(\ref{eqn4020402})).
 The distortion arises because $\expv{I_n/I_0} \neq \expv{I_n}/\expv{I_0}$,
 where the averaging is over many simulated experiments.
 The leading correction terms for small $\Qthre$ and not very large first bin can be found as
 (a detailed calculation by using Taylor expansion to second order
  will be given in App.~\ref{corr})
\beqn
       \delta\expv{v^n}
 \=    \alpha^n \calN_{\rm m}^2
       \cBiggl{(n+1) \bbigg{{\rm cov}(I_0,I_n)-\calN_{\rm m} I_n {\rm cov}(I_0,I_0)}}
       \non\\
\conti ~~~~~~~~~~~~ 
       \cBiggr{+2
                \bfrac{ \Qthre^{(n+1)/2}}{\FQthre} \bbigg{{\rm cov}(\rthre,I_0)
                       -\rthre \calN_{\rm m} {\rm cov}(I_0,I_0)}}
\~,
\label{eqn4020414}
\eeqn
 where the second line in Eq.(\ref{eqn4020410}) is significant only for $n = -1$.
 Note that this correction becomes very small
 if the statistical errors on the $I_n$ as well as on $\rthre$ become small.

 Meanwhile,
 according to some detailed numerical analyses \cite{DMDD},
 an ``error on the error'' should be added.
 The contribution to the diagonal entries of the covariance matrix
 given in Eq.(\ref{eqn4020405}) can be estimated as
\beq
   \sigma^2\abrac{{\rm cov}(I_n,I_n)}
 = \sum_a \frac{Q_a^{2n-2}}{F^8(Q_a)}
\~,
\label{eqn4020415}
\eeq
 the off-diagonal entries are then scaled up
 such that the correlation matrix remains unaltered.
 The numerical analyses show also that
 very rare events with large recoil energies
 contribute significantly more to the higher moments.
 Hence,
 an experiment with a small number of events
 will usually underestimate $\expv{v^n}$ and, especially, its error;
 the problem will become worse with increasing $n$.
 However,
 because this method uses whole experimental data together
 to determine the moments of $f_1(v)$,
 it has also been found that,
 based only on the first two or three moments,
 some non-trivial information can already be extracted from ${\cal O}(20)$ events.

 More details and discussions about
 the reconstruction of the velocity distribution
 and determination of its moments can be found in Ref.~\cite{DMDD}.
\section{Determining the WIMP mass}
\label{mchi}
 ~~~$\,$
 In the previous two sections
 I discussed how to use a recoil spectrum from direct Dark Matter detection
 as well as experimental data directly (i.e., the measured recoil energies)
 to reconstruct the velocity distribution function of WIMPs
 as well as to determine its moments.
 As noted earlier,
 for both of these reconstruction methods
 we need to know the mass of the incident WIMPs $\mchi$.
 In well-motivated WIMP models from elementary particle physics,
 $\mchi$ can be determined with high accuracy from future collider experiment data.
 However,
 one has to check experimentally that
 the particles produced at colliders are in fact the same ones
 seen in Dark Matter detection experiments which form the Galactic halo.
 In this section I present a method for (self-)determining the WIMP mass
 based on the determination of the moments of the velocity distribution function, $\expv{v^n}$,
 (presented in Sec.~\ref{expvvn})
 from two (or more) experimental data sets with different target materials.
\footnote{
 Note that the ansatz here is quite different
 from that used in Ref.~\cite{Green07},
 which assumes different WIMP velocity distributions
 with two input parameters:
 the WIMP mass and the WIMP-nucleon cross section,
 and then analyses with which precision
 in the usual WIMP mass-cross section plane
 the WIMP mass can be reproduced from the direct detection experiment.}
\subsection{Neglecting $\Qthre$}
 ~~~$\,$
 As mentioned in the end of Sec.~\ref{dRdQ},
 the basic idea for using two different detector materials
 to determine the WIMP mass is that,
 from independent direct WIMP detection experiments with different target nuclei,
 the measured recoil spectra should lead to
 the same (moments of the) velocity distribution function of incident WIMPs.

 For the case that the threshold energy $\Qthre$ can be neglected,
 the $n$-th moment of the velocity distribution function, $\expv{v^n}$,
 in Eq.(\ref{eqn4020402}) can be expressed simply as
\beq
   \expv{v^n}
 = \alpha^n (n+1) \afrac{I_n}{I_0} 
\~,
\label{eqn4030101}
\eeq
 where $I_n$ and $I_0$ can be estimated by Eq.(\ref{eqn4020403}).
 Suppose $X$ and $Y$ are two target nuclei.
 Eq.(\ref{eqn4030101}) implies
\beq
   \alpha_X^n \afrac{\InX}{\IzX}
 = \alpha_Y^n \afrac{\InY}{\IzY}
\~.
\eeq
 Note that
 the form factor $\FQ$ in Eq.(\ref{eqn4020403})
 for estimating $\InX$ and $\InY$ are different.
 Then,
 according to the definition of $\alpha$ in Eq.(\ref{eqn3010107})
 with the expression of the reduced mass $\mr$ in Eq.(\ref{eqn3010103})
 and using some simple algebra,
 one can find the WIMP mass as
\beq
   \mchi
 = \frac{\sqrt{\mX \mY}-\mX \calRn}{\calRn-\sqrt{\mX/\mY}}
\~,
\label{eqn4030102}
\eeq
 where I have defined
\beq
        \calRn
 \equiv \frac{\alpha_Y}{\alpha_X}
 =      \abrac{\frac{\InX}{\IzX} \cdot \frac{\IzY}{\InY}}^{1/n}
\~,
       ~~~~~~~~~~~~~~~~~~~~ 
        n
 \ne    0,~-1
.
\label{eqn4030103}
\eeq

\begin{figure}[t]
\begin{center}
\imageswitch{
\begin{picture}(14,10.2)
\put(0,0){\framebox(14,10.2){}}
\end{picture}}
{\includegraphics[width=13.5cm]{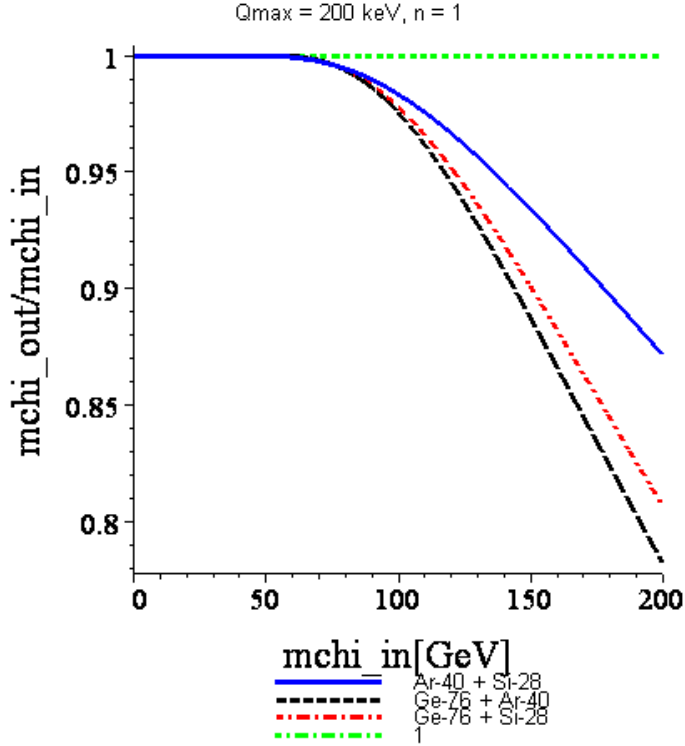}}
\end{center}
\caption{
 The curves show
 the ratios of the reproduced WIMP masses estimated by Eq.(\ref{eqn4030102})
 with different combinations of target nuclei
 to the input (true) one as functions of the input WIMP mass.
 $\calRn$ with $n = 1$ has been estimated
 by the integral form of $I_n$
 with a maximal measuring energy of 200 keV.
 The recoil energy spectrum for a shifted Maxwellian velocity distribution
 with the Woods-Saxon form factor has been used
 (parameters as in Fig.~\ref{fig4020101}).
 The solid (blue) line,
 the dashed (black) line,
 and the dash-dotted (red) line
 are for $\rmXA{Ar}{40}$ + $\rmXA{Si}{28}$,
 $\rmXA{Ge}{76}$ + $\rmXA{Ar}{40}$,
 and $\rmXA{Ge}{76}$ + $\rmXA{Si}{28}$ combination,
 respectively.
 The straight dash-dotted (green) line denotes 1.} 
\label{fig4030101}
\end{figure}
 Fig.~\ref{fig4030101} shows the ratios of the reproduced WIMP masses
 estimated by Eq.(\ref{eqn4030102}) with different combinations of target nuclei
 to the input (true) one as functions of the input WIMP mass.
 $\rmXA{Si}{28}$, $\rmXA{Ar}{40}$, and $\rmXA{Ge}{76}$ have been chosen
 as three target nuclei and
 thus three combinations for $\calRn$ defined in Eq.(\ref{eqn4030103})
 with $n = 1$ are shown.
 $\calRn$ has been estimated by the integral form of $I_n$ in Eq.(\ref{eqn4010012})
 with a maximal measuring energy of 200 keV.
 The theoretical predicted recoil spectrum
 for the shifted Maxwellian velocity distribution function,
 $(dR/dQ)_\sh$ in Eq.(\ref{eqn3020103}),
 with the Woods-Saxon form factor $F_{\rm WS}^2(Q)$ in Eq.(\ref{eqn3010202})
 has been used.
 In Fig.~\ref{fig4030101}
 one can see obviously a deviation of the reproduced WIMP mass
 from the input (true) one
 as input $\mchi~\amssyasymc{38}~60~{\rm GeV}/c^2$.
 The heavier the nuclear masses of two target nuclei,
 e.g., $\rmXA{Ge}{76}$ + $\rmXA{Si}{28}$,
 the larger the deviation from the true WIMP mass.
 This is caused by introducing the maximal measuring energy for estimating $I_n$.
 As discussed in Subsec.~\ref{expvvn},
 the heavier the nuclear mass $\mN$,
 or, equivalently,
 the larger $\alpha$,
 and the larger $n$,
 the more the contribution to $I_n$ comes from the high $Q$ region,
 and,
 for a fixed maximal measuring energy,
 the smaller the value for $\calRn$ and then for $\mchi$ will be estimated.
 As shown in Fig.~\ref{fig4030101},
 for $n = 1$ and input $\mchi = 200~{\rm GeV}/c^2$,
 the deviation with $Q_{\rm max} = 200$ keV is around 20\%.
 However,
 according to the numerical analysis,
 with $Q_{\rm max} = 250$ keV or 300 keV,
 this deviation will be reduced to around 10\% or even only 5\%.
 Later we will see,
 due to a quite large statistical error with very few events,
 a deviation around 10\% for input $\mchi = 200~{\rm GeV}/c^2$ is not very bad.
 Moreover,
 for input $\mchi~\amssyasymc{46}~120~{\rm GeV}/c^2$,
 the deviation should be less than 5\% or even 1\%
 for $Q_{\rm max} = 200$ keV or 250 keV.

 Furthermore,
 the statistical error on the reproduced WIMP mass
 can be obtained from Eq.(\ref{eqn4030102}) directly as
\beqn
        \sigma(\mchi)
 \=     \frac{\sqrt{\mX/\mY} \vbrac{\mX-\mY}}{\abrac{\calRn-\sqrt{\mX/\mY}}^2} \cdot
        \sigma(\calRn)
        \non\\
 \=     \frac{\calRn \sqrt{\mX/\mY} \vbrac{\mX-\mY}}{\abrac{\calRn-\sqrt{\mX/\mY}}^2}
        \non\\
 \conti ~~ \times
        \frac{1}{|n|}
        \bbrac{ \frac{\sigma^2\abrac{\InX}}{\InX^2}
               +\frac{\sigma^2\abrac{\IzX}}{\IzX^2}
               -\frac{2 {\rm cov}\abrac{\IzX,\InX}}{\IzX \InX}
               +(X \lto Y)}^{1/2} \!\!
,
\label{eqn4030104}
\eeqn
 where $\sigma^2\abrac{\InX} = {\rm cov}\abrac{\InX,\InX}$
 and ${\rm cov}\abrac{\IzX,\InX}$ and so on
 can be estimated from Eq.(\ref{eqn4020404}).

\begin{figure}[t]
\begin{center}
\imageswitch{
\begin{picture}(14,10.2)
\put(0,0){\framebox(14,10.2){}}
\end{picture}}
{\includegraphics[width=13.5cm]{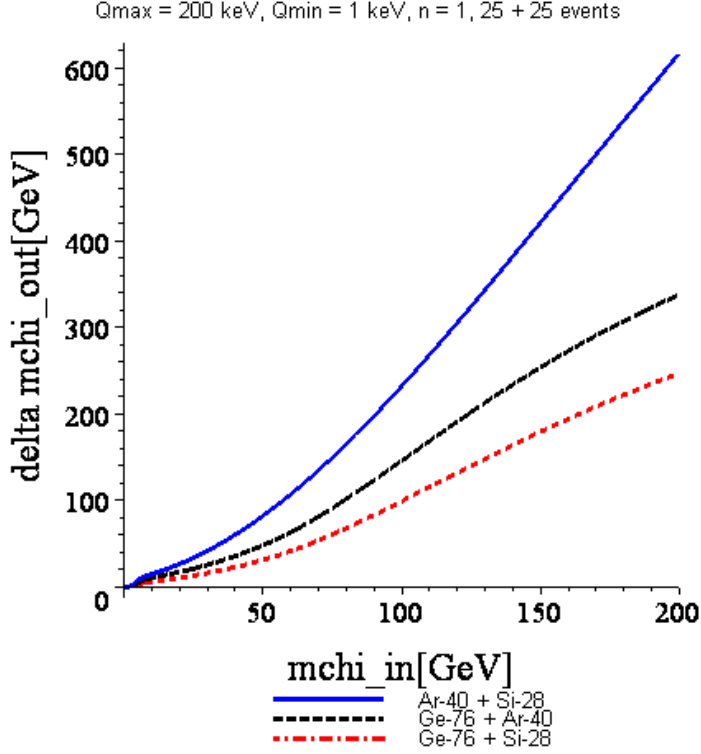}}
\end{center}
\caption{
 The curves show the statistical errors estimated by Eq.(\ref{eqn4030104})
 with different combinations of target nuclei
 as functions of the input WIMP mass.
 Each experiment has 25 events, i.e., totally 50 events.
 Parameters and indications of the lines as in Fig.~\ref{fig4030101}.} 
\label{fig4030102}
\end{figure}
 Fig.~\ref{fig4030102} shows the statistical errors estimated by Eq.(\ref{eqn4030104})
 with three different combinations of target nuclei
 as functions of the input (true) WIMP mass.
 Each experiment has 25 events, i.e., totally 50 events.
 Note that,
 in order to use the integral form of ${\rm cov}(I_n,I_m)$ in Eq.(\ref{eqn4020404}),
 a threshold energy $Q_{\rm min} = 1$ keV has been given.
 In Fig.~\ref{fig4030102}
 one can observe that
 {\em the larger the mass difference between two detector nuclei,
 the smaller the statistical error will be}.
 Hence,
 the combination with the largest mass difference,
 e.g., $\rmXA{Ge}{76}$ + $\rmXA{Si}{28}$
 will have the smallest statistical error.
 In principle
 one other combination: $\rmXA{Xe}{131}$ + $\rmXA{Ar}{40}$
 has larger mass difference and should have an even smaller statistical error.
 However,
 because the Woods-Saxon form factor has been used here,
 the integral form of ${\rm cov}(I_n,I_m)$ in Eq.(\ref{eqn4020404})
 has a pole at $Q \sim 100$ keV!
 Thus $\rmXA{Xe}{131}$ has been not used for this simulation.
 On the other hand,
 despite of the factor $1/|n|$ in Eq.(\ref{eqn4030104}),
 it has been found that
 the statistical errors increase with increasing $n$,
 except the $\rmXA{Ar}{40}$ + $\rmXA{Si}{28}$ combination.
 For this combination,
 the statistical error with $n = 2$ is a little smaller than with $n = 1$;
 but, with $n = 3$, the statistical error is significantly larger
 (and,
  as discussed above,
  the deviation of the reproduced WIMP mass should also be larger).
 Hence,
 $n = 1$ should be the best choice for $\mchi$ and $\sigma(\mchi)$
 in Eqs.(\ref{eqn4030102}) and (\ref{eqn4030104}),
 respectively.

\begin{figure}[p]
\begin{center}
\imageswitch{
\begin{picture}(15,20.5)
\put(0  , 0  ){\framebox(15,20.5){}}
\put(1  ,10.5){\framebox(13,10  ){}} 
\put(1  , 0  ){\framebox(13,10  ){}}
\end{picture}}
{\includegraphics[width=13cm]{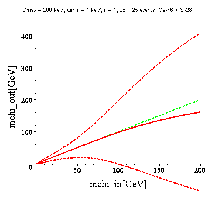} \\ \vspace{0.8cm}
 \includegraphics[width=13cm]{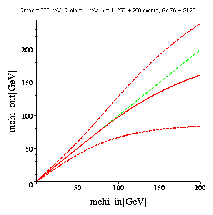}}
\end{center}
\caption{
 The reproduced WIMP mass with the statistical error
 by using $\rmXA{Ge}{76}$ and $\rmXA{Si}{28}$ as two target nuclei
 as a function of the input WIMP mass.
 The solid (red) line indicates the reproduced WIMP mass estimated by Eq.(\ref{eqn4030102}),
 the dashed (red) lines indicate the 1-$\sigma$ statistical error
 estimated by Eq.(\ref{eqn4030104}).
 The straight dash-dotted (green) line indicates the input (true) WIMP mass.
 Each experiment has 25 (250) events,
 i.e., totally 50 (500) events,
 in the upper (lower) frame.
 Parameters as in Fig.~\ref{fig4030101}.}
\label{fig4030103}
\end{figure}
 Figs.~\ref{fig4030103} show the reproduced WIMP mass with the statistical error
 by using $\rmXA{Ge}{76}$ and $\rmXA{Si}{28}$ as two target nuclei
 as a function of the input (true) WIMP mass.
 From the upper frame,
 it can be found that,
 despite of the very few (25 + 25, totally 50) events
 and correspondingly very large statistical error,
 for $\mchi \le 100~{\rm GeV}/c^2$,
 one can already extract some meaningful information on the WIMP mass.
 For example,
 for $\mchi = 25~{\rm GeV}/c^2$ and $\mchi = 50~{\rm GeV}/c^2$,
 we will reproduce $\mchi \simeq (25 \pm 13)~{\rm GeV}/c^2$
 and $\mchi \simeq (50 \pm 31)~{\rm GeV}/c^2$.
 For the case with 500 (250 + 250) total events,
 the statistical error will be reduced to less than 5 and 10 GeV$/c^2$,
 respectively!
 Certainly,
 as shown in the lower frame of Figs.~\ref{fig4030103},
 for the case with 500 total events,
 the deviation of the reproduced WIMP mass from the input one becomes important.
 Nevertheless,
 in practice,
 an experiment with more than 200 events
 should have a larger maximal measuring energy,
 and,
 as discussed above,
 the deviation can (should) be strongly reduced.

 For the simplified simulations with the integral form of $I_n$ presented above,
 the event numbers from both experiments have been considered to be equal.
 Practically,
 as described in Subsecs.~\ref{SIcrosssection} to \ref{SISDcrosssection},
 experiments with the higher mass nuclei, e.g., Ge or Xe,
 are expected to measure (much) more signal events.
 However,
 according to the expression for $\sigma(\mchi)$ in Eq.(\ref{eqn4030104})
 and the definition of $\calRn$ in Eq.(\ref{eqn4030103}),
 it can be found that
 only the terms in the brackets depends on the event number
 and the contributions from the two experiments are independent of each other.
 Moreover,
 a detailed analysis of contributions from different terms of $\sigma(\mchi)$ shows that
 the prefactor
\beq
 \frac{\calRn \sqrt{\mX/\mY} \vbrac{\mX-\mY}}{\abrac{\calRn-\sqrt{\mX/\mY}}^2}
\eeq
 which depends practically only on
 the choice of the combination of the two target nuclei
 is very large for every combination,
 while the terms in the brackets with the factor $1/|n|$
 are actually quite small.
 This implies that
 one {\em can not} reduce the statistical error of $\mchi$ estimated by Eq.(\ref{eqn4030104})
 by {\em improving only one experiment} with even very large event number,
 since the contribution from the other (poor) experiment will dominate the error.
\subsection{With $\Qthre > 0$}
 ~~~$\,$
 For the case that $\Qthre$ in Eq.(\ref{eqn4020402}) can not be neglected,
 $\calRn$ defined in Eq.(\ref{eqn4030103})
 should be modified to the following general form:
\beqn
        \calRn(\Qthre)
 \=     \bfrac{2 \QthreX^{(n+1)/2} \rthreX+(n+1) \InX \FQthreX}
              {2 \QthreX^{    1/2} \rthreX+      \IzX \FQthreX}^{1/n}
        \non\\
 \conti ~~~~~~~~ \times 
        \bfrac{2 \QthreY^{    1/2} \rthreY+      \IzY \FQthreY}
              {2 \QthreY^{(n+1)/2} \rthreY+(n+1) \InY \FQthreY}^{1/n}
\~,
\label{eqn4030201}
\eeqn
 where $\rthreX$ and $\rthreY$ should be determined by Eq.(\ref{eqn4020401})
 (practically) with different $r_1$, $k_1$, $Q_{s,1}$, and $\Qthre$.
 In this general form of $\calRn(\Qthre)$
 there are totally six variables:
 $\rthreX$, $\InX$, $\IzX$ and the other three for nucleus $Y$.
 This should generally produce a larger statistical error
 than that estimated by Eq.(\ref{eqn4030104})
 due to the contribution from $\rthre$
 (the statistical error of $\calRn(\Qthre)$ will be given in App.~\ref{sigmacalRn}).
 However,
 one can practically reduce the number of variables by choosing $n = -1$:
\beq
   \calRma(\Qthre)
 = \frac{\rthreY}{\rthreX}
   \bfrac{2 \QthreX^{1/2} \rthreX+\IzX \FQthreX}{2 \QthreY^{1/2} \rthreY+\IzY \FQthreY}
\~.
\label{eqn4030202}
\eeq
 Then $\sigma(\calRn)$ in the first line of Eq.(\ref{eqn4030104}) should be replaced by
\beqn
 \conti \sigma\aBig{\calRma(\Qthre)}
        \non\\
 \=     {\cal R}_{-1}(\Qthre)
        \cBiggl{ \Bfrac{\IzX \FQthreX}{2 \QthreX^{1/2} \rthreX+\IzX \FQthreX}^2}
        \non\\
 \conti ~~~~~~~~~~~~~~~~~~~~~~ \times 
                 \bbrac{ \frac{\sigma^2\abrac{\rthreX}}{\rthreX^2}
                        +\frac{\sigma^2\abrac{\IzX}}{\IzX^2}
                        -\frac{2 {\rm cov}\abrac{\rthreX,\IzX}}{\rthreX \IzX}}
        \non\\
 \conti ~~~~~~~~~~~~~~~~~~~~~~~~~~~~~~ 
        \cBiggr{+(X \lto Y)}^{1/2}
\~,
\label{eqn4030203}
\eeqn
 where $\sigma^2\abrac{\rthreX}$ and ${\rm cov}\abrac{\rthreX,\IzX}$ and so on
 can be calculated by Eqs.(\ref{eqn4020405}) and (\ref{eqn4020410}).

%% file: Doktorarbeit-Ch5.tex
\chapter{Annual Modulated Event Rate}
 ~~~$\,$
 In the previous chapter
 I have presented methods for reconstructing the velocity distribution function and its moments
 from the {\em time-averaged} recoil energy spectrum fitted to experimental data
 as well as from data directly.
 The annual modulation of the event rate discussed in Sec.~\ref{annualmodulation}
 has been ignored.
 As shown in Sec.~\ref{Qni},
 in the foreseeable future with rare signal events,
 the statistical errors will remain large
 and thus this is a reasonable first approximation.
 However,
 for the future detectors with strongly improved sensitivity and (very) large target mass
 (large exposure),
 the formulae and methods have to be extended
 to allow for an annual modulation of the event rate.

 In the first section of this chapter
 I extend the method developed in the previous chapter by considering
 {\em an arbitrary, but cosine-like time-dependent recoil spectrum with a one-year period}.
 In the second section
 I present the method for {\em reconstructing
 the amplitude of the (possible) annual modulation of the velocity distribution function}.
 An alternative, better way for {\em checking the annual modulation of the event rate}
 will also be described.
\section{Taking into account the annual modulation}
\label{dRdQt}
 ~~~$\,$
 For simplicity,
 in this chapter I take $t_p = 0$ in Eq.(\ref{eqn3020102}) and rewrite it as
\cheqnrefp{eqn3020102}
\beq
   v_e(t)
 = v_0 \bBig{1.05+0.07 \cos(\omega t)}
\~,
\label{eqn3020102p}
\eeq
\cheqnCN{-1}
 with
\beq
        \omega
 \equiv \frac{2 \pi}{365}
\~.
\label{eqn5010001}
\eeq
 This means that
 in the following analyses
 experiments (data) have been assumed to be started (collected), i.e., $t = 0$,
 when $v_e$ is maximal (around June 2nd, theoretically predicted)
 and the time $t$ will be measured in unit of ``day''.

\begin{figure}[p]
\begin{center}
\imageswitch{
\begin{picture}(15,20)
\put(0,0){\framebox(15,20){}}
\put(1  ,10.5){\framebox(13,9.5){}} 
\put(1  , 0  ){\framebox(13,9.5){}}
\end{picture}}
{\includegraphics[width=12.5cm]{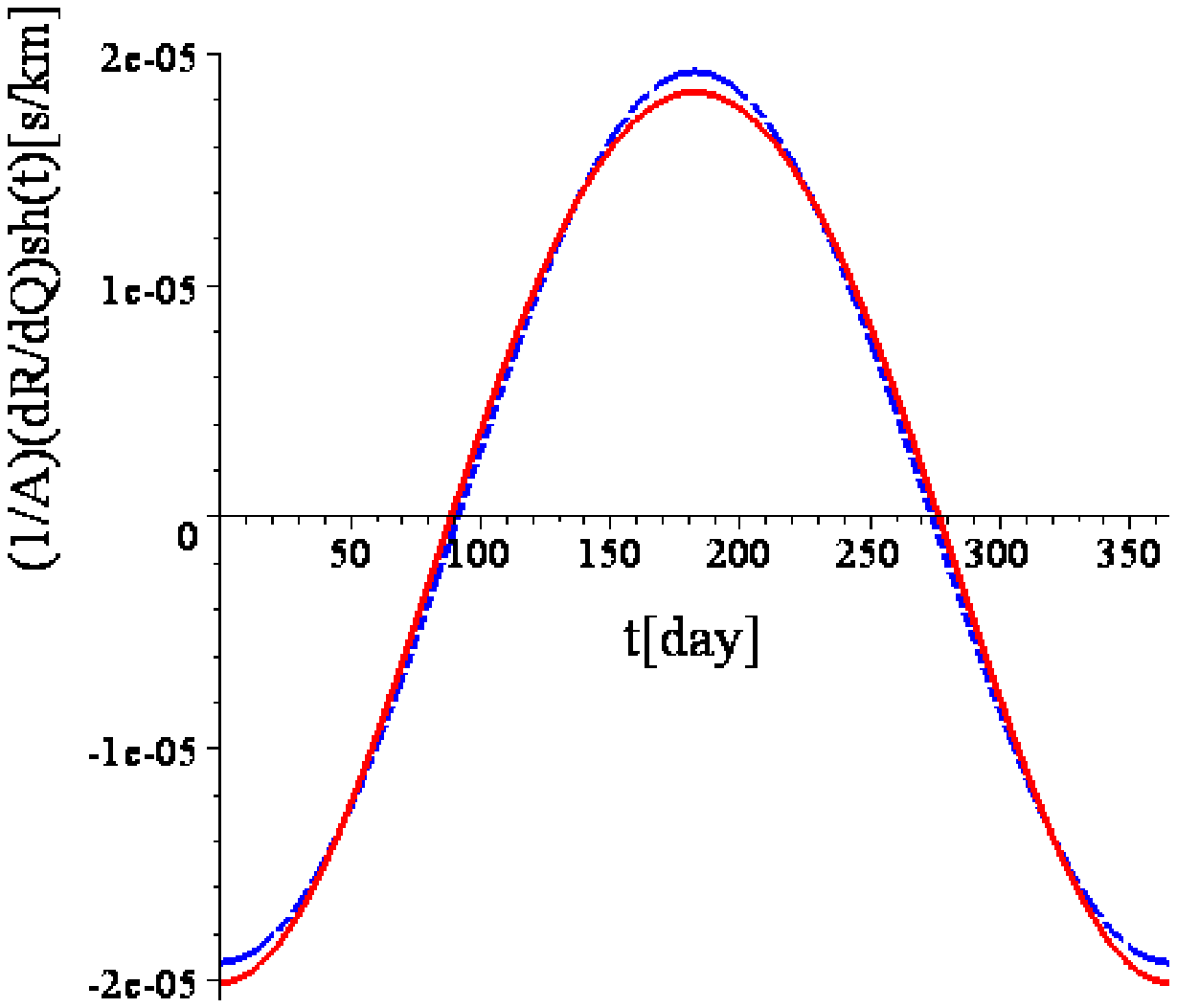} \\ \vspace{1cm}
 \includegraphics[width=12.5cm]{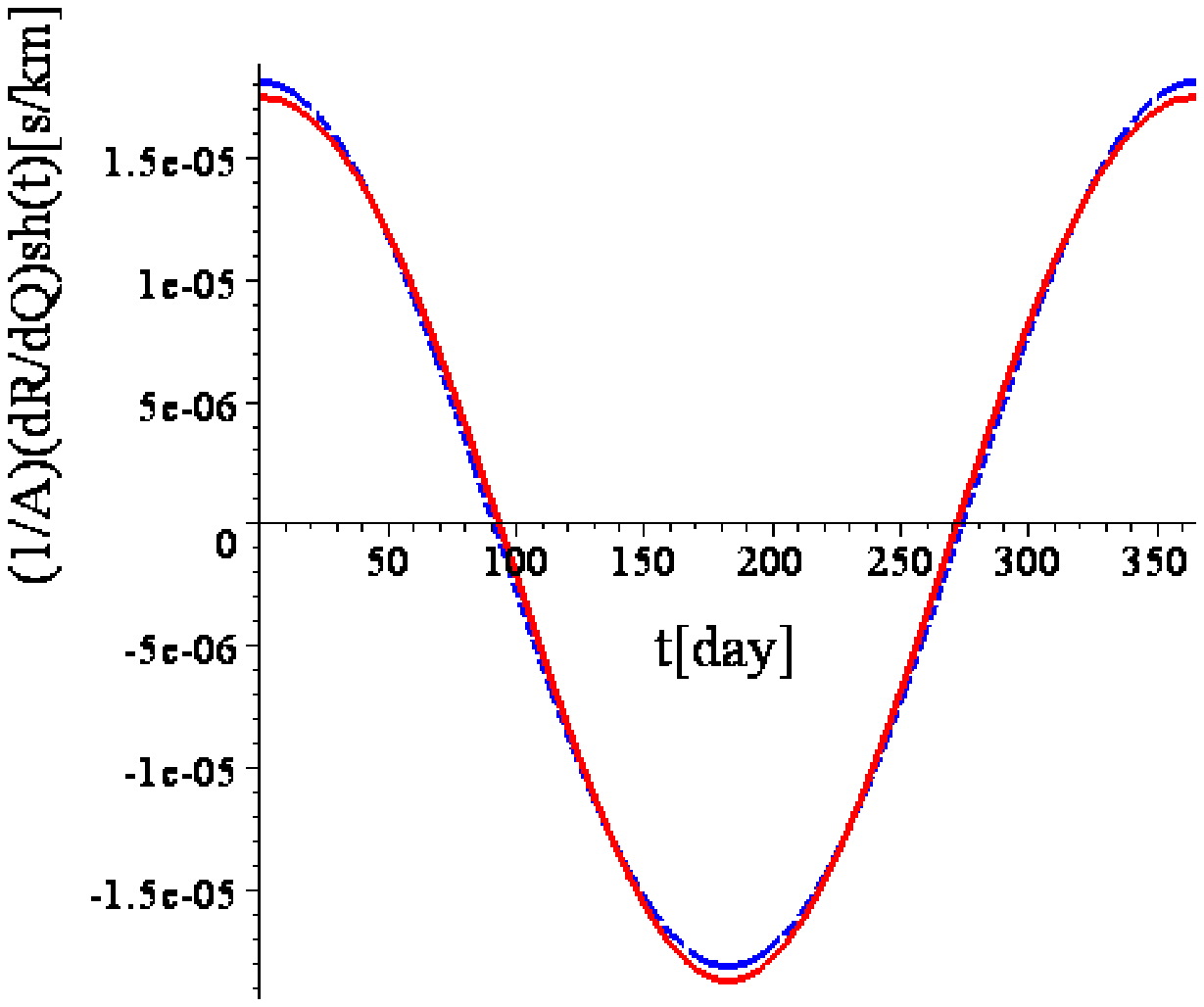}}
\end{center}
\caption{
 The solid (red) curves are
 the predicted modulations of the recoil energy spectrum
 for the shifted Maxwellian WIMP velocity distribution,
 $(dR/dQ)_\sh$ in Eq.(\ref{eqn3020103}),
 with the Woods-Saxon form factor $F_{\rm WS}^2(Q)$ in Eq.(\ref{eqn3010302});
 the dashed (blue) curves are cosine functions with an amplitude
 $[(dR/dQ)_\sh(t=0)-(dR/dQ)_\sh((t=\pi/2)]/2$.
 Here I have used $\mchi = 100~{\rm GeV}/c^2$,
 $\mN = 70.6~{\rm GeV}/c^2$ for $\rmXA{Ge}{76}$,
 $v_0 = 220~{\rm km/s}$.
 The upper and lower frames are drawn for $Q = 15$ keV and $Q = 30$ keV,
 respectively.}
\label{fig5010001}
\end{figure}

 As discussed in Secs.~\ref{elasticscattering} and \ref{annualmodulation},
 roughly speaking,
 the differential event rate for direct WIMP detection is proportional to the WIMP flux,
 or, equivalently, the velocity of the Earth relative to the WIMP halo.
 And,
 due to the motion of the Earth on an elliptical orbit around the Sun,
 the projection of the Earth's orbital speed
 on the orbital speed of the Sun around the Galactic center
 is {\em approximately} a cosine function.
 Therefore,
 as shown in Eqs.(\ref{eqn3020102}) and (\ref{eqn3020102}') above,
 the differential event rate should theoretically
 be a cosinusoidal function (c.f.~the DAMA results in Figs.~\ref{fig3070301}).
 On the other hand,
 substituting $v_e(t)$ in Eq.(\ref{eqn3020102}') into Eq.(\ref{eqn3020103}),
 it can be found that
 $(dR/dQ)_\sh$ are not exact cosine but {\em cosine-like} functions
 with a period of 365 days (shown in Figs.~\ref{fig5010001}).
 According to this observation,
 I assume generally
 {\em an arbitrary, but cosine-like time-dependent recoil spectrum with a one-year period}
 and then expand this spectrum and its corresponding velocity distribution function
 as {\em Fourier cosine series} as:
\beq
   \adRdQt
 = \adRdQz+\adRdQa \cos(\omega t)+\adRdQ_{(2)} \cos(2 \omega t)+\cdots
\~,
\label{eqn5010002}
\eeq
 and
\beq
   f_1(v,t)
 = \fzv+\fav \cos(\omega t)+f_{1,(2)}(v) \cos(2 \omega t)+\cdots
\~.
\label{eqn5010003}
\eeq
 According to Eq.(\ref{eqn3010108}),
 $\TadRdQt$ and $f_1(v,t)$ must satisfy the equation
 for the {\em time-dependent} WIMP-nucleus scattering spectrum:
\beq
   \adRdQt
 = \calA \FQ \intvmin \bfrac{f_1(v,t)}{v} dv
\~,
\label{eqn5010004}
\eeq
 and,
 consequently,
 each pair of their Fourier coefficients must satisfy a time-independent equation:
\beq
   \adRdQm
 = \calA \FQ \intvmin \bfrac{\fmv}{v} dv
\~,
   ~~~~~~~~~~~~~~~~ 
   m
 = 0,~1,~2,~\cdots
.
\label{eqn5010005}
\eeq
 Moreover,
 if we neglect
 (due to the very low detection rate discussed in Sec.~\ref{elasticscattering}
  and the tiny difference shown in Figs.~\ref{fig5010001},
  we can practically neglect)
 the terms with $m \ge 2$ in Eqs.(\ref{eqn5010002}) and (\ref{eqn5010003}),
 $\TadRdQz$ and $\fzv$ above are the time-averaged scattering spectrum
 and the time-averaged velocity distribution function of WIMPs,
 which we considered in Chap.~4,
 and $\TadRdQa$ and $\fav$ are the amplitudes of the annual modulations
 of the scattering spectrum and its corresponding velocity distribution.
 In addition,
 since $\TadRdQm$ are Fourier coefficients of $\TadRdQt$,
 we have
\beq
   \adRdQz
 = \frac{1}{365} \intyr \adRdQt dt
\~,
\label{eqn5010006}
\eeq
 and
\beq
   \adRdQm
 = \frac{2}{365} \intyr \adRdQt \cos(m \omega t) \~ dt
\~,
   ~~~~~~~~~~~~~~~~ 
   m
 = 1,~2,~\cdots
.
\label{eqn5010007}
\eeq

 Now,
 as mentioned in Subsec.~\ref{windowing},
 the important elements needed for the reconstruction of $f_{1,r}$ in Eq.(\ref{eqn4020301})
 are the number of events $N_{\mu}$ in the $\mu$-th window given in Eq.(\ref{eqn4020202}),
 as well as the averages $\Bar{Q-Q_{\mu}}|_{\mu}$ given in Eq.(\ref{eqn4020203}),
 which are theoretically defined as,
 respectively,
\beq
        N_{\mu}
 \equiv \int_{Q_{\mu}-w_{\mu}/2}^{Q_{\mu}+w_{\mu}/2} \adRdQz dQ
\~,
\label{eqn5010008}
\eeq
 and

\beq
        \Bar{(Q-Q_{\mu})^{\lambda}}|_{\mu}
 \equiv \frac{1}{N_{\mu}}
        \int_{Q_{\mu}-w_{\mu}/2}^{Q_{\mu}+w_{\mu}/2} (Q-Q_{\mu})^{\lambda} \adRdQz dQ
\~,
\label{eqn5010009}
\eeq
 where I have used generally the $\lambda$-th moment of the averaged recoil spectrum $\TadRdQz$.
 Substituting Eq.(\ref{eqn5010006}) into Eqs.(\ref{eqn5010008}) and (\ref{eqn5010009}),
 it can be found easily that,
 for a time-dependent recoil spectrum with a one-year period,
\beq
   N_{\mu}
 = \frac{1}{365} \intyr \int_{Q_{\mu}-w_{\mu}/2}^{Q_{\mu}+w_{\mu}/2} \adRdQt dQ \~ dt
 = \frac{N_{\mu,1~{\rm yr}}}{365}
\~,
\label{eqn5010010}
\eeq
 and
\beqn
    \Bar{(Q-Q_{\mu})^{\lambda}}|_{\mu}
 \= \frac{1}{N_{\mu}}
    \bbrac{\frac{1}{365} \intyr
           \int_{Q_{\mu}-w_{\mu}/2}^{Q_{\mu}+w_{\mu}/2} (Q-Q_{\mu})^{\lambda} \adRdQt dQ \~ dt}
    \non\\
 \= \frac{1}{N_{\mu,1~{\rm yr}}}
    \sum_{i=1}^{N_{\mu,1~{\rm yr}}} \abrac{Q_{\mu,i}-Q_{\mu}}^{\lambda}
\~,
\label{eqn5010011}
\eeqn
 where $Q_{\mu,i}$, $i = 1,~2,~\cdots,~N_{\mu,1~{\rm yr}}$,
 are the measured recoil energies from the direct WIMP detection experiment
 in the $\mu$-th window {\em in one year}.
 Note that the ``='' sign in the second line of Eq.(\ref{eqn5010011})
 denotes not mathematically equal
 but an experimental estimator for $\Bar{(Q-Q_{\mu})^{\lambda}}|_{\mu}$.

 Comparing these results with the expressions
 in Eqs.(\ref{eqn4020112}) and (\ref{eqn4020114}),
 Eqs.(\ref{eqn5010010}) and (\ref{eqn5010011}) show that,
 for an {\em arbitrary time-dependent} recoil spectrum with a one-year period
 (even though it is not cosine-like),
 we just have to take the experimental data {\em over some whole years} to find out
 the average event number (per day)
 and the annual average value of the energy transfer $(Q-Q_{\mu})^{\lambda}$
 in the $\mu$-th window (or bin).
 Then we can reconstruct the time-averaged velocity distribution
 by means of the method presented in the previous chapter directly.
 Moreover,
 the results above show also that
 it is actually {\em not necessary} to know when $v_e$ is maximal
 but only use all events collected in these (whole) years.
\section{Reconstructing the modulated amplitude of $f_1(v)$}
 ~~~$\,$
 In this section
 I follow the trick used with $\TadRdQz$ in the previous section
 and develop a method for reconstructing the (annual) modulated amplitude of $f_1(v)$.
 Meanwhile,
 I will also introduce two criteria for checking the annual modulation of the event rate.
\subsection{Criteria for the annual modulation}
\label{DeltaQt}
\begin{figure}[t]
\begin{center}
\imageswitch{
\begin{picture}(12,9)
\put(0,0){\framebox(12,9){}}
\end{picture}}
{\includegraphics[width=12cm]{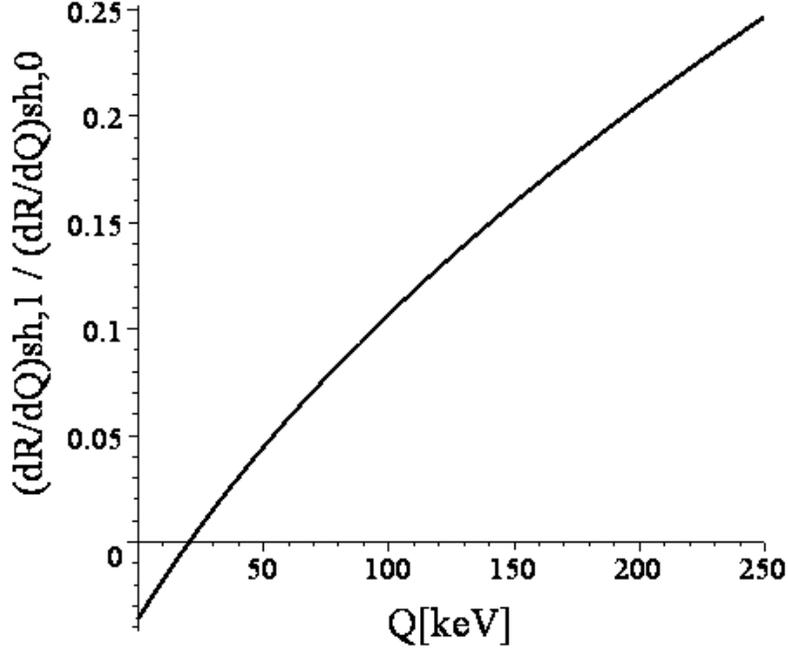}}
\end{center}
\caption{
 The curve shows the ratio of the modulated amplitude of the recoil spectrum, $\TadRdQa$,
 to the time-averaged recoil spectrum, $\TadRdQz$, as a function of the recoil energy $Q$.
 Parameters as in Figs.~\ref{fig5010001}.} 
\label{fig5020101}
\end{figure}

 ~~~$\,$
 Replacing $\TadRdQz$ in Eq.(\ref{eqn5010009}) by $\TadRdQa$ in Eq.(\ref{eqn5010007}),
 it can be found that
\footnote{
 For simplicity and clarity,
 I discuss in this section the case with bins,
 thus $\mu$ and $w_{\mu}$ in Eq.(\ref{eqn5010009})
 have been replaced here by $n$ and $b_n$,
 respectively.
 However,
 all formulae given in this section can be used for the case with windows
 by substituting $n$ and $b_n$ by $\mu$ and $w_{\mu}$.}
\beqn
 \conti \frac{1}{N_n} \intQnbn (Q-Q_n)^{\lambda} \adRdQa dQ
        \non\\
 \=     \frac{1}{N_n}
        \bbrac{\frac{2}{365} \intyr \intQnbn (Q-Q_n)^{\lambda} \cos(\omega t) \adRdQt dQ \~ dt}
        \non\\
 \=     \frac{2}{\Nnyr} \sumiNnyr \abrac{\Qni-Q_n}^{\lambda} \cos(\omega\tni)
\~,
\label{eqn5020101}
\eeqn
 where $\tni$, $i=1,~2,~\cdots,~\Nnyr$, $n=1,~2,~\cdots,~B$,
 are the ``measuring times''
 at which we measure the recoil energies $\Qni$ in the $n$-th bin in one year.
 Note that,
 first,
 the ``='' sign in the second line of Eq.(\ref{eqn5020101})
 denotes again an experimental estimator;
 second,
 the factor $\cos(\omega\tni)$ comes from the integral in Eq.(\ref{eqn5010007})
 (not from the second term of $v_e(t)$ in Eq.(\ref{eqn3020102}')!).
 On the other hand,
 by substituting $v_e(t)$ in Eq.(\ref{eqn3020102}')
 into $(dR/dQ)_\sh$ in Eq.(\ref{eqn3020103}),
 it can be found that
 the ratio of the modulated amplitude of the recoil spectrum, $\TadRdQa$,
 to the time-averaged recoil spectrum, $\TadRdQz$,
 increases monotonically with the recoil energy $Q$ and
 is approximately a linear function of $Q$
 (shown in Fig.~\ref{fig5020101}).
 Hence,
 I introduce an ansatz for the modulated amplitude of the recoil spectrum
 in the $n$-th bin:
\beq
   \adRdQan
 = \adRdQz \cdot \bBig{l_n (Q-Q_n)+h_n}
\~,
   ~~~~~~~~~~~~~~~~ 
   n
 = 1,~2,~\cdots,~B
.
\label{eqn5020102}
\eeq
 Note that $\TadRdQz$ here indicates generally a time-averaged recoil spectrum,
 not specified to the exponential ansatz in Eq.(\ref{eqn4020105}).
 Substituting the ansatz for $\TadRdQan$
 into the left-hand side of the Eq.(\ref{eqn5020101}),
 it can be found that
\beq
   \frac{1}{N_n} \intQnbn (Q-Q_n)^{\lambda} \adRdQan dQ
 = l_n \~ \bQxn{\lambda+1}+h_n \~ \bQxn{\lambda}
\~.
\label{eqn5020103}
\eeq
 Setting $\lambda = 0$ and 1
 and combining Eqs.(\ref{eqn5020101}) and (\ref{eqn5020103}),
 $l_n$ and $h_n$ in Eq.(\ref{eqn5020102}) can be solved as
\beq
   l_n
 = \frac{2 \~ \bQtn-2 \~ \btn \~ \bQn}{\bQQn-\bQn^2}
\~,
\label{eqn5020104}
\eeq
 and
\beq
   h_n
 = \frac{2 \~ \btn \~ \bQQn-2 \~ \bQtn \~ \bQn}{\bQQn-\bQn^2}
\~,
\label{eqn5020105}
\eeq
 where I have defined
\beq
        \bQxtyn{\lambda}{\rho}
 \equiv \frac{1}{\Nnyr} \sumiNnyr \abrac{\Qni-Q_n}^{\lambda} \cos^{\rho}(\omega\tni)
\~.
\label{eqn5020106}
\eeq

 Due to the annual modulation effect,
 around $t = 0$ we should get more events than around $t = \pi$.
 Recall that I have assumed that experiments start when $v_e$ is maximal.
 Thus,
 even though $\int_0^{2 \pi} \cos(\omega t) \~ dt = 0$,
 $\btn$ above are generally not equal to 0.
 Moreover,
 $l_n$ in Eq.(\ref{eqn5020104}) can be rewritten as
\cheqnrefp{eqn5020104}
\beq
   l_n
 = \frac{2 \~ {\rm cov}\abrac{\bQn,\btn}}{\sigma^2\abrac{\bQn}}
\~,
\label{eqn5020104p}
\eeq
\cheqnCN{-1}
 where ${\rm cov}(\bQn,\btn)$ is the covariance between
 the average value of the measured recoil energies $\Qni-Q_n$
 and that of $\cos(\omega\tni)$ in the $n$-th bin.
 According to Fig.~\ref{fig5020101} and the ansatz in Eq.(\ref{eqn5020103}),
 $l_n$ should be generally $>$ 0 and
 and this means that $\bQn$ and $\btn$ should be positively correlated.
 This result offers a better way to test the annual modulation effect!
 Traditionally,
 in order to confirm the annual modulation of the event rate,
 one has to collect the recoil events in a given energy range
 in several short time intervals (a few days to a couple weeks)
 and than compare the numbers of collected events
 in the different time intervals in one year
 (e.g., the DAMA 4-year and 7-year results shown in Figs.~\ref{fig3070301}).
 As mentioned in Sec.~\ref{annualmodulation},
 the annual modulation of event rate is expected to be only a few percent
 (about $-4\% \sim 5\%$ in the energy range between 0 and 50 keV,
  see Fig.~\ref{fig5020101})
 and this method can be used
 once more than one hundred events (in a few days!) have been accumulated.
 However,
 according to the discussion above,
 one can now collect all recoil events in a relatively larger energy range
 (since the calculation with bins can be extended directly to that with windows)
 in one year (or even several years)
 and then only has to check the following quantities:
\beq
        \Delta_t
 \equiv \Bar{\cos(\omega t)}
 =      \frac{1}{N_{tot}} \sum_a \cos(\omega t_a)
\~,
\label{eqn5020107}
\eeq
 and
\beq
        \Delta_{Q,t}
 \equiv {\rm cov}\abrac{\Bar{Q},\Bar{\cos(\omega t)}}
 =      \frac{1}{N_{tot}-1} \bbigg{\Bar{Q \cos(\omega t)}-\Bar{Q}~\Bar{\cos(\omega t)}}
\~,
\label{eqn5020108}
\eeq
 where the averages are over all events in the energy range which one concerns
 and $N_{tot}$ is the total event number in this energy range.
 If the annual modulation of event rate exists,
 one should than get $\Delta_t \ne 0$ and $\Delta_{Q,t} > 0$.
 Note that,
 for the case that some time-independent background events mixed into the true signals,
 the two quantities above will be {\em underestimated} through the averaging;
 or even worse,
 if most of the background events have been discriminated,
 then the contribution from the rest events can not cancel each other any more.
 However,
 for the case that
 the time-independent background events dominate the whole data set,
 one can use a quantity modified from $\Delta_t$ in Eq.(\ref{eqn5020107}):
\beq
        \Delta_t'
 \equiv \sum_a \cos(\omega t_a)
\~.
\label{eqn5020109}
\eeq
 The quantity $\Delta_t'$ defined here
 is not the average but the sum of the cosine values of the measuring times.
 Hence,
 the contributions from background events
 will cancel each other and not change the value of $\Delta_t'$ (too much).
 But in the statistical error of $\Delta_t'$:
\footnote{
 Here I have assumed that $N_{tot} \gg 1$ and
 then used
 $  \sigma^2\abig{\Bar{\cos(\omega t)}}
  = \bbrac{\Bar{\cos^2(\omega t)}-\Bar{\cos(\omega t)} ^2}/N_{tot}$.}
\beq
   \sigma^2\abrac{\Delta_t'}
 = \sum_a \cos^2(\omega t_a)
\~,
\label{eqn5020110}
\eeq
 there can not be any cancellation,
 i.e., $\sigma^2\abrac{\Delta_t'}$ will increase with the total event number $N_{tot}$.
 In addition,
 for a ``time-dependent'' background
 the quantity in Eq.(\ref{eqn5020109}) can not be used any more.

 Furthermore,
 for the check of the quantities in Eqs.(\ref{eqn5020107}) and (\ref{eqn5020108}),
 it is important to know when $v_e$ is maximal.
 This offers in practice a possibility to determine
 (to check theoretically predicted) $t_p$ in Eq.(\ref{eqn3020102}).
 One sets the starting date of the experiment on January 1st,
 inserts a phase $\varphi$ into Eqs.(\ref{eqn5020107}) and (\ref{eqn5020108}),
 and then finds out when the quantity
\cheqnrefp{eqn5020107}
\beq
   \Delta_{t-\varphi}
 = \Bar{\cos \omega (t-\varphi)}
\label{eqn5020107p}
\eeq
 is (almost) equal to 0,
 which corresponds to $t_p \pm \pi/2 \omega$,
 and when the quantity
\cheqnrefp{eqn5020108}
\beq
   \Delta_{Q,t-\varphi}
 = {\rm cov}\abrac{\Bar{Q},\Bar{\cos \omega (t-\varphi)}}
\label{eqn5020108p}
\eeq
\cheqnCN{-2}
 has a maximal value (positive), a minimal value (negative), or is (almost) equal to 0,
 which correspond to $t_p$, $t_p \pm \pi/\omega$, or $t_p \pm \pi/2 \omega$,
 respectively.
 Certainly,
 for the case that such annual modulation of the event rate does not exist,
 one will find that
 $\Delta_t$ and $\Delta_{Q,t}$ are independent of $\varphi$
 and always (approximately) equal to 0.
\subsection{Reconstructing the modulated amplitude of $f_1(v)$}
\label{f1un1(v)}
 ~~~$\,$
 According to Eq.(\ref{eqn4010008}),
 each coefficient of the time-dependent velocity distribution function,
 $f_1(v,t)$, in Eq.(\ref{eqn5010003}),
 can be solved from Eq.(\ref{eqn5010005}) as
\beq
   f_{1,(m)}(Q)
 = \calN \cbrac{-2 Q \cdot \dd{Q}\bbrac{\frac{1}{\FQ} \adRdQm}}
\~,
   ~~~~~~~~ 
   m
 = 0,~1,~2,~\cdots
,
\label{eqn5020201}
\eeq
 since, for all $m$,
\beq
     f_{1,(m)}(v \to \infty)
 \to 0
\~,
\label{eqn5020202}
\eeq
 see Eq.(\ref{eqn4010003}).
 Substituting the ansatz for $\TadRdQan$ in Eq.(\ref{eqn5020103}) at first
 and then the ansatz for $\TadRdQzn$ in Eq.(\ref{eqn4020105}) into Eq.(\ref{eqn5020201}),
 the ratio of the modulated amplitude of the velocity distribution function
 to the time-averaged one at the point $Q = Q_n$ (not at $v = v_n$!) can be obtained as
 (a detailed derivation will be given in App.~\ref{etan})
\beq
        \eta_n
 \equiv \frac{f_{1,(1),n}(Q_n)}{f_{1,(0),n}(Q_n)}
 =      h_n-l_n \bBigg{\dd{Q} \ln \FQ \bigg|_{Q = Q_n}-k_n}^{-1}
\~.
\label{eqn5020203}
\eeq
 Note that the first term in the brackets has been evaluated at $Q = Q_n$
 (not at $Q = Q_{s,n}$!).

 The result in Eq.(\ref{eqn5020203}) has three advantages.
 First,
 since $\eta_n$ here
 is the {\em ratio} of the modulated amplitude of the velocity distribution function
 to the time-average one in the $n$-th $Q$-bin,
 it is not necessary to combine the data from all bins to get the normalization constant,
 each one of these $\eta_n$ is independent of the others.
\footnote{
 For the case with windows,
 the $\eta_n$ are no more independent of each other.
 But the off-diagonal entries of the correlation matrix
 of the statistical error of the $\eta_n$
 are practically not important.}
 Second,
 due to the same reason,
 one can estimate the $\eta_n$
 even if he can not obtain (enough) events in the high energy range ($>$ 100 keV).
 It is,
 in contrast,
 necessary to collect enough events until high energy
 ($\approx$ 200 keV for a WIMP mass $\sim 100~{\rm GeV}/c^2$
  and $\rmXA{Ge}{76}$ as target nucleus)
 in order to determine the normalization constant $\calN$ in Eq.(\ref{eqn4020303}).
 Third,
 the $\eta_n$ are independent of $\alpha$ defined in Eq.(\ref{eqn3010107}),
 or, equivalently, independent of the WIMP mass $\mchi$.
 Certainly,
 one can use the method described in Sec.~\ref{mchi} to determine the WIMP mass.
 But for the case with very rare total events
 or too few events in the energy range higher than e.g., 100 keV,
 the deviation of the estimation of the WIMP mass from the true one
 and the statistical error will be very large.
 However,
 due to the independence of the $\eta_n$ of $\alpha$,
 the reconstruction by Eq.(\ref{eqn5020203}) will not be affected
 by the (large) uncertainty of the estimation of the WIMP mass.

 The statistical errors of $\eta_n$ in Eq.(\ref{eqn5020203}) can be expressed as
\beq
   \sigma^2(\eta_n)
 = \sum_{\nu=1}^{4} \aPp{\eta_n}{\ynun}^2 \sigma^2(\ynun)
  +\sumd{\nu,\tau=1}{\tau\neq\nu}^{4}
   \aPp{\eta_n}{\ynun} \aPp{\eta_n}{y_{\tau,n}} {\rm cov}\abrac{\ynun,y_{\tau,n}}
\~.
\label{eqn5020204}
\eeq
 Here I have defined
\beq
   \ynun
 = \bQn,~\bQQn,~\btn,~\bQtn
\~,
\label{eqn5020205}
\eeq
 and one has
\beq
   \sigma^2\abrac{\ynun}
 = {\rm cov}\abrac{\ynun,\ynun}
\~,
\label{eqn5020206}
\eeq
 with
\beqn
 \conti {\rm cov}\abrac{\bQxtyn{\lambda}{\rho},\bQxtyn{\nu}{\tau}}
        \non\\
 \=     \frac{1}{N_n-1}
        \bbiggl{ \bQxtyn{\lambda+\nu}{\rho+\tau}}
        \non\\
 \conti ~~~~~~~~~~~~~~~~ 
        \bbiggr{-\bQxtyn{\lambda}{\rho} \bQxtyn{\nu}{\tau}}
\~.
\label{eqn5020207}
\eeqn
 According to Eqs.(\ref{eqn4020113}),
\footnote{
 Since $l_n$ and $h_n$ are functions of both $\bQn$ and $\bQQn$,
 for simplicity,
 I have used here the expression for $k_n$ in Eq.(\ref{eqn4020113}).}
 (\ref{eqn5020104}), (\ref{eqn5020105}), and (\ref{eqn5020203}),
 the derivatives of $\eta_n$ with respect to each of the $\ynun$ in Eq.(\ref{eqn5020205})
 can be found easily as
\cheqnCa
\beqn
        \Pp{\eta_n}{\bQn}
 \=     \frac{h_n}{\sigma_n} \abigg{\bQn+K_n}
       -\frac{l_n}{\sigma_n} \bbigg{\bQQn+\bQn \~ K_n}
        \non\\
 \conti ~~~~~~~~~~~~~~~~ 
       -k_n \afrac{l_n K_n^2}{\bQn}
\~,
\eeqn
\cheqnCb
\beq
   \Pp{\eta_n}{\bQn}
 = \frac{l_n}{\sigma_n} \aBig{\bQn+K_n}-\frac{k_n^2}{2} \afrac{l_n K_n^2}{\bQn}
\~,
\eeq
\cheqnCc
\beq
   \Pp{\eta_n}{\btn}
 = \frac{2}{\sigma_n} \bbigg{\bQQn+\bQn \~ K_n}
\~,
\eeq
\cheqnCNx{-1}{d}
\beq
   \Pp{\eta_n}{\bQtn}
 =-\frac{2}{\sigma_n} \aBig{\bQn+K_n}
\~,
\eeq
\cheqnC
 where I have defined
\beq
        \sigma_n
 \equiv \sigma^2\abrac{\bQn}
 =      \frac{1}{N_n-1} \bbigg{\bQQn-\abrac{\bQn}^2}
\~,
\eeq
 and
\beq
        K_n
 \equiv \bbrac{\dd{Q} \ln \FQ \bigg|_{Q = Q_n}-k_n}^{-1}
\~.
\label{eqn5020208}
\eeq

%% file: Doktorarbeit-Ch6.tex
\chapter{Summary and Conclusions}
 ~~~$\,$
 In this thesis
 I have presented methods
 which allow to extract information
 on the WIMP velocity distribution as well as on the WIMP mass
 from the recoil energy spectrum $dR/dQ$
 measured in elastic WIMP-nucleus scattering experiments.
 In the long term
 the information on the WIMP velocity distribution
 can be used to test or constrain models of the dark halo of our Galaxy;
 this information would complement the information on the density distribution of WIMPs,
 which can be derived e.g., from measurements of the Galactic rotation curve.
 Meanwhile,
 the information on the WIMP mass
 can be used to constrain e.g., SUSY models in the elementary particle physics
 and compare with information from future collider experiments.

 In Sec.~\ref{dRdQ}
 I have derived the expression that allow to reconstruct
 the normalized one-dimensional velocity distribution function of WIMPs, $f_1(v)$,
 given an expression (e.g., a fit to data) for the recoil spectrum.
 I have also derived formulae for determining the moments of $f_1(v)$.
 All these expressions are independent of
 the as yet unknown WIMP density near the Earth
 as well as of the WIMP-nucleus cross section.
 The only information about the nature of WIMPs which one needs is the WIMP mass.

 Then,
 in Sec.~\ref{Qni},
 I have presented methods
 that allow to apply the expressions derived in Sec.~\ref{dRdQ}
 directly to experimental data,
 without the need to fit the recoil spectrum to a functional form.
 A good variable that allows direct reconstruction of $f_1(v)$
 is the average recoil energy in a given bin
 (or ``window'', introduced in Subsec.~\ref{windowing}).
 This average energy is sensitive to the slope of the recoil spectrum,
 which is the quantity one needs to reconstruct $f_1(v)$.
 The statistical error of the reconstruction of $f_1(v)$ has been analyzed.
 Unfortunately it has been found that
 several hundred events will be needed for the method
 to be able to extract meaningful information on $f_1(v)$.
 This is partly due to the fact that $f_1(v)$ is normalized,
 i.e., only the shape of this distribution contains meaningful information,
 and partly because this shape depends on the slope of the recoil spectrum,
 which is intrinsically difficult to determine.

 Meanwhile,
 a method for determining the moments of $f_1(v)$
 has also been presented in Subsec.~\ref{expvvn}.
 Numerical simulation shows that
 very rare events with large recoil energies
 contribute significantly more to the higher moments.
 Nevertheless,
 because this method uses the whole experimental data together
 to determine the moments of $f_1(v)$,
 it has been found that,
 based only on the first two or three moments,
 some non-trivial information can already be extracted from ${\cal O}(20)$ events.

 As noted earlier,
 one needs to know the WIMP mass $\mchi$
 for the reconstruction of (the moments of) the velocity distribution.
 Moreover,
 although in well-motivated WIMP models
 $\mchi$ can be determined with high accuracy from future collider data,
 we will have to check experimentally that
 the particles produced at colliders are in fact the same ones
 seen in direct Dark Matter detection experiments
 which form the Galactic halo.

 In Sec.~\ref{mchi}
 I have developed a method for (self-)determining $\mchi$
 based on the determination of the moments of $f_1(v)$
 by combining two (or more) experiments with different detector materials.
 The numerical analysis shows that,
 the larger the mass difference between two target nuclei,
 the smaller the statistical error will be.
 Hence,
 the combinations of two semiconductor detectors: Si and Ge
 and of two liquid noble gas detectors: Ar and Xe
 should be good choices.
 Meanwhile,
 due to the maximal measuring energy of the detector,
 there will be a deviation of the reproduced WIMP mass from the true one
 as WIMP masses $\amssyasymc{38}~60~{\rm GeV}/c^2$.
 However,
 the numerical analysis shows also that,
 for WIMP masses $\le 100~{\rm GeV}/c^2$
 some meaningful information on the WIMP mass can already be extracted
 from ${\cal O}(50)$ total (each experiment ${\cal O}(25)$) events.

 At the first step I have ignored the annual modulation of the WIMP flux.
 Given the large statistical errors expected in the foreseeable future,
 this is a reasonable first approximation.
 However,
 for the future detectors with strongly improved sensitivity
 and (very) large target mass (large exposure),
 the formulae and methods have to be extended
 to allow for an annual modulation of the event rate.
 Hence,
 in Sec.~\ref{dRdQt}
 I have discussed the extension
 of the reconstruction of the velocity distribution
 by taking into account the time dependence of the recoil spectrum.
 The analysis shows that
 the two important observables for reconstructing the velocity distribution function:
 the event number and the average recoil energy measured in a given bin (or window),
 can be obtained as the annual average of the total event number (per day)
 and the average value of the recoil energies
 measured in experiments operated over some whole years.

 Moreover,
 in Subsec.~\ref{f1un1(v)}
 I have presented a method for reconstructing
 the ratio of the modulated amplitude of the velocity distribution
 to the time-averaged one.
 The only information which one needs
 is the measured recoil energies and their measuring times.
 This reconstruction is independent of the WIMP mass
 and can be done even if we can not obtain (enough) events in the high energy range.
 Hence,
 before the sensitivity of detectors can be improved
 to offer enough data until high energy range,
 reconstructing this ratio directly from experimental data
 and comparing it with the theoretical predictions
 might be the best possibility to test
 the different models of the halo Dark Matter.

 Furthermore,
 in Subsec.~\ref{DeltaQt},
 I have given an alternative, and also better way
 to check whether the annual modulation of the event rate exists
 and thereby test models of the Dark Matter halo.
 The main advantage of this test is that,
 instead of (traditionally) comparing the numbers of collected signal events
 in different, short time intervals in one year,
 one can now use information,
 i.e., the measured recoil energies and their measuring times,
 from all signal events collected in one or even several years together.
 For the case that the background events dominate the whole data set,
 this test might be still useful,
 if one expects that the background is (almost) time independent.

 The analyses of this work are based on several simplifying assumptions.
 First,
 all experimental systematic uncertainties,
 as well as the uncertainty on the measurement of the recoil energy $Q$
 have been ignored.
 This should be a quite good approximation,
 given that we will have to live with quite large statistical uncertainties
 in the foreseeable future.
 Recall that,
 as shown in Secs.~\ref{cyrogenic} to \ref{SDD},
 not a single WIMP event has as yet been unambiguously recorded.

 I have also assumed that the detector consists of a single isotope.
 This is quite realistic for the current semiconductor (Si or Ge) detectors.
 On the other hand,
 for detectors containing more than one nucleus,
 by simultaneously measuring two signals,
 one might be able to tell on an event-by-event basis
 which kind of nucleus has been struck (see Subsec.~\ref{CRESST}).
 In this case,
 the methods can be applied straightforwardly to the separate sub-spectra.

 The analyses treat each recorded event as signal,
 i.e., background has been ignored altogether.
 At least after introducing a lower cut $\Qthre$ on the recoil energy,
 this may in fact not be unrealistic for modern detectors,
 which contain cosmic ray veto and neutron shielding systems
 (described in Subsec.~\ref{background}).
 Background subtraction should be relatively straightforward
 when fitting some function to the data,
 which would allow to use the expressions given in Sec.~\ref{dRdQ}.
 It should also be feasible in the method described in Sec.~\ref{Qni},
 if its effect on the average $Q$-values in the bins can be determined;
 in particular,
 an approximately flat ($Q$-independent) background
 would not change the slope of the recoil spectrum.

 In summary,
 a theoretical exploration of studying
 what direct Dark Matter detection experiments can teach us
 about the properties of Dark Matter particles in our Galactic neighborhood,
 e.g., their velocity distribution and their mass,
 the so-called ``WIMP astronomy'',
 has been started.
 However,
 the analyses show that
 this will require substantial data samples.
 Hopefully this work will encourage our experimental colleagues to plan future experiments
 well beyond the stage of ``merely'' detecting Dark Matter.
 On the other hand,
 due to the significantly reduced condition (less than 100 events)
 for extracting meaningful information on the WIMP mass
 by means of data from direct Dark Matter detection experiments,
 a championship for finding new particle(s)
 between the collaborations of direct Dark Matter detection
 and that of collider experiments has also been started.

%% file: Doktorarbeit-AppA.tex
\chapter{Expression of the Velocity Distribution of WIMPs}
 ~~~$\,$
 In this chapter
 I discuss at first some properties of the auxiliary function $F_1(v)$
 defined in Eq.(\ref{eqn4010001}).
 Then I show two different approaches
 to find out the expression of $f_1(v)$ in Eq.(\ref{eqn4010008}),
 and derive the normalization constant $\calN$ in Eq.(\ref{eqn4010009})
 as well as the expression of moments of $f_1(v)$, $\expv{v^n}$, in Eq.(\ref{eqn4010010}).
\section{Properties of $F_1(v)$ defined in Eq.(\ref{eqn4010001})}
\label{F1(v)}
\begin{figure}[t]
\begin{center}
\imageswitch{
\begin{picture}(7.5,5.5)
\put(0,0){\usebox{\FigureAa}}
\end{picture}}
{\includegraphics[width=10.5cm]{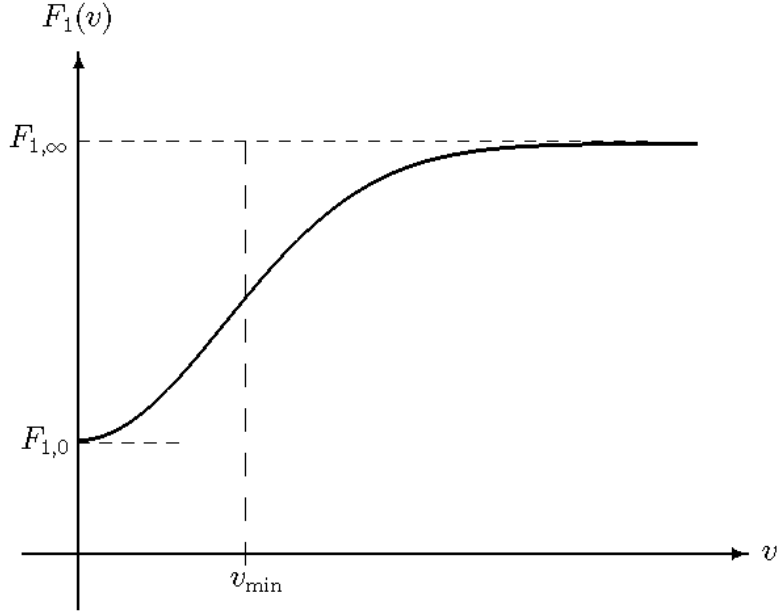}}
\end{center}
\caption{Sketch of the auxiliary function $F_1(v)$ defined in Eq.(\ref{eqn4010001}).}
\label{figA010001}
\end{figure}

 ~~~$\,$
 First,
 according to the definition of $F_1(v)$ in Eq.(\ref{eqn4010001})
 and noting that the velocity distribution function $f_1(v)$ can not be negative:
\beq
     f_1(v)
 \ge 0
\~,
\eeq
 I have
\beq
     \Dd{F_1(v)}{v}
 =   \frac{f_1(v)}{v}
 \ge 0
\~.
\eeq
 This means that $F_1(v)$ increases monotonically with $v$.
 Second,
 $f_1(v)$ must vanish as $v$ approaches infinity:
\cheqnref{eqn4010003}
\beq
     f_1(v \to \infty)
 \to 0
\~,
\eeq
 since WIMPs (as candidate for CDM) in today's Universe move quite slowly,
 then I have
\cheqnref{eqn4010004}
\beq
     \Dd{F_1(v)}{v}   \Bigg|_{v \to \infty}
 =   \frac{f_1(v)}{v} \Bigg|_{v \to \infty}
 \to 0
\~.
\eeq
\cheqnCN{-2}
 This means that $F_1(v)$ must approach a finite constant $F_{1,\infty}$ as $v \to \infty$.
 On the other hand,
 the three-dimensional velocity distribution function of WIMPs, $f(v)$, must be bounded:
\beq
         f(v)
 \propto \frac{f_1(v)}{v^2}
 \neq    \infty
\~,
\eeq
 where I have used
\beqN
   \int f(v) \~ d^3v
 = \int f(v) \~ v^2 \~ dv \~ d\Omega
 = \int f_1(v) \~ dv
\~.
\eeqN
 Hence,
\beq
   \Dd{F_1(v)}{v}   \Bigg|_{v=0}
 = \frac{f_1(v)}{v} \Bigg|_{v=0}
 = \bbrac{v \cdot \frac{f_1(v)}{v^2}}_{v=0}
 = 0
\~.
\eeq
 This means that $F_1(v)$ also approaches a constant $F_{1,0}$ as $v \to 0$.
 Actually,
 $F_{1,0}$ can be set to 0 without loss of generality,
 since Eq.(\ref{eqn4010001}) defines $F_1(v)$ only up to an additional constant.
 A sketch of the auxiliary function $F_1(v)$ is given in Fig.~\ref{figA010001}.
\section{Derivations for $f_1(v)$ in Eq.(\ref{eqn4010008})}
\label{f1(v)}
 ~~~$\,$
 According to Eq.(\ref{eqn3010106}),
 I have
\beq
   \Dd{\vmin}{Q}
 = \frac{\alpha}{2 \sqrt{Q}}
 = \frac{\vmin}{2 Q}
\~,
\eeq
 namely,
\beq
   \Dd{Q}{\vmin}
 = \frac{2 Q}{\vmin}
\~.
\eeq
 Differentiating both sides of Eq.(\ref{eqn4010002}) and using Eq.(\ref{eqn4010001}),
 one can obtain that
\cheqnref{eqn4010005}
\beqn
    \frac{f_1(\vmin)}{\vmin}
  = \Dd{F_1(\vmin)}{\vmin}
 \=-\frac{1}{\calA} \cbrac{\dd{\vmin} \bdRdQoFQ\Qvmina}
    \non\\
 \=-\frac{1}{\calA} \cbrac{\dd{Q}\bdRdQoFQ \cdot \aDd{Q}{\vmin}}\Qvmina
    \non\\
 \= \frac{1}{\vmin} \cdot \frac{1}{\calA} \cbrac{-2 Q \cdot \dd{Q}\bdRdQoFQ}\Qvmina
\~,
\eeqn
\cheqnCN{-1}
$\!\!\!\!\!\!$
 namely,
\beqn
    f_1(\vmin)
 \= \frac{1}{\calA} \cbrac{-2 Q \cdot \dd{Q}\bdRdQoFQ}\Qvmina
    \non\\
 \= \frac{1}{\calA} \cbrac{\frac{2 Q}{\FQ} \bbrac{\dFdQoF \adRdQ-\dd{Q}\adRdQ}}\Qvmina
\~.
\label{eqnA020001}
\eeqn

 On the other hand,
 according to {\it Leibnitz's Rule for Differentiation of Integrals}:
\beq
   \dd{t} \bbrac{\int_{a(t)}^{b(t)} F(x,t) \~ dx}
 = \int_{a(t)}^{b(t)} \bPp{F(x,t)}{t} dx + \bbrac{F(b,t) \aDd{b}{t}-F(a,t) \aDd{a}{t}}
\~,
\label{eqnA020002}
\eeq
 one can also differentiate both sides in Eq.(\ref{eqn3010108})
 with respect to $Q$ directly
 and obtain
\beqnN
 \conti \dd{Q}\adRdQ
        \non\\
 \=     \dd{Q} \cbrac{\calA \FQ \intvmin \bfrac{f_1(v)}{v} dv}
        \non\\
 \=     \calA \bDd{\FQ}{Q} \intvmin \bfrac{f_1(v)}{v} dv
       +\calA \FQ \bbrac{-\frac{f_1(\vmin)}{\vmin} \aDd{\vmin}{Q}}_{\vmin=\alpha \sqrt{Q}}
        \non\\
 \=     \bbrac{2 F(Q) \aDd{F}{Q}} \bdRdQoFQ
       -\calA \FQ \bbrac{\frac{f_1(\vmin)}{\vmin} \afrac{\vmin}{2 Q}}_{\vmin=\alpha \sqrt{Q}}
        \non\\
 \=     \dFdQoF \adRdQ-\calA \bfrac{\FQ}{2 Q} f_1\abrac{\vmin=\alpha \sqrt{Q}}
\~.
\eeqnN
 Therefore, it can be also found that
\cheqnref{eqnA020001}
\beq
   f_1(\vmin)
 = \frac{1}{\calA}
   \cbrac{\frac{2 Q}{\FQ} \bbrac{\dFdQoF \adRdQ-\dd{Q}\adRdQ}}\Qvmina
\~.
\eeq
\cheqnCN{-1}
\section{Normalization constant and moments of $f_1(v)$}
\label{calN}
 ~~~$\,$
 Since
\beq
   v
 = \alpha \sqrt{Q}
\~,
\label{eqnA030001}
\eeq
 I have
\beq
   dv
 = \afrac{\alpha}{2 \sqrt{Q}} dQ
\~.
\label{eqnA030002}
\eeq
 From Eq.(\ref{eqn4010008}) and according to the normalization condition in Eq.(\ref{eqn4010007}),
 I can get
\beqn
 \conti \intz f_1(v) \~ dv
        \non\\
 \=     \calN \intz \cbrac{-2 Q \cdot \ddRdQoFQdQ} \afrac{\alpha}{2 \sqrt{Q}} dQ 
        \non\\
 \=     \calN \cdot \abrac{-\alpha} \intz \sqrt{Q} \cdot \ddRdQoFQdQ dQ
        \non\\
 \=     \calN \cdot \abrac{-\alpha}
        \cbrac{\sqrt{Q} \bdRdQoFQ_0^{\infty}-\frac{1}{2} \intz \frac{1}{\sqrt{Q}} \bdRdQoFQ dQ}
        \non\\
 \=     \calN \afrac{\alpha}{2} \intz \frac{1}{\sqrt{Q}} \bdRdQoFQ dQ
        \non\\
 \=     1
\~,
\label{eqnA030003}
\eeqn
 where I have used the conditions:
\beq
     \dRdQ\Bigg|_{Q \to \infty}
 \to 0
\~,
\eeq
 and
\beq
      \dRdQ\Bigg|_{Q \to 0}
 \neq \infty
\~.
\eeq
 Eq.(\ref{eqn4010009}) follows immediately from Eq.(\ref{eqnA030003}). 

 Using Eqs.(\ref{eqnA030001}), (\ref{eqnA030002}) and integration by parts,
 I can also find the moments of $f_1(v)$,
 defined in Eq.(\ref{eqn4010010}) with a lower cut-off $\Qthre$ on the energy transfer,
 as follows:
\beqn
        \expv{v^n}
 \=     \int_{\vmin(\Qthre)}^\infty v^n f_1(v) \~ dv
        \non\\ 
 \=     \calN \int_{\Qthre}^\infty
        \aBig{\alpha \sqrt{Q}}^n \cbrac{-2 Q \cdot \ddRdQoFQdQ} \afrac{\alpha}{2 \sqrt{Q}} dQ
        \non\\
 \=     \calN \cdot \abrac{-\alpha^{n+1}} \int_{\Qthre}^\infty Q^{(n+1)/2} \cdot \ddRdQoFQdQ dQ
        \non\\
 \=     \calN \alpha^{n+1}
        \cBiggl{ \frac{\Qthre^{(n+1)/2}}{\FQthre} \adRdQ_{Q = \Qthre}}
        \non\\
 \conti ~~~~~~~~~~~~~~~~~~~~ 
        \cBiggr{ +\frac{n+1}{2} \int_{\Qthre}^\infty  Q^{(n-1)/2} \bdRdQoFQ dQ}
\~.
\eeqn
 This reproduces Eqs.(\ref{eqn4010011}) and (\ref{eqn4010012}) in Sec.~\ref{dRdQ}.

%% file: Doktorarbeit-AppB.tex
\chapter{Moments of the Velocity Distribution of WIMPs}
 ~~~$\,$
 In this chapter I derive at first the first and second moments of $f_1(v)$,
 i.e., the mean velocity and the velocity dispersion of WIMPs,
 for both of the two simplest semi-realistic halo models
 discussed in Subsecs.~\ref{fGau} and \ref{fsh}.
 Then,
 as tests of the formulae for reconstructing $f_1(v)$ and determining $\expv{v^n}$,
 I use the reduced spectra given in Eqs.(\ref{eqn3010302}') and (\ref{eqn3020103}')
 in Sec.~\ref{dRdQ}
 and the expressions for $f_1(v)$ in Eqs.(\ref{eqn4010008}) and (\ref{eqn4010009})
 as well as for the moments of $f_1(v)$ in Eqs.(\ref{eqn4010011}) and (\ref{eqn4010012})
 to obtain the same results.
\section{Calculating $\expv{v}$ and $\expv{v^2}$ from $f_1(v)$}
\subsection{From $f_{1,\Gau}(v)$ given in Eq.(\ref{eqn3010301})}
\label{vGau}
 ~~~$\,$
 For a simple isothermal Maxwellian halo,
 the normalized one-dimensional velocity distribution function has been given as
\cheqnref{eqn3010301}
\beq
   f_{1,\Gau}(v)
 = \frac{4}{\sqrt{\pi}} \afrac{v^2}{v_0^3} e^{-v^2/v_0^2}
\~.
\eeq
\cheqnCN{-1}
 One can find directly that
\beqn
    \expv{v^n}_\Gau
 \= \intz v^n f_{1,\Gau}(v) \~ dv
    \non\\
 \= \intz v^n \bbrac{\frac{4}{\sqrt{\pi}} \afrac{v^2}{v_0^3} e^{-v^2/v_0^2}} dv
    \non\\
 \= \frac{4}{\sqrt{\pi} v_0^3} \intz v^{n+2} e^{-v^2/v_0^2} \~ dv
    \non\\
 \= \frac{4}{\sqrt{\pi} v_0^3}
    \cbrac{\frac{\Gamma\bbrac{\frac{1}{2} (n+3)}}{2} \cdot \abrac{v_0^2}^{(n+3)/2}}
    \non\\
 \= \afrac{2}{\sqrt{\pi}} \Gamma\bbrac{\T \frac{1}{2} (n+1)+1} v_0^n
    \non\\
 \= \afrac{n+1}{\sqrt{\pi}} \Gamma\bbrac{\T \frac{1}{2} (n+1)} v_0^n
\~,
\label{eqnB010101}
\eeqn
 where I have used
\beqN
   \intz x^n e^{-ax^2} \~ dx
 = \frac{\Gamma\bbrac{\frac{1}{2} (n+1)}}{2 a^{(n+1)/2}}
\~,
\eeqN
 and
\beqN
   \Gamma(m+1)
 = m \~ \Gamma(m)
\~,
\eeqN
 for $m = 0,~\frac{1}{2},~1,~\frac{3}{2},~2,~\cdots$.
 Hence,
 using
\beqN
   \Gamma(n+1)
 = n!
\~,
\eeqN
 and
\beqN
   \Gamma\abrac{\T n+\frac{1}{2}}
 = \frac{1 \cdot 3 \cdots (2 n-1) \sqrt{\pi}}{2^n}
\~,
\eeqN
 for $n=0,~1,~2,~3,~\cdots$,
 Eqs.(\ref{eqn3010303}) and (\ref{eqn3010304}) can be obtained directly.
\subsection{From $f_{1,\sh}(v)$ given in Eq.(\ref{eqn3020101})}
\label{vsh}
 ~~~$\,$
 When we take into account
 the orbital motion of the Solar system around the Galaxy,
 the velocity distribution function should be modified to
\cheqnref{eqn3020101}
\beq
   f_{1,\sh}(v,v_e)
 = \frac{1}{\sqrt{\pi}} \afrac{v}{v_e v_0} \bbigg{e^{-(v-v_e)^2/v_0^2}-e^{-(v+v_e)^2/v_0^2}}
\~.
\eeq
\cheqnCN{-1}
 First, I have
\beqn
    \expv{v}_\sh
 \= \intz v f_{1,\sh}(v,v_e) \~ dv
    \non\\
 \= \frac{1}{\sqrt{\pi} v_e v_0}
    \intz v \cdot v \bbigg{e^{-(v-v_e)^2/v_0^2}-e^{-(v+v_e)^2/v_0^2}} dv
    \non\\
 \= \frac{1}{\sqrt{\pi} v_e}
    \bBigg{ \frac{1}{v_0} \intz v^2 e^{-(v-v_e)^2/v_0^2} \~ dv
           -\frac{1}{v_0} \intz v^2 e^{-(v+v_e)^2/v_0^2} \~ dv}
    \non\\
 \= \frac{1}{\sqrt{\pi} v_e} \aBig{{\bf V}_{1,-}-{\bf V}_{1,+}}
\~.
\label{eqnB010201}
\eeqn
 Define
\cheqnCa
\beq
        u_{\pm}
 \equiv \frac{v \pm v_e}{v_0}
\~,
\label{eqnB010202a}
\eeq
 i.e.,
\cheqnCb
\beq
   v
 = v_0 u_{\pm} \mp v_e
\~,
\label{eqnB010202b}
\eeq
\cheqnC
 it can be found that
\cheqnCa
\beqn
           {\bf V}_{1,-}
 \eqnequiv \frac{1}{v_0} \intz v^2 e^{-(v-v_e)^2/v_0^2} \~ dv
           \non\\
 \=        \frac{1}{v_0} \int_{-\tve}^{\infty} (v_0 u_{-}+v_e)^2 e^{-u_{-}^2}~(v_0 \~ du_{-})
           \non\\
 \=             v_0^2 \int_{-\tve}^{\infty} u_{-}^2 e^{-u_{-}^2} \~ du_{-}
          +2 v_e   v_0   \int_{-\tve}^{\infty} u_{-}   e^{-u_{-}^2} \~ du_{-}
          +  v_e^2       \int_{-\tve}^{\infty}         e^{-u_{-}^2} \~ du_{-}
           \non\\
 \=                v_0^2 \abrac{ 2 \int_0     ^{\tve}   u_{-}^2 e^{-u_{-}^2} \~ du_{-}
                                +  \int_{\tve}^{\infty} u_{-}^2 e^{-u_{-}^2} \~ du_{-}}
          +2 v_e   v_0             \int_{\tve}^{\infty} u_{-}   e^{-u_{-}^2} \~ du_{-}
           \non\\
 \conti    ~~~~~~~~~~~~~~~~~~~~ 
          +  v_e^2       \abrac{ 2 \int_0     ^{\tve}   e^{-u_{-}^2} \~ du_{-}
                                +  \int_{\tve}^{\infty} e^{-u_{-}^2} \~ du_{-}}
\~,
\label{eqnB010203a}
\eeqn
 and
\cheqnCb
\beqn
           {\bf V}_{1,+}
 \eqnequiv \frac{1}{v_0} \intz v^2 e^{-(v+v_e)^2/v_0^2} \~ dv
           \non\\
 \=        \frac{1}{v_0} \int_{\tve}^{\infty} (v_0 u_{+}-v_e)^2 e^{-u_{+}^2}~(v_0 \~ du_{+})
           \non\\
 \=                v_0^2 \int_{\tve}^{\infty} u_{+}^2 e^{-u_{+}^2} \~ du_{+}
          -2 v_e   v_0   \int_{\tve}^{\infty} u_{+}   e^{-u_{+}^2} \~ du_{+}
          +  v_e^2       \int_{\tve}^{\infty}         e^{-u_{+}^2} \~ du_{+}
\~,
\label{eqnB010203b}
\eeqn
\cheqnC
 where I have defined
\beq
        \tve
 \equiv \frac{v_e}{v_0}
\~.
\label{eqnB010204}
\eeq
 Combining Eqs.(\ref{eqnB010203a}) and (\ref{eqnB010203b}),
 one can get
\beqn
 \conti {\bf V}_{1,-}-{\bf V}_{1,+}
        \non\\
 \=     2       v_0^2 \int_0     ^{\tve}   u^2 e^{-u^2} \~ du
       +2 v_e   v_0   \int_{\tve}^{\infty}     e^{-u^2} \~ du^2
       +2 v_e^2       \int_0     ^{\tve}       e^{-u^2} \~ du
        \non\\
 \=     2       v_0^2
        \bbrac{-\frac{1}{2} \abrac{\tve e^{-\tve^2}}+\afrac{\sqrt{\pi}}{4} \erf(\tve)}
       +2 v_e   v_0   e^{-\tve^2}
       +2 v_e^2       \bbrac{\afrac{\sqrt{\pi}}{2} \erf(\tve)}
        \non\\
 \=     v_e v_0 e^{-v_e^2/v_0^2}+\sqrt{\pi} \abrac{\frac{v_0^2}{2}+v_e^2} \erf\afrac{v_e}{v_0}
\~,
\eeqn
 where I have used the definition of error function
\beqN
   \erf(x)
 = \frac{2}{\sqrt{\pi}} \int_0^{x} e^{-t^2} dt
\~,
\eeqN
 and
\beqN
    \int_0^x t^2 \~ e^{-t^2} dt
 =-\frac{1}{2} \abrac{x \~ e^{-x^2}}+\afrac{\sqrt{\pi}}{4} \erf(x)
\~.
\eeqN
 Hence,
 the mean velocity of WIMPs for a shifted Maxwellian halo discussed in Subsec.~\ref{fsh}
 can be found as
\cheqnref{eqn3020104}
\beq
   \expv{v}_\sh
 = \afrac{v_0}{\sqrt{\pi}} e^{-v_e^2/v_0^2}+\abrac{\frac{v_0^2}{2 v_e}+v_e} \erf\afrac{v_e}{v_0}
\~.
\eeq
\cheqnCN{-1}
 Meanwhile,
\beqn
    \expv{v^2}_\sh
 \= \intz v^2 f_{1,\sh}(v,v_e) \~ dv
    \non\\
 \= \frac{1}{\sqrt{\pi} v_e v_0}
    \intz v^2 \cdot v \bbigg{e^{-(v-v_e)^2/v_0^2}-e^{-(v+v_e)^2/v_0^2}} dv
    \non\\
 \= \frac{1}{\sqrt{\pi} v_e} \bbrac{ \frac{1}{v_0} \intz v^3 e^{-(v-v_e)^2/v_0^2} \~ dv
   -\frac{1}{v_0} \intz v^3 e^{-(v+v_e)^2/v_0^2} \~ dv}
    \non\\
 \= \frac{1}{\sqrt{\pi} v_e} \aBig{{\bf V}_{2,-}-{\bf V}_{2,+}}
\~.
\label{eqnB010205}
\eeqn
 Using Eqs.(\ref{eqnB010202a}), (\ref{eqnB010202b}), and (\ref{eqnB010204}),
 it can be found that
\cheqnCa
\beqn
           {\bf V}_{2,-}
 \eqnequiv \frac{1}{v_0} \intz v^3 e^{-(v-v_e)^2/v_0^2} \~ dv
           \non\\
 \=        \frac{1}{v_0} \int_{-\tve}^{\infty} (v_0 u_{-}+v_e)^3 e^{-u_{-}^2}~(v_0 \~ du_{-})
           \non\\
 \=                v_0^3 \int_{-\tve}^{\infty} u_{-}^3 e^{-u_{-}^2} \~ du_{-}
          +3 v_e   v_0^2 \int_{-\tve}^{\infty} u_{-}^2 e^{-u_{-}^2} \~ du_{-}
           \non\\
 \conti    ~~~~~~~~~~~~~~~~ 
          +3 v_e^2 v_0   \int_{-\tve}^{\infty} u_{-}   e^{-u_{-}^2} \~ du_{-}
          +  v_e^3       \int_{-\tve}^{\infty}         e^{-u_{-}^2} \~ du_{-}
           \non\\
 \=                v_0^3 \int_{\tve}^{\infty}  u_{-}^3 e^{-u_{-}^2} \~ du_{-}
          +3 v_e   v_0^2 \abrac{ \int_0^{\tve} u_{-}^2 e^{-u_{-}^2} \~ du_{-}
                                +\intz         u_{-}^2 e^{-u_{-}^2} \~ du_{-}}
           \non\\
 \conti    ~~~~~~~~ 
          +3 v_e^2 v_0   \int_{\tve}^{\infty}  u_{-}   e^{-u_{-}^2} \~ du_{-}
          +  v_e^3       \abrac{ \int_0^{\tve}         e^{-u_{-}^2} \~ du_{-}
                                +\intz                 e^{-u_{-}^2} \~ du_{-}}
\~,   
\eeqn
 and
\cheqnCb
\beqn
           {\bf V}_{2,+}
 \eqnequiv \frac{1}{v_0} \intz v^3 e^{-(v+v_e)^2/v_0^2} \~ dv
           \non\\
 \=        \frac{1}{v_0} \int_{\tve}^{\infty} (v_0 u_{+}-v_e)^3 e^{-u_{+}^2}~(v_0 \~ du_{+})
           \non\\
 \=                v_0^3 \int_{\tve}^{\infty} u_{+}^3 e^{-u_{+}^2} \~ du_{+}
          -3 v_e   v_0^2 \int_{\tve}^{\infty} u_{+}^2 e^{-u_{+}^2} \~ du_{+}
           \non\\
 \conti    ~~~~~~~~~~~~~~~~ 
          +3 v_e^2 v_0   \int_{\tve}^{\infty} u_{+}   e^{-u_{+}^2} \~ du_{+}
          -  v_e^3       \int_{\tve}^{\infty}         e^{-u_{+}^2} \~ du_{+}
\~.
\eeqn
\cheqnC
 Combining them, one can get
\beqn
    {\bf V}_{2,-}-{\bf V}_{2,+}
 \= 6 v_e   v_0^2 \intz u^2 e^{-u^2} \~ du
   +2 v_e^3       \intz     e^{-u^2} \~ du
    \non\\
 \= 6 v_e   v_0^2 \afrac{\Gamma\abrac{\frac{3}{2}}}{2}
   +2 v_e^3       \afrac{\sqrt{\pi}}{2}
    \non\\
 \= \afrac{3 \sqrt{\pi}}{2} v_e v_0^2+\sqrt{\pi} \~ v_e^3
\~.
\eeqn
 Therefore,
 the mean velocity of WIMPs for a shifted Maxwellian halo can be found as
\cheqnref{eqn3020105}
\beq
   \expv{v^2}_\sh
 = \afrac{3}{2} v_0^2+v_e^2
\~.
\eeq
\cheqnCN{-1}
 Finally,
 according to Leibnitz's Rule for Differentiation of Integrals
 given in Eq.(\ref{eqnA020002}),
 one has
\beq
   \dd{x}\bBig{\erf(x)}
 = \frac{2}{\sqrt{\pi}} \bbrac{\dd{x} \int_0^{x} e^{-t^2} dt}
 = \frac{2}{\sqrt{\pi}}~e^{-x^2}
\~.
\label{eqnB010206}
\eeq
 Then it is easily to prove that,
 for $v_e \ll v_0$,
 i.e., $\tve \ll 1$,
 Eqs.(\ref{eqn3020104}) and (\ref{eqn3020105}) will reduce to
 Eqs.(\ref{eqn3010303}) to (\ref{eqn3010304}).
\section{Calculating $f_1(v)$, $\expv{v}$, and $\expv{v^2}$ from $dR/dQ$}
\label{vdRdQ}
\subsection{From $(dR/dQ)_\Gau$ in Eq.(\ref{eqn3010302})}
\label{vGaudRdQ}
 ~~~$\,$
 Substituting Eq.(\ref{eqn3010302}') in Sec.~\ref{dRdQ} into Eq.(\ref{eqn4010008}),
 I have
\beqn
    f_{1,\Gau}(v)
 \= \calN_\Gau \cbrac{-2Q \cdot \dd{Q}\bbrac{\frac{1}{\FQ} \adRdQ_\Gau}}\Qva
    \non\\
 \= \calN_\Gau \bbrac{-2 Q \cdot \dd{Q}\abrac{e^{-\alpha^2 Q/v_0^2}}}\Qva
    \non\\
 \= \calN_\Gau \bbrac{2 \afrac{v}{v_0}^2 e^{-v^2/v_0^2}}
\~.
\label{eqnB020101}
\eeqn
 Meanwhile,
 according to Eq.(\ref{eqn4010009}),
 the normalization constant $\calN_\Gau$ can be found as
\beqn
    \calN_\Gau
 \= \frac{2}{\alpha}
    \cbrac{\intz \frac{1}{\sqrt{Q}} \bbrac{\frac{1}{\FQ} \adRdQ_\Gau} dQ}^{-1}
    \non\\
 \= \frac{1}{\alpha}
    \cbrac{\intz \bbrac{\frac{1}{\FQ} \adRdQ_\Gau}_{Q=q^2} dq}^{-1}
    \non\\
 \= \frac{1}{\alpha} \abrac{\intz e^{-\alpha^2 q^2/v_0^2} \~ dq}^{-1}
    \non\\
 \= \frac{1}{\alpha} \abrac{\frac{v_0}{\alpha} \cdot \frac{\sqrt{\pi}}{2}}^{-1}
    \non\\
 \= \frac{2}{\sqrt{\pi} v_0}
\~.
\label{eqnB020102}
\eeqn
 Here,
 for simplicity,
 I have defined
\footnote{
 I will use this definition in this section and the next chapter.
 Please do not confuse with the transferred 3-momentum in Eq.(\ref{eqn3010104}).}:
\cheqnCa
\beq
   Q
 = q^2
\~,
\label{eqnB020103a}
\eeq
 and then
\cheqnCb
\beq
   dQ
 = 2 q \~ dq
\~.
\label{eqnB020103b}
\eeq
\cheqnC
 Substituting Eq.(\ref{eqnB020102}) into Eq.(\ref{eqnB020101}),
 the normalized one-dimensional velocity distribution function $f_{1,\Gau}(v)$
 in Eq.(\ref{eqn3010301}) can be obtained directly.
 Moreover,
 substituting Eq.(\ref{eqn3010302}') into Eqs.(\ref{eqn4010012}) and (\ref{eqn4010011})
 and using the normalization constant in Eq.(\ref{eqnB020102}),
 one can get
\cheqnref{eqnB010101}
\beqn
    \expv{v^n}_\Gau
 \= \calN_\Gau \afrac{\alpha^{n+1}}{2} \cdot (n+1)
    \intz Q^{(n-1)/2} \bbrac{\frac{1}{\FQ} \adRdQ_\Gau} dQ
    \non\\
 \= \calN_\Gau (n+1) \afrac{\alpha^{n+1}}{2}
    \intz Q^{(n-1)/2} e^{-\alpha^2 Q/v_0^2} \~ dQ
    \non\\
 \= \calN_\Gau (n+1) \afrac{\alpha^{n+1}}{2}
    \intz q^{n-1} e^{-\alpha^2 q^2/v_0^2} \abrac{2 q \~ dq}
    \non\\
 \= \afrac{2}{\sqrt{\pi} v_0} (n+1) \alpha^{n+1}
    \cbrac{\frac{1}{2} \afrac{v_0}{\alpha}^{n+1} \Gamma\bbrac{\T \frac{1}{2}\abrac{n+1}}}
    \non\\
 \= \afrac{n+1}{\sqrt{\pi}} \Gamma\bbrac{\T \frac{1}{2}\abrac{n+1}} v_0^n 
\~.
\eeqn
\cheqnCN{-1}
 $\!\!\!\!\!\!$
 Note that I have set $\Qthre = 0$ here.
\subsection{From $(dR/dQ)_\sh$ in Eq.(\ref{eqn3020103})}
\label{vshdRdQ}
 ~~~$\,$
 According to Eq.(\ref{eqnB010206}),
 one can obtain that
\beq
   \dd{Q} \bbigg{\erf{\T \afrac{\alpha \sqrt{Q} \pm v_e}{v_0}}}
 = \frac{1}{\sqrt{\pi}} \afrac{\alpha}{v_0}
   \bbrac{\frac{1}{\sqrt{Q}}~e^{-\bbig{\abig{\alpha \sqrt{Q} \pm v_e}/v_0}^2}}
\~.
\label{eqnB020201}
\eeq
 Then,
 substituting Eq.(\ref{eqn3020103}') in Sec.~\ref{dRdQ} into Eq.(\ref{eqn4010008}),
 I have
\beqn
 \conti f_{1,_\sh}(v,v_e)
        \non\\
 \=     \calN_\sh \cbrac{-2 Q \cdot \dd{Q}\bbrac{\frac{1}{\FQ} \adRdQ_\sh}}\Qva
        \non\\
 \=     \calN_\sh
        \cbrac{-2 Q \cdot
                    \dd{Q}
                    \bbigg{ \erf{\T \afrac{\alpha \sqrt{Q}+v_e}{v_0}}
                           -\erf{\T \afrac{\alpha \sqrt{Q}-v_e}{v_0}}}}\Qva
        \non\\
 \=     \calN_\sh
        \cbrac{-2 Q \cdot
                    \frac{1}{\sqrt{\pi}} \afrac{\alpha}{v_0} \frac{1}{\sqrt{Q}}
                    \cbigg{ e^{-\bbig{\abig{\alpha \sqrt{Q}+v_e}/v_0}^2}
                           -e^{-\bbig{\abig{\alpha \sqrt{Q}-v_e}/v_0}^2}}}\Qva
        \non\\
 \=     \calN_\sh \cdot \frac{2}{\sqrt{\pi}} \afrac{v}{v_0}
        \bbigg{e^{-(v-v_e)^2/v_0^2}-e^{-(v+v_e)^2/v_0^2}}
\~.
\label{eqnB020202}
\eeqn
 Meanwhile,
 as done for $\calN_\Gau$ in Eq.(\ref{eqnB020102}),
 one can use Eqs.(\ref{eqnB020103a}) and (\ref{eqnB020103b}) and find that
\beqn
    \calN_\sh
 \= \frac{1}{\alpha} \cbrac{\intz \bbrac{\frac{1}{\FQ} \adRdQ_\sh}_{Q=q^2} dq}^{-1}
    \non\\
 \= \frac{1}{\alpha}
    \cBigg{\intz \bbrac{\erf\afrac{\alpha q+v_e}{v_0}-\erf\afrac{\alpha q-v_e}{v_0}} dq}^{-1}
    \non\\
 \= \frac{1}{\alpha} \bBig{{\bf V}_0(\infty)-{\bf V}_0(0)}^{-1}
\~,
\label{eqnB020203}
\eeqn
 where I have defined
\beq
        {\bf V}_0(q)
 \equiv \int \bbrac{\erf\afrac{\alpha q+v_e}{v_0}-\erf\afrac{\alpha q-v_e}{v_0}} dq
 \equiv {\bf V}_{0,+}(q)-{\bf V}_{0,-}(q)
\~.
\label{eqnB020204}
\eeq
 Define
\cheqnCa
\beq
        s_{\pm}
 \equiv \frac{\alpha q \pm v_e}{v_0}
\~,
\label{eqnB020205a}
\eeq
 i.e.,
\cheqnCb
\beq
   q
 = \frac{v_0 s_{\pm} \mp v_e}{\alpha}
\~,
\label{eqnB020205b}
\eeq
\cheqnC
 it can be found that
\beqn
           {\bf V}_{0,\pm}(q)
 \eqnequiv \int \erf\afrac{\alpha q \pm v_e}{v_0} dq
           \non\\
 \=        \int \erf(s_{\pm}) \bbrac{\afrac{v_0}{\alpha} ds_{\pm}}
           \non\\
 \=        \frac{v_0}{\alpha} \int \erf(s_{\pm}) \~ ds_{\pm}
           \non\\
 \=        \frac{v_0}{\alpha} \bbrac{s_{\pm} \erf(s_{\pm})+\frac{1}{\sqrt{\pi}}~e^{-s_{\pm}^2}}
           \non\\
 \=        \abrac{q \pm \frac{v_e}{\alpha}} \erf(s_{\pm})
          +\frac{1}{\sqrt{\pi}} \afrac{v_0}{\alpha} e^{-s_{\pm}^2}
\~,
\label{eqnB020206}
\eeqn
 where I have used
\beqN
   \int \erf(x) \~ dx
 = x \~ \erf(x)+\frac{1}{\sqrt{\pi}}~e^{-x^2}
\~.
\eeqN
 Substituting ${\bf V}_{0,\pm}(q)$ in Eq.(\ref{eqnB020206}) into Eq.(\ref{eqnB020204}),
 I can get
\beqn
    {\bf V}_0(q)
 \= q \bBig{\erf(s_{+})-\erf(s_{-})}
   +\frac{v_e}{\alpha} \bBig{\erf(s_{+})+\erf(s_{-})}
   +\frac{1}{\sqrt{\pi}} \afrac{v_0}{\alpha} \abrac{e^{-s_{+}^2}-e^{-s_{-}^2}}
\~.
    \non\\
\label{eqnB020207}
\eeqn
 Now note that, as $q \to \infty$,
\beq
   {\bf V}_0(q \to \infty)
 = \afrac{2}{\alpha} v_e
\~,
\label{eqnB020208}
\eeq
 since
\beq
     s_{\pm}(q \to \infty)
 \to \infty
\~,
\label{eqnB020209}
\eeq
 and
\beqN
   \erf(\infty)
 = \frac{2}{\sqrt{\pi}} \intz e^{-t^2} dt
 = 1
\~.
\eeqN
 While, since as $q = 0$,
\beq
   s_{\pm}(0)
 = \pm \frac{v_e}{v_0}
 = \pm \tve
\~,
\label{eqnB020210}
\eeq
 where I have used the definition in Eq.(\ref{eqnB020205a}),
 and
\beqN
   \erf(-x)
 =-\erf(x)
\~,
\eeqN
 it can be found that
\beq
   {\bf V}_0(0)
 = 0
\~.
\label{eqnB020211}
\eeq
 Substituting Eqs.(\ref{eqnB020209}) and (\ref{eqnB020211}) into Eq.(\ref{eqnB020203}),
 the normalization constant $\calN_\sh$ can be found as
\beq
   \calN_\sh
 = \frac{1}{\alpha} \bbrac{\afrac{2}{\alpha} v_e}^{-1}
 = \frac{1}{2 v_e}
\~.
\label{eqnB020212}
\eeq
 Then I can obtain the normalized velocity distribution function
 in Eq.(\ref{eqn3020101}) directly.
 Meanwhile,
 substituting Eq.(\ref{eqn3020103}') into Eq.(\ref{eqn4010011}) ($\Qthre = 0$)
 and using Eqs.(\ref{eqnB020103a}) and (\ref{eqnB020103b}),
 I have
\beqn
    \expv{v}_\sh
 \= \calN_\sh \cdot \alpha^2 \intz \bbrac{\frac{1}{\FQ} \adRdQ_\sh}_{Q=q^2} (2 q \~ dq)
    \non\\
 \= \frac{1}{2 v_e} \cdot 2 \alpha^2
    \intz q \bbrac{ \erf\afrac{\alpha q+v_e}{v_0}-\erf\afrac{\alpha q-v_e}{v_0}} dq
    \non\\
 \= \frac{\alpha^2}{v_e} \bBig{{\bf V}_1(\infty)-{\bf V}_1(0)}
\~,
\label{eqnB020213}
\eeqn
 where I have defined
\beq
        {\bf V}_1(q)
 \equiv \int q \bbrac{\erf\afrac{\alpha q+v_e}{v_0}-\erf\afrac{\alpha q-v_e}{v_0}} dq
 \equiv {\bf V}_{1,+}(q)-{\bf V}_{1,-}(q)
\~.
\label{eqnB020214}
\eeq
 Using Eqs.(\ref{eqnB020205a}) and (\ref{eqnB020205b}), it can be found that
\beqn
 \conti    {\bf V}_{1,\pm}(q)
           \non\\
 \eqnequiv \int q \~ \erf\afrac{\alpha q \pm v_e}{v_0} dq
           \non\\
 \=        \frac{v_0}{\alpha} \int \afrac{v_0 s_{\pm} \mp v_e}{\alpha} \erf(s_{\pm}) \~ ds_{\pm}
           \non\\
 \=            \afrac{v_0}{\alpha}^2    \int s_{\pm} \erf(s_{\pm}) \~ ds_{\pm}
           \mp \frac{v_e v_0}{\alpha^2} \int         \erf(s_{\pm}) \~ ds_{\pm}
           \non\\
 \=        \frac{1}{2} \afrac{v_0}{\alpha}^2
           \bbrac{ \abrac{s_{\pm}^2-\frac{1}{2}} \erf(s_{\pm})
                  +\frac{1}{\sqrt{\pi}} \~ s_{\pm} e^{-s_{\pm}^2}}
           \mp \frac{v_e v_0}{\alpha^2}
               \bbrac{s_{\pm} \erf(s_{\pm})+\frac{1}{\sqrt{\pi}} \~ e^{-s_{\pm}^2}}
           \non\\
 \=        \frac{1}{2} \afrac{v_0}{\alpha}^2
           \cbrac{ \abrac{s_{\pm} \mp \frac{2 v_e}{v_0}}
                   \bbrac{s_{\pm} \erf(s_{\pm})+\frac{1}{\sqrt{\pi}} \~ e^{-s_{\pm}^2}}
                  -\frac{1}{2} \~ \erf(s_{\pm})}
           \non\\
 \=        \frac{1}{2} \afrac{v_0}{\alpha}^2
           \bbrac{ \abrac{s_{+} s_{-}-\frac{1}{2}} \erf(s_{\pm})
                  +\frac{1}{\sqrt{\pi}} \~ s_{\mp} e^{-s_{\pm}^2}}
\~,
\label{eqnB020215}
\eeqn
 where I have used
\beqN
   \int x \~ \erf(x) \~ dx
 = \frac{1}{2} \bbrac{\abrac{x^2-\frac{1}{2}} \erf(x)+\frac{1}{\sqrt{\pi}} \~ x e^{-x^2}}
\~,
\eeqN
 and
\beq
   v_0 s_{\pm} \mp 2 v_e
 = (\alpha q \pm v_e) \mp 2 v_e
 = \alpha q \mp v_e
 = v_0 s_{\mp}
\~.
\eeq
 Hence, I can get
\beqn
    {\bf V}_1(q)
 \= \frac{1}{2} \afrac{v_0}{\alpha}^2
    \cBigg{ \abrac{s_{+} s_{-}-\frac{1}{2}} \bbigg{\erf(s_{+})-\erf(s_{-})}
           +\frac{1}{\sqrt{\pi}} \abrac{s_{-} e^{-s_{+}^2}-s_{+} e^{-s_{-}^2}}}
\~.
    \non\\
\label{eqnB020216}
\eeqn
 From Eq.(\ref{eqnB020209}), it can be found easily that
\beq
   {\bf V}_1(q \to \infty)
 = 0
\~,
\label{eqnB020217}
\eeq
 and, from Eq.(\ref{eqnB020210}), 
\beq
   {\bf V}_1(0)
 =-\afrac{v_0}{\alpha}^2
   \bbrac{\abrac{\tve^2+\frac{1}{2}} \erf(\tve)+\frac{1}{\sqrt{\pi}} \abigg{\tve e^{-\tve^2}}}
\~.
\eeq
 Therefore,
 substituting these results into Eq.(\ref{eqnB020213}),
 I can obtain that
\cheqnref{eqn3020104}
\beqn
    \expv{v}_\sh
 \= \frac{\alpha^2}{v_e}
    \cbrac{\afrac{v_0}{\alpha}^2
           \bbrac{ \abrac{\tve^2+\frac{1}{2}} \erf(\tve)
                  +\frac{1}{\sqrt{\pi}} \abigg{\tve e^{-\tve^2}}}}
    \non\\
 \= \abrac{v_e+\frac{v_0^2}{2 v_e}} \erf\afrac{v_e}{v_0}
   +\afrac{v_0}{\sqrt{\pi}} e^{-v_e^2/v_0^2}
\~.
\eeqn
\cheqnCN{-1}
 $\!\!\!\!\!\!$
 Similarly,
\beqn
    \expv{v^2}_\sh
 \= \calN_\sh \cdot \afrac{3}{2} \alpha^3
    \intz q \bbrac{\frac{1}{\FQ} \adRdQ_\sh}_{Q=q^2} (2 q \~ dq)
    \non\\
 \= \frac{1}{2 v_e} \cdot 3 \alpha^3
    \intz q^2 \bbrac{ \erf\afrac{\alpha q+v_e}{v_0}-\erf\afrac{\alpha q-v_e}{v_0}} dq
    \non\\
 \= \frac{3}{2} \afrac{\alpha^3}{v_e} \bBig{{\bf V}_2(\infty)-{\bf V}_2(0)}
\~,
\label{eqnB020218}
\eeqn
 where I have defined
\beq
        {\bf V}_2(q)
 \equiv \int q^2 \bbrac{\erf\afrac{\alpha q+v_e}{v_0}-\erf\afrac{\alpha q-v_e}{v_0}} dq
 \equiv {\bf V}_{2,+}(q)-{\bf V}_{2,-}(q)
\~.
\eeq
 Using Eqs.(\ref{eqnB020205a}) and (\ref{eqnB020205b}), it can be found that
\beqn
           {\bf V}_{2,\pm}(q)
 \eqnequiv \int q^2 \erf\afrac{\alpha q \pm v_e}{v_0} dq
           \non\\
 \=        \frac{v_0}{\alpha}
           \int \afrac{v_0 s_{\pm} \mp v_e}{\alpha}^2 \erf(s_{\pm}) \~ ds_{\pm}
           \non\\
 \=        \afrac{v_0}{\alpha^3}
           \int \aBig{v_0^2 s_{\pm}^2 \mp 2 v_e v_0 s_{\pm}+v_e^2} \~ \erf(s_{\pm}) \~ ds_{\pm}
           \non\\
 \=        \afrac{v_0}{\alpha}^3
           \bBigg{             \int s_{\pm}^2 \erf(s_{\pm}) \~ ds_{\pm}
                  \mp 2 \tve   \int s_{\pm}   \erf(s_{\pm}) \~ ds_{\pm}
                  +     \tve^2 \int           \erf(s_{\pm}) \~ ds_{\pm}}
           \non\\
 \=        \afrac{v_0}{\alpha}^3
           \cBiggl{ \frac{1}{3} \bbrac{ s_{\pm}^3 \erf(s_{\pm})
                   +\frac{1}{\sqrt{\pi}} \aBig{s_{\pm}^2+1} e^{-s_{\pm}^2}}
           \non\\
 \conti    ~~~~~~~~~~~~~~~~~~~~ 
                    \mp \tve
                        \bbrac{ \abrac{s_{\pm}^2-\frac{1}{2}} \erf(s_{\pm})
                               +\frac{1}{\sqrt{\pi}} \~ s_{\pm} e^{-s_{\pm}^2}}}
           \non\\
 \conti    ~~~~~~~~~~~~~~~~~~~~~~~~~~~~~~ 
           \cBiggr{+\tve^2 \bbrac{s_{\pm} \erf(s_{\pm})+\frac{1}{\sqrt{\pi}} \~ e^{-s_{\pm}^2}}}
           \non\\
 \=        \afrac{v_0}{\alpha}^3
           \cleft{ \bbrac{\frac{1}{3} \afrac{\alpha}{v_0}^3 q^3
                          \pm \abrac{\frac{\tve^3}{3}+\frac{\tve}{2}}}
                   \erf(s_{\pm})}
           \non\\
 \conti    ~~~~~~~~~~~~~~~~~~~~ 
           \cright{+\frac{1}{\sqrt{\pi}}
                    \abrac{\frac{s_{\pm}^2}{3} \mp \tve s_{\pm}+\tve^2+\frac{1}{3}} e^{-s_{\pm}^2}}
\~,
\label{eqnB020219}
\eeqn
 where I have used
\beqN
   \int x^2 \erf(x) \~ dx
 = \frac{1}{3} \bbrac{x^3 \erf(x)+\frac{1}{\sqrt{\pi}} \aBig{x^2+1} e^{-x^2}}
\~,
\eeqN
 and
\beqn
    \frac{s_{\pm}^3}{3} \mp \tve s_{\pm}^2+\tve^2 s_{\pm}
 \= \frac{1}{3}
    \bbigg{\abrac{s_{\pm}^3 \mp 3 \tve s_{\pm}^2+3 \tve^2 s_{\pm} \mp \tve^3} \pm \tve^3}
    \non\\
 \= \frac{1}{3} \afrac{v_0 s_{\pm} \mp v_e}{v_0}^3 \pm \frac{\tve^3}{3}
    \non\\
 \= \frac{1}{3} \afrac{\alpha}{v_0}^3 q^3 \pm \frac{\tve^3}{3}
\~.
\eeqn
 Hence, I can get
\beqn
        {\bf V}_2(q)
 \=     \afrac{v_0}{\alpha}^3
        \cleft{ \frac{1}{3} \afrac{\alpha}{v_0}^3 q^3 \bbigg{\erf(s_{+})-\erf(s_{-})}
               +\abrac{\frac{\tve^3}{3}+\frac{\tve}{2}} \bbigg{\erf(s_{+})+\erf(s_{-})}}
        \non\\
 \conti ~~~~~~~~~~~~~~~~~~~~ 
               +\frac{1}{\sqrt{\pi}}
                 \bbrac{ \abrac{\frac{s_{+}}{3}-\tve} s_{+} e^{-s_{+}^2}
                        -\abrac{\frac{s_{-}}{3}+\tve} s_{-} e^{-s_{-}^2}}
        \non\\
 \conti ~~~~~~~~~~~~~~~~~~~~~~~~~~~~~~ 
        +\cright{\frac{1}{\sqrt{\pi}} \abrac{\tve^2+\frac{1}{3}} \abrac{e^{-s_{+}^2}-e^{-s_{-}^2}}}
\~.
\label{eqnB020220}
\eeqn
 From Eq.(\ref{eqnB020209}), it can be found easily that 
\beq
   {\bf V}_2(q \to \infty)
 = \afrac{v_0}{\alpha}^3 \bbrac{2 \abrac{\frac{\tve^3}{3}+\frac{\tve}{2}}}
 = \frac{v_e}{\alpha^3} \bbrac{\afrac{2}{3} v_e^2+v_0^2}
\~,
\eeq
 and, from Eq.(\ref{eqnB020209}), 
\beqn
   {\bf V}_2(0)
 = 0
\~.
\eeqn
 Therefore,
 substituting these results into Eq.(\ref{eqnB020218}),
 I can obtain that
\cheqnref{eqn3020105}
\beq
   \expv{v^2}_\sh
 = \frac{3}{2} \afrac{\alpha^3}{v_e} \cBigg{\frac{v_e}{\alpha^3} \bbrac{\afrac{2}{3} v_e^2+v_0^2}}
 = v_e^2+\afrac{3}{2} v_0^2
\~.
\eeq
\cheqnCN{-1}

%% file: Doktorarbeit-AppC.tex
\chapter{Differential and Total Event Rates}
 ~~~$\,$
 In this chapter
 I derive the differential and total event rates
 for the simple and shifted isothermal Maxwellian halo models
 from their velocity distribution functions given
 in Eqs.(\ref{eqn3010301}) and (\ref{eqn3020101}).
 The case for $\FQ \approx 1$ and
 the case with the exponential form factor $F_{\rm ex}^2(Q)$ given in Eq.(\ref{eqn3010201})
 will be considered.
\section{Setting $\FQ \approx 1$}
\subsection{Starting with $f_{1,\Gau}(v)$ given in Eq.(\ref{eqn3010301})}
\label{dRdQGau}
 ~~~$\,$
 For a simple isothermal Maxwellian halo,
 the normalized one-dimensional velocity distribution function has been given as
\cheqnref{eqn3010301}
\beq
   f_{1,\Gau}(v)
 = \frac{4}{\sqrt{\pi}} \afrac{v^2}{v_0^3} e^{-v^2/v_0^2}
\~.
\eeq
\cheqnCN{-1}
 I can get directly that
\beqn
    \intvmin \bfrac{f_{1,\Gau}(v)}{v} dv
 \= \intvmin \frac{1}{v} \bbrac{\frac{4}{\sqrt{\pi}} \afrac{v^2}{v_0^3} e^{-v^2/v_0^2}} dv
    \non\\
 \= \frac{4}{\sqrt{\pi} v_0^3} \intvmin v e^{-v^2/v_0^2} \~ dv
    \non\\
 \= \frac{2}{\sqrt{\pi} v_0^3} \intvmin e^{-v^2/v_0^2} \~ dv^2
    \non\\
 \= \afrac{2}{\sqrt{\pi} v_0} e^{-\vmin^2/v_0^2}
\~.
\eeqn
 Using this result and Eqs.(\ref{eqn3010106}) and (\ref{eqn3010108}),
 one can obtain $(dR/dQ)_\Gau$ in Eq.(\ref{eqn3010302})
 and then $R_\Gau(\Qthre)$ in Eq.(\ref{eqn3010305}) easily
 when $\FQ$ has been neglected.
\subsection{Starting with $f_{1,\sh}(v)$ given in Eq.(\ref{eqn3020101})}
\label{dRdQsh}
 ~~~$\,$
 When we take into account
 the orbital motion of the Solar system around the Galaxy,
 the velocity distribution function should be modified to
\cheqnref{eqn3020101}
\beq
   f_{1,\sh}(v,v_e)
 = \frac{1}{\sqrt{\pi}} \afrac{v}{v_e v_0} \bbigg{e^{-(v-v_e)^2/v_0^2}-e^{-(v+v_e)^2/v_0^2}}
\~.
\eeq
\cheqnCN{-1}
 First,
 using Eqs.(\ref{eqnB010202a}) and (\ref{eqnB010202b}),
 it can be found that
\beqn
    \intvmin \bfrac{f_{1,\sh}(v)}{v} dv
 \= \frac{1}{\sqrt{\pi} \~ v_e v_0}
    \bbrac{\intvmin e^{-(v-v_e)^2/v_0^2} \~ dv -\intvmin e^{-(v+v_e)^2/v_0^2} \~ dv}
    \non\\
 \= \frac{1}{\sqrt{\pi} \~ v_e v_0}
    \bbrac{ \int_{u_{-,\rm min}}^{\infty} e^{-u_{-}^2} (v_0~du_{-})
           -\int_{u_{+,\rm min}}^{\infty} e^{-u_{+}^2} (v_0~du_{+})}
    \non\\
 \= \frac{1}{\sqrt{\pi} \~ v_e} \cdot
    \frac{\sqrt{\pi}}{2} \bbigg{\erfc(u_{-,\rm min})-\erfc(u_{+,\rm min})}
    \non\\
 \= \frac{1}{2 v_e} \bigg[\erf(u_{+,\rm min})-\erf(u_{-,\rm min})\bigg]
\~,
\label{eqnC010201}
\eeqn
 where I have used the definition
\beqN
   \erfc(x)
 = \frac{2}{\sqrt{\pi}} \int_{x}^{\infty} e^{-t^2} dt
 = 1-\erf(x)
\~,
\eeqN
 and
\beq
        u_{\pm,\rm min}
 \equiv \frac{\vmin \pm v_e}{v_0}
\~.
\label{eqnC010202}
\eeq
 Combining Eqs.(\ref{eqnC010201}) and (\ref{eqnC010202})
 with Eqs.(\ref{eqn3010106}) and (\ref{eqn3010108}),
 one can obtain $(dR/dQ)_\sh$ in Eq.(\ref{eqn3020103}).
 Moreover,
 by using Eqs.(\ref{eqnB020103a}) and (\ref{eqnB020103b}),
 one can find that
\beqn
 \conti \int_{\Qthre}^{\infty}
        \bbigg{\erf{\T\afrac{\alpha \sqrt{Q}+v_e}{v_0}}-\erf{\T\afrac{\alpha \sqrt{Q}-v_e}{v_0}}}
        dQ
        \non\\
 \=     \int_{q_{\rm thre}}^{\infty}
        \bbigg{\erf{\afrac{\alpha q+v_e}{v_0}}-\erf{\afrac{\alpha q-v_e}{v_0}}} (2 q \~ dq)
        \non\\
 \=     2 \bBig{{\bf V}_1(q \to \infty)-{\bf V}_1(q_{\rm thre})}
        \non\\
 \=     \afrac{v_0}{\alpha}^2
        \cBigg{ \abrac{\frac{1}{2}-S_{+} S_{-}} \bbigg{\erf(S_{+})-\erf(S_{-})}
               +\frac{1}{\sqrt{\pi}} \abrac{S_{+} e^{-S_{-}^2}-S_{-} e^{-S_{+}^2}}}
\~.
\eeqn
 Here I have defined
\beq
        q_{\rm thre}
 \equiv \sqrt{\Qthre}
\~,
\eeq
 and used Eqs.(\ref{eqnB020214}), (\ref{eqnB020217}), (\ref{eqnB020216}), (\ref{eqnB020205a}),
 and (\ref{eqn3020107}).
 Hence,
 for $\FQ \approx 1$,
 one can get $R_\sh(\Qthre)$ in Eq.(\ref{eqn3020106}) directly.
\section{Using $F_{\rm ex}^2(Q)$ given in Eq.(\ref{eqn3010201})}
\subsection{Starting with $(dR/dQ)_\Gau$ given in Eq.(\ref{eqn3010302})}
\label{RexGau}
 ~~~$\,$
 Substituting the exponential form factor $F_{\rm ex}^2(Q)$ given in Eq.(\ref{eqn3010201})
 into Eq.(\ref{eqn3010302}),
 one can get
\beq
   \adRdQ_{\Gau,\rm ex}
 = \calA \afrac{2}{\sqrt{\pi} v_0} e^{-\abrac{\alpha^2/v_0^2+1/Q_0} Q}
 = \calA \afrac{2}{\sqrt{\pi} v_0} e^{-\alpha^2 Q/v_0^2 \beta^2}
\~,
\eeq
 where $\beta$ has been defined in Eq.(\ref{eqn3010309}).
 Then it is easy to find that
\cheqnref{eqn3010307}
\beqn
    R_{\Gau,\rm ex}(\Qthre)
 \= \calA \afrac{2}{\sqrt{\pi} v_0}
    \int_{\Qthre}^{\infty} e^{-\alpha^2 Q/v_0^2 \beta^2} \~ dQ
    \non\\
 \= \frac{\rho_0 \sigma_0 \expv{v}_\Gau}{\mchi \mN}
    \abigg{\beta^2 \~ e^{-\alpha^2 \Qthre/v_0^2 \beta^2}}
\~.
\eeqn
\subsection{Starting with $(dR/dQ)_\sh$ given in Eq.(\ref{eqn3020103})}
\label{Rexsh}
 ~~~$\,$
 Substituting the exponential form factor $F_{\rm ex}^2(Q)$ into Eq.(\ref{eqn3020103}),
 one can get
\cheqnCN{-1}
\beq
   \adRdQ_{\sh,\rm ex}
 = \calA \afrac{1}{2 v_e} e^{-Q/Q_0}
   \bbigg{\erf{\T\afrac{\alpha \sqrt{Q}+v_e}{v_0}}-\erf{\T\afrac{\alpha \sqrt{Q}-v_e}{v_0}}}
\~.
\label{eqnC020201}
\eeq
 Consider
\beqn
 \conti \int e^{-Q/Q_0} \~ \erf{\T\afrac{\alpha \sqrt{Q} \pm v_e}{v_0}} \~ dQ
        \non\\
 \=    -Q_0 \~ e^{-Q/Q_0} \~ \erf{\T \afrac{\alpha \sqrt{Q} \pm v_e}{v_0}}
       +\frac{Q_0}{\sqrt{\pi}} \afrac{\alpha}{v_0}
        \int \frac{1}{\sqrt{Q}} \~ e^{-Q/Q_0-\bbig{\abig{\alpha \sqrt{Q} \pm v_e}/v_0}^2} dQ
\~,
\label{eqnC020202}
\eeqn
 where I have used integration by parts and Eqs.(\ref{eqnB020201}).
 Using Eqs.(\ref{eqnB020103a}) and (\ref{eqnB020103b}),
 the integral of the second term on the right-hand side above can be found as
\beqn
    \int \frac{1}{\sqrt{Q}} \~ e^{-Q/Q_0-\bbig{\abig{\alpha \sqrt{Q} \pm v_e}/v_0}^2} dQ
 \= 2
    \int
    e^{-\bbrac{    \abig{\alpha^2/v_0^2+1/Q_0} q^2
               \pm \abig{2 \alpha v_e/v_0^2} q
                  +v_e^2/v_0^2}} \~ dq
    \non\\
 \= \sqrt{\pi} \afrac{v_0 \beta}{\alpha} e^{-\abrac{1-\beta^2} \tve^2} \~
    \erf\abrac{\T \frac{\alpha \sqrt{Q}}{v_0 \beta} \pm \beta \tve}
\~,
\label{eqnC020203}
\eeqn
 where I have used
\beqN
   \int e^{-\abrac{a x^2+b x+c}} \~ dx
 = \frac{1}{2} \sfrac{\pi}{a}~e^{\abrac{b^2/4 a-c}} \~
   \erf\abrac{\T \sqrt{a} \~ x+\frac{b}{2 \sqrt{a}}}
\~,
\eeqN
 and the definition in Eq.(\ref{eqnB010204}).
 Combining Eqs.(\ref{eqnC020201}) to (\ref{eqnC020203}),
 one can get
\cheqnref{eqn3020109}
\beqn
 \conti R_{\sh,\rm ex}(\Qthre)
        \non\\
 \=     \calA \Afrac{Q_0}{2 v_e}
        \cBiggl{ e^{-\Qthre/Q_0}
                 \bbigg{ \erf{\T \afrac{\alpha \sqrt{\Qthre}+v_e}{v_0}}
                        -\erf{\T \afrac{\alpha \sqrt{\Qthre}-v_e}{v_0}}}}
        \non\\
 \conti ~~~~~~~~~~~~~~~~~~~~ 
        \cBiggr{-\beta e^{-\abrac{1-\beta^2} \tve^2}
                 \bbigg{ \erf{\T \afrac{\alpha \sqrt{\Qthre}+\beta^2 v_e}{v_0 \beta}}
                        -\erf{\T \afrac{\alpha \sqrt{\Qthre}-\beta^2 v_e}{v_0 \beta}}}}
        \non\\
 \=     \frac{\rho_0 \sigma_0}{\mchi \mN} \afrac{v_0^2}{2 v_e} \afrac{\beta^2}{1-\beta^2}
        \non\\
 \conti ~~~~~~ \times 
        \cBiggl{ e^{-\abrac{1-\beta^2} \alpha^2 \Qthre/v_0^2 \beta^2} 
                 \bbigg{\erf(S_{+})-\erf(S_{-})}}
        \non\\
 \conti ~~~~~~~~~~~~~~~~~~~~~~~~~~~ 
        \cBiggr{-\beta e^{-\abrac{1-\beta^2} v_e^2/v_0^2}
                 \bbigg{\erf(T_{+})-\erf(T_{-})}}
\~,
\eeqn
 where I have used the definitions in Eqs.(\ref{eqn3020107}) and (\ref{eqn3020110}) as well as
\cheqnrefp{eqn3010309}
\beq
   Q_0
 = \frac{v_0^2}{\alpha^2} \afrac{\beta^2}{1-\beta^2}
\~.
\eeq
\cheqnCN{-1}

%% file: Doktorarbeit-AppD.tex
\chapter{Some Old Attempts}
 ~~~$\,$
 In this chapter
 I present some old attempts for reconstructing the velocity distribution function,
 eventually also for determining its moments.
 I describe also their disadvantages and problems.
 However,
 these unsuccessful attempts could perhaps inspire some new ideas. 
\section{Binning the data set}
\label{binning}
 ~~~$\,$
 The usually used choice for binning a data set is that every bin has that same width:
\beq
   b_n
 = b
 = \frac{Q_{\rm max}-Q_{\rm min}}{B}
\~,
\eeq
 and thus
\beq
   Q_n
 = Q_{\rm min}+\abrac{n-\frac{1}{2}} b
\~.
\eeq
 However,
 as discussed in Subsec.~\ref{windowing},
 using bins with linearly increasing widths can make the errors roughly equal:
\cheqnref{eqn4020200a}
\beq
   b_n
 = b_1+(n-1) \delta
\~,
\eeq
 here the increment $\delta$ satisfies
\cheqnref{eqn4020200c}
\beq
   \delta
 = \frac{2}{B (B-1)} \aBig{Q_{\rm max}-Q_{\rm min}-B b_1}
\~.
\eeq
\cheqnCN{-2}
 Hence,
 for the $n$-th $Q$-bin,
 one has
\cheqnCa
\beq
   Q_{n,\rm min}
 = Q_{\rm min}+(n-1) b_1+\bfrac{(n-1) (n-2)}{2} \delta
\~,
\eeq
 and
\cheqnCb
\beq
   Q_{n,\rm max}
 = Q_{\rm min}+n b_1+\bfrac{n (n-1)}{2} \delta
\~.
\eeq
\cheqnC
 This means that
\cheqnref{eqn4020200b}
\beq
   Q_n
 = Q_{\rm min}+\abrac{n-\frac{1}{2}} b_1+\bfrac{(n-1)^2}{2} \delta
\~.
\eeq
\cheqnCN{-1}
 Moreover,
 one other choice for binning the data set is
\beq
   b_n
 = b_1 \delta^{n-1}
\~.
\eeq
 It is more comfortable
 if we choose $\delta$ as the input parameter and then determine $b_1$ as
\beq
   b_1
 = \afrac{\delta-1}{\delta^B-1} \aBig{Q_{\rm max}-Q_{\rm min}}
\~.
\eeq
 Hence,
 for the $n$-th $Q$-bin,
 one has
\cheqnCa
\beq
   Q_{n,\rm min}
 = Q_{\rm min}+\afrac{\delta^{n-1}-1}{\delta-1} b_1
\~,
\eeq
 and
\cheqnCb
\beq
   Q_{n,\rm max}
 = Q_{\rm min}+\afrac{\delta^n-1}{\delta-1} b_1
\~.
\eeq
\cheqnC
 This means that
\beq
   Q_n
 = Q_{\rm min}+\bfrac{\delta^n+\delta^{n-1}-2}{2 \abrac{\delta-1}} b_1
\~.
\eeq
\section{Reconstructing $f_1(v)$ without derivatives}
 ~~~$\,$
 According to the expression of the differential event rate in Eq.(\ref{eqn3010108}),
 I have
\beq
   \adRdQ_{Q = Q_n}
 = \calA F^2(Q_n) \int_{v_n}^{\infty} \bfrac{f_1(v)}{v} dv
\~,
\eeq
 where, from Eq.(\ref{eqn3010106}),
\beq
   v_n
 = \alpha \sqrt{Q_n}
\~.
\label{eqnD020001}
\eeq
 Then it can be found that
\beqn
           \int_{v_{n}}^{v_{n+1}} \bfrac{f_1(v)}{v} dv
 \=        \frac{1}{\cal A}
           \bbrac{ \frac{1}{F^2(Q_{n  })} \adRdQ_{Q = Q_{n  }}
                  -\frac{1}{F^2(Q_{n+1})} \adRdQ_{Q = Q_{n+1}}}
           \non\\
 \eqnequiv \Delta_n
\~.
\label{eqnD020002}
\eeqn
 The mean value theorem of calculus implies
\beq
   \frac{\Delta_{n}}{v_{n+1}-v_{n}}
 = \bfrac{f_1(v)}{v}_{v = \td{v}_{n}}
\~,
\label{eqnD020003}
\eeq
 where
\(
     v_{n}
 \le \Td{v}_{n}
 \le v_{n+1}
\).
 Hence, I can let
\beq
   \Td{v}_{n}
 = \alpha_{n} v_{n+1}+(1-\alpha_{n}) v_{n}
 = v_{n}+\alpha_{n} (v_{n+1}-v_{n})
\~,
   ~~~~~~~~~~~~~ 
   (0 \le \alpha_n \le 1)
,
\eeq
 and rewrite Eq.(\ref{eqnD020003}) to
\beq
   f_1(\Td{v}_{n})
 = \afrac{\Td{v}_{n}}{v_{n+1}-v_{n}} \Delta_{n}
 = \bbrac{\frac{1}{\abrac{v_{n+1}/v_{n}}-1}+\alpha_{n}} \Delta_{n}
\~,
   ~~~~~ 
   (0 \le \alpha_n \le 1)
.
\eeq
 Therefore,
 the error of $f_1(\Td{v}_{n})$ can be given as
\beq
   \sigma\aBig{f_1(\Td{v}_{n})}
 = \bbrac{\frac{1}{\abrac{v_{n+1}/v_{n}}-1}+\alpha_{n}} \sigma(\Delta_{n})
\~.
\eeq
 Usually, one sets
\(
   \alpha_n
 = \frac{1}{2}
\)
 and then it can be reduced to
\beq
        \sigma\aBig{f_1(\Td{v}_{n},\alpha_n=1/2)}
 \equiv \sigma\aBig{f_1(v_{n+1/2})}
 =      \frac{1}{2} \afrac{v_{n+1}+v_{n}}{v_{n+1}-v_{n}} \sigma(\Delta_{n})
\~.
\eeq
 Here, from Eq.(\ref{eqnD020002}),
\beqn
    \sigma(\Delta_{n})
 \= \frac{1}{\cal A}
    \cbrac{ \frac{1}{F^4(Q_{n  })}~\sigma^2\bbrac{\T\adRdQ_{Q = Q_n  }}
           +\frac{1}{F^4(Q_{n+1})}~\sigma^2\bbrac{\T\adRdQ_{Q = Q_n+1}}}^{1/2}
    \non\\
 \= \frac{1}{\cal A}
    \cbrac{ \frac{1}{F^4(Q_{n  })} \afrac{N_{n  }}{b_{n  }^2}
           +\frac{1}{F^4(Q_{n+1})} \afrac{N_{n+1}}{b_{n+1}^2}}^{1/2}
\~,
\eeqn
 where I have used the standard estimator
 for for $dR/dQ$ at the point $Q = Q_n$ in Eq.(\ref{eqn4020102})
 and then its statistical error in Eq.(\ref{eqn4020103}).

 This method is straightforward.
 However,
 neither $f_1(\Td{v}_n)$ nor its statistical error
 is independent of the unknown constant $\calA$.
 Moreover,
 this method has an {\em anti-correlation} problem:
 An upward fluctuation of the counting rate in the $n$-th $Q$-bin
 will lead to too small $f_1$ in the $n-1$-st $v$-bin,
 but tends to give too large $f_1$ in the $n$th $v$-bin.
\section{Average logarithmic slope}
\label{knave}
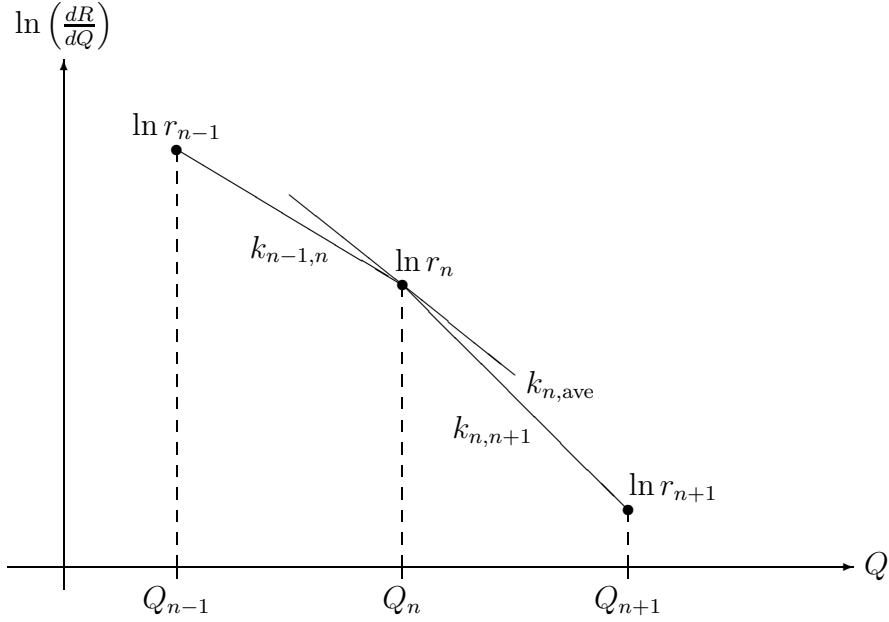
\begin{figure}[t]
\begin{center}
\begin{picture}(9,6)
\put(0,0){\usebox{\Figurekave}}
\end{picture}
\end{center}
\caption{
 Sketch of the average slope for the function $\ln(dR/dQ)_{Q \simeq Q_n}$ at $Q = Q_n$,
 $k_{n,\rmave}$,
 defined in Eq.(\ref{eqnD030002a}).}
\label{figD030001}
\end{figure}

 ~~~$\,$
 As shown in Fig.~\ref{fig4020101},
 the theoretically predicted recoil spectrum is approximately exponential.
 And,
 as discussed in Subsec.~\ref{dRdQkn},
 an exponential approximation can approximate the recoil spectrum for a wider bin.
 Hence,
 I have considered the exponential ansatz in Eq.(\ref{eqn4020105}),
 but at beginning only naively combined with
 the standard estimator for $dR/dQ$ at the point $Q = Q_n$ as
\beq
   \adRdQ_{Q \simeq Q_n}
 = r_n e^{k_n (Q-Q_n)}
\~,
\eeq
 where $r_n = N_n/b_n$ is the standard estimator given in Eq.(\ref{eqn4020102}).
 Define the slope of the straight line
 with two endpoints $\abrac{Q_{n},\ln(dR/dQ)_{Q = Q_{n}}}$ and
 $\abrac{Q_{n+1},\ln(dR/dQ)_{Q = Q_{n+1}}}$ as
\beq
        \knnp
 \equiv \frac{\ln r_{n+1}-\ln r_{n}}{Q_{n+1}-Q_{n}}
\~,
        ~~~~~~~~~~~~~~~~~~ 
        n
 =      1,~2,~\cdots,~B-1
.
\label{eqnD030001}
\eeq
 Then I can define an {\em average slope} for the function $\ln(dR/dQ)_{Q \simeq Q_n}$
 at the point $Q = Q_n$, $n=2,~3,~\cdots,~B-1$, as (see Fig.~\ref{figD030001})
\cheqnCa
\beq
        k_{n,\rmave}
 \equiv \frac{k_{n-1,n}+k_{n,n+1}}{2}
 =      \frac{1}{2}
        \abrac{ \frac{\ln r_{n  }-\ln r_{n-1}}{Q_{n  }-Q_{n-1}}
               +\frac{\ln r_{n+1}-\ln r_{n  }}{Q_{n+1}-Q_{n  }}}
\~,
\label{eqnD030002a}
\eeq
 but, at the point $Q = Q_1$, I have defined
\cheqnCb
\beq
        k_{1,\rmave}
 \equiv k_{1,2}
 =      \frac{\ln r_2-\ln r_1}{Q_2-Q_1}
\~.
\label{eqnD030002b}
\eeq
\cheqnC
 The statistical errors on $k_{n,\rmave}$
 can be obtained directly from Eqs.(\ref{eqnD030002a}) and (\ref{eqnD030002b}) as
\cheqnCa
\beqn
        \sigma^2\abrac{k_{n,\rmave}}
 \=     \frac{1}{4}
        \bleft{ \abrac{\frac{1}{Q_{n}-Q_{n-1}}-\frac{1}{Q_{n+1}-Q_{n}}}^2 \frac{1}{N_n}}
        \non\\
 \conti ~~~~~~~~~~~~ 
        \bright{+\afrac{1}{Q_{n  }-Q_{n-1}}^2 \frac{1}{N_{n-1}}
                +\afrac{1}{Q_{n+1}-Q_{n  }}^2 \frac{1}{N_{n+1}}}
\~,
\label{eqnD030003a}
\eeqn
 for $n=2,~3,~\cdots,~B-1$,
 and
\cheqnCb
\beq
   \sigma^2\abrac{k_{1,\rmave}}
 = \afrac{1}{Q_2-Q_1}^2 \aBigg{\frac{1}{N_1}+\frac{1}{N_2}}
\~,
\label{eqnD030003b}
\eeq
\cheqnC
 where I have used Eqs.(\ref{eqn4020102}) and (\ref{eqn4020103}) to get
\beq
   \frac{\sigma^2(r_n)}{r_n^2}
 = \frac{1}{N_n}
\~.
\eeq
 Moreover,
 the $k_{n,\rmave}$ given in Eqs.(\ref{eqnD030002a}) and (\ref{eqnD030002b}) are correlated.
 Hence,
 for $k_{n,\rmave}$ in Eq.(\ref{eqnD030002a}),
 I have \\
\cheqnCa
\beqn
 \conti {\rm cov}\abrac{k_{n,\rmave},k_{n+1,\rmave}}
        \non\\
 \=     \frac{1}{4}
        \bleft{-\abrac{\frac{1}{Q_{n}-Q_{n-1}}-\frac{1}{Q_{n+1}-Q_{n}}}
                \afrac{1}{Q_{n+1}-Q_{n}}
                \frac{1}{N_{n}}}
        \non\\
 \conti ~~~~~~~~~~~~ 
        \bright{+\afrac{1}{Q_{n+1}-Q_{n}}
                 \abrac{\frac{1}{Q_{n+1}-Q_{n}}-\frac{1}{Q_{n+2}-Q_{n+1}}}
                 \frac{1}{N_{n+1}}}
\~,
\label{eqnD030004a}
\eeqn
 and
\cheqnCb
\beq
   {\rm cov}\abrac{k_{n,\rmave},k_{n+2,\rmave}}
 =-\frac{1}{4}
   \afrac{1}{Q_{n+1}-Q_{n}}
   \afrac{1}{Q_{n+2}-Q_{n+1}}
   \frac{1}{N_{n+1}}
\~,
\label{eqnD030004b}
\eeq
 while,
 for $k_{1,\rmave}$ in Eq.(\ref{eqnD030002b}),
 I have
\cheqnCc
\beqn
    {\rm cov}\abrac{k_{1,\rmave},k_{2,\rmave}}
 \= \frac{1}{2}
    \bbrac{ \afrac{1}{Q_2-Q_1}^2 \frac{1}{N_1}
           +\frac{1}{Q_2-Q_1} \abrac{\frac{1}{Q_2-Q_1}-\frac{1}{Q_3-Q_2}} \frac{1}{N_2}}
\~,
    \non\\
\label{eqnD030004c}
\eeqn
 and
\cheqnCNx{-1}{d}
\beq
   {\rm cov}\abrac{k_{1,\rmave},k_{3,\rmave}}
 =-\frac{1}{2} \afrac{1}{Q_2-Q_1} \afrac{1}{Q_3-Q_2} \frac{1}{N_2}
\~.
\label{eqnD030004d}
\eeq
\cheqnC

\begin{figure}[t]
\begin{center}
\begin{picture}(9,6)
\put(0,0){\usebox{\FigurelnrnQ}}
\end{picture}
\end{center}
\caption{
 Sketch of the reconstructed segment of the function $\ln(dR/dQ)$
 between $(Q_{n-1}+Q_n)/2$ and $(Q_n+Q_{n+1})/2$,
 $\ln r_{n,\rmave}(Q)$.}
\label{figD030002}
\end{figure}
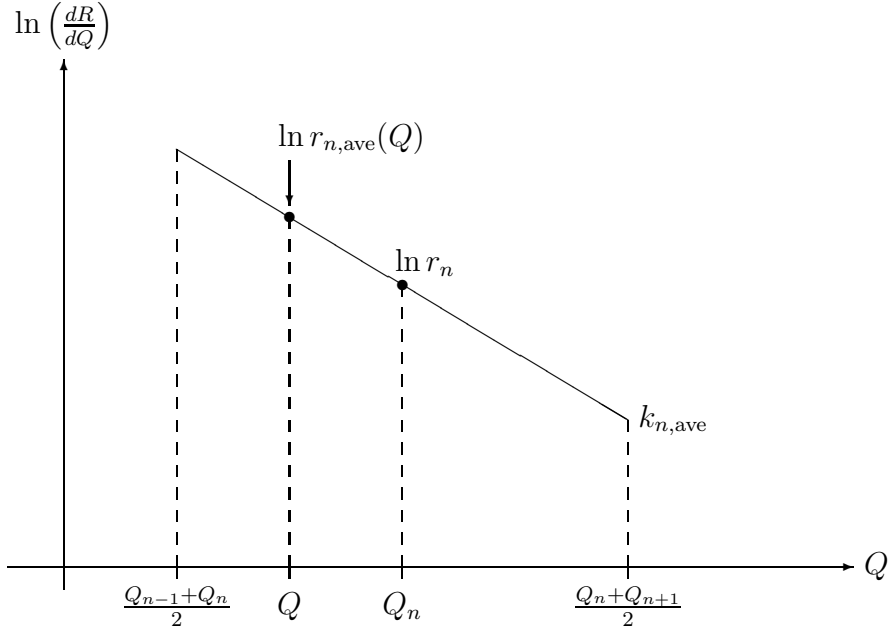

 Now I can begin to reconstruct the recoil spectrum.
 The basic idea is that
 I approximate the function $\ln(dR/dQ)$ in each bin by a straight line $\ln r_{n,\rmave}(Q)$
 which has the slope $k_{n,\rmave}$ and passes through the point $\abrac{Q_n,\ln r_n}$ 
 (see Fig.~\ref{figD030002}):
\beq
   \frac{\ln r_{n,\rmave}(Q)-\ln r_n}{Q-Q_n}
 = k_{n,\rmave}
\~.
\eeq
 Hence,
 I have
\cheqnCa
\beq
   \adRdQ_{Q \simeq Q_n}
 = r_{n,\rmave}(Q)
 = r_n \~ e^{k_{n,\rmave} (Q-Q_n)}
\~,
   ~~~~~~~~~~ 
   n
 = 2,~3,~\cdots,~B-1
,
\label{eqnD030005a}
\eeq
 in the $n$-th $Q$-bin:
\beq
        {\T \frac{Q_{n-1}+Q_{n  }}{2}}
 \equiv Q_{n-}
 \le    Q
 \le    Q_{n+}
 \equiv {\T \frac{Q_{n  }+Q_{n+1}}{2}}
\~,
\label{eqnD030006a}
\eeq
 and
\cheqnCNx{-2}{b}
\beq
   \adRdQ_{Q \simeq Q_1}
 = r_{1,\rmave}(Q)
 = r_1 \~ e^{k_{1,\rmave} (Q-Q_1)}
\~,
\label{eqnD030005b}
\eeq
 in the first $Q$-bin:
\beq
        \Qthre
 \equiv Q_{1-}
 \le    Q
 \le    Q_{1+}
 \equiv {\T \frac{Q_1+Q_2}{2}}
\~,
\label{eqnD030006b}
\eeq
\cheqnC
 where $\Qthre$ is the threshold energy
 and the $k_{n,\rmave}$, $n=1,~2,~\cdots,~B-1$,
 are given in Eqs.(\ref{eqnD030002a}) and (\ref{eqnD030002b}),
 respectively.
 Then,
 similar to Eq.(\ref{eqn4020301}),
 the velocity distribution function $f_1(v)$ given in Eq.(\ref{eqn4010008})
 can be reconstructed as
\beq
   f_{1,\rmave}(v_n)
 = \calN_\rmave \bfrac{2 Q_n r_n}{\FQn}
   \bbrac{\dd{Q} \ln\FQ \bigg|_{Q = Q_n}-k_{n,\rmave}}
\~,
\label{eqnD030007}
\eeq
 for $n=1,~2,~\cdots,~B-1$.
 Here $v_n$ is given in Eq.(\ref{eqnD020001}).

 The first problem with the expression in Eq.(\ref{eqnD030007}) is
 estimating the normalization constant $\calN_\rmave$.
 One possibility is
 inserting $r_{n,\rmave}(Q)$ given in Eqs.(\ref{eqnD030005a}) and (\ref{eqnD030005b})
 into Eq.(\ref{eqn4010009}) directly:
\beq
   \calN_\rmave
 = \frac{2}{\alpha}
   \cbrac{\sum_{i=1}^{B-1}
          \int_{Q_{i-}}^{Q_{i+}} \frac{1}{\sqrt{Q}} \bfrac{r_{i,\rmave}(Q)}{\FQ} dQ}^{-1}
\~,
\label{eqnD030008}
\eeq
 where $Q_{i\pm}$ are given in Eqs.(\ref{eqnD030006a}) and (\ref{eqnD030006b}).
 Similarly,
 $I_n$ defined in Eq.(\ref{eqn4010012})
 and $(dR/dQ)_{Q = \Qthre}$ in Eq.(\ref{eqn4010011})
 can also be estimated as
\beq
   I_n
 = \sum_{i=1}^{B-1} \int_{Q_{i-}}^{Q_{i+}} Q^{(n-1)/2} \bfrac{r_{i,\rmave}(Q)}{\FQ} dQ
\~,
\label{eqnD030009}
\eeq
 and
\beq
   \adRdQ_{\rmave, Q = \Qthre}
 = r_{1,\rmave}(\Qthre)
 = r_1 \~ e^{k_{1,\rmave} (\Qthre-Q_1)}
\~.
\label{eqnD030010}
\eeq
 However,
 it should be pretty complicated
 to estimate the statistical errors of $f_{1,\rmave}(v_n)$ given in Eq.(\ref{eqnD030007})
 with $\calN_\rmave$ estimated in Eq.(\ref{eqnD030008}).

\begin{figure}[t]
\begin{center}
\imageswitch{
\begin{picture}(9,6)
\put(0,0){\usebox{\Figuredeltarn}}
\end{picture}}
{\includegraphics[width=12.5cm]{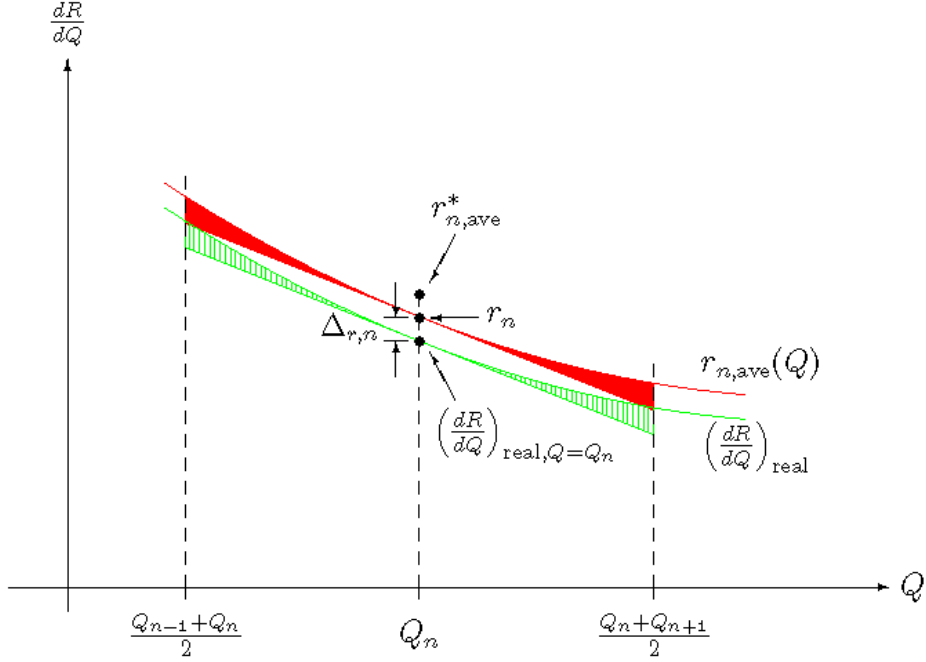}}
\end{center}
\caption{
 Sketch of the elevation from $(dR/dQ)_{{\rm real},Q = Q_n}$ to $r_n$
 and from $r_n$ to $r_{n,\rmave}^{\ast}$
 due to the concavity of the recoil curve, $(dR/dQ)_{\rm real}$,
 and that of the reconstructed spectrum, $r_{n,\rmave}(Q)$,
 between $(Q_{n-1}+Q_n)/2$ and $(Q_n+Q_{n+1})/2$,
 respectively.}
\label{figD030003}
\end{figure}
 The other serious problem
 with the ansatz in Eqs.(\ref{eqnD030005a}) and (\ref{eqnD030005b}) is that
 one must also consider a systematic error
 caused by using the exponential ansatz with the standard estimator $r_n = N_n/b_n$.
 Suppose that
 $(dR/dQ)_{\rm real}$ is the real recoil spectrum
 and passes through the point $\abrac{Q_n,(dR/dQ)_{{\rm real},Q = Q_n}}$
 (see Fig.~\ref{figD030003}).
 During the experiment
 we measure deposited energies and count the event rate,
 which is proportional to the area under the real recoil spectrum
 (see Eq.(\ref{eqn4020107})),
 in the $n$-th $Q$-bin,
 and then estimate $r_n = N_n/b_n$.
 However,
 because the recoil spectrum is concave,
 the estimator $r_n$ is a little larger than the real value,
 $(dR/dQ)_{{\rm real},Q = Q_n}$
 (see $\trn$ given in Eq.(\ref{eqn4020109})).
 Define this elevation from $(dR/dQ)_{{\rm real},Q = Q_n}$ to $r_n$ as
\beq
        \Delta_{r,n}
 \equiv r_{n,\rmave}(Q_n)-\adRdQ_{{\rm real},Q = Q_n}
 \equiv r_n-r_{n,{\rm real}}
\~,
        ~~~~
        n
 =      1,~2,~\cdots,~B-1
.
\label{eqnD030011}
\eeq
 On the other hand,
 it is plausible to suppose that
 the reconstructed recoil spectrum $r_{n,\rmave}(Q)$
 in Eq.(\ref{eqnD030003a}) and (\ref{eqnD030003b})
 are approximately parallel to the real one $(dR/dQ)_{\rm real}$,
 thus I can estimate the elevation by
\beq
         \Delta_{r,n}
 \approx r_{n,\rmave}^{\ast}-r_n
\~,
        ~~~~~~~~~~~~~~~~~~~~~~~~ 
        n
 =      1,~2,~\cdots,~B-1
.
\label{eqnD030012}
\eeq
 Here $r_{n,\rmave}^{\ast}$ can be calculated from $r_{n,\rmave}(Q_n)$ as
\cheqnCa
\beqn
    r_{n,\rmave}^{\ast}
 \= \frac{2}{Q_{n+1}-Q_{n-1}}
    \int_{(Q_{n-1}+Q_{n})/2}^{(Q_{n}+Q_{n+1})/2} r_{n,\rmave}(Q) \~ dQ
    \non\\
 \= \frac{2}{Q_{n+1}-Q_{n-1}}
    \int_{(Q_{n-1}+Q_{n})/2}^{(Q_{n}+Q_{n+1})/2} \bbrac{r_n \~ e^{k_{n,\rmave} (Q-Q_n)}} dQ
    \non\\
 \= r_n \bfrac{2}{k_{n,\rmave} (Q_{n+1}-Q_{n-1})}
    \bbrac{e^{k_{n,\rmave} \afrac{Q_{n+1}-Q_{n}}{2}}-e^{-k_{n,\rmave} \afrac{Q_{n}-Q_{n-1}}{2}}}
\~,
\label{eqnD030013a}
\eeqn
 for $n=2,~3,~\cdots,~B-1$, and for $n = 1$,
\cheqnCb
\beqn
    r_{1,\rmave}^{\ast}
 \= \frac{2}{Q_1+Q_2-2 \Qthre} \int_{\Qthre}^{(Q_1+Q_2)/2} r_{1,\rmave}(Q) \~ dQ
    \non\\
 \= \frac{2}{Q_1+Q_2-2 \Qthre}
    \int_{\Qthre}^{(Q_1+Q_2)/2} \bbrac{r_1 \~ e^{k_{1,\rmave} (Q-Q_1)}} dQ
    \non\\
 \= r_1 \bfrac{2}{k_{1,\rmave} (Q_1+Q_2-2 \Qthre)}
    \bBigg{e^{k_{1,\rmave} \afrac{Q_2-Q_1}{2}}-e^{-k_{1,\rmave} (Q_1-\Qthre)}}
\~.
\label{eqnD030013b}
\eeqn
\cheqnC
 Combining Eqs.(\ref{eqnD030011}) to (\ref{eqnD030013b}),
 the real value of the recoil spectrum $(dR/dQ)_{\rm real}$
 at the point $Q = Q_n$, $n=2,~3,~\cdots,~B-1$, can be obtained (approximately) as
\cheqnCa
\beqn
            r_{n,{\rm real}}
 \eqnapprox 2 r_n-r_{n,\rmave}^{\ast}
            \non\\
 \=         2 r_n
              \cleft{ 1
                     -\bfrac{1}{k_{n,\rmave} (Q_{n+1}-Q_{n-1})}}
            \non\\
 \conti     ~~~~~~~~~~~~~~~~~~~~ \times 
              \cright{\bbrac{ e^{ k_{n,\rmave} \afrac{Q_{n+1}-Q_{n  }}{2}}
                             -e^{-k_{n,\rmave} \afrac{Q_{n  }-Q_{n-1}}{2}}}}
\~,
\label{eqnD030014a}
\eeqn
 and at the point $Q = Q_1$,
\cheqnCb
\beqn
            r_{1,{\rm real}}
 \eqnapprox 2 r_1-r_{1,\rmave}^{\ast}
            \non\\
 \=         2 r_1
            \cleft{ 1
                   -\bfrac{1}{k_{1,\rmave} (Q_1+Q_2-2 \Qthre)}}
            \non\\
 \conti     ~~~~~~~~~~~~~~~~~~~~ \times 
            \cright{\bBigg{e^{k_{1,\rmave} \afrac{Q_2-Q_1}{2}}-e^{-k_{1,\rmave} (Q_1-\Qthre)}}}
\~.
\label{eqnD030014b}
\eeqn
\cheqnC
 Note that
 the correction of $r_n$ here is essentially the same as
 the expression of $\trn$ in Eq.(\ref{eqn4020109}).

 Now one can replace $r_n$ in Eqs.(\ref{eqnD030002a}) and (\ref{eqnD030002b})
 by $r_{n,{\rm real}}$ estimated by Eqs.(\ref{eqnD030014a}) and (\ref{eqnD030014b})
 to get $k_{n,\rmave}$,
 and then substitute $r_{n,{\rm real}}$ and $k_{n,\rmave}$
 into Eqs.(\ref{eqnD030007}) to (\ref{eqnD030010})
 to reconstruct $f_{1,\rmave}(v_n)$ and so on.
 However,
 due to the dependence of $r_{n,{\rm real}}$ on $r_n$ and $k_{n,\rmave}$, 
 it is very complicated to modify
 even the statistical errors in Eqs.(\ref{eqnD030003a}) to (\ref{eqnD030004d})!

 Moreover,
 this ``average-logarithmic-slope'' method
 has also the same ``anti-correlation'' problem
 as the method described in the previous section.
 An upward fluctuation in the $n$-th $Q$-bin leads to a too small slope $k_{n-1,n}$
 and a too large slope $k_{n,n+1}$,
 even though the fluctuation of the ``average'' slope
 $k_{n,\rmave}=(k_{n-1,n}+k_{n,n+1})/2$
 could be more or less decreased and
 the value of $k_{n,\rmave}$ should be not very bad.

 Furthermore,
 as shown in e.g., Eqs.(\ref{eqnD030001}), (\ref{eqnD030005a}), and (\ref{eqnD030011}),
 from $B$ bins one can get $B-1$ $k_{n,\rmave}$, $r_{n,\rmave}(Q)$, and $r_{n,{\rm real}}$.
 Then,
 after one replaces $r_n$ in Eq.(\ref{eqnD030001}) by $r_{n,{\rm real}}$
 and runs the whole process from Eq.(\ref{eqnD030001}) to Eq.(\ref{eqnD030010}),
 one can have only $B-2$ $f_{1,\rmave}(v_n)$ given by Eq.(\ref{eqnD030007}).
 However,
 as shown in Figs.~\ref{fig4020301},
 with 500 (or even 5000) events,
 one has only 4 (or 8) bins to use.
\footnote{
 Not 5 or 10 bins,
 because the last one or two bin are almost empty,
 see the discussion in Subsec.~\ref{f1r(v)}.}
 Hence,
 it can not be allowed to lose 2 bins (points) more!
\section{Linear approximations of $(dQ/dR)_{Q \simeq Q_n}$}
\label{lineardRdQn}
 ~~~$\,$
 As noted in the beginning of Sec.~\ref{Qni},
 according to the expression in Eq.(\ref{eqn4010008}),
 one needs not only an estimator for $dR/dQ$ at $Q=Q_n$
 but also one for the {\em slope} of the recoil spectrum
 to reconstruct the velocity distribution.
 A rather crude estimator of this slope is
\beq
        s_{1,n}
 \equiv \bbrac{\dd{Q}\adRdQ}_{Q = Q_n}
 =      \frac{N_{n,Q > Q_n}-N_{n,Q < Q_n}}{(b_n/2)^2}
\~,
\label{eqnD040001}
\eeq
 where $N_{n,Q > Q_n}$ and $N_{n,Q < Q_n}$ are the numbers of events in bin $n$
 which have measured recoil energy $Q$ larger and smaller than $Q_n$,
 respectively.
 This estimator is rather crude,
 since it only uses the information in which half of its bin a given event falls.

 It is clear intuitively that
 an estimator that makes use of the exact $Q$-value of each event should be better.
 This can e.g., be obtained from the average $Q$-value in a given bin.
 Taylor-expanding $dR/dQ$ around $Q = Q_n$,
 keeping terms up to linear order,
 gives
\beqn
           \adRdQ_{Q \simeq Q_n}
 \eqnsimeq \adRdQ_{Q = Q_n} + (Q-Q_n) \bbrac{\dd{Q}\adRdQ}_{Q=Q_n}
 =         r_n+(Q-Q_n) s_n
\~.
           \non\\
\label{eqnD040002}
\eeqn
 Using this linear approximation for the recoil spectrum,
 one can find
\beq
   N_n
 = \intQnbn \bBig{r_n+(Q-Q_n) s_n} dQ
 = r_n b_n
\~,
\label{eqnD040003}
\eeq
 (of course, this reproduces the standard estimator in Eq.(\ref{eqn4020102}))
 and the average value of the recoil energies in the $n$-th $Q$-bin:
\beq
   \bQn
 = \frac{1}{N_n} \intQnbn (Q-Q_n) \bBig{r_n+(Q-Q_n) s_n} dQ
 = \afrac{b_n^2}{12 r_n} s_n
\~.
\label{eqnD040004}
\eeq
 Hence, an improved estimator of the slope of $dR/dQ$ at $Q = Q_n$ is
\beq
    s_{2,n}
 = \frac{12 r_n \bQn}{b_n^2}
\~.
\label{eqnD040005}
\eeq
 According to the definition of $s_{1,n}$ in Eq.(\ref{eqnD040001})
 it can be found that
\beqn
    \sigma^2(s_{1,n})
 \= \bfrac{1}{(b_n/2)^2}^2 \bbigg{\sigma^2\aBig{N_{n,Q > Q_n}}+\sigma^2\aBig{N_{n,Q < Q_n}}}
    \non\\
 \= \frac{16}{b_n^4} \aBig{N_{n,Q > Q_n}+N_{n,Q < Q_n}}
    \non\\
 \= \frac{16 r_n}{b_n^3}
\~,
\label{eqnD040006}
\eeqn
 where I have used Eqs.(\ref{eqn4020104}) and (\ref{eqn4020102}).
 On the other hand,
 according to the expression of $s_{2,n}$ in Eq.(\ref{eqnD040005}),
 and treating the number of events and the average $Q$-value in a given bin
 as two independent variables,
 one can obtain that
\beq
   \sigma^2(s_{2,n})
 = \afrac{12 \bQn}{b_n^2}^2 \sigma^2(r_n)+\afrac{12 r_n}{b_n^2}^2 \sigma^2\abrac{\bQn}
 = \frac{12 r_n}{b_n^3}
\~.
\label{eqnD040007}
\eeq
 Here I have used Eq.(\ref{eqn4020103}),
 the definition in Eq.(\ref{eqn4020111})
 with the linear approximation in Eq.(\ref{eqnD040002}),
\beq
   \bQQn
 = \frac{1}{N_n} \intQnbn (Q-Q_n)^2 \bBig{r_n+(Q-Q_n) s_n} dQ
 = \frac{b_n^2}{12}
\~,
\eeq
 and
\footnote{
 Strictly speaking,
 the denominator should be $N_n-1$ as I used in Eq.(\ref{eqn4020119}).}
\beq
   \sigma^2\abrac{\bQn}
 = \frac{1}{N_n} \bbigg{\bQQn-\bQn^2}
\~.
\label{eqnD040008}
\eeq
 This simple calculation shows that
 the estimator $s_{2,n}$ given in Eq.(\ref{eqnD040005})
 indeed has a smaller statistical error
 than the crude estimator $s_{1,n}$ in Eq.(\ref{eqnD040001})
 by a factor of $\sqrt{3/4}$.
\section{Using the exponential ansatz in Eq.(\ref{eqn4020105})}
 ~~~$\,$
 In App.~\ref{knave}
 I have used an exponential approximation
 with the standard estimator $r_n$
 to reconstruct the recoil spectrum.
 A correction due to the approximately exponential form of the recoil spectrum
 has also been discussed.
 The use of the average logarithmic slope $k_{n,\rmave}$ combined with the correction
 is very complicated,
 especially for the error analysis.
 However,
 it is clear that
 an exponential approximation can approximate the recoil spectrum much better
 than a linear one.
 On the other hand,
 in the previous section
 I have introduced the use of the exact $Q$-value of each event.
 The analysis done with two linear approximations has shown that
 the statistical error can be strongly reduced.
 Hence,
 it is pretty straightforwardly to combine
 these two techniques and their advantages together.

 By using an exponential ansatz for the recoil spectrum in each $Q$-bin,
 combining with a prefactor which can be adjusted by the event number in this bin,
 and then estimating the logarithmic slope
 by the average value of the recoil energies measured in this bin,
 one can already obtain the expressions given in Subsec.~\ref{dRdQkn}.
 Substituting the first expression of the exponential ansatz in Eq.(\ref{eqn4020105})
 into Eq.(\ref{eqn4010008})
 with the logarithmic slope $k_n$ estimated by Eq.(\ref{eqn4020110}),
 the velocity distribution function can be reconstructed as
\beq
   f_{1,\Bar{Q}}(v_n)
 = \calN_{\Bar{Q}} \bfrac{2 Q_n \trn}{\FQn} \bbrac{\dd{Q} \ln\FQ \bigg|_{Q = Q_n}-k_n}
\~,
\label{eqnD050001}
\eeq
 where $v_n$ is given in Eq.(\ref{eqnD020001}).
 This expression is already almost the same as the expression given in Eq.(\ref{eqn4020301}),
 except the central point $Q_n$ in the $n$-th $Q$-bin has been used here
 instead of the shifted point $Q_{s,\mu}$ in the $\mu$-th $Q$-window,
 and thus I have used $\trn$ instead of $r_{\mu}$ here.
 Moreover,
 I have to determine the normalization constant $\calN_{\Bar{Q}}$ here.
 It can be done by Eq.(\ref{eqnD030008})
 with replacing $r_{i,\rmave}(Q)$ by $\trn e^{k_n (Q-Q_n)}$:
\beq
   \calN_{\Bar{Q}}
 = \frac{2}{\alpha}
   \cbrac{\sum_{i=1}^{B}
          \int_{Q_{i-}}^{Q_{i+}} \frac{1}{\sqrt{Q}} \bfrac{\trn e^{k_n (Q-Q_n)}}{\FQ} dQ}^{-1}
\~,
\eeq
 where $Q_{n\pm}$ have been given as
\beq
        \T{Q_n-\frac{b_n}{2}}
 \equiv Q_{n-}
 \le    Q
 \le    Q_{n+}
 \equiv \T{Q_n+\frac{b_n}{2}}
\~.
\eeq
 However,
 in order to estimate the statistical error of $f_{1,\Bar{Q}}(v_n)$ in Eq.(\ref{eqnD050001})
 more easily,
 I have defined
\beq
        f_{1,n,\Bar{Q}}(v)
 \equiv \calN_{\Bar{Q}}
        \cbrac{\frac{2 Q \trn e^{k_n (Q-Q_n)}}{\FQ} \bbrac{\Dd{\ln\FQ}{Q}-k_n}}\Qva
\~,
\label{eqnD050002}
\eeq
 and
\beq
        \td{f}_{1,n,\Bar{Q}}(v)
 \equiv 2 \cbrac{\frac{Q \trn e^{k_n (Q-Q_n)}}{\FQ} \bbrac{\Dd{\ln\FQ}{Q}-k_n}}\Qva
 =      \frac{f_{1,n,\Bar{Q}}(v)}{\calN_{\Bar{Q}}}
\~,
\label{eqnD050003}
\eeq
 in the $n$-th $v$-bin:
\beq
        \T{\alpha \sqrt{Q_{n-}}}
 \equiv v_{n-}
 \le    v
 \le    v_{n+}
 \equiv \T{\alpha \sqrt{Q_{n+}}}
\~,
\eeq
 for $n=1,~2,~3,~\cdots,~B$.
 The normalization condition in Eq.(\ref{eqn4010007}) can be rewritten as
\beq
   \sum_{i=1}^{B} \intvi f_{1,i,\Bar{Q}}(v) \~ dv
 = \calN_{\Bar{Q}} \bbrac{\sum_{i=1}^{B} \intvi \td{f}_{1,i,\Bar{Q}}(v) \~ dv}
 = 1
\~.
\eeq
 Then the normalization constant can be obtained directly by 
\beq
   \calN_{\Bar{Q}}
 = \bbrac{\sum_{i=1}^{B} \intvi \td{f}_{1,i,\Bar{Q}}(v) \~ dv}^{-1}
\~,
\label{eqnD050004}
\eeq
 and $f_{1,n,\Bar{Q}}(v)$ in Eq.(\ref{eqnD050002}) can be rewritten as
\beq
   f_{1,n,\Bar{Q}}(v)
 = \bbrac{\sum_{i=1}^{B} \intvi \td{f}_{1,i,\Bar{Q}}(v) \~ dv}^{-1} \td{f}_{1,n,\Bar{Q}}(v)
 = \frac{\td{f}_{1,n,\Bar{Q}}(v)}{S_0}
\~.
\label{eqnD050005}
\eeq
 Here I have defined
\beq
        S_{\lambda}
 \equiv \sum_{i=1}^{B} \intvi v^{\lambda} \td{f}_{1,i,\Bar{Q}}(v) \~ dv
\~.
\label{eqnD050006}
\eeq
 Moreover,
 it is reasonable to define the $n$-th moment of $f_{1,n,\Bar{Q}}(v)$ as
\beqn
        \expv{v^n}_{\Bar{Q}}
 \equiv \sum_{i=1}^{B} \intvi v^n f_{1,i,\Bar{Q}}(v) \~ dv
 =      \frac{S_n}{S_0}
\~.
\label{eqnD050007}
\eeqn

 Furthermore,
 in order to estimate the statistical errors of
 $f_{1,n,\Bar{Q}}(v)$ and $\expv{v^n}_{\Bar{Q}}$
 given in Eqs.(\ref{eqnD050005}) and (\ref{eqnD050007}),
 I have denoted first
 the independent variables of $\td{f}_{1,n,\Bar{Q}}(v)$
 defined in Eq.(\ref{eqnD050003}) as $x_{\nu,j}$,
 where the subscript $\nu$ stands for different species of variable,
 and $j=1,~2,~3,~\cdots,~B$ stands for the bins.
 Meanwhile,
 I have assumed that,
 in the $j$-th $Q$-bin (and then also in the $j$-th $v$-bin),
 the error of each of these variables $x_{\nu,j}$ is approximately equal.
 Hence,
 I can use its value at the point $Q=Q_j$, defined as $\delta x_{\nu,j}$,
 for the whole $j$-th $Q$- or $v$-bin.
 From the expression of $f_{1,n,\Bar{Q}}(v)$ in Eq.(\ref{eqnD050005}),
 its statistical error can be found directly as
\beqn
    \sigma^2\aBig{f_{1,n,\Bar{Q}}(v)}
 \= \sumnu \sum_{j=1}^{B} \bPp{f_{1,n,\Bar{Q}}(v)}{x_{\nu,j}}^2 \sigma^2(x_{\nu,j})
    \non\\
 \= \frac{1}{S_0^4}
    \sumnu \sum_{j=1}^{B}
    \bbigg{ S_0 \abrac{\p_{\nu,j} \td{f}_{1,n,\Bar{Q}}(v)}
           -S_{\nu,j;0} \~ \td{f}_{1,n,\Bar{Q}}(v)}^2
    \sigma^2(x_{\nu,j})
\~.
\label{eqnD050008}
\eeqn
 Here I have defined
\beq
        \p_{\nu,j} \td{f}_{1,n,\Bar{Q}}(v)
 \equiv \Pp{\td{f}_{1,n,\Bar{Q}}(v)}{x_{\nu,j}}
\~,
\eeq
 and
\beq
        S_{\nu,j;\lambda}
 \equiv \sum_{i=1}^{B} \intvi v^{\lambda} \bbrac{\p_{\nu,j} \td{f}_{1,i,\Bar{Q}}(v)} dv
\label{eqnD050009}
\~.
\eeq
 Similarly,
 from the definition of $\expv{v^n}_{\Bar{Q}}$ in Eq.(\ref{eqnD050007}),
 it can be found that
\beqn
    \sigma^2\abrac{\expv{v^n}_{\Bar{Q}}}
 \= \sumnu \sum_{j=1}^{B} \aPp{\expv{v^n}_{\Bar{Q}}}{x_{\nu,j}}^2 \sigma^2(x_{\nu,j})
    \non\\
 \= \frac{1}{S_0^4}
    \sumnu \sum_{j=1}^{B} \aBig{S_0 \~ S_{\nu,j;n}-S_n \~ S_{\nu,j;0}}^{2} \sigma^2(x_{\nu,j})
\~.
\label{eqnD050010}
\eeqn
 According to the expression of $\td{f}_{1,n,\Bar{Q}}(v)$ in Eq.(\ref{eqnD050003})
 and $\trn$ and $k_n$ given in Eqs.(\ref{eqn4020109}) and (\ref{eqn4020110}),
 the independent variables of $\td{f}_{1,n,\Bar{Q}}(v)$ should be chosen as
\beq
   x_{1,n}
 = N_n
\~,
   ~~~~~~~~~~~~~~~~~~ 
   x_{2,n}
 = k_n
\~,
\eeq
 with $\sigma^2(N_n)$ and $\sigma^2(k_n)$ given in Eqs.(\ref{eqn4020104}) and (\ref{eqn4020116}).
 Since $\td{f}_{1,n,\Bar{Q}}(v)$ depends only on $N_n$ and $k_n$,
 Eqs.(\ref{eqnD050008}) and (\ref{eqnD050010}) can be reduced to
\beq
   \sigma^2\aBig{f_{1,n,\Bar{Q}}(v)}
 = \frac{1}{S_0^4}
   \sumnu \sum_{j=1}^{B}
   \bbigg{\delta_{nj} S_0 \abrac{\p_{\nu} \tfQnv}-I_{\nu,j;0} \~ \tfQnv}^{2}
   \sigma^2(x_{\nu,j})
\~,
\label{eqnD050011}
\eeq
 and
\beq
   \sigma^2\abrac{\expv{v^n}_{\Bar{Q}}}
 = \frac{1}{S_0^4}
   \sumnu \sum_{j=1}^{B} \aBig{S_0 \~ I_{\nu,j;n}-S_n \~ I_{\nu,j;0}}^{2} \sigma^2(x_{\nu,j})
\~.
\label{eqnD050012}
\eeq
 Here I have defined
\beq
        \p_N \td{f}_{1,n,\Bar{Q}}(v)
 \equiv \Pp{\td{f}_{1,n,\Bar{Q}}(v)}{N_n}
 =      \frac{\td{f}_{1,n,\Bar{Q}}(v)}{N_n}
\~,
\label{eqnD050013}
\eeq
\beqn
           \p_k \td{f}_{1,n,\Bar{Q}}(v)
 \eqnequiv \Pp{\td{f}_{1,n,\Bar{Q}}(v)}{k_n}
           \non\\
 \=       -\td{f}_{1,n,\Bar{Q}}(v) \cbrac{\bQn+\bbrac{\Dd{\ln \FQ}{Q}-k_n}^{-1}}\Qva
\~,
\label{eqnD050014}
\eeqn
 and, from Eq.(\ref{eqnD050009}),
\beq
        I_{\nu,j;\lambda}
 \equiv \intvj v^{\lambda} \bbrac{\p_{\nu} \td{f}_{1,j,\Bar{Q}}(v)} dv
\~.
\label{eqnD050015}
\eeq
\section{Introducing the average value of $Q^{\lambda}/\FQ$}
 ~~~$\,$
 The method presented in the previous section
 has two disadvantages.
 First,
 the estimator of $\calN_{\Bar{Q}}$ in Eq.(\ref{eqnD050004})
 is the sum of several {\em integrals},
 this makes the estimation complicated.
 Second,
 and also the worse disadvantage,
 by using $\expv{v^n}_{\Bar{Q}}$ given in Eq.(\ref{eqnD050007})
 with $S_{\lambda}$ defined in Eq.(\ref{eqnD050006}),
 one has to know $\td{f}_{1,n,\Bar{Q}}(v)$ defined in Eq.(\ref{eqnD050003}),
 i.e., $f_{1,n,\Bar{Q}}(v)$ defined in Eq.(\ref{eqnD050002}).
 It is not only complicated
 but also loses the advantage of the expressions
 in Eqs.(\ref{eqn4010011}) and (\ref{eqn4010012}),
 by which one can evaluate the moments of $f_1(v)$
 without knowing the functional form of $f_1(v)$.
 This problem comes essentially from
 the estimator of $\calN_{\Bar{Q}}$ in Eq.(\ref{eqnD050004})
 obtained from the normalization condition in Eq.(\ref{eqn4010007}).
 Hence,
 one needs a new estimator for the normalization constant.

 Similar to the use of the moments of the recoil spectrum
 in Eqs.(\ref{eqn4020110}) and (\ref{eqn4020111}),
 I have defined
 an average value of $Q^{\lambda}/\FQ$ for all events in the $n$-th $Q$-bin:
\beq
        \frac{1}{N_n} \intQnbn \frac{Q^{\lambda}}{\FQ} \adRdQ dQ
 =      \frac{1}{N_n} \sumiNn \frac{\Qni^{\lambda}}{\FQni}
 \equiv \bSbxn{\lambda}
\~.
\label{eqnD060001}
\eeq
 Then, for all recorded events in the sample, I can use
\beq
     \intz \frac{Q^{\lambda}}{\FQ} \adRdQ dQ
 \to \sum_{n=1}^B \sumiNn \frac{\Qni^{\lambda}}{\FQni}
 =   \sum_{n=1}^B N_n \bSbxn{\lambda}
 \equiv \Sbxtot{\lambda}
\~.
\label{eqnD060002}
\eeq
 Note that
 the recoil spectrum $dR/dQ$ here is not specified
 to the exponential ansatz $(dR/dQ)_n$ in Eq.(\ref{eqn4020105}).
 Using the definition in Eq.(\ref{eqnD060002}),
 $\calN$ in Eq.(\ref{eqn4010009}) and $I_n$ in Eq.(\ref{eqn4010012})
 can be estimated by
\beq
   \calN
 = \frac{2}{\alpha} \afrac{1}{\Sbxtot{-1/2}}
 = \frac{2}{\alpha} \abrac{\sum_{n=1}^B N_n \bSbxn{-1/2}}^{-1}
\~,
\label{eqnD060003}
\eeq
 and
\beq
   I_n
 = \Sbxtot{(n-1)/2}
 = \sum_{n=1}^B N_n \bSbxn{(n-1)/2}
\~.
\label{eqnD060004}
\eeq
 The expressions in Eqs.(\ref{eqnD060003}) and (\ref{eqnD060004})
 are already essentially the same as the expressions
 given in Eqs.(\ref{eqn4020303}) and (\ref{eqn4020404}),
 respectively.

 Now replacing $\calN_{\Bar{Q}}$ in Eq.(\ref{eqnD050001})
 by $\calN$ given in Eq.(\ref{eqnD060003}),
 the reconstructed velocity distribution function at point $v = v_n$,
 $f_{1,n,\Bar{Q}}(v_n)$,
 in Eq.(\ref{eqnD050001}) can be expressed simply as
\beq
   f_1(v_n)
 = \calN \bfrac{2 Q_n \trn}{\FQn} \bbrac{\dd{Q} \ln\FQ \bigg|_{Q = Q_n}-k_n}
 = \frac{2}{\alpha} \afrac{\tfn}{\Sbxtot{-1/2}}
\~.
\label{eqnD060005}
\eeq
 Here,
 similar to $\td{f}_{1,n,\Bar{Q}}(v)$ defined in Eq.(\ref{eqnD050003}),
 I have defined
\beq
        \tfn
 \equiv \frac{2 Q_n \trn}{\FQn} \bbrac{\dd{Q} \ln\FQ\bigg|_{Q = Q_n}-k_n}
\~.
\label{eqnD060006}
\eeq
 Moreover,
 the $n$-th moment of the velocity distribution function, $\expv{v^n}$,
 determined by Eq.(\ref{eqn4010011}) can now be expressed as (with $\Qthre = 0$)
\beq
   \expv{v^n}
 = \alpha^n (n+1) \afrac{\Sbxtot{(n-1)/2}}{\Sbxtot{-1/2}}
\~.
\label{eqnD060007}
\eeq
 By means of this expression,
 one can finally estimate the moments of the velocity distribution function
 directly from the experimental data given in Eq.(\ref{eqn4020101})
 without knowing the exact form of $f_1(v_n)$.
 Actually,
 according to Eq.(\ref{eqnD060004}),
 the expression of $\expv{v^n}$ given in Eq.(\ref{eqnD060007})
 is exactly the same as that given in Eq.(\ref{eqn4030101}),
 or the general form in Eq.(\ref{eqn4020402}) with $\Qthre = 0$.

 On the other hand,
 before beginning to calculate the statistical errors of
 $f_1(v_n)$ and $\expv{v^n}$ in Eqs.(\ref{eqnD060005}) and (\ref{eqnD060007}),
 one must pay some special attention with the variables involved in their expressions.
 According to the expression of $\tfn$ in Eq.(\ref{eqnD060006})
 and of $\trn$ and $k_n$ in Eqs.(\ref{eqn4020109}) and (\ref{eqn4020110}),
 $\tfn$ is a function of only two variables: $N_n$ and $\bQn$;
 while,
 according to the definition in Eq.(\ref{eqnD060004}),
 $\Sbxtot{\lambda}$ is a function of $2B$ variables:
 $N_n$ and $\bSbxn{\lambda}$ for all $n=1,~2,~\cdots, B$.
 Hence,
 $f_1(v_n)$ and $\expv{v^n}$ depend on $2B+2$ and $3B$ variables,
 respectively,

 Then,
 from Eq.(\ref{eqnD060005}),
 it can be found that
\cheqnCa
\beq
   \Pp{f_1(v_n)}{\bQn}
 =-f_1(v_n) \abigg{\bQn+K_n} \bfrac{k_n^2}{g(\kappa_n)}
\~,
\label{eqnD060008a}
\eeq
 where $K_n$ has been defined in Eq.(\ref{eqn5020208})
 and I have used $d\bQn/d k_n$ in Eq.(\ref{eqn4020117});
 for $m=1,~2,~\cdots,~B$,
 one has
\cheqnCb
\beq
   \Pp{f_1(v_n)}{N_m}
 = f_1(v_n) \abrac{\frac{\delta_{nm}}{N_m}-\frac{\bSbxm{-1/2}}{\Sbxtot{-1/2}}}
\~,
\label{eqnD060008b}
\eeq
 and
\cheqnCc
\beq
   \Pp{f_1(v_n)}{\bSbxm{-1/2}}
 =-f_1(v_n) \afrac{N_m}{\Sbxtot{-1/2}}
\~.
\label{eqnD060008c}
\eeq
\cheqnC
 Then the statistical error of $f_1(v_n)$ estimated in Eq.(\ref{eqnD060005})
 can be expressed as
\beqn
        \sigma^2\aBig{f_1(v_n)}
 \=     f_1^2(v_n)
        \cleft{    \bfrac{\abrac{\bQn+\Kn} k_n^2}{g(\kappa_n)}^2 \sigma^2\aBig{\bQn}}
                +  \frac{1}{N_n}
        \non\\
 \conti ~~~~~~~~~~~~~~~~ 
                +2 \bfrac{\abrac{\bQn+\Kn} k_n^2}{g(\kappa_n)}
                   \Afrac{\bSbxn{1/2}-Q_n \bSbxn{-1/2}}{\Sbxtot{-1/2}}
        \non\\
 \conti ~~~~~~~~~~~~~~~~~~~~~~ 
        \cright{+  \frac{\Sdxtot{-1}}{\Sbxtot{-1/2}^2}
                -  \frac{2 \bSbxn{-1/2}}{\Sbxtot{-1/2}}}
\~.
\label{eqnD060009}
\eeqn
 Here I have used Eq.(\ref{eqn4020104}) for $\sigma^2(N_n)$,
 and, for simplicity,
 set $N_m \gg 1$ for all $m$ in order to use
\beq
   {\rm cov}\abrac{\bQn,\bSbxn{\lambda}}
 = \frac{1}{N_n} \aBig{\bSbxn{\lambda+1}-Q_n \bSbxn{\lambda}}
\~,
\label{eqnD060010}
\eeq
 and
\beq
   {\rm cov}\abrac{\bSbxn{\lambda},\bSbxn{\rho}}
 = \frac{1}{N_n} \abrac{\bSdxn{\lambda+\rho}-\bSbxn{\lambda}~\bSbxn{\rho}}
\~,
\label{eqnD060011}
\eeq
 with the definition
\beq
        \bSdxn{\lambda}
 \equiv \frac{1}{N_n} \sumiNn \frac{\Qni^{\lambda}}{F^4(\Qni)}
\~,
\label{eqnD060012}
\eeq
 and then
\beq
        \Sdxtot{\lambda}
 \equiv \sum_{n=1}^{B} N_n \bSdxn{\lambda}
\~,
\label{eqnD060013}
\eeq
 see Eqs.(\ref{eqnD060001}) and (\ref{eqnD060002}).
 Meanwhile,
 from the expression of $\expv{v^n}$ given in Eq.(\ref{eqnD060007}),
 it can be found that
\cheqnCa
\beq
   \Pp{\expv{v^n}}{\bSbxm{(n-1)/2}}
 = \expv{v^n} \afrac{N_m}{\Sbxtot{(n-1)/2}}
\~,
\label{eqnD060014a}
\eeq
\cheqnCb
\beq
   \Pp{\expv{v^n}}{\bSbxm{-1/2}}
 =-\expv{v^n} \afrac{N_m}{\Sbxtot{-1/2}}
\~,
\label{eqnD060014b}
\eeq
 and
\cheqnCc
\beq
   \Pp{\expv{v^n}}{N_m}
 = \expv{v^n}
   \abrac{\frac{\bSbxm{(n-1)/2}}{\Sbxtot{(n-1)/2}}-\frac{\bSbxm{-1/2}}{\Sbxtot{-1/2}}}
\~.
\label{eqnD060014c}
\eeq
\cheqnC
 Hence,
 the statistical error of $\expv{v^n}$ can be expressed as
\beq
   \sigma^2\aBig{\expv{v^n}}
 = \expv{v^n}^2
   \abrac{ \frac{\Sdxtot{n-1}}{\Sbxtot{(n-1)/2}^2}
          +\frac{\Sdxtot{ -1}}{\Sbxtot{   -1/2}^2}
          -\frac{2 \Sdxtot{(n-2)/2}}{\Sbxtot{(n-1)/2} \Sbxtot{-1/2}}}
\~,
\label{eqnD060015}
\eeq
 where I have used Eq.(\ref{eqnD060011}).

 Finally,
 if one considers the reconstruction only in bins,
 the diagonal entries of the covariance matrix in Eq.(\ref{eqn4020304}) can be reduced to
\beqn
    \sigma^2\aBig{f_{1,r}^2(v_{s,n})}
 \= \frac{f_{1,r}^2(v_{s,n})}{N_n}+\calN^2 \bfrac{2 Q_{s,n} r_n}{F^2(Q_{s,n})}^2 \sigma^2(k_n)
    \non\\
 \= f_{1,r}^2(v_{s,n})
    \bbrac{\frac{1}{N_n}+K_{s,n}^2 \sigma^2(k_n)}
\~,
\label{eqnD060016}
\eeqn
 since $r_n$ and $k_n$ are now two independent variables,
 and,
 similar to Eq.(\ref{eqn5020208}),
 I have defined here
\beq
        K_{s,n}
 \equiv \bbrac{\dd{Q} \ln \FQ \bigg|_{Q = Q_{s,n}}-k_n}^{-1}
\~.
\eeq
 The expression in Eq.(\ref{eqnD060016}) is essentially
 the same as that in Eq.(\ref{eqnD060009})
 without the last three terms involving $\Sbxtot{-1/2}$,
 which correspond to the statistical error of the estimator for $\calN$
 and have been neglected in Eq.(\ref{eqn4020304}).
 Note that
 $f_1(v_n)$ in Eq.(\ref{eqnD060005}) is obtained
 from the first expression of the exponential ansatz in Eq.(\ref{eqn4020105})
 and estimated at $v = v_n$,
 while $f_{1,r}^2(v_{s,\mu})$ in Eq.(\ref{eqn4020301}) is obtained
 from the second expression in Eq.(\ref{eqn4020105})
 and estimated at $v = v_{s,\mu}$,
 which is not a fixed value like $v_n$ but actually depends on $k_n$
 through Eqs.(\ref{eqn4020115}).
 Hence,
 since the two expressions in Eq.(\ref{eqn4020105}) are equivalent,
 when one takes into account
 the uncertainty of the determination of $Q_{s,\mu}$ by Eq.(\ref{eqn4020115}),
 the statistical error of $f_{1,r}^2(v_{s,\mu})$
 will be identical to the first two terms of $\sigma^2(f_1(v_n))$
 in Eq.(\ref{eqnD060009}).

 Similarly,
 if one neglects $\Qthre$ and thus all terms involving $\rthre$,
 the diagonal entries of the general form of the covariance matrix
 given in Eq.(\ref{eqn4020411}) can be reduced to
\beqn
    \sigma^2\aBig{\expv{v^n}}
 \= \frac{1}{I_0^2}
    \bbigg{ \expv{v^n}^2 \sigma^2(I_0)
           +\alpha^{2n} (n+1)^2 \sigma^2(I_n)
           -2 \alpha^n (n+1) \expv{v^n} {\rm cov}(I_0,I_n)}
    \non\\
 \= \expv{v^n}^2
    \bbrac{ \frac{\sigma^2(I_0)}{I_0^2}
           +\frac{\sigma^2(I_n)}{I_n^2}
           -\frac{2 {\rm cov}(I_n,I_0)}{I_n I_0}}
\~,
\label{eqnD060017}
\eeqn
 where I have used $\expv{v^n}$ in Eq.(\ref{eqn4030101}).
 Comparing the definition of $\Sdxtot{\lambda}$ in Eq.(\ref{eqnD060013})
 with the expression of ${\rm cov}(I_n,I_m)$ in Eq.(\ref{eqn4020404})
 and using Eq.(\ref{eqnD060004}),
 the expression given in Eq.(\ref{eqnD060015})
 is exactly identical to that in Eq.(\ref{eqnD060017}).

%% file: Doktorarbeit-AppE.tex
\chapter{Some Detailed Calculations}
 ~~~$\,$
 In this chapter
 I give some detailed derivations.
\section{Derivations of covariances in Sec.~\ref{Qni}}
\subsection{Covariances in Subsec.~\ref{windowing}}
\label{covbQnrn}
 ~~~$\,$
 From Eq.(\ref{eqn4020202}), one has
\beq
   \Pp{N_{\mu}}{N_{n^{\ast}}}
 = 1
\~,
\label{eqnE010101}
\eeq
 where $n^{\ast}$ denotes a given bin between the $n_{\mu-}$-th and the $n_{\mu+}$-th bins.
 Then,
 from (\ref{eqn4020203}),
 one has
\beq
   \Pp{\Bar{Q-Q_{\mu}}|_{\mu}}{N_{n^{\ast}}}
 = \frac{\Bar{Q}|_{n^{\ast}}}{N_{\mu}}
  -\frac{1}{N_{\mu}^2}\abrac{\sum_{n=n_{\mu-}}^{n_{\mu+}} N_n \Bar{Q}|_{n}}
 = \frac{\Bar{Q}|_{n^{\ast}}-\Bar{Q}|_{\mu}}{N_{\mu}}
\~,
\label{eqnE010102}
\eeq
 and
\beq
   \Pp{\Bar{Q-Q_{\mu}}|_{\mu}}{\Bar{Q}|_{n^{\ast}}}
 = \frac{N_{n^{\ast}}}{N_{\mu}}
\~.
\label{eqnE010103}
\eeq
 Combining Eqs.(\ref{eqnE010101}) to (\ref{eqnE010103}),
 it can be found that
\cheqnref{eqn4020204}
\beqn
 \conti {\rm cov}\abrac{\Bar{Q-Q_{\mu}}|_{\mu},\Bar{Q-Q_{\nu}}|_{\nu}}
        \non\\
 \=     \sum_{n^{\ast}}
        \bleft{ \aPp{\Bar{Q-Q_{\mu}}|_{\mu}}{N_{n^{\ast}}}
                \aPp{\Bar{Q-Q_{\nu}}|_{\nu}}{N_{n^{\ast}}}
                \sigma^2(N_{n^{\ast}})}
        \non\\
 \conti ~~~~~~~~~~~~~~~~ 
        \bright{+\aPp{\Bar{Q-Q_{\mu}}|_{\mu}}{\Bar{Q}|_{n^{\ast}}}
                \aPp{\Bar{Q-Q_{\nu}}|_{\nu}}{\Bar{Q}|_{n^{\ast}}}
                \sigma^2\abrac{\Bar{Q}|_{n^{\ast}}}}
        \non\\
 \=     \frac{1}{N_{\mu} N_{\nu}}
        \sum_{n^{\ast}}
        \bbigg{ N_{n^{\ast}}
                \abrac{\Bar{Q}|_{n^{\ast}}-\Bar{Q}|_{\mu}}
                \abrac{\Bar{Q}|_{n^{\ast}}-\Bar{Q}|_{\nu}}
               +N_{n^{\ast}}^2 \sigma^2\abrac{\Bar{Q}|_{n^{\ast}}}}
\~.
\eeqn
\cheqnCN{-1}
 Meanwhile,
 from Eqs.(\ref{eqn4020202}) and (\ref{eqn4020205}),
 one has
\beq
   r_{\mu}
 = \frac{1}{w_{\mu}} \sum_{n=n_{\mu-}}^{n_{\mu+}} N_n
\~.
\eeq
 Thus,
\beq
   \Pp{r_{\mu}}{N_{n^{\ast}}}
 = \frac{1}{w_{\mu}}
\~,
\label{eqnE010104}
\eeq
 and then
\cheqnref{eqn4020206}
\beq
   {\rm cov}(r_\mu, r_\nu)
 = \sum_{n^{\ast}} \aPp{r_{\mu}}{N_{n^{\ast}}} \aPp{r_{\nu}}{N_{n^{\ast}}} \sigma^2(N_{n^{\ast}})
 = \frac{1}{w_{\mu} w_{\nu}} \sum_{n^{\ast}} N_{n^{\ast}}
\~,
\eeq
\cheqnCN{-1}
 Combining Eqs.(\ref{eqnE010102}) and (\ref{eqnE010104}),
 it can be found that
\cheqnref{eqn4020207}
\beqn
    {\rm cov}\abrac{r_{\mu},\Bar{Q-Q_{\nu}}|_{\nu}}
 \= \sum_{n^{\ast}}
    \aPp{r_{\mu}}{N_{n^{\ast}}} \aPp{\Bar{Q-Q_{\nu}}|_{\nu}}{N_{n^{\ast}}}
    \sigma^2(N_{n^{\ast}})
    \non\\
 \= \frac{1}{w_{\mu} N_{\nu}}
    \sum_{n^{\ast}} N_{n^{\ast}} \abrac{\Bar{Q}|_{n^{\ast}}-\Bar{Q}|_{\nu}}
\~.
\eeqn
\subsection{Covariance in Eq.(\ref{eqn4020411})}
\label{covexpvvn}
 ~~~$\,$
 The expression of $\expv{v^n}$ in Eq.(\ref{eqn4020402}) can be rewritten as
\cheqnrefp{eqn4020402}
\beq
   \expv{v^n}
 = \alpha^n \calN_{\rm m} \bbrac{\frac{2 \Qthre^{(n+1)/2} \rthre}{\FQthre}+(n+1) I_n}
\~,
\eeq
 with
\cheqnrefp{eqn4020412}
\beq
   \calN_{\rm m}
 = \bbrac{\frac{2 \Qthre^{1/2} \rthre}{\FQthre}+I_0}^{-1}
\~.
\eeq
\cheqnCN{-3}
 Hence, it can be found that
\beq
   \Pp{\expv{v^n}}{I_n}
 = \calN_{\rm m} \alpha^n (n+1)
\~,
\label{eqnE010201}
\eeq
 and
\beq
   \Pp{\expv{v^n}}{I_0}
 =-\alpha^n \calN_{\rm m}^2 \bbrac{\frac{2 \Qthre^{(n+1)/2} \rthre}{\FQthre}+(n+1) I_n}
 =-\calN_{\rm m} \expv{v^n}
\~.
\label{eqnE010202}
\eeq
 Moreover,
\beqn
    \Pp{\expv{v^n}}{\rthre}
 \= \alpha^n \calN_{\rm m}   \bfrac{2 \Qthre^{(n+1)/2}}{\FQthre}
   -\alpha^n \calN_{\rm m}^2 \bbrac{\frac{2 \Qthre^{(n+1)/2} \rthre}{\FQthre}+(n+1) I_n}
    \bfrac{2 \Qthre^{1/2}}{\FQthre}
    \non\\
 \= \calN_{\rm m} \bfrac{2}{\FQthre}
    \abrac{\alpha^n \Qthre^{(n+1)/2}-\sqrt{\Qthre} \~ \expv{v^n}}
\~.
\label{eqnE010203}
\eeqn
 This leads to the definition of $D_n$ in Eq.(\ref{eqn4020413}).
 Combining Eqs.(\ref{eqnE010201}), (\ref{eqnE010202}), and (\ref{eqn4020413}),
 Eq.(\ref{eqn4020411}) can be obtained directly.
\section{Derivation of the correction terms in Eq.(\ref{eqn4020414})}
\label{corr}
 ~~~$\,$
 Starting point is the observation that we wish to compute the ratio of two integrals,
\beq
     \frac{G_1}{G_2}
 =   \frac{\int g_1(x) \~ dx}{\int g_2(x) \~ dx}
 \to \frac{\sum_i n_i g_1(x_i)} {\sum_j n_j g_2(x_j)}
\~.
\eeq
 In the second step the integrals have been discretized,
 i.e., replaced by sums over bins $i$ with $n_i$ events per bin.
 $n_i$ can be written as sum of average value $\bar{n}_i$ and fluctuation $\delta n_i$:
\beq
   \frac{G_1}{G_2}
 = \frac{\sum_i (\bar{n}_i+\delta n_i) g_1(x_i)}
        {\sum_j \bar{n}_j g_2(x_j)+\sum_j \delta n_j g_2(x_j)}
\~.
\eeq
 Introducing the notation
\beq
   \Bar{G}_a
 = \sum_i \bar{n}_i g_a(x_i)
\~,
\eeq
 for $a=1,~2$,
 and expanding up to second order in the $\delta n_i$,
 one has
\beqn
           \frac{G_1}{G_2}
 \eqnsimeq \frac{\Bar{G}_1+\sum_i \delta n_i g_1(x_i)}{\Bar{G}_2}
           \bbrac{ 1
                   -\frac{\sum_j \delta n_j g_2(x_j)}{\Bar{G}_2}
                   +\afrac{\sum_j \delta n_j g_2(x_j)}{\Bar{G}_2}^2}
           \non\\
 \eqnsimeq \frac{\Bar{G}_1}{\Bar{G}_2}
          +\frac{1}{\Bar{G}_2} \abrac{\sum_i \delta n_i g_1(x_i)}
          -\frac{\Bar{G}_1}{\Bar{G}_2^2} \abrac{\sum_i \delta n_i g_2(x_i)}
           \non\\ 
 \conti    ~~~~~~ 
          -\frac{1}{\Bar{G}_2^2}
           \abrac{\sum_i \delta n_i g_1(x_i)}
           \aBigg{\sum_j \delta n_j g_2(x_j)}
          +\frac{\Bar{G}_1}{\Bar{G}_2^3}
           \abrac{\sum_i \delta n_i g_2(x_i)}^2
\~.
\eeqn
 Now consider the average over many experiments.
 Of course,
 $\delta n_i$ averages to zero,
 but the product $\delta n_i \delta n_j$ averages to $\bar{n}_i \delta_{ij}$,
 i.e., it is non-zero for $i = j$.
 Hence:
\beq
        \Expv{\frac{G_1}{G_2}}
 \simeq \frac{\Bar{G}_1}{\Bar{G}_2} 
       -\frac{1}{\Bar{G}_2^2} \abrac{\sum_i \bar{n}_i g_1(x_i) g_2(x_i)}
       +\frac{\Bar{G}_1}{\Bar{G}_2^3} \abrac{\sum_i \bar{n}_i g_2^2(x_i)}
\~.
\eeq
 The sums appearing in the two correction terms
 also appear in the definition of the covariance matrix between $G_1$ and $G_2$.
 Note that we wish to compute the first term on the right-hand side,
 since in this case the estimators for $G_1$ and $G_2$ indeed average to the correct values.
 This then leads to the final result
\beq
   \frac{\Bar{G}_1}{\Bar{G}_2}-\Expv{\frac{G_1}{G_2}} 
 = \afrac{1}{\Bar{G}_2^2} {\rm cov}(G_1,G_2)-\afrac{\Bar{G}_1}{\Bar{G}_2^3} {\rm cov}(G_2,G_2)
\~.
\eeq
 Applying this result to Eq.(\ref{eqn4020402})
 then immediately leads to Eq.(\ref{eqn4020414}).
\section{Statistical error of $\calRn(\Qthre)$ in Eq.(\ref{eqn4030201})}
\label{sigmacalRn}
 ~~~$\,$
 From Eq.(\ref{eqn4030201}),
 it can be found directly that
\cheqnCa
\beqn
        \Pp{\calRn(\Qthre)}{\rthreX}
 \=     \frac{2}{n}
        \bfrac{  \QthreX^{(n+1)/2} \IzX-(n+1) \QthreX^{1/2} \InX}
              {2 \QthreX^{(n+1)/2} \rthreX+(n+1) \InX \FQthreX}
        \non\\
 \conti ~~~~~~~~~~~~ \times 
        \bfrac{\FQthreX}{2 \QthreX^{1/2} \rthreX+\IzX \FQthreX} \calRn(\Qthre)
\~,
\eeqn
\cheqnCb
\beqn
    \Pp{\calRn(\Qthre)}{\InX}
 \= \frac{n+1}{n}
    \bfrac{\FQthreX}{2 \QthreX^{(n+1)/2} \rthreX+(n+1) \InX \FQthreX} \calRn(\Qthre)
\~,
    \non\\
\eeqn
 and
\cheqnCc
\beq
   \Pp{\calRn(\Qthre)}{\IzX}
 =-\frac{1}{n} \bfrac{\FQthreX}{2 \QthreX^{1/2} \rthreX+\IzX \FQthreX} \calRn(\Qthre)
\~.
\eeq
\cheqnC
 By first exchanging $\QthreX^{(n+1)/2}$ and $(n+1) \InX$
 with $\QthreX^{1/2}$ and $\IzX$,
 respectively,
 and then replacing $X$ by $Y$,
 one can get
\cheqnCa
\beqn
        \Pp{\calRn(\Qthre)}{\rthreY}
 \=    -\frac{2}{n}
        \bfrac{  \QthreY^{(n+1)/2} \IzY-(n+1) \QthreY^{1/2} \InY}
              {2 \QthreY^{(n+1)/2} \rthreY+(n+1) \InY \FQthreY}
        \non\\
 \conti ~~~~~~~~~~~~ \times 
        \bfrac{\FQthreY}{2 \QthreY^{1/2} \rthreY+\IzY \FQthreY} \calRn(\Qthre)
\~,
\eeqn
\cheqnCb
\beqn
    \Pp{\calRn(\Qthre)}{\InY}
 \=-\frac{n+1}{n}
    \bfrac{\FQthreY}{2 \QthreY^{(n+1)/2} \rthreY+(n+1) \InY \FQthreY} \calRn(\Qthre)
\~,
    \non\\
\eeqn
 and
\cheqnCc
\beq
   \Pp{\calRn(\Qthre)}{\IzY}
 = \frac{1}{n} \bfrac{\FQthreY}{2 \QthreY^{1/2} \rthreY+\IzY \FQthreY} \calRn(\Qthre)
\~.
\eeq
\cheqnC
 By setting $\Qthre = 0$,
 the above expressions can be reduced to $\sigma(\mchi)$ given in Eq.(\ref{eqn4030104})
 directly.
\section{Derivation of $\eta_n$ in Eq.(\ref{eqn5020203})}
\label{etan}
 ~~~$\,$
 From Eq.(\ref{eqn5020201}), I have
\beqn
    \frac{f_{1,(m)}(Q)}{2 \calN}
 \= \frac{Q}{\FQ} \adRdQm \cbrac{\dFdQoF-\frac{1}{\TadRdQm} \bbrac{\dd{Q}\adRdQm}}
\~.
    \non\\
\eeqn
 Substituting the ansatz for $\TadRdQan$ in Eq.(\ref{eqn5020102}), it can be found that
\beqn
 \conti \frac{1}{(dR/dQ)_{(1),n}} \bbrac{\dd{Q} \adRdQan}
        \non\\
 \=     \frac{1}{(dR/dQ)_{(0)  }} \bbrac{\dd{Q} \adRdQz }+\frac{l_n}{l_n (Q-Q_n)+h_n}
\~.
\eeqn
 Thus I can obtain that
\beqn
 \conti \frac{f_{1,(m)}(Q)}{2 \calN}
        \non\\
 \=     \frac{Q}{\FQ} \adRdQz \bBig{l_n (Q-Q_n)+h_n}
        \non\\
 \conti ~~~~~~~~ \times 
        \cbrac{ \dFdQoF
               -\frac{1}{(dR/dQ)_{(0)}} \bbrac{\dd{Q}\adRdQz}
               -\frac{l_n}{l_n (Q-Q_n)+h_n}}
        \non\\
 \=     \frac{f_{1,(0)}(Q)}{2 \calN} \bBig{l_n (Q-Q_n)+h_n}
       -l_n \bbrac{\frac{Q}{\FQ} \adRdQz}
\~,
\eeqn
 namely,
\beqn
 \conti \frac{f_{1,(1),n}(Q)}{f_{1,(0)}(Q)}
        \non\\
 \=     \bBig{l_n (Q-Q_n)+h_n}
       -l_n \cbrac{\dFdQoF-\frac{1}{(dR/dQ)_{(0)}} \bbrac{\dd{Q}\adRdQz}}^{-1} \!\!
.
\eeqn
 Now I use the ansatz for $\TadRdQzn$ in Eq.(\ref{eqn4020105}) to get
\beq
   \frac{f_{1,(1),n}(Q)}{f_{1,(0),n}(Q)}
 = \bBig{l_n (Q-Q_n)+h_n}-l_n \bbrac{\dFdQoF-k_n}^{-1}
\~.
\eeq
 Let $Q = Q_n$,
 the expression of $\eta_n$ in Eq.(\ref{eqn5020203}) can be obtained directly.